# Projection-operator methods for classical transport in magnetised plasmas. II. Nonlinear response and the Burnett equations

**John A. Krommes†**

Princeton University, Plasma Physics Laboratory
P. O. Box 451, MS 28, Princeton, New Jersey 08543–0451 USA

The time-independent projection-operator formalism of Brey *et al.* [Physica **109A**, 425–444 (1981)] for the derivation of Burnett equations is extended and considered in the context of multispecies and magnetised plasmas. The procedure provides specific formulas for the transport coefficients in terms of two-time correlation functions involving both two and three phase-space points. It is shown how to calculate those correlation functions in the limit of weak coupling. The results are used to demonstrate, with the aid of a particular nontrivial example, that the Chapman–Enskog methodology employed by Catto & Simakov (CS) [Phys. Plasmas **11**, 90–102 (2004)] to calculate the contributions to the parallel viscosity driven by temperature gradients is consistent with formulas previously derived from the two-time formalism by J. J. Brey [J. Chem. Phys. **79**, 4585–4598 (1983)]. The work serves to unify previous work on plasma kinetic theory with formalism usually applied to turbulence. Additional contributions include discussions of (i) Braginskii-order interspecies momentum exchange from the point of view of two-time correlations; and (ii) a simple stochastic model, unrelated to many-body theory, that exhibits Burnett effects. Insights from that model emphasize the role of non-Gaussian statistics in the evaluation of Burnett transport coefficients, including the effects calculated by CS that stem from the nonlinear collision operator. Together, Parts I and II of this series provide an introduction to projection-operator methods that should be broadly useful in theoretical plasma physics.

† Email address for correspondence: krommes@princeton.edu

CONTENTS

## 1. Introduction

The collisional fluid equations for weakly coupled and magnetised plasma, often referred to as the *Braginskii equations*, are very well known (Braginskii 1965). In Part I of this series (Krommes 2018*b*), I discussed the derivation of the linearized Braginskii equations by means of the projection-operator formalism (Mori 1965) that has been frequently used in the context of neutral gases. That formalism is appealing on both heuristic and technical levels. The use of projection operators fosters a clean separation between the hydrodynamic and orthogonal subspaces, and it provides an efficient construction of the subtracted fluxes whose autocorrelations define the transport coefficients. It also leads to new results. For example, I showed in Part I that for linear response it leads to a straightforward derivation of fluctuating hydrodynamics, a topic not explicitly discussed by Braginskii.

Of course, the true fluid equations are nonlinear. The Braginskii equations are really a hybrid: the Euler parts of the equations, which close without approximation in terms of the fluid variables (density $n$, flow velocity $\boldsymbol{u}$, and temperature $T$), are nonlinear and valid to all orders in the macroscopic gradients, while the dissipative corrections are valid only to lowest order in the gradients or, in multicomponent systems, the momentum and energy exchange terms involving (in a quasineutral plasma with one ion species) $\Delta\boldsymbol{u} \doteq \boldsymbol{u}_e - \boldsymbol{u}_i$ and $\Delta T \doteq T_e - T_i$. (I use $\doteq$ for definitions.) That is known as *Navier–Stokes order* in nonequilibrium statistical mechanics; one obtains standard effects such as heat fluxes ($\boldsymbol{q} = -n\kappa\boldsymbol{\nabla}T$, where $\kappa$ is the thermal conductivity) or, in a plasma, frictional drag and collisional temperature equilibration. Equations valid to next order in the gradients are called the *Burnett equations*; a review is by García-Colín *et al.* (2008). They contain new linear dissipative fluxes such as $\kappa'\nabla^2 T$, nonlinear gradient–gradient interactions such as $\boldsymbol{\nabla}T\boldsymbol{\cdot}\boldsymbol{\nabla}T$ and, in a multicomponent system, additional terms such as $\Delta\boldsymbol{u}\boldsymbol{\cdot}\boldsymbol{\nabla}T$ or $|\Delta\boldsymbol{u}|^2$. Terms at Burnett order could be important in situations involving strong fluctuations and/or gradients, such as may obtain in tokamak edges. In the present paper, I discuss some aspects of the calculation of Burnett transport coefficients for plasmas. My focus is on the basic principles and technical methodology; I shall not consider practical implications, for which further research is required.

Although it is clearly desirable to have a description that is superior to Braginskii's when the gradients and/or fluctuations are strong, there are daunting technical obstacles. Of course, when the nonlinearities are extremely large perturbation theory loses its validity and a renormalized description involving terms of all orders becomes necessary. Such a theory is beyond the scope of this paper. But even when the gradients are modest, so that one might expect that regular perturbation theory truncated at second order would give useful corrections to the Braginskii equations, there is a deeply troubling issue. It is unfortunately the case that the second-order (Burnett) description may not actually exist; coefficients such as $\kappa'$ can be infinite, signifying an embarrassing breakdown of the (perturbative and Markovian) approach. The difficulty arises because of the collisional excitation of long-lived and long-ranged hydrodynamic fluctuations that lead to slowly decaying tails on correlation functions and violate basic assumptions of locality. There is a large literature devoted to this topic, which regrettably is too extensive to review here. A fundamental paper was by Alder & Wainwright (1970), who discovered long-time tails in molecular-dynamics computer simulations of neutral fluids. The issue arises as well in plasma kinetic theory, where it was investigated by Krommes & Oberman (1976*b*); see that paper for earlier references. The topic is also discussed in the books by Balescu (1975, Sec. 21.5), Reichl (1998, Sec. S11.A), and Zwanzig (2001, Chap. 9). Brey (1983) showed that the Burnett coefficients for neutral fluids were infinite. Although the specific generalization of Brey's calculations to plasma kinetic theory at the Burnett level has not been done, there is no reason to believe that the situation is any better in plasmas. Why, then, should one proceed? This question was addressed by Wong *et al.* (1978) as follows:[1]

> It is known that persistent correlations (Alder & Wainwright 1970, and the additional references cited by Wong *et al.* in their Ref. 11) lead to a divergence of the linear Burnett-order coefficients, which is to say that the expansion used is not justified; a well-defined expansion involving fractional powers of the wavenumber has been obtained by Ernst & Dorfman (1972). For the nonlinear case a nonanalytic dependence of stress on strain rate has been found (Kawasaki & Gunton 1973, and further references cited by Wong *et al.* in their Refs. 9 and 13) such that the corresponding nonlinear Burnett coefficients are also divergent. Thus an expansion of Chapman–Enskog type is actually not legitimate. However, it is possible to make a separation into a regular part (for which a Chapman–Enskog expansion is valid) plus a singular part, and in the cases considered the singular part is found to be small (Kawasaki & Gunton 1973; Ernst & Dorfman 1972; Ernst *et al.* 1978). Hence the formulas ... will give accurate results if evaluated in a way which suppresses persistent correlations .... Our point of view is therefore to ignore the effects of persistent correlations, but the results nevertheless shed some light on the nature of the divergences.

A summary of that same point was given slightly later by Brey *et al.* (1981):

> We want to point out that in spite of the problem of the divergencies, an expansion in gradients of the transport equations may still be useful. First, it may allow the calculation of the 'regular' or nonsingular part of these coefficients and this may be interesting since the regular part is probably dominant for small, but not asymptotically small, gradients.

For those readers used to calculations based on the Landau (or Balescu–Lenard) collision operator, it is necessary to emphasize that no divergences arise in that context; those operators produce finite results for transport coefficients at both Navier–Stokes and Burnett orders. [See the discussion of the calculations of Catto & Simakov (2004) in §6.] The point is that those operators omit crucial many-body physics related to long-lived correlations. Although the strength of those effects naively scales with $\epsilon_{\mathrm{p}}^n$, where $n \geqslant 2$ ($\epsilon_{\mathrm{p}} \doteq 1/n\lambda_{\mathrm{D}}^3$ is the plasma discreteness parameter), so they are ignored in the derivations of the usual operators, such fluctuations are excited for each of an infinity of long-wavelength modes, the superposition of which need not be negligible and can even lead to infinities at Burnett order. In the language of the above two quotes, calculations based on the Landau operator lead to the regular parts of the transport coefficients.

Divergences of the transport coefficients aside, there is another well-known difficulty with the Burnett equations: they may be mathematically ill-posed in the sense that singularities in the predicted nonlinear fluid motions may arise for sufficiently large gradients. Furthermore, the proper choice of boundary conditions is an issue. There has been a great deal of work on these topics, which can be located through literature searches. Although they are obviously important, I shall not pursue them here.

In spite of these serious concerns, I shall proceed in the present paper to discuss the formal structure of the Burnett equations for multispecies and magnetised plasmas (ignoring any issues with long-time tails and nonlocality); then I shall compare the results to earlier work for weakly coupled plasma. In plasma physics, equations that contain (some) Burnett terms have been derived by Mikhaĭlovskiĭ (1967), Mikhaĭlovskiĭ

---

[1] The quotation has been modified to accommodate the author-year style of referencing used in the present article.

& Tsypin (1971, 1984), and Catto & Simakov (2004). None of those authors remarked on connections to the extensive work on Burnett equations in neutral gases. In particular, neither Mikhaĭlovskiĭ & Tsypin (1984) nor Catto & Simakov (2004) seem to have been aware of the earlier work of Wong *et al.* (1978) and the more technically efficient subsequent work of Brey *et al.* (1981), so they did not attempt to demonstrate consistency between their calculations and previously known general formulas. One goal of the present research is to show that the results of Catto & Simakov are, in fact, consistent[2] with a subset of the general formulas of Brey *et al.* as specialized to a one-component fluid by Brey (1983) and calculated to the lowest nontrivial order in weak coupling. Another motivation, quite apart from the possible practical implications of Burnett-level effects (which are not addressed in this paper), is the elucidation of the proper treatment of plasma kinetic theory beyond linear order, which is generally instructive and has implications that transcend the specific application to second-order plasma transport. The discussion serves to unify several research threads, including traditional plasma kinetic theory and the results of Rose (1979) on non-Gaussian statistical closures that embrace both collective turbulence and discrete-particle effects.

The formulas recorded by Brey (1983) apply to an unmagnetised, one-component, neutral fluid. Adding a background magnetic field $\boldsymbol{B}$ to the general formalism is formally trivial. However, the resulting symmetry breaking leads to many additional transport coefficients relative to the unmagnetised case. I shall not give specific formulas for those coefficients at Burnett order. Although the way to do so is straightforward, in the limit of large $B$ those coefficients are very small; Catto & Simakov did not calculate them. The magnetised problem at Navier–Stokes order was treated from the projection-operator point of view in Part I.

In a multicomponent system, additional interspecies momentum and energy *exchange effects* emerge. At Navier–Stokes order, those were treated in Part I, and I shall give some further discussion in this paper. At Burnett order, the presence of exchange effects leads to yet more terms. Although I shall indicate where those effects arise in the general formalism, I shall not derive specific formulas for them because the first-order exchange effects are already small.

Given the restrictions listed in the last two paragraphs, one sees that the present paper does not present a complete description of Burnett effects in plasmas. Instead, the focus is on developing the basic ideas, illustrating the use of projection operators in a nontrivial context, and discussing the equivalence between the two-time formalism and Chapman–Enskog theory. Various other insights emerge along the way.

### 1.1. *Introduction to Burnett effects*

At first order in a weakly coupled, unmagnetised, one-component plasma, there are just two nonvanishing dissipative coefficients, namely the kinematic viscosity $\mu$ and the thermal conductivity $\kappa$; see §I:2 and §I:A.[3] For strong coupling, the bulk viscosity $\zeta$ is also required. At second order, there is an explosion of terms. The general theory of an unmagnetised, one-component plasma with arbitrary coupling involves at Burnett order 13 additional $\mu$'s and 8 additional $\kappa$'s, collectively described in terms of 23 nontrivial integrals over two-time correlation functions involving two or three phase-space points (see §4). The details are tedious. However, the basic idea is clear.

[2] I say 'consistent' because in order to be quantitative Catto & Simakov calculate the corrections to the distribution function approximately (inversion of the collision operator is involved), thus obscuring the general structure of the theory. However, if one manipulates the basic equations solved by Catto & Simakov, the equivalence of the formalisms can be demonstrated. That is done in §6. It can then be seen that the calculations of Catto & Simakov of the parallel viscosity amount to approximate evaluations of some of the formulas of Brey (1983), and that straightforward generalization of those unmagnetised formulas would lead to the magnetic-field effects evaluated by Catto & Simakov.

[3] Section and equation numbers that refer to Part I or the Supplement to Part II are prefaced by 'I:' or 'II-S:', respectively.

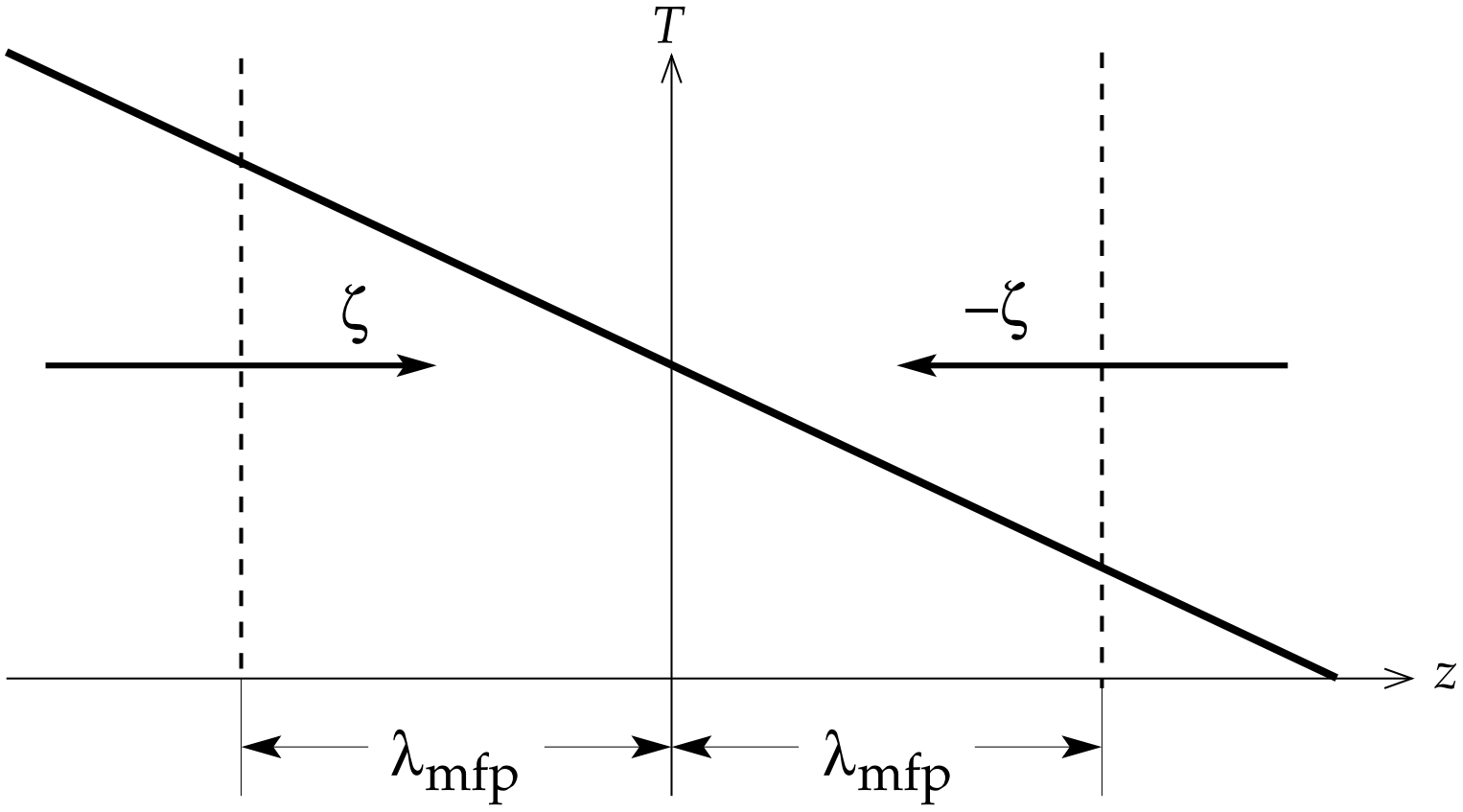

FIGURE 1. One mechanism leading to a Burnett contribution to the momentum flux that is proportional to $|\boldsymbol{\nabla}T|^2$. The thick solid line depicts a temperature profile with constant gradient. The net second-order momentum flux across the $z = 0$ plane arises from the unbalanced portion of the first-order viscous forces exerted on the velocity streams arriving from a mean free path $\lambda_{\rm mfp}$ away. See the text for further discussion.

I shall illustrate by considering the scaling of the contribution of the unmagnetised (or parallel) momentum flux proportional to $(\nabla T)^2$ (here I ignore tensorial properties, i.e., the distinction between $\boldsymbol{\nabla}T\,\boldsymbol{\nabla}T$ and $|\boldsymbol{\nabla}T|^2\boldsymbol{I}$). One mechanism is illustrated in figure 1, which is in the spirit of some diagrams used by Braginskii (1965).[4] Assume a constant temperature gradient in the $-z$ direction and zero net flow velocity, $u_z = 0$. Divide the distribution function into positive- and negative-going streams whose velocities are $\zeta_+ \equiv \zeta$ and $\zeta_- = -\zeta$; those streams arrive from distances of the order of the mean free path $\lambda_{\rm mfp} \doteq v_{\rm t}/\nu$, where $v_{\rm t} \doteq (T/m)^{1/2}$ is the thermal velocity and $\nu$ is the collision frequency. Arguing heuristically, one can assert that at first order each stream experiences a viscous force $-nm\mu\nabla\zeta$. Because the streams arrive from regions of differing temperatures, that force is unbalanced; the net momentum flux $\pi$ across the plane $z = 0$ is

$$\pi = -nm\mu_+\nabla\zeta_+ - nm\mu_-\nabla\zeta_- \approx -nm(\mu_+ - \mu_-)\nabla\zeta. \qquad (1.1)$$

One has $\nabla\zeta = (\partial_T\zeta)\nabla T$ and $\mu \sim \langle\delta v^2\rangle/\nu$, $\delta v$ being a thermal velocity fluctuation. Temporarily assuming that $\nu$ is constant (the temperature dependence of $\nu$ is properly taken into account in the detailed calculations described later), one has

$$\mu(z) \approx \mu_0 + z(\nu^{-1}\partial_T\langle\delta v^2\rangle)\nabla T. \qquad (1.2)$$

The $\mu_0$ contributions cancel in (1.1). If one assumes that the streams arrive from distances $\mp\lambda_{\rm mfp}$, the remainder is

$$\pi \approx -nm[-2\lambda_{\rm mfp}(\nu^{-1}\partial_T\langle\delta v^2\rangle)\nabla T](\partial_T\zeta)\nabla T \sim nm\mu\left(\frac{\lambda_{\rm mfp}}{L}\right)\left(\frac{v_{\rm t}}{L}\right). \qquad (1.3a)$$

[4]This picture cannot possibly be new, but I am unable to cite a reference where it is presented in the present form, including the subsequent discussion of non-Gaussian statistics.

The small ratio $\lambda_{\mathrm{mfp}}/L$ signifies that this Burnett effect is of one higher order in the gradients than is the usual first-order viscous stress.

The combination $m\langle\delta v^2\rangle v_{\mathrm{t}}$, which appears in the above estimate, arises microscopically from the kinetic energy flux $\boldsymbol{J}^E = \frac{1}{2}mv^2\boldsymbol{v}$. In detail, it is shown later that the present effect arises from the cross correlation between (subtracted versions of[5]) the microscopic momentum flux $m\boldsymbol{v}\boldsymbol{v}$ and $\boldsymbol{J}^E$; see formula (4.15*e*), which according to (4.8*e*) gives one contribution to the Burnett momentum flux proportional to $\boldsymbol{\nabla}T\,\boldsymbol{\nabla}T$.

It is easy to see that an extension of the above argument to include a nonconstant temperature gradient will lead to another Burnett term involving $\nabla^2 T$. Various effects involving flow gradients follow as well. Detailed formulas for all of those are derived in the subsequent sections.

The concepts of velocity streams with definite speed, and of a mean free path of constant value, are useful for simple heuristic arguments. However, they provide a poor description of the actual particle probability density function (PDF) and cannot be used if systematic and quantitative results are desired. In reality, both the particle velocities and the mean free path are random variables at the microscopic level. Consider, for example, the classical (first-order) viscosity $\mu \sim v_{\mathrm{t}}^2/\nu = v_{\mathrm{t}}\lambda_{\mathrm{mfp}}$. More properly, with $\widetilde{\lambda} = \widetilde{v}/\nu$ and $\langle\widetilde{v}\rangle = 0$, viscosity is expressed as the statistical average $\mu \sim \langle\widetilde{v}\widetilde{\lambda}\rangle = \langle\delta v^2\rangle/\nu = v_{\mathrm{t}}^2/\nu$. (In reality, $\nu$ also fluctuates, but that does not affect the basic scaling.) At second order, we know from the above physical picture that one needs to consider differences in the statistics of fluctuations in regions separated by a mean free path. This suggests a Taylor expansion of the velocity fluctuations and implies a second-order contribution to momentum flux of the form

$$\Delta_2(nm\langle\delta v\,\delta v\rangle) \sim nm\langle[\widetilde{\lambda}(\partial_T\delta v)\nabla T]\delta v\rangle. \tag{1.4}$$

This involves a triplet correlation function, essentially $\langle\delta v\,\delta v\,\delta v\rangle$. Whereas at first order in the gradients it is adequate to assume that $\delta v$ is Gaussian (only the second cumulant enters the above expression for $\mu$), one must do better at second order because triplet correlations vanish for Gaussian statistics. Indeed, the presence of a temperature gradient distorts the distribution from Gaussian form, and on dimensional grounds it is reasonable to guess that

$$\langle\delta v\,\delta v\,\delta v\rangle \sim \langle\delta v^2\rangle\langle\delta v^2\rangle^{1/2}\lambda_{\mathrm{mfp}}\nabla T. \tag{1.5}$$

Upon inserting this estimate into (1.4), one recovers (1.3*a*).

More quantitatively, the significance of non-Gaussian triplet correlations for Burnett-order effects can be demonstrated by a simple stochastic model discussed in appendix A. That model has nothing to do with many-body physics *per se*, but it demonstrates the role of symmetry breaking in producing non-Gaussian statistics. Study of that appendix is not necessary for the understanding of the remainder of this paper, but it adds additional perspective and may be of interest to people with backgrounds in statistical turbulence theory.

Ultimately, all of the Burnett coefficients will be expressed in terms of two-time correlation functions involving the Klimontovich phase-space microdensity [see (2.6) for the definition of that quantity]. Although from the above argument those coefficients appear to be related to certain three-point correlation functions (which describe non-Gaussian symmetry-breaking effects), this (perhaps paradoxically) does not mean that in all cases one needs to evaluate Klimontovich correlations involving three phase-space points. The formalism to be described does a preliminary processing that expresses the non-Gaussian effects in terms of two-time Klimontovich correlations that can involve either two or three phase-space points. For example, it turns out that the mechanism estimated above by (1.3*a*) can be calculated from a two-point correlation function. [This is plausible because (1.5) is written in terms of $\langle \delta v^2 \rangle$, which is a second-order cumulant.] On the other hand, other non-Gaussian effects cannot be so expressed, and some contributions to the Burnett coefficients require calculation of three-point phase-space correlations. See, for example, (4.8*e*), which shows that a variety of effects contribute to the coefficient of $\boldsymbol{\nabla} T\, \boldsymbol{\nabla} T$ in the momentum equation. Roughly speaking, effects calculable from two-point Klimontovich correlations are related to the linearized collision operator, while ones calculable from three-point Klimontovich correlations involve the nonlinear collision operator. This point will become clearer as we work through the details.

[5] Subtracted fluxes are discussed in §2.6.2 and calculated in appendix D.

### 1.2. *The relationship between two-time formalism and Chapman–Enskog theory*

In its most general form, the present procedure (based on time correlation functions) produces a plethora of transport effects, not all of which appear in the work of Catto & Simakov. It must be stated clearly that the absence of certain terms in that work is not related to the authors' formalism or to algebraic mistakes; rather, it has to do with their ordering choices (made for relevance to drift-wave physics in magnetised plasmas). First, note that in a linear analysis (such as is presented in Part I) it does not make sense to discuss the size of, say, the perturbed fluid velocity; it is formally infinitesimal. Thus, there is confusion already with the first sentence of Catto & Simakov (2004), which reads

> The short mean free path description of magnetised plasma as originally formulated by Braginskii ... and Robinson and Bernstein ... assumes an ordering in which the ion mean flow is on the order of the ion thermal speed.

While this is a familiar, often-repeated statement, it does not apply to linearized hydrodynamics. Instead, that emerges by working to first order in $\Delta \doteq k_\parallel \lambda_{\mathrm{mfp}} \ll 1$ and $\delta \doteq k_\perp \rho \ll 1$. (Here I use the notation of Catto & Simakov for $\Delta$ and $\delta$; $k_\parallel \doteq |\nabla_\parallel| \equiv L_\parallel^{-1}$ and $k_\perp \doteq |\boldsymbol{\nabla}_\perp| \equiv L_\perp^{-1}$.) If one pursues an expansion to second order in $\Delta$ and $\delta$ with no further approximations, the complete set of Burnett effects will appear. For example, if at first (Navier–Stokes) order in the gradients one finds dissipative fluxes proportional to $\nabla \boldsymbol{u}$ (momentum diffusion due to viscosity) or $\nabla T$ (heat diffusion due to thermal conductivity), at second (Burnett) order cross effects in the fluxes such as $(\nabla \boldsymbol{u}) \boldsymbol{\cdot} (\nabla T)$ will emerge. However, Catto & Simakov order $u$ itself to be small, $u \sim \nabla T$. Then $(\nabla u) \boldsymbol{\cdot} (\nabla T)$ becomes formally of third order in the gradients and is neglected in their analysis. As they state, their ordering is appropriate 'for most magnetic confinement and fusion devices in general, and the edge of many tokamaks in particular', and I am not arguing that they have made an error with this ordering. Of course, it is possible that the complete set of terms is required for some esoteric situations. If not, however, it is straightforward to apply the ordering of Catto & Simakov to the general results and attempt to reduce them to the smaller set of terms retained by those authors. I shall do that later in the paper.

Apart from subsidiary orderings, one can ask whether the end points of the several possible approaches should agree in general. The distinction is between (i) Chapman–Enskog analysis of a $\mu$-space kinetic theory that includes a nonlinear collision operator (the approach used by Catto & Simakov); and (ii) a $\mathit{\Gamma}$-space formalism based on multipoint, two-time correlation functions (the approach used by Brey *et al.*). This question was answered in the affirmative by Wong *et al.* (1978) for the dilute neutral gas. I shall argue that the analogous correspondence holds as well for weakly coupled plasmas, and I shall illustrate by working out a certain example in detail. While it is clear on general principles that the procedures should be equivalent, a detailed demonstration is decidedly nontrivial, as it requires careful attention to the calculation and subsequent manipulation of certain correlation functions involving two times and as many as three phase-space points.

It is interesting to focus on the specific advance made by Catto & Simakov over the work of previous authors such as Mikhailovskii & Tsypin, which was to calculate heat-flow contributions to the plasma stress tensor due to the nonlinearity of the plasma collision operator. I shall show where those terms arise in the theory based on time correlation functions. That provides an interesting perspective on the role of so-called nonlinear noise terms that are relevant not only in the many-body theory of discrete particles but also in continuum turbulence theory.

### 1.3. *Gaussian and non-Gaussian statistics in kinetic theory*

Before I delve into formal mathematics, I shall give a brief prelude to motivate the relationship between single-time, $\mu$-space kinetic theory and the $\mathit{\Gamma}$-space theory of two-time correlations. Consider a simple spatial random walk (Rudnick & Gaspari 2004) and, for example, the associated density diffusion equation, a simplified version of which is $\partial_t n = D\nabla^2 n$. Similar equations for flow velocity $\boldsymbol{u}$ (involving viscosity $\mu$) and temperature $T$ (involving thermal conductivity $\kappa$) can be derived by Chapman–Enskog expansion of the one-time Landau kinetic equation for plasmas. On the other hand, Taylor (1921) showed in the context of turbulence theory that the spatial diffusion coefficient $D$ (for a test particle or fluid element) can be obtained as the time integral of the two-time Lagrangian velocity correlation function:[6]

$$D = \int_0^\infty \mathrm{d}\overline{\tau}\, C_{vv}(\overline{\tau}). \tag{1.6}$$

Similar integrals (Kubo formulas) can be written for $\mu$ and $\kappa$. Thus, there is an intimate relationship between one- and two-time theory.

The technical problem with the Taylor formula (1.6) is that Lagrangian correlation functions are very difficult to calculate in the general case. However, it is well known that $D \sim \Delta x^2/\Delta t$, where $\Delta x$ is the magnitude of the characteristic spatial step taken during the step time $\Delta t$. The steps $\Delta x$ and $\Delta t$ represent two-point *Eulerian* information: two distinct spatial points are required in order to measure a spatial difference $\Delta x$, and two distinct time points are required for $\Delta t$. Of course, these are statistical characteristics, so the relevant quantities are certain two-point Eulerian correlation functions in space and time that I shall here denote generically by $C(\boldsymbol{x}, t, \boldsymbol{x}', t') \equiv C(\boldsymbol{\rho}, \tau)$, where $\boldsymbol{\rho} \doteq \boldsymbol{x} - \boldsymbol{x}'$ and $\tau \doteq t - t'$. (I assume homogeneous, stationary statistics.) One goal of theories of turbulence or nonequilibrium statistical mechanics is to establish the specific relationships between Lagrangian representations of transport coefficients and the relevant Eulerian correlation functions that enable one to make quantitative calculations.

It is the right-hand side of the equation $\partial_\tau C(\boldsymbol{\rho}, \tau) = \cdots$ from which $C$ acquires its characteristic space and time scales. Since in the present paper we are considering classical spatial transport, those scales are $\Delta x \sim \lambda_{\mathrm{mfp}}$ and $\Delta t \sim \nu^{-1}$. However, in the usual weakly coupled plasma kinetic theory, those 'kinetic' scales do not appear in

---

[6]The Lagrangian velocity correlation function is defined by

$$C_{vv}(\tau) \doteq \langle \delta v(\widetilde{x}(\tau), \tau)\delta v(\widetilde{x}(0), 0)\rangle,$$

where $\delta v$ denotes the velocity fluctuation from the mean and $\widetilde{x}(t)$ is the random trajectory of a fluid element (or test particle).

the lowest-order equation for two-point correlations, which follows from the linearized Klimontovich equation. That equation contains only Vlasov physics; it describes the formation over a timescale of $\omega_p^{-1}$ ($\omega_p$ is the plasma frequency) of plasma shielding clouds whose spatial extent is the Debye length $\lambda_D$. This is adequate for discussions of velocity-space diffusion and test-particle polarization effects (the content of the Balescu–Lenard and Landau collision operators), but it does not capture the kinetic scales required for spatial transport. This implies that evaluation of spatial transport coefficients from two-time theory requires one to consider collisional corrections to the evolution equation for two-point correlations. By standard arguments, those corrections are represented by triplet correlation functions,[7] which describe (some) non-Gaussian effects. To repeat, those effects are not included in the derivation of the Landau collision operator, as they are of higher order in the plasma parameter $\epsilon_p$.

It is important to be clear about what 'non-Gaussian' means in this context. I am not referring to the fact that the shape of the one-particle distribution function $f_s(\boldsymbol{x}, \boldsymbol{v}, t) \equiv f(\mu, t)$ ($s$ is a species label and $\mu \doteq \{\boldsymbol{x}, \boldsymbol{v}, s\}$) usually differs from a Maxwellian in velocity. In general, the statistics of a Gaussian time series $\widetilde{\psi}(t)$ are completely specified by the mean $\langle \widetilde{\psi}(t) \rangle$ and the two-time correlation function $\langle \delta\widetilde{\psi}(t)\delta\widetilde{\psi}(t') \rangle$, otherwise known as the first and second cumulants.[8] In the present discussion, the underlying random variable is the Klimontovich microdensity $\widetilde{f}(\mu, t)$, defined by (2.6) below. Its statistical mean is the (in general, non-Maxwellian) $f(\mu, t)$, and its two-point correlation function is $C(\mu, t, \mu', t') \doteq \langle \delta\widetilde{f}(\mu, t)\delta\widetilde{f}(\mu', t') \rangle$. At equal times, one has

$$C(\mu, \mu', t) = \overline{n}^{-1}\delta(\mu - \mu')f(\mu', t) + g(\mu, \mu', t), \tag{1.8}$$

where $\overline{n}$ is the mean density and $g$ is the pair correlation function. It is the approximate calculation of the long-time limit of that $g$ to first order in $\epsilon_p$, producing a functional $g[f]$, that ultimately leads to the weakly coupled plasma collision operator. From this point of view, one might argue that the plasma kinetic equation with the nonlinear Landau collision operator is a Gaussian theory, since it involves only (special cases of) the first two cumulants of $\widetilde{f}$, namely the equal-time $f$ and $g$.

The reader may recognize a possible paradox here. I have argued that determination of the characteristic random-walk steps from two-point correlations requires non-Gaussian corrections. Yet in Part I I showed from several points of view — classical Chapman–Enskog theory and projection-operator methods — that it is possible to calculate the correct Braginskii transport coefficients from the linearized Landau kinetic equation, which does not seem to be aware of non-Gaussian effects. How can this be reconciled?

The first part of the answer involves the reminder (see footnote 8) that a random process is Gaussian only if all *multiple-time* cumulants vanish beyond second order. This is not the case for plasma kinetic theory due to the intrinsic nonlinearity of the Coulomb interaction.

---

[7] This kinetic-theory problem is completely analogous to the problem of Navier–Stokes turbulence, which has been discussed extensively. A highly incomplete list of discussions of statistical closure theory includes Kraichnan (1961, 1962), Leslie (1973), McComb (1990), Krommes (2002), McComb (2014), and Krommes & Parker (2018, and references therein).

[8] The Gaussian probability density functional has the form

$$P[\psi] = \mathscr{N} \exp\left(-\frac{1}{2}\int_{-\infty}^{\infty} \mathrm{d}t \int_{-\infty}^{\infty} \mathrm{d}t' \, [\psi(t) - \langle\psi\rangle(t)]C^{-1}(t, t')[\psi(t') - \langle\psi\rangle(t')]\right), \tag{1.7}$$

where $\mathscr{N}$ is a normalization factor and $C(t, t') \doteq \langle \delta\psi(t)\delta\psi(t') \rangle$. The inverse $C^{-1}$ is defined by the equation $\int \mathrm{d}\overline{t}\, C(t, \overline{t})C^{-1}(\overline{t}, t') = \delta(t - t')$.

Nevertheless, apparently it is possible to calculate transport effects from either a one-time theory involving just $f(t)$ and $g[f(t)]$ or a two-time theory of correlation functions. For consistency, there must therefore be a deep relationship between one- and two-time theory. This follows as a statistical generalization of the well-known fact that a linear ordinary differential equation (ODE) of first order in time can be solved by means of a two-time Green's function. Namely, the ODE

$$\frac{\mathrm{d}\psi}{\mathrm{d}t} + \mathrm{i}\mathrm{L}\psi = s(t), \tag{1.9}$$

where L is a linear operator and $s(t)$ is a given source, has the solution

$$\psi(t) = R(t;0)\psi(0) + \int_0^t \mathrm{d}\overline{t}\, R(t;\overline{t})s(\overline{t}) \quad (t > 0), \tag{1.10}$$

where Green's function $R$ obeys

$$\partial_t R(t;t') + \mathrm{i}\mathrm{L}R(t;t') = \delta(t-t'), \quad R(t'-\epsilon;t') = 0. \tag{1.11}$$

Formally,

$$R(t;t') = \frac{\delta\psi(t)}{\delta\widehat{\eta}(t')}, \tag{1.12}$$

where $\widehat{\eta}(t)$ is an arbitrary function[9] that replaces $s(t)$ in (1.9). That is, the functional variation of a one-time field with respect to a time-dependent perturbation is described by a two-time dynamics. The details of the statistical generalization will be described later (see §5), but the basic result is that, in essence, an $n$-point cumulant function can be obtained from the appropriate functional derivative of an $(n-1)$-point cumulant. Thus, functionally differentiating the (one-time) nonlinear Landau kinetic equation both linearizes that equation and produces a two-time dynamics that includes collisional corrections involving the linearized collision operator. In a direct derivation of the equation for the two-time correlation function (say from the Klimontovich equation), the collisional corrections arise from triplet correlations. But at first order the forms of the linearized equations in the $t$ variable are the same in both the two-time theory[10] and the one-time Chapman–Enskog approach (or the equivalent projection-operator approach described in Part I).

A more difficult question is what happens at second order. Upon pursuing the one-time Chapman–Enskog approach to that order, Catto & Simakov found contributions to the Burnett stress tensor from the nonlinear (in fact, bilinear when the Landau operator is used) collision operator evaluated with the first-order correction to the one-particle distribution function (i.e., $\mathrm{C}[f_1, \overline{f}_1]$).[11] For those to be compatible with the general two-time theory, it has to be the case that certain triplet correlations are driven in a definite way by the nonlinear collision operator. Technically, it is by no means immediately obvious that this is so. In order to demonstrate it, we shall have to proceed systematically.

---

[9] The hat on $\widehat{\eta}$ is used for consistency with the notation of Martin *et al.* (1973). See Krommes (2002) for a review of the MSR formalism and for further discussion of source functions.

[10] The fact that at linear order transport coefficients can be represented in terms of two-time correlation functions is well known from the work of Kubo on linear response theory. Selected references include Kubo (1957), Kubo (1959), and Kubo (1974).

[11] Conventionally, the bilinear Landau collision operator is written as $\mathrm{C}[f, f]$. I shall instead write $\mathrm{C}[f, \overline{f}]$, where the overline indicates dependence on the integration variable $\overline{\boldsymbol{v}}$ associated with the field particles. Strictly speaking, the overline is unnecessary because the square-bracket notation already indicates general functional dependence. However, the notation will make some subsequent manipulations easier to follow.

Representations of the unmagnetised Burnett transport coefficients in terms of time correlation functions were obtained by Wong *et al.* (1978). In the subsequent discussion, I shall instead closely follow the later paper of Brey *et al.* (1981) because it fits more naturally into the projection-operator formalism that is explicated in this series of articles, and because it is in some ways more technically concise. This choice is in no way intended to detract from the importance of the earlier work by Wong *et al.*, whose general results were correct and were reproduced by Brey *et al.*

### 1.4. *Liouvillian dynamics vs kinetic equations; $\mathit{\Gamma}$ space vs $\mu$ space*

The Chapman–Enskog formalism is explicitly a one-time theory. The following remarks instead pertain to the representation of transport coefficients in terms of two-time correlations. Two-time theory can be developed within either a $\mathit{\Gamma}$-space or a $\mu$-space formalism.

The work of Brey *et al.* proceeds directly from the Liouville equation

$$\partial_t P_{\mathscr{N}}(\mathit{\Gamma}, t) + \mathrm{i}\mathscr{L} P_{\mathscr{N}} = 0, \tag{1.13}$$

where $\mathit{\Gamma}$ denotes the complete set of phase-space variables for the $\mathscr{N}$-particle system,[12] $P_{\mathscr{N}}$ is the $\mathscr{N}$-particle PDF, and $\mathscr{L}$ is the (first-order, differential) Liouville operator:[13]

$$\mathrm{i}\mathscr{L} \doteq \sum_{i=1}^{\mathscr{N}} \boldsymbol{v}_i \cdot \boldsymbol{\nabla}_i + \sum_{i=1}^{\mathscr{N}} \left(\frac{q}{m}\right)_i (\boldsymbol{E}_i + c^{-1}\boldsymbol{v}_i \times \boldsymbol{B}_i^{\mathrm{ext}}) \cdot \frac{\partial}{\partial \boldsymbol{v}_i}, \tag{1.14}$$

where $\boldsymbol{E}_i \doteq \sum_{j\neq i}^{\mathscr{N}} \boldsymbol{\epsilon}_{ij} q_j$ and $\boldsymbol{\epsilon}_{ij} \doteq -\boldsymbol{\nabla}_i |\boldsymbol{x}_i - \boldsymbol{x}_j|^{-1}$. I consider only the electrostatic approximation (no magnetic perturbations are allowed), so $\boldsymbol{B}_i^{\mathrm{ext}} \equiv \boldsymbol{B}^{\mathrm{ext}}(\boldsymbol{x}_i)$ is a given external or 'background' magnetic field. (Brey *et al.* and others did not include a magnetic field.) The final formulas are valid, in principle, for arbitrarily strong coupling, although I shall not evaluate any terms for strongly coupled systems. For weak coupling, an alternative procedure that was used in Part I is to first derive the kinetic equation (I shall follow many authors, including Catto & Simakov, and use the Landau form), then process that equation to obtain two-time correlation functions and, ultimately, transport coefficients. That is, while Brey *et al.* work in $\mathit{\Gamma}$ space, one could in principle proceed in $\mu$ space. The differences between the two approaches manifest in several ways.

#### 1.4.1. Fluxes in $\mathit{\Gamma}$ space and $\mu$ space

In the work of Brey *et al.*, the transport coefficients involve microscopic fluxes that appear in formulas that are averaged over a $\mathit{\Gamma}$-space distribution function. Those formulas are difficult to evaluate because of the presence of potential-energy terms. For weak coupling those terms may be neglected,[14] but one must still perform a $\mathit{\Gamma}$-space average.

---

[12] It is much more conventional to use $N$ rather than $\mathscr{N}$ for the total number of particles (and $V$ rather than $\mathscr{V}$ for the system volume). The script font is used in this paper in order to be consistent with a notation of Brey's for the total (volume-integrated) amount of a quantity. See §2.2, especially (2.3), for further discussion.

[13] Brey *et al.* do not include the i in the definition of the Liouville operator. I prefer to retain it as a reminder that the operator is time-reversible and so that the i disappears in $\boldsymbol{k}$ space. For example, under Fourier transformation the operator $\boldsymbol{v} \cdot \boldsymbol{\nabla} \to \mathrm{i}\boldsymbol{k} \cdot \boldsymbol{v}$, leading to the streaming operator $\mathscr{L}_{\boldsymbol{v}} = \sum_{i=1}^{\mathscr{N}} \boldsymbol{k} \cdot \boldsymbol{v}_i$. Dissipative effects in the many-body system arise from a generalization of the fact that the Maxwellian velocity average of the reversible streaming effect $\exp(-\mathrm{i}\boldsymbol{k} \cdot \boldsymbol{v}\tau)$ decays in $\tau$.

[14] Of course, potential-energy interactions cannot be neglected everywhere, since the dissipation arising from Coulomb collisions depends on them. The relevant correlations are contained in the Liouville operator.

If one instead works with a $\mu$-space description as was done in Part I, the kinetic fluxes ($m\boldsymbol{v}\boldsymbol{v}$ and $\frac{1}{2}mv^2\boldsymbol{v}$) appear directly, and the weakly coupled transport coefficients can be expressed as simple matrix elements in velocity space. For weak coupling, this would seem to argue in favour of a $\mu$-space approach to the theory of two-time correlations that begins with the irreversible kinetic equation; however, see the next subsection.

### 1.4.2. Reversible vs irreversible operators

Because the Liouville operator $\mathscr{L}$ is time-reversible whereas the kinetic-equation approach involves the time-irreversible collision operator, there arise technical differences in the analyses relating to the absence or presence of a dissipative collision operator in the kinetic equation. For example, $\mathrm{e}^{-\mathrm{i}\mathscr{L}\tau}$ is a propagator that moves the phase-space variables back in time by an amount $\tau$,[15] so if $A(\mathit{\Gamma})$ and $B(\mathit{\Gamma})$ are two arbitrary phase functions, one has the useful result

$$\mathrm{e}^{-\mathrm{i}\mathscr{L}\tau}(AB) = \left(\mathrm{e}^{-\mathrm{i}\mathscr{L}\tau}A\right)\left(\mathrm{e}^{-\mathrm{i}\mathscr{L}\tau}B\right). \tag{1.15}$$

This property is used by Brey *et al.* in the course of various manipulations that lead to the final form of the Burnett coefficients. However, if $\widehat{\mathrm{C}}$ is the linearized collision operator and if $a$ and $b$ are two functions that live in $\mu$ space, it is not true that[16] $(\mathrm{e}^{-\widehat{\mathrm{C}}\tau}a)(\mathrm{e}^{-\widehat{\mathrm{C}}\tau}b) = \mathrm{e}^{-\widehat{\mathrm{C}}\tau}(ab)$. Due to this difficulty, I shall abandon a $\mu$-space approach in favour of the $\mathit{\Gamma}$-space one followed by Brey *et al.* That has the added advantage of leading to formulas that are applicable to arbitrary coupling. Although in general those formulas are difficult to evaluate, it is not difficult to reduce them to the weakly coupled limit.

## 1.5. *Modified and unmodified propagators*

The interspecies coupling in a multispecies plasma introduces some complications that are absent in a one-component neutral gas. Thus, some generalizations and modifications of the pure-fluid formulas are required. Were it not for those modifications, the formulas of Brey (1983) could be used with only the straightforward addition of the Lorentz force term to the momentum equation and some generalizations of various symmetry arguments to allow for the anisotropy induced by a background magnetic field. In the multispecies case, however, one must be careful about the handling of the orthogonal projector Q, so one must repeat and extend the derivation, paying close attention at every step to the assumptions that Brey *et al.* make. To be more specific, I note that in linear response theory a crucial role is played by the modified propagator $\mathrm{G_Q} \doteq \exp(-\mathrm{QiLQ}t)$, where L is the linear operator in the perturbed kinetic equation. As was discussed in §I:G.3, an identity relates $\mathrm{G_Q}$ to the unmodified propagator $\mathrm{G} \doteq \exp(-\mathrm{iL}t)$, and for the

[15] This property follows as a generalization of the basic properties

$$\mathrm{i}\mathscr{L}\boldsymbol{x}_i = \boldsymbol{v}_i = \frac{\mathrm{d}\boldsymbol{x}_i}{\mathrm{d}t}, \quad \mathrm{i}\mathscr{L}\boldsymbol{v}_i = \left(\frac{q}{m}\right)_i (\boldsymbol{E}_i + c^{-1}\boldsymbol{v}_i \times \boldsymbol{B}_i^{\mathrm{ext}}) = \frac{\mathrm{d}\boldsymbol{v}_i}{\mathrm{d}t},$$

or for arbitrary phase function $f(\mathit{\Gamma})$

$$\mathrm{i}\mathscr{L}f(\mathit{\Gamma}) = \frac{\mathrm{d}}{\mathrm{d}t}f(\mathit{\Gamma}(t))|_{t=0}.$$

An equivalent discussion is given by Piccirelli (1968); see his equations (23) and (24).

[16] This result would be true if $\widehat{\mathrm{C}}$ were a first-order differential operator with a constant coefficient, such as $\partial_v$. However, $\widehat{\mathrm{C}}$ is, at the very least, a second-order differential operator in velocity. (In fact, it is an integro-differential operator.)

neutral-fluid case transport coefficients follow as the double (ordered) limit

$$\boldsymbol{D} = \lim_{\omega \to 0} \lim_{k \to 0} k^{-2} \widehat{\boldsymbol{\Phi}}_k(\omega), \tag{1.16}$$

where $\widehat{\boldsymbol{\Phi}}$ is the hydrodynamic transport matrix defined with G instead of $\mathrm{G_Q}$. At linear order, Brey *et al.* (1981) express the transport matrix $\widehat{\Sigma}$ (defined with $\mathrm{G_Q}$) in terms of the equivalent quantity $\widehat{\boldsymbol{\Phi}}$ defined with G, then argue that to second order in the gradient expansion the correction is negligible. That argument fails for the multispecies case because there the collision operator does not satisfy $\mathrm{Q}\widehat{\mathrm{C}}\mathrm{Q} = \widehat{\mathrm{C}}$ and because $\widehat{\mathrm{C}}$ does not vanish with the gradients. The implication is that both the Navier–Stokes and Burnett transport coefficients for a multispecies plasma are determined by matrix elements involving $(\mathrm{Q}\widehat{\mathrm{C}}\mathrm{Q})^{-1}$ rather than $\widehat{\mathrm{C}}^{-1}$, and the equations will also contain extra terms due to the interspecies collisional coupling. The Navier–Stokes interspecies coupling terms were already treated in Part I for the special case of small electron-to-ion mass ratio. Because for standard orderings those exchange terms are already small, I shall eschew discussion of the analogous Burnett effects in this work.

### 1.6. *Plan of the paper*

The remainder of the paper is organized as follows. In §2 I introduce the general time-independent projection-operator formalism of Brey *et al.* (1981). In §3 I repeat their reduction of the general formulas through Burnett order, paying careful attention to modifications introduced by the multispecies and magnetised nature of the plasma. In §4 I follow Brey (1983) and record the explicit formulas, written in terms of two-time correlation functions, for the special case of an unmagnetised one-component fluid. In §5 I discuss some aspects of the theory of those correlation functions, and I show how to calculate them in the limit of weak coupling. In §6 I compare the predictions of the two-time formalism to those of (the one-time) Chapman–Enskog theory; I conclude that they are equivalent. I summarize and discuss the paper in §7. Several appendixes are included. In appendix A I present a simple stochastic model that shows that Burnett effects arise in contexts more general than classical many-body theory and emphasizes the importance of non-Gaussian statistics. In appendix B I discuss the time evolution of the microscopic momentum density, paying particular attention to modifications introduced by the long-range nature of the Coulomb force and the presence of multiple species. In appendix C I discuss the choice of conjugate variables, whose gradients serve as thermodynamic forces. In appendix D I describe the evaluation of the fundamental subtracted fluxes from which the transport coefficients are constructed. In appendix E I evaluate the hydrodynamic portion of the first-order momentum exchange term from its time-correlation representation and, in a nontrivial calculation, show that one obtains Braginskii's result. Some technical manipulations leading to one of the non-Gaussian terms are described in appendix F. Derivations of both the nonlinear and linearized plasma collision operators are described in appendix G. Also discussed there is the relationship between certain so-called nonlinear noise terms and the nonlinear Balescu–Lenard operator. In appendix H I address some issues with one-sided correlations and the interpretation of nonlinear noise. Finally, I summarize the principal notation in appendix I. Some further details and supporting calculations are given in the online Part II Supplement (Krommes 2018*c*). Both Part II and its Supplement rely on a basic understanding of the material in Part I.

## 2. The time-independent projection-operator formalism of Brey *et al.*

In this section I shall review the time-independent projection-operator formalism of Brey *et al.* (1981), paying particular attention to the generalizations needed to include a background magnetic field and especially multiple species.

### 2.1. *Statistical ensembles and distribution functions*

We shall consider several kinds of statistical ensembles. Most fundamentally, it is adequate for present purposes to work in a very large, fixed box of volume $\mathscr{V}$ that is in thermal contact with a heat reservoir at temperature $T$ but is impervious to particles. Thus, the total particle number $\mathscr{N}$ does not change and the mean density $\overline{n} \doteq \mathscr{N}/\mathscr{V}$ is spatially constant. This system is described in thermal equilibrium by the standard canonical ensemble. I shall consider the thermodynamic limit $\mathscr{N} \to \infty$, $\mathscr{V} \to \infty$, $\overline{n} = \text{const}$. I shall not consider an external electric field $\boldsymbol{E}^{\text{ext}}$, but I shall allow for an external constant magnetic field $\boldsymbol{B}^{\text{ext}}$.

In principle, all statistical properties of the previous system follow from solution of the Liouville equation (1.13) (or of multiple-time generalizations of that equation). However, when the goal is to develop local transport equations in which fluxes are driven by gradients of thermodynamic forces, direct solution of the Liouville equation is difficult, particularly when one wants to perturbatively develop expressions to higher order in the gradients. One cannot impose a constant temperature gradient across the entire box because the boundaries of the box are supposed to be at the constant bath temperature $T$. One can attempt to circumvent that problem by imposing periodic boundary conditions and considering long-wavelength sinusoidal perturbations. Then one can imagine dividing $\mathscr{V}$ into a large collection of cubic cells of fixed volume $\Delta\mathscr{V}$, where $(\Delta\mathscr{V})^{1/3}$ is much larger than a collisional correlation length $\lambda_{\text{mfp}}$ but much smaller than a macroscopic gradient scale length $L$. Each cell will contain the (variable) number of particles $\widetilde{\mathscr{N}}_{\Delta\mathscr{V}}(\boldsymbol{r},t)$, so will be of random density $\widetilde{n}(\boldsymbol{r},t) \doteq \widetilde{\mathscr{N}}_{\Delta\mathscr{V}}(\boldsymbol{r},t)/\Delta\mathscr{V}$. Thus, on the average and at lowest order a cell has a local particle density $n(\boldsymbol{r},t)$, is moving with a local flow velocity $\boldsymbol{u}(\boldsymbol{r},t)$, and has a local temperature $T(\boldsymbol{r},t)$ (i.e., it is in local thermodynamic equilibrium).[17] At next order, velocity and temperature differences between the cell boundaries induce gradients that lead to small corrections to the locally Gibbsian statistical distribution. It is those corrections that determine the transport coefficients.

There are technical difficulties with such a procedure. First, a division into intermediate-sized cells fails if there are long-ranged correlations at the microscopic level. But even if that issue is ignored, as I shall do (see the previous quotes from Wong *et al.* and Brey *et al.* in §1), one must face the problem of connecting the cells together in a way that is compatible with the solution of the Liouville equation in the entire box. At lowest order, one popular way of doing that begins with the introduction of the so-called *local equilibrium distribution*[18] $F_B(\mathit{\Gamma};t)$. This will be defined and further discussed in §2.4. Here, it is necessary to clarify the use of the word 'local', which may be confusing. In fact, the local equilibrium ensemble applies to the entire box; it is not parameterized by any particular spatial location. Rather, 'local' refers to the fact that each cell is approximately in a local equilibrium. Corrections to $F_B$ define the dissipative

---

[17] Note that if the cells are arranged in a fixed Eulerian tiling, particles can cross cell boundaries and the lowest-order local cell distribution is a grand canonical ensemble.

[18] The significance of the $B$ subscript on $F_B$ will become clear later. In this context, $B$ denotes a conjugate variable, not the magnetic field.

fluxes and, through the transport equations, provide the means of connecting the cells together.

Unfortunately, direct use of the local equilibrium distribution as a basis for a local gradient expansion turns out to be technically somewhat unwieldy because $F_B$ contains a nonlocal spatial integral over the entire box. Brey *et al.* therefore employ a *reference distribution* $F_0(\Gamma; \boldsymbol{r}, t)$ that is intermediate between the two distributions introduced above. It is essentially a canonical ensemble, but with local parameters $\beta(\boldsymbol{r}, t)$ and $(\beta\mu)(\boldsymbol{r}, t)$. (Here $\beta$ is the inverse temperature and $\mu$ is the chemical potential per particle.) The reference distribution will be defined and further discussed in §2.5.

Following Brey *et al.*, I shall employ two sets of space-time arguments: the position and time at which the evolution of the hydrodynamic variables is desired are denoted as above by $(\boldsymbol{r}, t)$, while dependence on space and time at an arbitrary space-time point is denoted by $(\boldsymbol{x}, s)$. The distinction between dependence on species $s$ (a subscript) and arbitrary time $s$ (an argument) should be clear. Thus, for example, one can discuss the number density $n_s(\boldsymbol{x}, s)$; its hydrodynamic evolution equation will be written $\partial_t n_s(\boldsymbol{r}, t) = \cdots$.

Although the paper of Brey *et al.* is fairly self-contained, it is succinct and might not serve as a suitable starting point for someone who is not familiar with the long history and extensive technical developments of the field of nonequilibrium statistical mechanics. A very clear explanation of the basic issues is given by Piccirelli (1968), who cites earlier fundamental papers. In particular, he discusses the philosophy of the Bogoliubov–Chapman–Enskog asymptotic methods for the development of transport equations, he explains the interpretation of the local equilibrium ensemble, and he shows how an appropriate projection operation can be employed for usefully rearranging the content of the full Liouville distribution. Some of his appendixes and technical footnotes may also be useful for the following discussion.

### 2.2. *Fundamental microscopic variables*

I shall denote a random variable by a tilde. Given a random time series $\widetilde{\psi}(t)$ and a PDF $P(\psi, t)$, the average of a function $g(\widetilde{\psi}(t))$ is given by $\langle g(\widetilde{\psi}(t))\rangle \doteq \int \mathrm{d}\psi\, g(\psi) P(\psi, t)$ (Schrödinger representation). The PDF itself is given by

$$P(\psi, t) = \langle \delta(\psi - \widetilde{\psi}(t))\rangle. \tag{2.1}$$

The nonrandom variable $\psi$ is called the observer coordinate. The representation (2.1) is trivial if the average is taken with the PDF at time $t$, but is nontrivial if $\widetilde{\psi}(t)$ has evolved from random variables at some earlier time $t_0$ and the average is evaluated with $P(\psi_0, t_0)$ (Heisenberg representation). It is a common abuse of notation to drop the tildes on random variables inside expectations and thus to write $\langle g(\widetilde{\psi}(t))\rangle \equiv \langle g(\psi)\rangle_t \equiv \langle g(\psi)\rangle$, the evaluation time $t$ being understood implicitly in the last form.

Brey *et al.* considered only a single species of particle (i.e., a one-component fluid). I shall allow $S$ species ($S = 2$ being the most popular case for plasma physicists), but shall consider interspecies coupling effects only at lowest (Braginskii) order. The fundamental random variables are then chosen to be

$$\widetilde{\boldsymbol{A}}_s(\boldsymbol{r}, t) = (\widetilde{N}_s \;\; \widetilde{\boldsymbol{P}}_s \;\; \widetilde{E}_s)^{\mathrm{T}} \tag{2.2}$$

(T denotes transpose), where $\widetilde{N}_s(\boldsymbol{r}, t)$ is the microscopic number density, $\widetilde{\boldsymbol{P}}_s(\boldsymbol{r}, t)$ is the microscopic momentum density,[19] and $\widetilde{E}_s(\boldsymbol{r}, t)$ is the microscopic energy density. (All of these are defined below.) A script notation will be used for the total (volume-integrated)

[19] Brey *et al.* used $\widetilde{\boldsymbol{G}}$ instead of $\widetilde{\boldsymbol{P}}$ and $\boldsymbol{g}$ instead of $\boldsymbol{p}$). The boldfaced $\widetilde{\boldsymbol{P}}$ for microscopic

amounts of these and other quantities in the system. That is, for an arbitrary variable $\widetilde{A}$, the total amount of $\widetilde{A}$ is

$$\widetilde{\mathscr{A}}(t) \doteq \int \mathrm{d}\boldsymbol{r}\, \widetilde{A}(\boldsymbol{r}, t). \tag{2.3}$$

Such total quantities will be seen to appear in the reference distribution $F_0$ [(2.37) below], which is a spatially dependent generalization of the Gibbs distribution.

The averages of the microscopic densities will be denoted by lower-case quantities:

$$\boldsymbol{a}_s(\boldsymbol{r}, t) \doteq \langle \widetilde{\boldsymbol{A}}_s \rangle = (n_s \;\; \boldsymbol{p}_s \;\; e_s)^{\mathrm{T}}. \tag{2.4}$$

Lower case will also be used for other intensive quantities such as the pressure $p$. (Temperature $T$ is an exception, as lower case would conflict with time $t$.) The mean internal energy density will be denoted by $u$, which in principle conflicts with $u \doteq |\boldsymbol{u}|$, where $\boldsymbol{u}$ is the fluid velocity. However, there should be no confusion in context.

I shall use Greek letters to denote the components of the hydrodynamic column vectors: $\widetilde{\boldsymbol{A}} \to \widetilde{A}^\mu$ and $\boldsymbol{a} \to a^\mu$, where $\mu \in \{n,\ \boldsymbol{p},\ e\}$. (Brey *et al.* used upper case for those $\mu$'s.) The Einstein summation convention will apply for repeated indices. Species labels will be subsumed into the field indices unless they are written explicitly for emphasis.

### 2.2.1. Number density $\widetilde{N}$

The microscopic number density is

$$\widetilde{N}_s(\boldsymbol{r}, t) \doteq \sum_{i \in s} \delta(\boldsymbol{r} - \widetilde{\boldsymbol{x}}_i(t)), \tag{2.5}$$

where $\sum_{i \in s}$ denotes summation over all of the $\mathscr{N}_s$ particles of species $s$. Obviously, $\widetilde{N}_s$ depends on (some of) the random phase-coordinates $\widetilde{\Gamma}(t)$, but that dependence will be indicated only by the tilde. In terms of the Klimontovich microdensity $\widetilde{f}_s$ as usually defined (such that its velocity integral is dimensionless),

$$\widetilde{f}_s(\boldsymbol{r}, \boldsymbol{v}, t) \doteq \frac{1}{\overline{n}_s} \sum_{i \in s} \delta(\boldsymbol{r} - \widetilde{\boldsymbol{x}}_i(t)) \delta(\boldsymbol{v} - \widetilde{\boldsymbol{v}}_i(t)), \tag{2.6}$$

where $\overline{n}_s \doteq \mathscr{N}_s / \mathscr{V}$ is the mean density of species $s$, one has

$$\widetilde{N}_s(\boldsymbol{r}, t) = \overline{n}_s \int \mathrm{d}\boldsymbol{v}\, \widetilde{f}_s(\boldsymbol{r}, \boldsymbol{v}, t). \tag{2.7}$$

The average over a pure (statistically homogeneous) thermal-equilibrium ensemble is

$$\langle \widetilde{N}_s \rangle = \sum_{i \in s} \int \frac{\mathrm{d}\boldsymbol{x}_i}{\mathscr{V}} \delta(\boldsymbol{r} - \boldsymbol{x}_i) = \mathscr{N}_s / \mathscr{V} = \overline{n}_s = \mathrm{const.} \tag{2.8}$$

In an arbitrary ensemble, one has that the mean particle density is

$$n_s(\boldsymbol{r}, t) = \langle \widetilde{N}_s(\boldsymbol{r}, t) \rangle = \overline{n}_s \int \mathrm{d}\boldsymbol{v}\, f_s(\boldsymbol{r}, \boldsymbol{v}, t), \tag{2.9}$$

---

momentum density should be distinguished from a projection operator P and a PDF $P$; boldfaced $\boldsymbol{p}$ (mean momentum density) should be distinguished from the pressure $p$.

where $f \doteq \langle \widetilde{f} \rangle$ is the one-particle distribution function in a convenient normalization.[20] The total particle number is

$$\widetilde{\mathscr{N}_s}(t) \doteq \int \mathrm{d}\boldsymbol{r} \sum_{i \in s} \delta(\boldsymbol{r} - \widetilde{\boldsymbol{x}}_i(t)) = \mathscr{N}_s. \tag{2.10}$$

Note that although in principle $\widetilde{\mathscr{N}_s}$ is random, it is constant in the ensemble considered here.

### 2.2.2. Momentum density $\widetilde{\boldsymbol{P}}$

The microscopic momentum density is

$$\widetilde{\boldsymbol{P}}_s(\boldsymbol{r}, t) \doteq \sum_{i \in s} m_s \widetilde{\boldsymbol{v}}_i(t) \delta(\boldsymbol{r} - \widetilde{\boldsymbol{x}}_i(t)). \tag{2.11}$$

The total momentum is

$$\widetilde{\boldsymbol{\mathscr{G}}}_s(t) \doteq \int \mathrm{d}\boldsymbol{r}\, \widetilde{\boldsymbol{P}}_s(\boldsymbol{r}, t) = \sum_{i \in s} m_s \widetilde{\boldsymbol{v}}_i(t) = \boldsymbol{0}. \tag{2.12}$$

The last equality is an assumption.[21] The mean momentum density is

$$\boldsymbol{p}_s(\boldsymbol{r}, t) \doteq \langle \widetilde{\boldsymbol{P}}_s(\boldsymbol{r}, t) \rangle = n_s(\boldsymbol{r}, t) m_s \boldsymbol{u}_s(\boldsymbol{r}, t), \tag{2.13}$$

which defines the mean flow velocity $\boldsymbol{u}_s$.

### 2.2.3. Energy density $\widetilde{E}$

The microscopic energy density is

$$\widetilde{E}_s(\boldsymbol{r}, t) \doteq \sum_{i \in s} \widetilde{E}_i(t) \delta(\boldsymbol{r} - \widetilde{\boldsymbol{x}}_i(t)), \tag{2.14}$$

where

$$\widetilde{E}_i \doteq \frac{1}{2} m \widetilde{v}_i^2(t) + \widetilde{U}_i(t) \tag{2.15}$$

and the microscopic potential energy is

$$\widetilde{U}_i(t) \doteq \frac{1}{2} \sum_{j \neq i}^{\mathscr{N}} \widetilde{U}_{ij}(t), \tag{2.16}$$

with $\widetilde{U}_{ij}(t) \doteq U(\widetilde{\boldsymbol{x}}_i(t) - \widetilde{\boldsymbol{x}}_j(t))$ being the two-particle potential energy. (By assumption, there is no external potential.) Note that the $j$ sum in (2.16) is extended over all particles, so $\widetilde{U}_i$ contains contributions from multiple species. Equation (2.14) can be written as

$$\widetilde{E}_s(\boldsymbol{r}, t) = \frac{1}{2} m_s u_s^2(\boldsymbol{r}, t) \widetilde{N}_s(\boldsymbol{r}, t) + \widetilde{E}_{0,s}(\boldsymbol{r}, t), \tag{2.17}$$

[20] In kinetic theory, it is convenient to define $n$-particle distribution functions $f_n$ by $f_n \doteq \mathscr{V}^n P_n$. I shall write $f_1 \equiv f$. Also, to make the notation more consistent with that of Brey *et al.*, I shall write $P_{\mathscr{N}} \equiv F$.

[21] The centre-of-mass velocity of the entire collection of particles is proportional to $\sum_s \widetilde{\boldsymbol{\mathscr{G}}}_s$, and that can be set to zero with no loss of generality. I make the more restrictive assumption that no species has bulk motion. This does not preclude a nonvanishing local fluid velocity $\boldsymbol{u}_s(\boldsymbol{r}, t)$ at a particular point in the fluid.

where, with $\widetilde{\boldsymbol{w}}_i \doteq \widetilde{\boldsymbol{v}}_i - \boldsymbol{u}_s(\boldsymbol{r}, t)$,

$$\widetilde{E}_{0,s}(\boldsymbol{r}, t) \doteq \sum_{i \in s} \left( \frac{1}{2} m_s \widetilde{w}_i^2(t) + \widetilde{U}_i(t) \right) \delta(\boldsymbol{r} - \widetilde{\boldsymbol{x}}_i(t)) \tag{2.18}$$

is the total thermal energy density with respect to the local moving frame. The total amount of particle energy (mechanical plus thermal) is[22]

$$\widetilde{\mathscr{E}}_s(t) \doteq \sum_{i \in s} \left( \frac{1}{2} m_s \widetilde{v}_i^2(t) + \widetilde{U}_i(t) \right). \tag{2.19}$$

The mean energy density, including mechanical motion, is

$$e_s(\boldsymbol{r}, t) = \langle \widetilde{E}_s(\boldsymbol{r}, t) \rangle = \left( \frac{1}{2} m_s u_s^2(\boldsymbol{r}, t) \right) n_s(\boldsymbol{r}, t) + \langle \widetilde{E}_{0,s}(\boldsymbol{r}, t) \rangle. \tag{2.20}$$

Upon performing the velocity average according to $\frac{1}{2} m \langle w^2 \rangle = \frac{3}{2} T$, one finds[23]

$$\langle \widetilde{E}_{0,s}(\boldsymbol{r}, t) \rangle = \frac{3}{2} n_s(\boldsymbol{r}, t) T_s(\boldsymbol{r}, t) + u_s(\boldsymbol{r}, t), \tag{2.21}$$

where $u$ is the internal energy density.

In spite of my convention that a tilde denotes a random variable, I shall frequently drop it for the particle variables $\boldsymbol{x}_i$ and $\boldsymbol{v}_i$, since the presence of a particle index also indicates a random nature. I shall also drop the time argument for those variables when there is no possibility of confusion between the Schrödinger and Heisenberg representations of phase-space averages. Thus, I shall write formulas such as $\widetilde{N}_s(\boldsymbol{r}) \doteq \sum_{i \in s} \delta(\boldsymbol{r} - \boldsymbol{x}_i)$.

## 2.3. *The microscopic fluxes*

For a one-component, unmagnetised system with short-ranged forces, it is well known that the time derivatives of the microscopic densities can be written as the divergences of microscopic fluxes or currents according to

$$\partial_t \widetilde{\boldsymbol{A}}(\boldsymbol{r}, t) = -\boldsymbol{\nabla} \cdot \widetilde{\boldsymbol{J}}(\boldsymbol{r}, t), \tag{2.22}$$

where $\widetilde{\boldsymbol{J}}$ is a column vector of random currents. Specifically,

$$\partial_t \widetilde{N} = -\boldsymbol{\nabla} \cdot (m^{-1} \widetilde{\boldsymbol{P}}), \tag{2.23$a$}$$

$$\partial_t \widetilde{\boldsymbol{P}} = -\boldsymbol{\nabla} \cdot \widetilde{\boldsymbol{\tau}}, \tag{2.23$b$}$$

$$\partial_t \widetilde{E} = -\boldsymbol{\nabla} \cdot \widetilde{\boldsymbol{J}}^E, \tag{2.23$c$}$$

[22] Particle energy is not conserved in a finite-volume canonical ensemble. However, it is well known that the relative energy fluctuations scale as $\mathscr{N}^{-1/2}$, so they are negligible in the thermodynamic limit.

[23] Strictly speaking, if the many-particle system is in contact with a heat bath at temperature $T$, then the equilibrium temperature of all species should be $T$. However, because interspecies temperature relaxation is slow, I follow the standard assumption of allowing for different temperatures $T_s$ (Braginskii 1965). In any event, perturbations from a common temperature can clearly be species-dependent.

where the Fourier transforms of the microscopic stress tensor[24] $\widetilde{\boldsymbol{\tau}}$ and energy current $\widetilde{\boldsymbol{J}}^E$ are

$$\widetilde{\boldsymbol{\tau}}(\boldsymbol{k}) \doteq \sum_{i=1}^{\mathscr{N}} [m\boldsymbol{v}_i\boldsymbol{v}_i + \Delta\widetilde{\boldsymbol{\tau}}_i(\boldsymbol{k})]\mathrm{e}^{-\mathrm{i}\boldsymbol{k}\cdot\boldsymbol{x}_i}, \tag{2.24$a$}$$

$$\widetilde{\boldsymbol{J}}^E(\boldsymbol{k}) \doteq \sum_{i=1}^{\mathscr{N}} [E_i\boldsymbol{v}_i + \Delta\widetilde{\boldsymbol{\tau}}_i(\boldsymbol{k})\cdot\boldsymbol{v}_i]\mathrm{e}^{-\mathrm{i}\boldsymbol{k}\cdot\boldsymbol{x}_i}, \tag{2.24$b$}$$

with (appendix B)

$$\Delta\widetilde{\boldsymbol{\tau}}_i(\boldsymbol{k}) \doteq -\frac{1}{2}\sum_{ij}^{\mathscr{N}} \frac{\partial \widetilde{U}_{ij}}{\partial \boldsymbol{x}_{ij}}\,\boldsymbol{x}_{ij}\left(\frac{1-\mathrm{e}^{\mathrm{i}\boldsymbol{k}\cdot\boldsymbol{x}_{ij}}}{-\mathrm{i}\boldsymbol{k}\cdot\boldsymbol{x}_{ij}}\right). \tag{2.25}$$

For multiple species, the density continuity equation (2.23$a$) holds separately for each species. However, for a plasma the momentum and energy equations must be revisited in order to account for the long-ranged nature of the Coulomb force, the Lorentz force, and random interspecies coupling or exchange terms. As shown in appendix B, one finds

$$\partial_t \widetilde{N}_s(\boldsymbol{r},t) = -\boldsymbol{\nabla}\cdot(m_s^{-1}\widetilde{\boldsymbol{P}}_s), \tag{2.26$a$}$$

$$\partial_t \widetilde{\boldsymbol{P}}_s(\boldsymbol{r},t) = (nq)_s\boldsymbol{E} + \omega_{cs}\widetilde{\boldsymbol{P}}_s\times\widehat{\boldsymbol{b}} - \boldsymbol{\nabla}\cdot\widetilde{\boldsymbol{\tau}}_s' + \dot{\widetilde{\boldsymbol{P}}}{}'_{\Delta,s}, \tag{2.26$b$}$$

$$\partial_t \widetilde{E}_s(\boldsymbol{r},t) = -\boldsymbol{\nabla}\cdot\widetilde{\boldsymbol{J}}_s^E + \dot{\widetilde{E}}_{\Delta,s}, \tag{2.26$c$}$$

where $\boldsymbol{E}(\boldsymbol{r},t)$ is the macroscopic electric field determined via Poisson's equation (an external electric field is assumed to vanish), $\widehat{\boldsymbol{b}}$ is the unit vector in the direction of the magnetic field, $\omega_{cs} \doteq (qB/mc)_s$, a prime denotes removal of any contribution from the long-ranged mean potential, $\widetilde{\boldsymbol{\tau}}_s'$ is discussed in appendix B, and $\dot{\widetilde{\boldsymbol{P}}}{}'_{\Delta,s}$ and $\dot{\widetilde{E}}_{\Delta,s}$ represent the random momentum and energy exchanges due to interactions with other species. Ultimately, those terms with a $\Delta$ subscript lead to collisional interspecies relaxation, which I shall often call *exchange effects*. For weakly coupled plasmas, the consequences of those effects are already well known to first order in the interspecies velocity and temperature differences $\Delta\boldsymbol{u}$ and $\Delta T$, as discussed in Part I; with standard orderings, they are small. At second order, a plethora of additional terms will arise involving $\Delta\boldsymbol{u}\,\Delta\boldsymbol{u}$, $\Delta\boldsymbol{u}\,\Delta T$, and $\Delta T^2$ as well as cross terms involving products of the differences and gradients, such as $\Delta\boldsymbol{u}\cdot\boldsymbol{\nabla}T$. I shall neglect all such effects (as did Catto & Simakov). In the general, strongly coupled case, analysis of those terms ultimately requires the derivation of a collision operator that goes beyond the Landau and Balescu–Lenard forms, but that is beyond the scope of this paper.

For general manipulations, it is useful to write (2.26) succinctly as

$$\partial_t\widetilde{\boldsymbol{A}}_s = \widetilde{\boldsymbol{F}}_{\mathrm{EM,s}} - \boldsymbol{\nabla}\cdot\widetilde{\boldsymbol{J}}_s + \dot{\widetilde{\boldsymbol{A}}}{}'_{\Delta,s}. \tag{2.27}$$

[24] In a consistent notation, the microscopic stress tensor would ideally be written in upper case. However, the symbol for Greek Tau is indistinquishable from $T$, which is reserved for temperature.

This involves the random electromagnetic force[25] $\widetilde{\boldsymbol{F}}_{\mathrm{EM,s}}$, which has only a momentum component; the generalized random currents $\widetilde{\boldsymbol{J}}_s$; and the random exchange terms $\dot{\widetilde{\boldsymbol{A}}}'_{\Delta,s}$.

## 2.4. *The local equilibrium distribution*

The local equilibrium distribution $F_B(\mathit{\Gamma};t)$ generalized to include multiple species is defined by

$$F_B(\mathit{\Gamma};t) \doteq [Z_B(t)]^{-1} \exp\left(\sum_{\overline{s}} \int \mathrm{d}\overline{\boldsymbol{r}}\, \boldsymbol{A}_{\overline{s}}(\mathit{\Gamma};\overline{\boldsymbol{r}}) \cdot \boldsymbol{B}_{\overline{s}}(\overline{\boldsymbol{r}},t)\right), \tag{2.28}$$

where

$$Z_B(t;[\boldsymbol{B}]) \doteq \int \mathrm{d}\mathit{\Gamma}\, \exp\left(\sum_{\overline{s}} \int \mathrm{d}\overline{\boldsymbol{r}}\, \boldsymbol{A}_{\overline{s}}(\mathit{\Gamma};\overline{\boldsymbol{r}}) \cdot \boldsymbol{B}_{\overline{s}}(\overline{\boldsymbol{r}},t)\right). \tag{2.29}$$

The choice of the *conjugate variables*[26] $\boldsymbol{B}_s$ will be discussed below and in appendix C. $F_B(\mathit{\Gamma};t)$ is the generalization of an equilibrium Gibbs distribution to the general case in which the thermodynamic variables vary with time and assume different values in different local regions of space. (Think of the $\overline{\boldsymbol{r}}$ integration as a Riemann sum.) A useful shorthand notation is to define a $\star$ operation[27] via

$$\sum_{\overline{s}} \int \mathrm{d}\overline{\boldsymbol{r}}\, \boldsymbol{A}_{\overline{s}}(\overline{\boldsymbol{r}},t) \cdot \boldsymbol{B}_{\overline{s}}(\overline{\boldsymbol{r}},t) \equiv \boldsymbol{A} \star \boldsymbol{B}; \tag{2.30}$$

then one can write

$$F_B(\mathit{\Gamma};t) = \frac{\mathrm{e}^{\boldsymbol{A}\star\boldsymbol{B}}}{\int \mathrm{d}\mathit{\Gamma}\, \mathrm{e}^{\boldsymbol{A}\star\boldsymbol{B}}}. \tag{2.31}$$

The local-equilibrium partition function $Z_B[t;\boldsymbol{B}]$, a time-independent functional of the $\boldsymbol{B}_s(\boldsymbol{r},t)$'s at a particular time $t$, serves as a generating function for the equal-time moments of the $\widetilde{\boldsymbol{A}}_s$'s *in the local equilibrium ensemble*, and its logarithm serves as a cumulant generating function. For example, upon suppressing the time argument,

$$\boldsymbol{a}_{s,B}(\boldsymbol{r}) \doteq \langle \widetilde{\boldsymbol{A}}_s(\boldsymbol{r}) \rangle_B = \frac{\delta \ln Z_B}{\delta \boldsymbol{B}_s(\boldsymbol{r})}, \tag{2.32$a$}$$

$$\langle \widetilde{\boldsymbol{A}}'_{s,B}(\boldsymbol{r}) \widetilde{\boldsymbol{A}}'_{s',B}(\boldsymbol{r}') \rangle_B = \frac{\delta^2 \ln Z_B}{\delta \boldsymbol{B}_s(\boldsymbol{r}) \delta \boldsymbol{B}_{s'}(\boldsymbol{r}')} = \frac{\delta \langle \widetilde{\boldsymbol{A}}_s(\boldsymbol{r}) \rangle_B}{\delta \boldsymbol{B}_{s'}(\boldsymbol{r}')}. \tag{2.32$b$}$$

where $\widetilde{\boldsymbol{A}}'_{s,\boldsymbol{B}} \doteq \widetilde{\boldsymbol{A}}_s - \langle \widetilde{\boldsymbol{A}}_s \rangle_B$ is the fluctuation with respect to the local mean. Note that the multipoint correlation functions so defined are not quite the true ones for the nonequilibrium $\mathscr{N}$-particle system because the expectations are calculated in the local equilibrium ensemble, not the exact one described by $F(\mathit{\Gamma},t)$. The latter contains the dissipative transport fluxes, while the former does not.

---

[25]In general, each of the terms in the product $n_s\boldsymbol{E}$ is random when the system is macroscopically turbulent. However, in this paper fluctuations of those terms are ignored in order to focus on classical transport.

[26]I also use $\boldsymbol{B}$ for the magnetic field, but there should be no confusion in context. For the vector field, I write $\boldsymbol{B}^{\mathrm{ext}}$, but I drop the superscript for $B = |\boldsymbol{B}^{\mathrm{ext}}|$ and $\widehat{\boldsymbol{b}} \doteq \boldsymbol{B}^{\mathrm{ext}}/B$.

Why are the nonrandom conjugate variables $\boldsymbol{B}$ written in upper case? I yield to the notation used by Brey *et al.*

[27]Brey *et al.* (1981) used the symbol $*$ for this operation, but I prefer to use $*$ to denote convolution. $A(\boldsymbol{r}) * B(\boldsymbol{r}) \doteq \int \mathrm{d}\overline{\boldsymbol{r}}\, A(\boldsymbol{r} - \overline{\boldsymbol{r}}) B(\overline{\boldsymbol{r}})$ depends on $\boldsymbol{r}$, but $A(\boldsymbol{r}) \star B(\boldsymbol{r})$ as defined by (2.30) is independent of $\boldsymbol{r}$. The $\star$ operation is a generalization of Einstein's convention for summation and/or integration over repeated indices.

***It is technically convenient to choose the $\boldsymbol{B}$'s such that the $\boldsymbol{a}$'s are the true fluid variables;*** thus, one enforces the constraint

$$\boldsymbol{a}_s(\boldsymbol{r},t) = \boldsymbol{a}_{s,B}(\boldsymbol{r},t). \tag{2.33}$$

This choice is discussed at length by Piccirelli (1968). The basic idea is that the local equilibrium ensemble is supposed to provide a good zeroth-order starting point for a subsequent expansion in the gradients of the $\boldsymbol{B}$'s (or, more generally, in differences $\Delta\boldsymbol{B}$, which include exchange effects). In fact, this choice is not unique, and Piccirelli points out that a variety of forms for $F_B$ will lead to similar expansions in terms of other quantities $\Delta\boldsymbol{B}'$. However, if the present choice is not made, various difficulties may ensue: $\Delta\boldsymbol{B}'$ may not have a correspondence to natural thermodynamic forces that can be measured physically, or there may be difficulty in the preparation of an appropriate initial state.[28]

For the one-component fluid, Piccirelli showed that the conjugate variables $\boldsymbol{B}$ are

$$\boldsymbol{B}(\boldsymbol{r},t) = \left(\beta(\mu - \tfrac{1}{2}mu^2) \quad \beta\boldsymbol{u} \quad -\beta\right)^{\mathrm{T}}, \tag{2.34}$$

where $\beta \doteq T^{-1}$ and all of the quantities on the right-hand side are functions of $\boldsymbol{r}$ and $t$. In appendix C, I give some details of the argument extended to the multispecies case and conclude that this choice remains valid with the mere addition of species labels to all quantities in (2.34).

The method to be described shortly will express the time evolution of the $\boldsymbol{a}$'s in terms of the $\boldsymbol{B}$'s; it does not depend on the definition of the $\boldsymbol{B}$'s. However, in order to obtain a closed set of transport equations, the $\boldsymbol{B}$'s must be expressed in terms of the $\boldsymbol{a}$'s. For example, in a weakly coupled system the energy density $e$ reduces to $\frac{3}{2}nT$. With the definition (2.34), the system $\{\partial_t n = f_n(\boldsymbol{B}),\ \partial_t\boldsymbol{u} = f_{\boldsymbol{u}}(\boldsymbol{B}),\ \partial_t T = f_T(\boldsymbol{B})\}$ (where the $f$'s are certain functions of $\boldsymbol{B}$ and/or $\boldsymbol{\nabla}\boldsymbol{B}$) is then partially closed in terms of the variables $\boldsymbol{u}$ and $T = \beta^{-1}$; the only remaining closure problem is to determine the chemical potential $\mu(n,T)$. That can be done by using thermodynamic relations valid in local thermal equilibrium. See §II-S:1.1 for further details.

[28] An extended quotation from Piccirelli (1968) may be useful. In his discussion of projection techniques, he says (using the notation $D$ instead of $P_{\mathscr{N}} \equiv F$, and $D_0$ instead of $F_B$),

> Since ... almost closed expressions for the irreversible currents are central to the present work, the following remarks should be made. The expressions are almost closed only because we have assumed that $D(\Gamma,t)$ is split into a differentiable time-independent functional $D_0$ of the densities and a remainder and not because of the particular form of $D_0$. [Several sentences of technical justification are omitted here.]
>
> From this point of view, our choice of $D_0$ would appear to be arbitrary; any of a whole class of choices will yield almost closed expressions for the currents. One should notice concerning this point: First, that a certain arbitrariness over various generalized equilibrium distributions is desirable to reflect differences in experimental circumstances. It will be argued in the concluding section that the final results are invariant to such choices. Second, we have just seen that the use of generalized equilibrium distributions for $D_0$ introduces the conjugate variables and their gradients in a very natural way. As will emerge presently, this yields a structural form of the equations which has a very strong and complete analogy to the classical ones which must be their limiting form. Finally, there is no reason to expect that arbitrary choices of $D_0$ will lead by any route to the classical limit, and in particular, one cannot expect the initial value terms to have the proper behavior for arbitrary choices of $D_0$ which may not correspond to physically realizable initial preparations ... .

## 2.5. *The reference distribution*

Consider a particular reference space-time point $(\boldsymbol{r}, t)$ and reference species $s$, then define[29]

$$\Delta\boldsymbol{B}_{\overline{s}s}(\overline{\boldsymbol{r}}, \boldsymbol{r}, t) \doteq \boldsymbol{B}_{\overline{s}}(\overline{\boldsymbol{r}}, t) - \boldsymbol{B}_s(\boldsymbol{r}, t). \tag{2.35}$$

Equation (2.28) can then be written as

$$F_B(\mathit{\Gamma}; t) = Z_B^{-1} \exp\left(\sum_{\overline{s}} \boldsymbol{\mathscr{A}}_{\overline{s}} \cdot \boldsymbol{B}_{\overline{s}}(\boldsymbol{r}, t) + \sum_{\overline{s}} \int \mathrm{d}\overline{\boldsymbol{r}}\, \boldsymbol{A}_{\overline{s}}(\overline{\boldsymbol{r}}) \cdot \Delta\boldsymbol{B}_{\overline{s}\,\overline{s}}(\overline{\boldsymbol{r}}, \boldsymbol{r}, t)\right). \tag{2.36}$$

(This quantity is independent of $\boldsymbol{r}$, although the individual terms depend on $\boldsymbol{r}$.) The reference distribution $F_0$ is defined by omitting the terms in $\Delta\boldsymbol{B}$. (The effects of the $\Delta\boldsymbol{B}$'s will be seen to be small when appropriate averages are taken, so they can be treated as perturbative corrections to $F_0$.) Thus,

$$F_0(\mathit{\Gamma}; \boldsymbol{r}, t) \doteq [Z_0(\boldsymbol{r}, t)]^{-1} \exp\left(\sum_{\overline{s}} \boldsymbol{\mathscr{A}}_{\overline{s}}(\mathit{\Gamma}) \cdot \boldsymbol{B}_{\overline{s}}(\boldsymbol{r}, t)\right), \tag{2.37}$$

where the *local partition function* is

$$Z_0(\boldsymbol{r}, t) \doteq \int \mathrm{d}\mathit{\Gamma} \exp\left(\sum_{\overline{s}} \boldsymbol{\mathscr{A}}_{\overline{s}}(\mathit{\Gamma}) \cdot \boldsymbol{B}_{\overline{s}}(\boldsymbol{r}, t)\right). \tag{2.38}$$

Here $\boldsymbol{\mathscr{A}}_s(\mathit{\Gamma}) \doteq \boldsymbol{\mathscr{A}}_s(\widetilde{\mathit{\Gamma}}(t))|_{\widetilde{\mathit{\Gamma}}(t)=\mathit{\Gamma}}$. The order of the arguments in $F_0(\mathit{\Gamma}; \boldsymbol{r}, t)$ emphasizes that $F_0$ is fundamentally a function of $\mathit{\Gamma}$ but is parameterized by $\boldsymbol{r}$ and $t$. By rearranging the dot product and using the definitions (2.34) of the $\boldsymbol{B}$'s, one can see that

$$F_0(\mathit{\Gamma}) = Z_0^{-1} \exp\left(\sum_{\overline{s}} N_{\overline{s}}\beta_{\overline{s}}\mu_{\overline{s}} - \beta_{\overline{s}}\mathscr{E}_{0\overline{s}}(\mathit{\Gamma})\right), \tag{2.39}$$

where

$$\mathscr{E}_{0s}(\mathit{\Gamma}) \doteq \sum_{i\in s}\left(\frac{1}{2}m_s w_i^2 + U_i\right) \tag{2.40}$$

is the total energy with respect to the local reference frame moving with the velocity $\boldsymbol{u}_s$. Note that although the parameters $\beta$, $\mu$, and $\boldsymbol{u}$ depend on $\boldsymbol{r}$ and $t$, the spatial phase-space dependence of $F_0$ enters only through the internal energy, so $F_0$ is invariant under translations of the particle positions.

Differentiations with respect to the parameters $(\beta\mu)_s$ and $-\beta_s$ generate the system-mean number and thermal-energy densities:

$$\left(\frac{\partial \ln Z_0}{\partial[(\beta\mu)_s(\boldsymbol{r}, t)]}\right)_\beta = \langle\widetilde{\mathscr{N}_s}\rangle = \mathscr{N}_s = \mathscr{V}\overline{n}_s, \tag{2.41$a$}$$

$$\left(\frac{\partial \ln Z_0}{\partial[-\beta_s(\boldsymbol{r}, t)]}\right)_{\beta\mu} = \langle\widetilde{\mathscr{E}}_0\rangle = \mathscr{V}\left(e_{0s}(\boldsymbol{r}, t) - \frac{1}{2}m_s\overline{n}_s u_s^2(\boldsymbol{r}, t)\right). \tag{2.41$b$}$$

Note that these are ordinary partial derivatives, not functional derivatives. I shall write, for example, $\partial/\partial\beta|_{\beta\mu} \equiv \partial/\partial\beta$ when it is clear which variable is being held fixed. One also holds $\boldsymbol{u}$ fixed when performing those derivatives. Note that $\overline{n}$ does not depend on $\boldsymbol{r}$ and $t$, while the reference energy density $e_0$ does (e.g., through its dependence on the local temperature).

---

[29] I deviate from the notation of Brey *et al.* and call this quantity $\Delta\boldsymbol{B}$ instead of $\boldsymbol{B}_0$ in order to emphasize with the $\Delta$ that $\Delta\boldsymbol{B}$ will be treated as small.

With

$$\Delta\boldsymbol{B}_{\overline{s}}(\overline{\boldsymbol{r}},\boldsymbol{r},t) \doteq \Delta\boldsymbol{B}_{\overline{s}\,\overline{s}}(\overline{\boldsymbol{r}},\boldsymbol{r},t) \tag{2.42}$$

(subsequently, $\Delta\boldsymbol{B}$'s without subscripts will denote this species-diagonal part), it can be seen from (2.36) that expectations of an arbitrary quantity $\widetilde{G}$ with the local equilibrium distribution $F_B$ can be reexpressed in terms of the reference distribution $F_0$ as

$$\langle\widetilde{G}\rangle_B(t) = \frac{\langle\widetilde{G}\,\mathrm{e}^{\widetilde{\boldsymbol{A}}\star\Delta\boldsymbol{B}}\rangle_0}{\langle\mathrm{e}^{\widetilde{\boldsymbol{A}}\star\Delta\boldsymbol{B}}\rangle_0}. \tag{2.43}$$

This important result will be used later. Note that the left-hand side of (2.43) does not depend on $\boldsymbol{r}$ even though both $F_0$ and $\Delta\boldsymbol{B}$ do. Clearly, the local equilibrium average $\langle\ldots\rangle_B$ and the reference average $\langle\ldots\rangle_0$ are equivalent to zeroth order in $\Delta\boldsymbol{B}$.

## 2.6. *Projection operators and subtracted fluxes*

The ultimate goal is the derivation of closed evolution equations for the hydrodynamic variables: $\partial_t\boldsymbol{a}_s = \cdots$. The strategy is to evaluate the full expectation $\boldsymbol{a}_s = \langle\widetilde{\boldsymbol{A}}_s\rangle$, which is taken with respect to the Liouville distribution $F$, by referring it to the reference distribution $F_0$:

$$\boldsymbol{a}_s(\boldsymbol{r},t) = \int \mathrm{d}\mathit{\Gamma}\,\boldsymbol{A}_s(\mathit{\Gamma};\boldsymbol{r})\left(\frac{F(\mathit{\Gamma},t)}{F_0(\mathit{\Gamma};\boldsymbol{r},t)}\right)F_0(\mathit{\Gamma};\boldsymbol{r},t). \tag{2.44}$$

The ratio $F/F_0$ will be expanded to second order with the aid of projection operators. Obviously, the utility of the manipulations depends on an apt choice of those operators.

### 2.6.1. The fundamental projection operator P

Given the definition (2.30) of the $\star$ operator, the form of the projection operator P used by Brey *et al.* holds as well for multiple species and is defined for arbitrary phase function $\widetilde{\chi}$ in terms of averages over the reference distribution by

$$\mathrm{P}\widetilde{\chi} \doteq \langle\widetilde{\chi}\rangle_0 + \widetilde{\boldsymbol{A}}'^{\mathrm{T}}\star\boldsymbol{\mathscr{M}}^{-1}\star\langle\widetilde{\boldsymbol{A}}'\,\widetilde{\chi}'\rangle_0, \tag{2.45}$$

where

$$\boldsymbol{A}'(\boldsymbol{r},t;\mathit{\Gamma}) \doteq \widetilde{\boldsymbol{A}}(\boldsymbol{r})|_{\widetilde{\mathit{\Gamma}}(t)=\mathit{\Gamma}} - \langle\widetilde{\boldsymbol{A}}\rangle_0 \equiv \widetilde{\boldsymbol{A}}'(\boldsymbol{r},t), \tag{2.46$a$}$$

$$\mathscr{M}^{\mu\mu'}_{ss'}(\boldsymbol{x},\boldsymbol{x}') \doteq \langle\widetilde{A}'^{\mu}_{s}(\boldsymbol{x},t)\widetilde{A}'^{\mu'}_{s'}(\boldsymbol{x}',t)\rangle_0 \equiv \mathscr{M}^{\mu\mu'}(\boldsymbol{x},\boldsymbol{x}'). \tag{2.46$b$}$$

In the last form, the species indices have been subsumed into the field indices. Note that $\boldsymbol{\mathscr{M}}$ is an equal-time cumulant and that averaging with respect to $F_0(\mathit{\Gamma};\boldsymbol{r},t)$ introduces a dependence on $\boldsymbol{r}$ and $t$, which is indicated implicitly through the subscript 0 or the prime that indicates a fluctuation from the reference state. Strictly speaking, each of P, Q $\doteq$ 1 − P, $\boldsymbol{A}'$, $\boldsymbol{\mathscr{M}}$, and various other quantities to be introduced below should be adorned with a subscript 0, as Brey *et al.* do. I shall eschew those subscripts with the goal of somewhat uncluttering the notation. In more detail, (2.45) means

$$\mathrm{P}\widetilde{\chi} = \langle\widetilde{\chi}\rangle_0 + \sum_{\overline{\mu},\overline{\mu}'}\int \mathrm{d}\overline{\boldsymbol{x}}\,\mathrm{d}\overline{\boldsymbol{x}}'\,A'^{\overline{\mu}}(\overline{\boldsymbol{x}},t;\mathit{\Gamma})\mathscr{M}^{-1}_{\overline{\mu}\,\overline{\mu}'}(\overline{\boldsymbol{x}},\overline{\boldsymbol{x}}')\langle\widetilde{A}'^{\overline{\mu}'}(\overline{\boldsymbol{x}}')\widetilde{\chi}\rangle_0(\boldsymbol{r},t), \tag{2.47}$$

with $\mu$ denoting both the index for the hydrodynamic column vector as well as species dependence. Note that P = P$(\mathit{\Gamma};\boldsymbol{r},t)$, where the $(\boldsymbol{r},t)$ arises from the various averages over the reference distribution.

The significance of P is that it projects onto the hydrodynamic subspace based on the reference state. Namely, upon suppressing $t$ arguments for brevity,

$$\langle \widetilde{A}^{\sigma}(\boldsymbol{r})\mathrm{P}\widetilde{\chi}\rangle_0 = \langle \widetilde{A}^{\sigma}(\boldsymbol{r})\rangle_0\langle\widetilde{\chi}\rangle_0 + \sum_{\overline{\mu},\overline{\mu}'}\int \mathrm{d}\overline{\boldsymbol{x}}\,\mathrm{d}\overline{\boldsymbol{x}}'\,\langle \widetilde{A}^{\sigma}(\boldsymbol{r})\widetilde{A}'^{\overline{\mu}}(\overline{\boldsymbol{x}})\rangle \mathscr{M}^{-1}_{\overline{\mu}\,\overline{\mu}'}(\overline{\boldsymbol{x}},\overline{\boldsymbol{x}}')\langle \widetilde{A}'^{\overline{\mu}'}(\overline{\boldsymbol{x}}')\widetilde{\chi}\rangle_0 \tag{2.48$a$}$$

$$= \langle \widetilde{A}^{\sigma}(\boldsymbol{r})\rangle_0\langle\widetilde{\chi}\rangle_0 + \langle \widetilde{A}'^{\sigma}(\boldsymbol{r})\widetilde{\chi}\rangle_0 \tag{2.48$b$}$$

$$= \langle \widetilde{A}^{\sigma}(\boldsymbol{r})\widetilde{\chi}\rangle_0. \tag{2.48$c$}$$

If $\chi \doteq F/F_0$, the average (2.48$c$) produces the hydrodynamic variables $a^{\sigma}$ [see (2.44); note that $\sigma$ incorporates both a species index and a field index].

The $\mathit{\Gamma}$-space projection operation $\mathrm{P} \equiv {}_{\Gamma}\mathrm{P}$ defined here generalizes the $\mu$-space one

$${}_{\mu}\mathrm{P} \doteq |\, A^{\mu}\rangle M_{\mu\,\mu'}\langle A^{\mu'}\,| \tag{2.49}$$

that was used in Part I to project the weakly coupled kinetic equation onto the hydrodynamic subspace [see (I:3.20)]. That operation is local in space (potential-energy terms are neglected in the definitions of the $\boldsymbol{A}$'s). It can be confusing to work with the $\mu$-space Dirac notation because the scalar product implied by the bra involves a species summation, but (as explained in Part I) that summation must be inhibited when a specific component $A^{\mu}$ appears under the expectation. No such difficulty arises with ${}_{\Gamma}\mathrm{P}$. The difference arises because ${}_{\Gamma}\mathrm{P}$ is built from expectations taken with the $\mathscr{N}$-particle distribution $F(\mathit{\Gamma})$, which has no preferred species dependence. On the other hand, ${}_{\mu}\mathrm{P}$ is built from expectations taken with the one-particle distribution $f_s(\boldsymbol{v})$, which intrinsically involves a particular species. Given the choice between the $\mathit{\Gamma}$-space and the $\mu$-space projection routes, the $\mathit{\Gamma}$-space one used in the present paper is the more general, the cleaner, and the easier to understand.

Nevertheless, both routes must lead to the same $\boldsymbol{a}_s$ for the same physics situation. It is therefore instructive to contemplate the subtle differences in the definitions (2.45) of ${}_{\Gamma}\mathrm{P}$ and (2.49) for ${}_{\mu}\mathrm{P}$, and of the associated $\boldsymbol{A}$'s. In the former, the mean is broken out separately from the remainder of the projection, defined with ${}_{\Gamma}\boldsymbol{A}'_s \doteq (\widetilde{N}'_s,\ \widetilde{\boldsymbol{P}}'_s,\ \widetilde{E}'_s)^{\mathrm{T}}$; in the latter, the mean is not broken out and the mixed vector ${}_{\mu}\boldsymbol{A}_s = (1,\ \boldsymbol{P}'_s(\boldsymbol{v}),\ K'_s(\boldsymbol{v}))^{\mathrm{T}}$ [see (I:2.25)] is used. It is left as an exercise to convince oneself that both of these projections produce equivalent results for linear response in a weakly coupled plasma.

### 2.6.2. The basic subtracted fluxes

The quantities that will be shown to appear in the expressions for the transport coefficients are the *subtracted fluxes* $\widehat{\boldsymbol{J}} \doteq \mathrm{Q}\widetilde{\boldsymbol{J}}$, a concept already familiar from the discussions in Part I. (Note that a hatted quantity is also random, and of course $\mathrm{PQ} = \mathrm{QP} = 0$.) The physical meaning of a subtracted flux is that it is the residual, gradient-driven portion of the total flux, over and above the microscopic part that already exists in local thermal equilibrium.

The projections of the basic fluxes are worked out in appendix D; one finds

$$\widehat{\boldsymbol{J}}^{N}_s = 0, \tag{2.50$a$}$$

$$\widehat{\boldsymbol{J}}^{\boldsymbol{P}}_s \equiv \widehat{\boldsymbol{\tau}}_s = \widetilde{\boldsymbol{\tau}}'_s - \boldsymbol{I}\left[p_s + \sum_{\overline{s}} N'_{\overline{s}}\left(\frac{\partial p_s}{\partial n_{\overline{s}}}\right)_e + \sum_{\overline{s}} E'_{\overline{s}}\left(\frac{\partial p_s}{\partial e_{\overline{s}}}\right)_n\right], \tag{2.50$b$}$$

$$\widehat{\boldsymbol{J}}^{E}_s = \widetilde{\boldsymbol{J}}^{E}_s - \left(\frac{h}{mn}\right)_s \widetilde{\boldsymbol{P}}_s. \tag{2.50$c$}$$

Here $h \doteq e + p$ is the enthalpy density. (For an ideal gas, $h = \frac{5}{2}nT$.) As is clear from appendix D, ***the subtracted fluxes are to be evaluated in the local frame with*** $\boldsymbol{u}_s = \mathbf{0}$.

### 2.7. *Formal solution of the Liouville equation using time-independent projection operators*

In this section I shall describe the process of constructing a formal solution to the Liouville equation by using projection operators. Because the P as defined above depends on time due to its dependence on the reference distribution, straightforward application of the projection procedures described in Part I leads to technical complications involving time-ordered propagators.[30] To circumvent that, Brey *et al.* use a clever trick. They describe their strategy as follows:

> The technical advantage of our method [over the ones used by Wong *et al.* (1978) and earlier workers] will consist of using a reference state that is determined by the local properties of the system at the position and time of interest. We refer the evolution of the system for all times $s$ prior to the chosen $t$ to the reference distribution function $F_0(\mathit{\Gamma};\boldsymbol{r},t)$; i.e., the reference state is the same for the whole past evolution of the system. The motivation for this choice is clear, if we remember that we expect $F_0(\mathit{\Gamma};\boldsymbol{r},t)$ to carry the main information about the system at $(\boldsymbol{r},t)$.

Following Brey *et al.*, I write the solution of the Liouville equation (1.13) as

$$F(\mathit{\Gamma},s) = \mathrm{e}^{W(\mathit{\Gamma},s;\boldsymbol{r},t)}F_0(\mathit{\Gamma};\boldsymbol{r},t) \quad (s<t). \tag{2.51}$$

This exact representation of $F$ is not unique, but it will turn out to be convenient. It is the $\mathit{\Gamma}$-space generalization of the decomposition of the one-particle distribution function $f$ used in Part I, $f_s = f_{\mathrm{M},s} + \chi_s f_{\mathrm{M},s} + \cdots$. Note that the $\boldsymbol{r}$ and $t$ dependence of $W$ must turn out to be such that $F(\mathit{\Gamma},s)$ does not depend on $\boldsymbol{r}$ and $t$. This is possible because the functions are related according to

$$\ln F(\mathit{\Gamma},s) = W(\mathit{\Gamma},s;\boldsymbol{r},t) + \ln F_0(\mathit{\Gamma};\boldsymbol{r},t)]. \tag{2.52}$$

To find an evolution equation for $W$, apply the linear operator $\partial_s + \mathrm{i}\mathscr{L}$ to (2.52) (note that the time derivative is with respect to $s$, not $t$) and use (1.13) and (2.37):

$$0 = (\partial_s + \mathrm{i}\mathscr{L})W + \sum_{\overline{s}}(\mathrm{i}\mathscr{L}\boldsymbol{\mathscr{A}}_{\overline{s}})\boldsymbol{\cdot}\boldsymbol{B}_{\overline{s}}. \tag{2.53}$$

---

[30]To eliminate the orthogonal projections, one needs to solve equations of the form

$$\partial_t\psi(t) + \mathrm{i}\widehat{\mathrm{L}}(t)\psi = s(t).$$

When $\widehat{\mathrm{L}}(t)$ is a scalar function, Green's function is easily seen to be

$$G(t;t') = H(t-t')\exp\left(-\int_{t'}^{t}\mathrm{d}\overline{t}\,\mathrm{i}\widehat{\mathrm{L}}(\overline{t})\right).$$

When $\widehat{\mathrm{L}}(t)$ is instead an operator such that $\widehat{\mathrm{L}}(t_1)\widehat{\mathrm{L}}(t_2) - \widehat{\mathrm{L}}(t_2)\widehat{\mathrm{L}}(t_2) \neq 0$, iterative solution shows that Green's function generalizes to

$$G(t;t') = H(t-t')\exp_{+}\left(-\int_{t'}^{t}\mathrm{d}\overline{t}\,\mathrm{i}\widehat{\mathrm{L}}(\overline{t})\right),$$

where the plus subscript indicates that in the series expansion of the exponential the operators must be ordered such that their time arguments increase from right to left.

From footnote 15 on page 14 and the spatial integrals of the microscopic evolution equations (2.26), one has

$$\mathrm{i}\mathscr{L}\boldsymbol{\mathscr{A}}_s = \partial_t\boldsymbol{\mathscr{A}}_s = \int \mathrm{d}\boldsymbol{r}\,[(nq)_s(\boldsymbol{E} + c^{-1}\boldsymbol{u}_s \times \boldsymbol{B}^{\mathrm{ext}})\delta^{\mu\boldsymbol{p}} - \boldsymbol{\nabla}\cdot\boldsymbol{J}_s + \dot{\boldsymbol{A}}'_{\Delta,s}]. \tag{2.54}$$

The divergence term integrates away. So does the $(nq\boldsymbol{u})_s$ term by virtue of the assumption (2.12). In the one-component case the $\dot{\boldsymbol{A}}_\Delta$ term is absent and the $\boldsymbol{E}$ term can be shown to vanish,[31] so one concludes that $W$ obeys the homogeneous Liouville equation (Brey *et al.* 1981). More generally, total conservation of momentum and energy gives $\sum_{\overline{s}}\dot{\boldsymbol{\mathscr{A}}}'_{\Delta,\overline{s}} = \boldsymbol{0}$, and $\sum_{\overline{s}}\int \mathrm{d}\boldsymbol{r}\,(nq)_{\overline{s}}\boldsymbol{E} = \int \mathrm{d}\boldsymbol{r}\,\rho\boldsymbol{E}$ vanishes by the same manipulation used in footnote 31. If one refers the $\boldsymbol{B}_{\overline{s}}$ to a fixed $\boldsymbol{B}_s$, one has

$$S_\Delta \doteq -\sum_{\overline{s}}\dot{\boldsymbol{\mathscr{A}}}_{\overline{s}}\cdot\boldsymbol{B}_{\overline{s}}(\boldsymbol{r},t) = -\left[\sum_{\overline{s}}\dot{\boldsymbol{\mathscr{A}}}_{\overline{s}}\cdot\Delta\boldsymbol{B}_{\overline{s}s}(\boldsymbol{r},t) + \left(\sum_{\overline{s}}\dot{\boldsymbol{\mathscr{A}}}_{\overline{s}}\right)\cdot\boldsymbol{B}_s\right] \tag{2.55a}$$

$$= -\sum_{\overline{s}}\dot{\boldsymbol{\mathscr{A}}}'_{\Delta,\overline{s}}\cdot\Delta\boldsymbol{B}_{\overline{s}s}(\boldsymbol{r},t). \tag{2.55b}$$

In obtaining the last result, I required $\int \mathrm{d}\boldsymbol{r}\,(nq)_s\boldsymbol{E} = \boldsymbol{0}$ in order to ensure the constraint (2.12). Thus, one needs to solve

$$\partial_s W + \mathrm{i}\mathscr{L}W = S_\Delta, \tag{2.56}$$

where the effect of $S_\Delta$ is no larger than first order. The source $S_\Delta$ will ultimately give rise to the momentum and energy exchange terms; for example, $\Delta\boldsymbol{B}^{\boldsymbol{u}_e}_{ie}(\boldsymbol{r},t) \doteq (\beta\boldsymbol{u})_i(\boldsymbol{r},t) - (\beta\boldsymbol{u})_e(\boldsymbol{r},t) \approx -T^{-1}[\boldsymbol{u}_e(\boldsymbol{r},t) - \boldsymbol{u}_i(\boldsymbol{r},t)]$. Note that although $\Delta\boldsymbol{B}_{\overline{s}s}$ depends on $s$, $S_\Delta$ is independent of $s$.

It may seem that one has merely traded one very difficult problem — solution of the original, homogeneous Liouville equation (1.13) — for one of at least equal difficulty, namely the inhomogeneous Liouville equation (2.56) for $W$. However, as we shall see, the fact that both the reference distribution $F_0$ and the projection operator P are built from the $\widetilde{\boldsymbol{A}}$'s and the $\boldsymbol{B}$'s enables one to make progress when an expansion in the gradients is desired. The strategy of Brey *et al.* is to solve for $W$ by calculating Q$W$, then adding the result to the formal representation of P$W$. Because the required projection operations depend only on $t$, not $s$, they can be passed through the $s$ derivative in (2.56) and one has

$$\partial_s \mathrm{P}W + \mathrm{P}\mathrm{i}\mathscr{L}\mathrm{P}W + \mathrm{P}\mathrm{i}\mathscr{L}\mathrm{Q}W = \mathrm{P}S_\Delta, \tag{2.57a}$$

$$\partial_s \mathrm{Q}W + \mathrm{Q}\mathrm{i}\mathscr{L}\mathrm{Q}W + \mathrm{Q}\mathrm{i}\mathscr{L}\mathrm{P}W = \mathrm{Q}S_\Delta. \tag{2.57b}$$

The solution of (2.57*b*) is

$$\mathrm{Q}W(s) = \mathrm{U}(s)W(0) - \int_0^s \mathrm{d}\overline{s}\,\mathrm{U}(s-\overline{s})\mathrm{i}\mathscr{L}\mathrm{P}W(\overline{s}) + \varUpsilon_\Delta(s), \tag{2.58}$$

where the *modified propagator* is

$$\mathrm{U}(s) \doteq \mathrm{Q}\mathrm{e}^{-\mathrm{Q}\mathrm{i}\mathscr{L}\mathrm{Q}s}\mathrm{Q} \tag{2.59}$$

[31] One has $\boldsymbol{E} = -\boldsymbol{\nabla}\phi$; Poisson's equation is $-\nabla^2\phi = 4\pi\rho$. In the one-component case, $\rho = (nq)_s + (\overline{n}q)_{s'}$, where $s$ is the active species and $s'$ represents the neutralizing background. Thus, $\int \mathrm{d}\boldsymbol{r}\,(nq)_s\boldsymbol{E} = \int \mathrm{d}\boldsymbol{r}\,[\rho - (\overline{n}q)_{s'}](-\boldsymbol{\nabla}\phi)$. The $\overline{n}$ term integrates away. From Poisson's equation, $-\rho\boldsymbol{\nabla}\phi = (4\pi)^{-1}(\nabla^2\phi)(\boldsymbol{\nabla}\phi) = (8\pi)^{-1}\boldsymbol{\nabla}|\boldsymbol{\nabla}\phi|^2$. This term integrates away as well.

and

$$\varUpsilon_\Delta(s) = \left(\int_0^s \mathrm{d}\overline{\tau}\, \mathrm{U}(\overline{\tau})\right) S_\Delta = \mathrm{Q}\left(\int_0^s \mathrm{d}\overline{\tau}\, \mathrm{e}^{-\mathrm{Qi}\mathscr{L}\overline{\tau}}\right) \mathrm{Q}S_\Delta. \tag{2.60}$$

From the definition (2.45) of the projection operation, one finds the representation

$$\mathrm{P}W(s) = \omega(s) + \boldsymbol{A}'^{\mathrm{T}} \star \boldsymbol{b}(s), \tag{2.61}$$

where $\omega(s) \doteq \langle W \rangle_0(s)$ and

$$\boldsymbol{b}_s(s) \doteq \boldsymbol{\mathscr{M}}^{-1} \star \langle \boldsymbol{A}'_s\, W(s) \rangle_0. \tag{2.62}$$

Equation (2.58) can thus be written as

$$\mathrm{Q}W(s) = \mathrm{U}(s)W(0) + \int_0^s \mathrm{d}\overline{s}\, \boldsymbol{F}^{\mathrm{T}}(s-\overline{s}) \star \boldsymbol{b}(\overline{s}) + \varUpsilon_\Delta(s), \tag{2.63}$$

where

$$\boldsymbol{F}_s(s) \doteq -\mathrm{U}(s)\mathrm{i}\mathscr{L}\boldsymbol{A}'_s. \tag{2.64}$$

Brey *et al.* argue that it is legitimate to assert the initial condition $\mathrm{Q}W(0) = 0$. That removes the initial-condition term in (2.58). Then, upon adding (2.61) and (2.63), one obtains

$$W(s) = \omega(s) + \boldsymbol{A}'^{\mathrm{T}} \star \boldsymbol{b}(s) + \int_0^s \mathrm{d}\overline{s}\, \boldsymbol{F}^{\mathrm{T}}(s-\overline{s}) \star \boldsymbol{b}(\overline{s}) + \varUpsilon_\Delta(s). \tag{2.65}$$

[Note that the P-projected equation (2.57*a*) has not been used at this point. Its role will ultimately be to determine the proper representations of the dissipative fluxes in terms of Q$W$.] Finally, define [for arbitrary vector $\boldsymbol{\gamma}(s)$]

$$\psi_{\boldsymbol{\gamma}}(s) \doteq \int_0^s d\overline{s}\, \boldsymbol{F}^{\mathrm{T}}(s-\overline{s}) \star \boldsymbol{\gamma}(\overline{s}). \tag{2.66}$$

Upon taking the limit $s \to t$, one then finds that (2.51) can be written as

$$F(\varGamma, t) = \exp[\omega(t) + \boldsymbol{A}'^{\mathrm{T}} \star \boldsymbol{b}(t) + \psi_{\boldsymbol{b}}(t) + \varUpsilon_\Delta(t)] F_0(\varGamma; \boldsymbol{r}, t). \tag{2.67}$$

This generalizes formula (36) of Brey *et al.* to include interspecies exchange effects (through the $\varUpsilon_\Delta$ term). [The value of $\omega(t)$, which plays the role of a normalization factor, will not be required.]

As Brey *et al.* emphasize, the representation (2.67) is an exact consequence of the Liouville equation, given the particular choice of initial condition. Note from (2.62) that $\boldsymbol{b}$ contains $W$, thus is still unknown. That is, (2.67) is an implicit representation of the phase-space distribution; it is not an explicit solution for it. The clever advance of Brey *et al.* was to show how to usefully exploit that representation in order to develop hydrodynamics as a gradient expansion. I now sketch that procedure. Again, my discussion closely follows that of Brey *et al.* except for the extra exchange term $\varUpsilon_\Delta$ and the implicit sum over species that resides in the $\star$ operation.

Consider the phase-space average of any random function $\widetilde{G}$. From (2.67), one has

$$\langle \widetilde{G} \rangle(t) = \int \mathrm{d}\varGamma\, G(\varGamma) F(\varGamma; t) = \frac{\langle \widetilde{G} \exp[\widetilde{\boldsymbol{A}}'^{\mathrm{T}} \star \boldsymbol{b}(t) + \widetilde{\psi}_{\boldsymbol{b}}(t) + \widetilde{\varUpsilon}_\Delta(t)] \rangle_0}{\langle \exp[\widetilde{\boldsymbol{A}}'^{\mathrm{T}} \star \boldsymbol{b}(t) + \widetilde{\psi}_{\boldsymbol{b}}(t) + \widetilde{\varUpsilon}_\Delta(t)] \rangle_0}. \tag{2.68}$$

[The normalization of $F$ has been used to eliminate the $\exp[\omega(t)]$ factor in (2.67).] As a special case of (2.68), the hydrodynamic variables are

$$\boldsymbol{a}_s(\boldsymbol{r}, t) = \frac{\langle \widetilde{\boldsymbol{A}}_s(\boldsymbol{r}, t) \exp[\widetilde{\boldsymbol{A}}'^{\mathrm{T}} \star \boldsymbol{b}(t) + \widetilde{\psi}_{\boldsymbol{b}}(t) + \widetilde{\varUpsilon}_\Delta(t)] \rangle_0}{\langle \exp[\widetilde{\boldsymbol{A}}'^{\mathrm{T}} \star \boldsymbol{b}(t) + \widetilde{\psi}_{\boldsymbol{b}}(t) + \widetilde{\varUpsilon}_\Delta(t)] \rangle_0}. \tag{2.69}$$

An independent expression for $\boldsymbol{a}_s(\boldsymbol{r},t)$ follows from the constraint (2.33) that the conjugate variables $\boldsymbol{B}_s(\boldsymbol{r},t)$ are supposed to be defined such that the mean hydrodynamic variables are obtained exactly by averaging over the local equilibrium distribution:

$$\boldsymbol{a}_s(t) = \int \mathrm{d}\Gamma\, \boldsymbol{A}_s(\Gamma;\boldsymbol{r}) F_B(\Gamma;t), \tag{2.70}$$

where $F_B$ is given by (2.28). From (2.43), one can reexpress (2.70) as an average over the reference distribution as

$$\boldsymbol{a}_s(\boldsymbol{r},t) = \frac{\langle \widetilde{\boldsymbol{A}}_s(\boldsymbol{r},t)\exp(\widetilde{\boldsymbol{A}}'^{\mathrm{T}} \star \Delta\boldsymbol{B})\rangle_0}{\langle \exp(\boldsymbol{A}'^{\mathrm{T}} \star \Delta\boldsymbol{B})\rangle_0}, \tag{2.71}$$

where again $\Delta\boldsymbol{B}_{\overline{s}}$ is defined by (2.42) as being the diagonal part of $\Delta\boldsymbol{B}_{\overline{s}s}$. Because the average limits the range of the $\overline{\boldsymbol{r}}$ in the $\star$ operation to lie within a correlation length of $\boldsymbol{r}$, $\Delta\boldsymbol{B}_{\overline{s}s}(\overline{\boldsymbol{r}},\boldsymbol{r},t)$ will be assumed to be small. Thus, when a Markovian approximation is invoked, one has

$$\Delta\boldsymbol{B}_{\overline{s}s}(\overline{\boldsymbol{r}},\boldsymbol{r},t) \approx \Delta\boldsymbol{B}_{\overline{s}s}(\boldsymbol{r},\boldsymbol{r},t) + (\overline{\boldsymbol{r}}-\boldsymbol{r}) \cdot \boldsymbol{\nabla}\boldsymbol{B}_{\overline{s}}(\boldsymbol{r},t) + \cdots . \tag{2.72}$$

The first term vanishes for $s = \overline{s}$, leaving only a small gradient contribution; that is the effect studied by Brey *et al.* through second order in the gradients. For $s \neq \overline{s}$, the first term does not vanish; it appears in $\Upsilon_\Delta$ [(2.60)] through the definition (2.55*b*) of $S_\Delta$ and will lead to exchange effects. ***As a fundamental assumption, I shall order the exchange effects to be small and of the same order as the gradient terms.*** If gradients are symbolically represented by $\nabla$ and exchange effects by $\Delta$, this ordering implies that a general theory of Burnett-order transport will include terms of order $\nabla^2$, $\Delta\nabla$, and $\Delta^2$. However, although I shall indicate where the latter two kinds of terms arise, *in this work I shall not calculate terms of order* $\Delta\nabla$ *and* $\Delta^2$.

It will be seen that the unknown $\boldsymbol{b}_s$'s are approximately equal to the $\Delta\boldsymbol{B}_s$'s and therefore are also small. Thus, one can expand (2.69) in powers of $\boldsymbol{b}$, expand (2.71) in powers of $\Delta\boldsymbol{B}$, then equate the two equivalent representations to find an expression for the $\boldsymbol{b}$'s in terms of the $\Delta\boldsymbol{B}$'s. The result of this exercise is

$$\boldsymbol{b}(t) = \Delta\boldsymbol{B}(t) + \Delta\boldsymbol{B}^{(2)}(t), \tag{2.73}$$

where

$$\begin{aligned}\Delta\boldsymbol{B}^{(2)}(t) \doteq -\boldsymbol{\mathscr{M}}^{-1} \star \left( \langle \boldsymbol{A}'\psi_{\Delta\boldsymbol{B}}(t)\boldsymbol{A}'^{\mathrm{T}}\rangle_0 \star \Delta\boldsymbol{B}(t) + \frac{1}{2}\langle \boldsymbol{A}'\psi^2_{\Delta\boldsymbol{B}}(t)\rangle_0 \right) \\ + O(\Upsilon_\Delta^2, \Upsilon_\Delta \Delta\boldsymbol{B}) + \text{third-order terms.}\end{aligned} \tag{2.74}$$

Since $\Upsilon_\Delta(t)$ is additive to $\psi_b$ in (2.68), the form of the unwritten second-order terms in (2.74) involving $\Upsilon_\Delta$ can be obtained from the explicit terms by replacing $\psi_{\Delta\boldsymbol{B}}$ by $\psi_{\Delta\boldsymbol{B}} + \Upsilon_\Delta$.

These results can now be used to expand (2.68) to the desired order. I shall retain gradient effects through second order but exchange effects only through first order. The result is then

$$\begin{aligned}\langle G\rangle(t) \approx \underbrace{\langle G\rangle_B(t)}_{\text{(i)}} + \underbrace{\langle G'\psi_{\Delta\boldsymbol{B}}(t)\rangle_0}_{(\mathrm{ii}_{\boldsymbol{\nabla}})} + \underbrace{\langle G'\Upsilon_\Delta(t)\rangle_0}_{(\mathrm{ii}_\Delta)} \\ + \underbrace{\langle G'\mathrm{Q}\psi_{\Delta\boldsymbol{B}}(t)\boldsymbol{A}'\rangle_0 \star \Delta\boldsymbol{B}(t)}_{(\mathrm{iii_a})} + \underbrace{\frac{1}{2}\langle G'\mathrm{Q}\psi^2_{\Delta\boldsymbol{B}}(t)\rangle_0}_{(\mathrm{iii_b})} + \underbrace{\langle G'\psi_{\Delta\boldsymbol{B}^{(2)}}\rangle_0}_{\text{(iv)}} .\end{aligned} \tag{2.75}$$

This generalizes formula (51) of Brey *et al.* to include the term (ii$_\Delta$). See Brey *et al.* for a discussion of how some terms have been combined into the local equilibrium average $\langle G\rangle_B$ [term (i)]. Terms (ii)–(iv) will ultimately lead to the dissipative fluxes and exchange effects.

## 3. The gradient expansion through Burnett order

I now discuss and simplify each of the terms in (2.75). The discussion follows Brey *et al.* closely, but generalizations are necessary at various steps in order to deal with the magnetic field and especially the interspecies coupling.

### 3.1. *Term (i)*

As was noted above, term (i) is the local equilibrium average. As a consistency check, one can see that when $G$ is replaced by $\boldsymbol{A}$ all terms except for term (i) vanish because of either explicit or implicit Q's. [$\psi_\gamma$ contains a factor of Q on the left because of definitions (2.66), (2.64), and (2.59).] In that case, term (i) generates the right-hand side of the Euler equations.

### 3.2. *Term (ii$_{\boldsymbol{\nabla}}$)*

Upon using the definitions (2.66) and (2.64), one has

$$\text{term (ii}_{\boldsymbol{\nabla}}) \doteq \langle G'\psi_{\Delta\boldsymbol{B}}(t)\rangle_0 = -\int_0^t \mathrm{d}s\,\langle G'\mathrm{U}(t-s)\mathrm{i}\mathscr{L}\boldsymbol{A}'^{\mathrm{T}}\rangle_0 \star \Delta\boldsymbol{B}(s). \tag{3.1}$$

This form is somewhat schematic because only the time dependence is displayed. In order to process it further, it is necessary to be more explicit. Recall that the transport equations will be evaluated at the reference variables $(\boldsymbol{r},t)$, while contributions to those equations (e.g., the transport coefficients) will involve integrals over the past state of the system at $(\boldsymbol{x},s)$. In the following, I shall adopt a somewhat more symmetrical notation in which the reference variables are unbarred while the integration variables are barred [i.e., $(\boldsymbol{x},s)\to(\overline{\boldsymbol{x}},\overline{t})$]. With $\mu\doteq\{\boldsymbol{r},s\}$ (here $s$ is a species index) and upon noting that the propagator U depends on the time difference $\overline{\tau}\doteq t-\overline{t}$, one can write term (ii$_{\boldsymbol{\nabla}}$) as

$$\langle G'(\mu)\psi_{\Delta\boldsymbol{B}}(t)\rangle_0 = -\int_0^t \mathrm{d}\overline{\tau}\int \mathrm{d}\overline{\mu}\,\langle G'(\mu)\mathrm{U}(\overline{\tau})\mathrm{i}\mathscr{L}A'^{\beta}(\overline{\mu})\rangle_0 \Delta B_\beta(\overline{\mu},t-\overline{\tau}), \tag{3.2}$$

where $\int\mathrm{d}\overline{\mu}\equiv\sum_{\overline{s}}\int\mathrm{d}\overline{\boldsymbol{x}}$, $\boldsymbol{A}'(\overline{\mu})\doteq\boldsymbol{A}_{\overline{s}}(\overline{\boldsymbol{x}})-\langle\boldsymbol{A}\rangle_{0\overline{s}}(\boldsymbol{r},t)$, and $\Delta\boldsymbol{B}(\overline{\mu},\overline{t})\doteq\boldsymbol{B}_{\overline{s}}(\overline{\boldsymbol{x}},\overline{t})-\boldsymbol{B}_{\overline{s}}(\boldsymbol{r},t)$. [The $(\boldsymbol{r},t)$ dependence on the reference state is implicit in the notation $\boldsymbol{A}'(\overline{\mu})$ and $\Delta\boldsymbol{B}(\overline{\mu},\overline{t})$.]

Expression (3.2) displays two difficulties: it involves the modified propagator U [see (2.59)], which is difficult to work with because of the Q projections in the operator Qi$\mathscr{L}$Q; and it is nonlocal in space and time. The nonlocality is relatively straightforward to deal with by means of a Markovian approximation. Thus, the characteristic correlation length introduced by U is, for the case of the weakly coupled plasma, either the collisional mean free path or the gyroradius; both are assumed to be of short range relative to the macroscopic gradient scale length. (It is at this point that one ignores the possibility of long-ranged correlations.) Then, with $\overline{\boldsymbol{\rho}}\doteq\boldsymbol{r}-\overline{\boldsymbol{x}}$ and upon noting that $\Delta\boldsymbol{B}_{\overline{s}\,\overline{s}}(\boldsymbol{r},t)=0$, one has

$$\begin{aligned}\Delta\boldsymbol{B}(\overline{\mu},t-\overline{\tau}) &= -\overline{\boldsymbol{\rho}}\cdot\boldsymbol{\nabla}\boldsymbol{B}_{\overline{s}}(\boldsymbol{r},t)-\overline{\tau}\,\partial_t\boldsymbol{B}^{(1)}_{\overline{s}}(\boldsymbol{r},t)\\ &\quad+\frac{1}{2}(\overline{\boldsymbol{\rho}}\,\overline{\boldsymbol{\rho}}):\boldsymbol{\nabla}\boldsymbol{\nabla}\boldsymbol{B}_{\overline{s}}(\boldsymbol{r},t)+\overline{\tau}\,\overline{\boldsymbol{\rho}}\cdot\boldsymbol{\nabla}\partial_t\boldsymbol{B}^{(1)}_{\overline{s}}(\boldsymbol{r},t)+\frac{1}{2}\overline{\tau}^2\partial_t^2\boldsymbol{B}^{(1)}_{\overline{s}}(\boldsymbol{r},t)+\cdots.\end{aligned} \tag{3.3}$$

For the terms involving time derivatives, note that it is adequate to calculate $\partial_t \boldsymbol{B}$ only to first order (i.e., from the Euler equations).

The Q projections are more problematical. For a one-component neutral fluid, Brey *et al.* argue that for small gradients the $\mathrm{Qi}\mathscr{L}\mathrm{Q}$ in U can be replaced merely by $\mathrm{i}\mathscr{L}$. The analogous statement for the one-component plasma at first order in the gradients (§I:2) is that $\mathrm{Q}\widehat{\mathrm{C}}\mathrm{Q}$ can be replaced merely by $\widehat{\mathrm{C}}$. I shall first describe the procedure followed by Brey *et al.*, then indicate a difficulty for the multicomponent case. Brey *et al.* invoke the identity[32]

$$\mathrm{U}(s) = \mathrm{Q}\mathrm{e}^{-\mathrm{i}\mathscr{L}s}\mathrm{Q} + \int_0^s \mathrm{d}\overline{s}\,\mathrm{Q}\mathrm{e}^{-\mathrm{i}\mathscr{L}(s-\overline{s})}\mathrm{P}\mathrm{i}\mathscr{L}\mathrm{U}(\overline{s}). \tag{3.4}$$

Because for any phase function $\widetilde{g}$ one has $\mathrm{i}\mathscr{L}\widetilde{g} = \mathrm{d}\widetilde{g}/\mathrm{d}t$ (see footnote 15 on page 14), when the form $\mathrm{i}\mathscr{L}\boldsymbol{A}'$ appears to the far right of any expression involving phase-space variables it can be replaced by the right-hand side of (2.27). The electric and magnetic force terms in the momentum equation do not enter the final expressions because they are invariably preceded by a Q. In the long-time limit, one is led, after a spatial integration by parts and with the aid of (3.3), to

$$\langle G'(\mu)\psi_{\Delta\boldsymbol{B}}(t)\rangle_0 = -\int_0^t \mathrm{d}\overline{\tau}\int \mathrm{d}\overline{\mu}\,\boldsymbol{K}_G^{\overline{\beta}}(\mu,\overline{\mu},\overline{\tau})\cdot\nabla_{\overline{\boldsymbol{x}}}\Delta B_{\overline{\beta}}(\overline{\mu},t-\overline{\tau}) \tag{3.5}$$

$$\begin{aligned} &\approx -\int_0^\infty \mathrm{d}\overline{\tau}\int \mathrm{d}\overline{\mu}\,\boldsymbol{K}_G^{\overline{\beta}}(\mu,\overline{\mu},\overline{\tau})\cdot\nabla B_{\overline{\beta}}(\boldsymbol{r},t) \\ &\quad -\int_0^\infty \mathrm{d}\overline{\tau}\int \mathrm{d}\overline{\mu}\,\boldsymbol{K}_G^{\overline{\beta}}(\mu,\overline{\mu},\overline{\tau})\overline{\tau}\cdot\nabla[\partial_t B_{\overline{\beta}}(\boldsymbol{r},t)]^{(1)} \\ &\quad +\int_0^\infty \mathrm{d}\overline{\tau}\int \mathrm{d}\overline{\mu}\,\boldsymbol{K}_G^{\overline{\beta}}(\mu,\overline{\mu},\overline{\tau})\overline{\boldsymbol{\rho}} : \nabla\nabla B_{\overline{\beta}}(\boldsymbol{r},t), \end{aligned} \tag{3.6}$$

where $\boldsymbol{K} = {}_a\boldsymbol{K} + {}_b\boldsymbol{K}$ and

$${}_a\boldsymbol{K}_G^{\overline{\beta}}(\mu,\overline{\mu},\overline{\tau}) \doteq \langle\widehat{G}(\mu)\mathrm{e}^{-\mathrm{i}\mathscr{L}\overline{\tau}}\widehat{\boldsymbol{J}}^{\overline{\beta}}(\overline{\mu})\rangle_0, \tag{3.7$a$}$$

$$\begin{aligned} {}_b\boldsymbol{K}_G^{\overline{\beta}}(\mu,\overline{\mu},\overline{\tau}) \doteq &-\int_0^{\overline{\tau}} \mathrm{d}\overline{s}\,\langle\widehat{G}(\mu)\mathrm{e}^{-\mathrm{i}\mathscr{L}(\overline{\tau}-\overline{s})}\boldsymbol{A}'^{\mathrm{T}}\rangle_0 \star \mathscr{M}^{-1}\star\langle(\mathrm{i}\mathscr{L}\boldsymbol{A}')U(\overline{s})\widehat{\boldsymbol{J}}^{\overline{\beta}}(\overline{\mu})\rangle_0 \\ &+ (\Delta \text{ correction}), \end{aligned} \tag{3.7$b$}$$

To obtain (3.7$b$), the Liouville operator was integrated by parts under the last $F_0$ average. The correction terms, not written explicitly, arise from the fact that in general $\mathrm{i}\mathscr{L}F_0 \neq 0$ [see the discussion after (2.56)].

---

[32]See the related discussion in footnote 25 of Part I. In detail,

$$\mathrm{U}(s) \doteq \mathrm{Q}\mathrm{e}^{-\mathrm{Qi}\mathscr{L}\mathrm{Q}s}\mathrm{Q} = \mathrm{Q}\mathrm{e}^{-\mathrm{Qi}\mathscr{L}s}\mathrm{Q} = \mathrm{e}^{-\mathrm{Qi}\mathscr{L}s}\mathrm{Q},$$

$$\partial_s\mathrm{U} = -\mathrm{Qi}\mathscr{L}\mathrm{U}(s) = -(1-\mathrm{P})\mathrm{i}\mathscr{L}\mathrm{U}(s), \quad \text{or} \quad \partial_s\mathrm{U} + \mathrm{i}\mathscr{L}\mathrm{U} = \mathrm{P}\mathrm{i}\mathscr{L}\mathrm{U},$$

Solution by means of a Green's function leads to

$$\mathrm{U}(s) = \mathrm{e}^{-\mathrm{i}\mathscr{L}s}\mathrm{U}(0) + \int_0^s \mathrm{d}\overline{s}\,\mathrm{e}^{-\mathrm{i}\mathscr{L}(s-\overline{s})}\mathrm{P}\mathrm{i}\mathscr{L}\mathrm{U}(\overline{s}).$$

One has $\mathrm{U}(0) = \mathrm{Q}^2 = \mathrm{Q}$. Since one knows from the original definition that $\mathrm{PU} = 0$, it must be the case that applying P to the right-hand side of the last result yields 0. That can be checked term by term upon expanding in $s$. Thus, the right-hand side can be multiplied by Q with no change in value, whereupon one obtains (3.4).

### 3.2.1. Two-time $\boldsymbol{x}$-space correlation functions as velocity integrals over Klimontovich correlations

Before I consider (3.7) in detail, I digress to discuss expectations of the form $M_0(\mu,\mu',\tau) \doteq \langle \widetilde{a}(\mu)\mathrm{e}^{-\mathrm{i}\mathscr{L}\tau}\widetilde{b}(\mu')\rangle_0$, which appear in each of those equations. Here

$$\widetilde{a}(\mu) \doteq \sum_{i\in s} a_s(\boldsymbol{x}_i,\boldsymbol{v}_i)\delta(\boldsymbol{r}-\boldsymbol{x}_i), \tag{3.8}$$

where $a_s(\boldsymbol{x},\boldsymbol{v})$ is a prescribed function, and similarly for $\widetilde{b}$. (At this point, there may also be implicit dependence of $a$ and $b$ on the particle coordinates $\boldsymbol{x}_{j\neq i}$; that is not indicated explicitly.) A standard manipulation (Klimontovich 1967; Krommes & Oberman 1976*a*) shows that such expectations can be written in terms of correlations of the Klimontovich microdensity $\widetilde{f}$. Namely, upon using the definition (2.6) of $\widetilde{f}$ and the fact that $\mathrm{e}^{-\mathrm{i}\mathscr{L}\tau}$ moves phase-space coordinates back in time by interval $\tau$, one has

$$M_0(\mu,\mu',\tau) = \left\langle \sum_{i\in s} a_s(\boldsymbol{x}_i,\boldsymbol{v}_i)\delta(\boldsymbol{r}-\boldsymbol{x}_i)\mathrm{e}^{-\mathrm{i}\mathscr{L}\tau}\sum_{j\in s'} b_{s'}(\boldsymbol{x}_j,\boldsymbol{v}_j)\delta(\boldsymbol{r}'-\boldsymbol{x}_j)\right\rangle_0 \tag{3.9a}$$

$$= \overline{n}_s\overline{n}_{s'}\left\langle \frac{1}{\overline{n}_s}\sum_{i\in s} a_s(\boldsymbol{x}_i(0),\boldsymbol{v}_i(0))\delta(\boldsymbol{r}-\boldsymbol{x}_i(0)) \times \frac{1}{\overline{n}_{s'}}\sum_{j\in s'} b_{s'}(\boldsymbol{x}_j(-\tau),\boldsymbol{v}_j(-\tau))\right\rangle_0 \tag{3.9b}$$

$$= \overline{n}_s\overline{n}_{s'}\left\langle \int \mathrm{d}\overline{\boldsymbol{v}}\, a_s(\boldsymbol{r},\overline{\boldsymbol{v}})\frac{1}{\overline{n}_s}\sum_{i\in s}\delta(\boldsymbol{r}-\boldsymbol{x}_i(0))\delta(\overline{\boldsymbol{v}}-\boldsymbol{v}_i(0)) \times \int \mathrm{d}\overline{\boldsymbol{v}}'\, b_{s'}(\boldsymbol{r}',\overline{\boldsymbol{v}}')\frac{1}{\overline{n}_{s'}}\sum_{j\in s'}\delta(\boldsymbol{r}'-\boldsymbol{x}_j(-\tau))\delta(\overline{\boldsymbol{v}}'-\boldsymbol{v}_j(-\tau))\right\rangle_0 \tag{3.9c}$$

$$= \overline{n}_s\overline{n}_{s'}\int \mathrm{d}\overline{\boldsymbol{v}}\,\mathrm{d}\overline{\boldsymbol{v}}'\, a_s(\boldsymbol{r},\overline{\boldsymbol{v}})\langle \widetilde{f}_s(\boldsymbol{r},\overline{\boldsymbol{v}},0)\widetilde{f}_{s'}(\boldsymbol{r}',\overline{\boldsymbol{v}}',-\tau)\rangle_0 b_{s'}(\boldsymbol{r}',\overline{\boldsymbol{v}}'), \tag{3.9d}$$

the last result holding when neither $a$ nor $b$ implicitly depends on any phase-space variable. The random currents on the right-hand side of (2.23) satisfy this property when the potential-energy contributions involving $\Delta\boldsymbol{\tau}_i$ are neglected; then $a$ and $b$ depend only on $\boldsymbol{v}$ and $s$. In that case (valid for the weakly coupled plasma), the $\varGamma$-space expectation has been reduced to integrals of the one-point distribution function $f$ and the two-point Klimontovich correlation function $C(\tau) \doteq \langle \delta f(\tau)\delta f(0)\rangle_0$:

$$M_0(\mu,\mu',\tau) = \left(\overline{n}_s\int \mathrm{d}\overline{\boldsymbol{v}}\, a_s(\boldsymbol{r},\overline{\boldsymbol{v}})f_s(\boldsymbol{r},\overline{\boldsymbol{v}})\right)\left(\overline{n}_{s'}\int \mathrm{d}\overline{\boldsymbol{v}}'\, b_{s'}(\boldsymbol{r}',\overline{\boldsymbol{v}}')f_{s'}(\boldsymbol{r}',\overline{\boldsymbol{v}}')\right) + \overline{n}_s\overline{n}_{s'}\int \mathrm{d}\overline{\boldsymbol{v}}\,\mathrm{d}\overline{\boldsymbol{v}}'\, a_s(\boldsymbol{r},\overline{\boldsymbol{v}})C_{ss'}(\boldsymbol{r},\overline{\boldsymbol{v}},\tau,\boldsymbol{r}',\overline{\boldsymbol{v}}',0)b_{s'}(\boldsymbol{r}',\overline{\boldsymbol{v}}'). \tag{3.10}$$

Here the assumption of stationary statistics was used to shift the $\tau$ argument.[33] In the uses I shall make of this formula, the mean-field terms [first line of (3.10)] will vanish. When potential-energy contributions to $a$ and $b$ are included, a generalization of the

[33] If the expectation were taken in a stationary, homogeneous ensemble, $M_0(\mu,\mu',\tau)$ would depend on only the spatial and temporal difference variables: $M_0(\mu,\mu',\tau) \to M_{ss'}(\boldsymbol{r}-\boldsymbol{r}',\tau)$. It could then be Fourier transformed with respect to that variable, giving the function $\widehat{M}_{ss',\boldsymbol{k}}(\tau)$. For transport theory, the relevant limit is $k\to 0$. In fact, however, the expectation is taken in the reference ensemble, which is weakly dependent on $\{\boldsymbol{r},t\} \equiv r$. One must then consider a

previous argument shows that the $\Gamma$-space expectation can be reduced to integrals of correlation functions with no more than four phase-space points.

The theory of the Klimontovich two-time correlation function $C(\tau)$ will be described in §5. It is shown there that to lowest order in a weakly coupled plasma the time dependence of the one-sided[34] function $C_+(\tau)$ is given for small $k$ by the linearized Vlasov response function [which lies at the heart of Rostoker's Test Particle Superposition Principle (Krommes 1976, and references therein)] modified to include collisional corrections given by the linearized collision operator $\widehat{\mathrm{C}}$. This information is sufficient to inform the following discussion, in which I shall consider each of (3.7*a*) and (3.7*b*) in turn.

### 3.2.2. Equation (3.7*a*)

Eventually, we shall see that the Braginskii (weakly coupled Navier–Stokes) transport coefficients that multiply gradients will arise from the $k \to 0$ limit of the $\overline{\tau}$ integrals of expression (3.7*a*) with $\widehat{G}$ replaced by $\widehat{\boldsymbol{J}}^\alpha$. The first-order gradient-driven transport fluxes are then

$$-\int_0^\infty \mathrm{d}\overline{\tau}\int \mathrm{d}\boldsymbol{v}\sum_{\overline{s}}\int \mathrm{d}\overline{\boldsymbol{v}}\,\widehat{\boldsymbol{J}}^\alpha_s(\boldsymbol{v})C_{s\overline{s},\boldsymbol{k}=\boldsymbol{0}}(\boldsymbol{v},\overline{\boldsymbol{v}},\overline{\tau})\widehat{\boldsymbol{J}}^{\overline{\beta}\mathrm{T}}_{\overline{s}}(\overline{\boldsymbol{v}})\boldsymbol{\cdot}\boldsymbol{\nabla}B_{\overline{\beta},\overline{s}}, \tag{3.11}$$

where both $\alpha$ and $\overline{\beta}$ may assume the values $\boldsymbol{p}$ or $e$. ($\widehat{\boldsymbol{J}}^n = \boldsymbol{0}$.) According to the discussion in the last paragraph, the time dependence is given by $\exp(-\widehat{\mathrm{C}}\overline{\tau})C(\tau=0)$, with $C(\tau=0)$ being given by (1.8). At $\boldsymbol{k}=\boldsymbol{0}$ the Fourier transform of the pair correlation function $g$ vanishes due to the normalization constraint, so one has

$$C_{s\overline{s},\boldsymbol{k}=\boldsymbol{0}}(\boldsymbol{v},\overline{\boldsymbol{v}},\tau=0) = \overline{n}_s^{-1}\delta(\boldsymbol{v}-\overline{\boldsymbol{v}})\delta_{s\overline{s}}f_{0,\overline{s}}(\overline{\boldsymbol{v}}); \tag{3.12}$$

note the appearance of the one-particle reference distribution in this expression. Because $\widehat{\mathrm{C}}$ is isotropic in velocity space, symmetry considerations restrict $\overline{\beta}$ to equal $\alpha$. Upon performing the $\overline{v}$ integration and the $\overline{s}$ and $\overline{\beta}$ summations in (3.11), one is led to the transport matrices

$$\boldsymbol{D}^\alpha_s = \int_0^\infty \mathrm{d}\overline{\tau}\,\langle \widehat{\boldsymbol{J}}^\alpha_{(s)}|\mathrm{e}^{-\widehat{\mathrm{C}}\overline{\tau}}|\widehat{\boldsymbol{J}}^{\alpha\mathrm{T}}_{(s)}\rangle_0, \tag{3.13}$$

where the presence of the vertical bar indicates the $\mu$-space scalar product (I:3.15) used in Part I (here averaged with the local equilibrium distribution). This reproduces the transport formulas discussed in Part I for the irreversible fluxes that are linearly proportional to gradients (now evaluated locally rather than with absolute equilibrium parameters).

### 3.2.3. Equation (3.7*b*)

Brey *et al.* argue that (3.7*b*) is negligible in a gradient expansion. They say,

---

function

$$M_0(r,\overline{r}) \equiv M(r-\overline{r}\mid \tfrac{1}{2}(r+\overline{r})) = \int\frac{\mathrm{d}k}{(2\pi)^4}\int\frac{\mathrm{d}q}{(2\pi)^4}\widehat{M}_{k,q}\mathrm{e}^{\mathrm{i}k(r-\overline{r})}\mathrm{e}^{\mathrm{i}q(r+\overline{r})/2}.$$

Transport coefficients will be given by

$$\int \mathrm{d}\overline{r}\,M_0(r,\overline{r}) = \int\frac{\mathrm{d}q}{(2\pi)^4}\,\widehat{M}_{q/2,q}\mathrm{e}^{\mathrm{i}qr} \approx \widehat{M}(k=0\mid r),$$

the latter following when the microscopic correlation scales are much smaller than the macroscopic hydrodynamic scales.

[34]See appendix H for a discussion of one-sided functions.

From the definition of P, it follows that the sequence $\mathrm{P}\mathscr{L}$ introduces factors of $\mathscr{L}\boldsymbol{A}$ and these yield ... gradient operators acting on $\boldsymbol{B}$. If we keep terms only up to second order in gradients of $\boldsymbol{B}$, we see that [(3.7*b*)] ... can be neglected since it is proportional to $\langle G(\boldsymbol{r})\mathrm{Q}\exp[-(\overline{\tau}-\overline{\tau}')\mathrm{i}\mathscr{L}]\boldsymbol{A}'\rangle_0$ evaluated to zeroth order, which vanishes. This may easily be seen, because to this order

$$\langle G(\boldsymbol{r})\mathrm{Q}\exp[-(\overline{\tau}-\overline{\tau}')\mathrm{i}\mathscr{L}]\boldsymbol{A}'\rangle_0 = \langle G(\boldsymbol{r})\mathrm{Q}\boldsymbol{A}'\rangle_0 = 0. \tag{3.14}$$

This argument must be revised in the multispecies case because of the exchange terms on the right-hand side of (2.26), which are not proportional to gradients. However, one does expect that the effects of those terms will be small, proportional to $\Delta\boldsymbol{u}$ or $\Delta T$, and I have assumed that those are of the same order of smallness as are the gradients. Therefore, the arguments of Brey *et al.* might appear to suggest that the effect of (3.7*b*), which to lowest order multiplies $-\boldsymbol{\nabla}\boldsymbol{B}$, is negligible.

There is, however, a subtlety, which is that the left-hand side of (3.14) appears under an infinite time integral; it is not clear that arguments based on truncated power-series expansion are adequate. To make the mathematics match the discussion in Part I as closely as possible, assume that it is legitimate to replace $\mathrm{i}\mathscr{L}$ by $\widehat{\mathrm{C}}$ (this ignores a $\boldsymbol{v}\cdot\boldsymbol{\nabla}$ streaming term) and treat the expectations and projections as the $\mu$-space ones used in Part I. The integral of (3.7*b*) then becomes

$$\lim_{t\to\infty}\int_0^t \mathrm{d}\overline{\tau}\; {}_b\boldsymbol{K}_G^{\beta}(\mu,\overline{\mu},\overline{\tau}) = \lim_{t\to\infty}\int_0^t \mathrm{d}\overline{\tau}\int_0^{\overline{\tau}} \mathrm{d}\overline{s}\,\langle\widehat{G}(\mu)|\mathrm{e}^{-\widehat{\mathrm{C}}(\overline{\tau}-\overline{s})}|\mathrm{P}\widehat{\mathrm{C}}\mathrm{U}(\overline{s})\widehat{\boldsymbol{J}}(\overline{\mu})\rangle_0 \tag{3.15a}$$

$$= \lim_{t\to\infty}\int_0^t \mathrm{d}\overline{s}\int_{\overline{s}}^{t} \mathrm{d}\overline{\tau}\,\langle\widehat{G}(\mu)|\mathrm{e}^{-\widehat{\mathrm{C}}(\overline{\tau}-\overline{s})}|\mathrm{P}\widehat{\mathrm{C}}\mathrm{U}(\overline{s})\widehat{\boldsymbol{J}}(\overline{\mu})\rangle_0 \tag{3.15b}$$

$$= \lim_{t\to\infty}\int_0^t \mathrm{d}\overline{s}\int_0^{t-\overline{s}} \mathrm{d}\overline{\tau}\,\langle\widehat{G}(\mu)|\mathrm{e}^{-\widehat{\mathrm{C}}\overline{\tau}}|\mathrm{P}\widehat{\mathrm{C}}\mathrm{U}(\overline{s})\widehat{\boldsymbol{J}}(\overline{\mu})\rangle_0. \tag{3.15c}$$

The modified propagator $\mathrm{U}(\overline{s})$ constrains $\overline{s}$ to be less than or of the order of a collision time, whereas $t\to\infty$. Thus, $\overline{s}$ is negligible in the second integral, which can then be performed to give

$$\lim_{t\to\infty}\int_0^t \mathrm{d}\overline{\tau}\; {}_b\boldsymbol{K}_G^{\beta}(\mu,\overline{\mu},\overline{\tau}) = \lim_{t\to\infty}\left\langle \widehat{G}(\mu)\,\middle|\,\widehat{\mathrm{C}}^{-1}\left(1-\mathrm{e}^{-\widehat{\mathrm{C}}t}\right)\,\middle|\,\mathrm{P}\widehat{\mathrm{C}}\left(\int_0^\infty \mathrm{d}\overline{s}\,\mathrm{U}(\overline{s})\right)\widehat{\boldsymbol{J}}(\overline{\mu})\right\rangle_0. \tag{3.16}$$

The properties of this expression differ for one-component and multicomponent systems. In a one-component system, the basis functions from which P is constructed are the null eigenfunctions of $\widehat{\mathrm{C}}$; thus, $\widehat{\mathrm{C}}\mathrm{P}=0$. One has

$$\mathrm{Q}\widehat{\mathrm{C}}^{-1}\left(1-\mathrm{e}^{-\widehat{\mathrm{C}}t}\right)\mathrm{P} = \mathrm{Q}\left(t-\frac{1}{2}\widehat{\mathrm{C}}\,t^2+\cdots\right)\mathrm{P} = \mathrm{Q}\,t\,\mathrm{P} = 0. \tag{3.17}$$

Here 'zero' here means that the effect is at least of first order in $k$, as can be seen by adding the streaming contribution $\mathrm{i}\boldsymbol{k}\cdot\boldsymbol{v}$ to $\widehat{\mathrm{C}}$. Upon similarly processing the factor $\mathrm{P}\widehat{\mathrm{C}}\int_0^\infty d\overline{s}\,\mathrm{U}(\overline{s})\widehat{\boldsymbol{J}}(\overline{\mu})$, one finds that that too is $O(k)$. Since term b multiplies $\boldsymbol{\nabla}\boldsymbol{B}$, one concludes that for a one-component system the net effect of term b is of third order in the gradients, thus is negligible to Burnett order. This recovers the conclusion of Brey *et al.*, which was obtained directly from the expression involving the Liouville operator. Brey *et al.* did not perform the $\overline{\tau}$ integral as was done above, but rather examined the power-series expansion of $\exp(-\mathrm{i}\mathscr{L}\overline{\tau})$, noting that the ultimate effect of the factor $\mathrm{i}\mathscr{L}\boldsymbol{A}'$ is a term proportional to a gradient. Similar arguments will be used later; the manipulations above justify that procedure for a one-component system.

The situation is different for a multicomponent system. In that case $\widehat{\mathrm{C}}\mathrm{P} \neq 0$ because momentum and energy can be exchanged between unlike species. Then, in the limit $t \to \infty$ one has $\exp(-\widehat{\mathrm{C}}t) \to 0$ and one is left with the construction $\widehat{\mathrm{C}}^{-1}\mathrm{P}\widehat{\mathrm{C}}$. Although I shall not work it out in detail, the ultimate value of expression (3.16) will generate a small exchange contribution that will ultimately multiply $\boldsymbol{\nabla B}$. That second-order term should be retained, in principle; however, I shall neglect all such effects. The $\Delta$ correction terms are neglected for the same reason.

Mathematically, the various behaviours are an example of an interchange of limits. That is, with $\widehat{\mathrm{C}} \to \nu$ and $\tau_{\mathrm{coll}} \doteq \nu^{-1}$, one has

$$\nu^{-1}(1 - \mathrm{e}^{-\nu t}) = \tau_{\mathrm{coll}}(1 - \mathrm{e}^{-t/\tau_{\mathrm{coll}}}) \to \begin{cases} t & \text{for } t \text{ fixed, } \tau_{\mathrm{coll}} \to \infty; \\ \tau_{\mathrm{coll}} & \text{for } \tau_{\mathrm{coll}} \text{ fixed, } t \to \infty. \end{cases} \tag{3.18}$$

In future manipulations of second-order terms, I shall follow Brey *et al.* in counting the powers of gradients as though the system contains just one component. That is, the first limit in (3.18) is used. This can be interpreted as a formal way of ordering out second-order exchange effects — they may happen, but only on a timescale longer than that of interest.

## 3.3. *Term (ii$_\Delta$)*

From (2.60) and (2.55*b*), one has

$$\text{term (ii}_\Delta) \doteq \langle G'(\mu)\Upsilon_\Delta(t)\rangle_0 = -\int_0^t \mathrm{d}\overline{\tau} \int \mathrm{d}\overline{\mu}\, \breve{K}_G^{\overline{\beta}}(\mu, \overline{\mu}, \overline{\tau}) B_{\overline{\beta}}(\boldsymbol{r}, t), \tag{3.19}$$

where

$$\breve{\boldsymbol{K}}_G(\mu, \overline{\mu}, \overline{\tau}) \doteq \langle G'(\mu)\mathrm{U}(\overline{\tau})\dot{\boldsymbol{A}}'_{\Delta,\overline{s}}(\overline{\mu})\rangle_0. \tag{3.20}$$

This formula may again be simplified by using the identity (3.4). Thus, one can write $\breve{\boldsymbol{K}} = {}_a\breve{\boldsymbol{K}} + {}_b\breve{\boldsymbol{K}}$, where

$${}_a\breve{\boldsymbol{K}}_G \doteq \langle G' \mathrm{e}^{-\mathrm{i}\mathscr{L}\overline{\tau}} \dot{\boldsymbol{A}}'_\Delta(\overline{\mu})\rangle_0, \tag{3.21$a$}$$

$${}_b\breve{\boldsymbol{K}}_G \doteq \int_0^{\overline{\tau}} \mathrm{d}\overline{s}\, \langle G' \ldots \rangle_0 + (\Delta \text{ correction}). \tag{3.21$b$}$$

### 3.3.1. **Equation (3.21*a*)**

I show in appendix E that ${}_a\breve{\boldsymbol{K}}$ gives the hydrodynamic part of the Navier–Stokes exchange effects as calculated in Part I.

### 3.3.2. **Equation (3.21*b*)**

If one generalizes the above quote that surrounds (3.14) by replacing 'gradient operators' with 'gradient or $\Delta$ operators', one might be led to conclude that (3.21*b*) is negligible, apparently giving a contribution that is at least of third order. However, the discussion in §I:3.2 of the multicomponent plasma shows that there is a difficulty with this argument for the multispecies case, for there it was shown that the nonhydrodynamic part of the momentum exchange is not negligible but is rather of order unity relative to the hydrodynamic part. Manipulations identical to those done for (3.7*b*) show that

$$\lim_{t\to\infty} \int_0^t \mathrm{d}\overline{\tau}\, {}_b\breve{K}_G^{\overline{\beta}}(\mu, \overline{\mu}, \overline{\tau}) = \left\langle \widehat{G}\widehat{\mathrm{C}}^{-1}\mathrm{P}\widehat{\mathrm{C}} \int_0^\infty \mathrm{d}\overline{s}\, U(\overline{s}) \widehat{A}'^{\overline{\beta}}_\Delta \right\rangle. \tag{3.22}$$

Upon comparing this result to the analogous calculations in Part I, one sees that the last integral is essentially the orthogonal projection $|\mathrm{Q}\chi\rangle$, and the construction $\widehat{\mathrm{C}}^{-1}\mathrm{P}\widehat{\mathrm{C}}|\mathrm{Q}\chi\rangle$

is the one that would follow from (I:3.35) by formally multiplying that equation through by $\widehat{\mathrm{C}}^{-1}$. Thus, the correct answer for the nonhydrodynamic part of the exchange effect follows from (3.21$b$).

### 3.4. *Term (iii)*

Term ($\mathrm{iii_b}$) is explicitly

$$\frac{1}{2}\langle G'\mathrm{Q}\psi^2_{\Delta\boldsymbol{B}}\rangle_0 = \frac{1}{2}\left\langle G'\mathrm{Q}\left(-\int_0^t \mathrm{d}\overline{\tau}\,\mathrm{U}(\overline{\tau})\mathrm{i}\mathscr{L}\boldsymbol{A}'\star\Delta\boldsymbol{B}(t-\overline{\tau})\right)\right.$$
$$\left.\times\left(-\int_0^t \mathrm{d}\overline{\tau}'\,\mathrm{U}(\overline{\tau}')\mathrm{i}\mathscr{L}\boldsymbol{A}'\star\Delta\boldsymbol{B}(t-\overline{\tau}')\right)\right\rangle_0. \tag{3.23}$$

Because this expression contains two factors of $\mathrm{i}\mathscr{L}\boldsymbol{A}'$, each of which generates either gradients or exchange effects, it is adequate through second order to use the time-local forms $\Delta\boldsymbol{B}(t)$. One then has a construction of the form

$$\frac{1}{2}\int_0^t \mathrm{d}\overline{\tau}\int_0^t \mathrm{d}\overline{\tau}'\,F(\overline{\tau},\overline{\tau}') = \int_0^t \mathrm{d}\overline{\tau}\int_0^{\overline{\tau}} \mathrm{d}\overline{\tau}'\,F(\overline{\tau},\overline{\tau}'), \tag{3.24}$$

upon using the fact that $F$ is a symmetric function. For the $\overline{\tau}'$ integral, invoke the identity[35]

$$\int_0^\tau \mathrm{d}\overline{\tau}\,\mathrm{U}(\overline{\tau})\mathrm{i}\mathscr{L}\boldsymbol{A} = \boldsymbol{A} - \mathrm{e}^{-\mathrm{Qi}\mathscr{L}\tau}\boldsymbol{A} \tag{3.25}$$

with $\tau$ being replaced by $\overline{\tau}$. The contribution from the $\boldsymbol{A}$ term (with $\boldsymbol{A}\to\boldsymbol{A}'$) cancels term ($\mathrm{iii_a}$). Furthermore, one has[36]

$$\mathrm{U}(\tau) \doteq \mathrm{Q}\mathrm{e}^{-\mathrm{Qi}\mathscr{L}\mathrm{Q}\tau}\mathrm{Q} = \mathrm{e}^{-\mathrm{Qi}\mathscr{L}\tau}\mathrm{Q} = \mathrm{R}_1(\tau)\mathrm{Q}. \tag{3.26}$$

In the last expression, I have used one instance of the shorthand notation

$$\mathrm{R}_0(\tau) \doteq \mathrm{e}^{-\mathrm{i}\mathscr{L}\tau},\quad \mathrm{R}_1(\tau) \doteq \mathrm{e}^{-\mathrm{Qi}\mathscr{L}\tau},\quad \mathrm{R}_2(\tau) \doteq \mathrm{e}^{-\mathrm{Qi}\mathscr{L}\mathrm{Q}\tau}. \tag{3.27}$$

Thus,

$$\text{term (iii)} = -\int_0^t \mathrm{d}\overline{\tau}\int \mathrm{d}\overline{\mu}\int \mathrm{d}\overline{\mu}'\,\langle\widehat{G}(\mu)[\mathrm{R}_1(\overline{\tau})\mathrm{Qi}\mathscr{L}A'^{\overline{\beta}}(\overline{\mu})]\mathrm{R}_1(\overline{\tau})A'^{\overline{\gamma}}(\overline{\mu}')\rangle_0$$
$$\times\,\Delta B_{\overline{\beta}}(\boldsymbol{r}-\overline{\boldsymbol{\rho}},t)\Delta B_{\overline{\gamma}}(\boldsymbol{r}-\overline{\boldsymbol{\rho}}',t). \tag{3.28}$$

The first $\Delta\boldsymbol{B}$ is operated on by $\mathrm{i}\mathscr{L}\boldsymbol{A}'$, so it is already of first order. Because $\mathrm{R}_1(\overline{\tau}) = 1 + O(\mathrm{Qi}\mathscr{L}\overline{\tau})$, use of the expansion (3.3) shows that the second $\Delta\boldsymbol{B}$ factor is also at least of first order.

---

[35] It is easy to verify that (3.25) is true by showing that the time derivatives of both sides are equal. The derivative of the left-hand side is $\mathrm{U}(\tau)\mathrm{i}\mathscr{L}A$, while the derivative of the right-hand side is $\mathrm{e}^{-\mathrm{Qi}\mathscr{L}\tau}\mathrm{Qi}\mathscr{L}A = \mathrm{Q}\mathrm{e}^{-\mathrm{Qi}\mathscr{L}\tau}\mathrm{Qi}\mathscr{L}A = \mathrm{U}(\tau)\mathrm{i}\mathscr{L}A$. The integration constant is zero because both sides vanish at $\tau = 0$. An explicit proof follows by performing the time integration:

$$\int_0^\tau \mathrm{d}\overline{\tau}\,\mathrm{Q}\mathrm{e}^{-\mathrm{Qi}\mathscr{L}\mathrm{Q}\overline{\tau}}\mathrm{Qi}\mathscr{L}A = \int_0^\tau \mathrm{d}\overline{\tau}\,\mathrm{e}^{-\mathrm{Qi}\mathscr{L}\overline{\tau}}\mathrm{Qi}\mathscr{L}A$$
$$= -(\mathrm{e}^{-\mathrm{Qi}\mathscr{L}\tau}-1)(\mathrm{Qi}\mathscr{L})^{-1}\mathrm{Qi}\mathscr{L}A = A - \mathrm{e}^{-\mathrm{Qi}\mathscr{L}\tau}A,$$

provided that $\mathrm{Q}\mathscr{L}$ is invertible. That can be argued from the properties of Q.

[36] The identity (3.26) can be verified by comparing the Taylor expansions of the exponentials and using the fact that $\mathrm{Q}^2 = \mathrm{Q}$.

With exchange effects neglected, one may replace $\mathrm{Qi}\mathscr{L}\boldsymbol{A}'(\overline{\mu})$ by $\mathrm{Q}\overline{\boldsymbol{\nabla}}\boldsymbol{\cdot}\boldsymbol{J}(\overline{\mu})$. Upon integrating the $\overline{\boldsymbol{\nabla}}$ by parts and using the definition $\overline{\boldsymbol{\rho}} \doteq \boldsymbol{r} - \overline{\boldsymbol{x}}$ and the expansion (3.3), one finds

$$\text{term (iii)} = -\int_0^t \mathrm{d}\overline{\tau} \int \mathrm{d}\overline{\mu} \int \mathrm{d}\overline{\mu}'\,\langle \widehat{G}(\mu)[\mathrm{R}_1(\overline{\tau})\widehat{\boldsymbol{J}}^{\overline{\beta}}(\overline{\mu})]\mathrm{R}_1(\overline{\tau})A'^{\overline{\gamma}}(\overline{\mu}')\rangle_0 \boldsymbol{\cdot} [\boldsymbol{\nabla} B_{\overline{\beta}}(\boldsymbol{r},t)]\Delta B_{\overline{\gamma}}(\boldsymbol{r}-\overline{\boldsymbol{\rho}}',t). \tag{3.29}$$

Although both factors of expression (3.29) involve $\mathrm{R}_1(\overline{\tau})$, the first $\mathrm{R}_1$ acts on the subtracted flux $\widehat{\boldsymbol{J}}$ (which lives in the orthogonal subspace) while the second one acts merely on $\boldsymbol{A}'$ (which lives in the hydrodynamic subspace). To lowest order in the gradients, one sees from (3.14) that the first $\mathrm{R}_1$ may be replaced by $\mathrm{R}_0$. Brey *et al.* show that it is fruitful to manipulate the second $\mathrm{R}_1$ by using the identity

$$\mathrm{e}^{-\mathrm{Qi}\mathscr{L}\tau} = \mathrm{e}^{-\mathrm{i}\mathscr{L}\tau} + \int_0^\tau \mathrm{d}\overline{\tau}\,\mathrm{e}^{-\mathrm{i}\mathscr{L}(\tau-\overline{\tau})}\mathrm{Pi}\mathscr{L}\mathrm{e}^{-\mathrm{Qi}\mathscr{L}\overline{\tau}}. \tag{3.30}$$

Thus, term (iii) = term (iii–c) + term (iii–d), where

$$\text{term (iii–c)} \doteq \int_0^t \mathrm{d}\overline{\tau} \int \mathrm{d}\overline{\mu} \int \mathrm{d}\overline{\mu}'\,\langle \widehat{G}(\mu)[\mathrm{R}_0(\overline{\tau})\widehat{\boldsymbol{J}}^{\overline{\beta}}(\overline{\mu})]\mathrm{R}_0(\overline{\tau})A'^{\overline{\gamma}'}(\overline{\mu}')\rangle_0 \boldsymbol{\cdot}\boldsymbol{\nabla} B_{\overline{\beta}}(\boldsymbol{r},t)\Delta B_{\overline{\gamma}'}(\boldsymbol{r}-\overline{\boldsymbol{\rho}}',t), \tag{3.31a}$$

$$\text{term (iii–d)} \doteq \int_0^t \mathrm{d}\overline{\tau} \int \mathrm{d}\overline{\mu} \int \mathrm{d}\overline{\mu}'\,\langle \widehat{G}(\mu)[\mathrm{R}_0(\overline{\tau})\widehat{\boldsymbol{J}}^{\overline{\beta}}(\overline{\mu})] \boldsymbol{\cdot} \int_0^\tau \mathrm{d}\overline{\tau}\,\mathrm{R}_0(\overline{\tau}-\overline{\tau}')\mathrm{Pi}\mathscr{L}\mathrm{R}_1(\overline{\tau}')A'^{\overline{\gamma}'}(\overline{\mu}')\rangle_0 \boldsymbol{\nabla} B_{\overline{\beta}}(\boldsymbol{r},t)\Delta B_{\overline{\gamma}'}(\boldsymbol{r}-\overline{\boldsymbol{\rho}}',t). \tag{3.31b}$$

In term (iii–c), Taylor expansion gives $\Delta B_{\overline{\gamma}'}(\boldsymbol{r}-\boldsymbol{\rho}',t) \approx \Delta B_{\overline{\gamma}',\overline{s}'s}(\boldsymbol{r}) - \overline{\boldsymbol{\rho}}'\boldsymbol{\cdot}\boldsymbol{\nabla} B_{\overline{\gamma}'}(\boldsymbol{r},t)$. $\Delta\boldsymbol{B}_{\overline{s}'s}$ is again neglected, as it generates an exchange effect. Thus, term (iii–c) reduces to

$$\text{term (iii–c)} \approx -\boldsymbol{N}_2^{\overline{\beta}\overline{\gamma}}[G](\mu,t) : \boldsymbol{\nabla} B_{\overline{\beta}}\boldsymbol{\nabla} B_{\overline{\gamma}}, \tag{3.32}$$

where

$$\boldsymbol{N}_2^{\overline{\beta}\overline{\gamma}}[G](\mu,t) \doteq -\int_0^\infty \mathrm{d}\overline{\tau} \int \mathrm{d}\overline{\boldsymbol{x}}\,\boldsymbol{N}_2^{\overline{\beta}\overline{\gamma}}[G](\mu,\overline{\mu},\overline{\tau};t)\overline{\boldsymbol{\rho}} \tag{3.33}$$

and

$$\boldsymbol{N}_2^{\overline{\beta}\overline{\gamma}}[G](\mu,\overline{\mu},\overline{\tau};t) \doteq \langle[\mathrm{e}^{\mathrm{i}\mathscr{L}\overline{\tau}}\widehat{G}(\mu)][\widehat{\boldsymbol{\mathscr{J}}}^{\overline{\beta}}A'^{\overline{\gamma}}(\overline{\mu})]\rangle_0. \tag{3.34}$$

When only kinetic contributions are used in $\widehat{G}$, $\widehat{\boldsymbol{\mathscr{J}}}$, and $\boldsymbol{A}'$, it is easy to see that (3.34) involves the Klimontovich correlation function for three phase-space points.

The further reduction of term (iii–d) is described in appendix B of Brey *et al.* (1981), who also use results from their appendix A, in which a representation of $(\partial_t\boldsymbol{B})^{(1)}$ is derived. The final result, correct to second order in the gradients and in the absence of exchange effects, is

$$\text{term (iii–d)} = \mathscr{M}_2^{\overline{\beta}\overline{\gamma}}[G](\mu,t)\boldsymbol{\cdot}\boldsymbol{\nabla} B_{\overline{\beta}}(\boldsymbol{r},t)[\partial_t B_{\overline{\gamma}}(\boldsymbol{r},t)]^{(1)}, \tag{3.35}$$

where

$$\mathscr{M}_2^{\beta\gamma}[G](\mu,t) \doteq \int_0^\infty \mathrm{d}\overline{\tau} \int \mathrm{d}\overline{\boldsymbol{x}}\,\boldsymbol{N}_2^{\beta\gamma}[G](\mu,\overline{\mu},\overline{\tau};t)\overline{\tau}. \tag{3.36}$$

## 3.5. *Term (iv)*

We have

$$\text{term (iv)} \doteq \langle G'\psi_{\Delta\boldsymbol{B}^{(2)}}\rangle_0, \tag{3.37}$$

where $\Delta\boldsymbol{B}^{(2)}$ is given by (2.74). Although both of the explicit terms in (2.74) are of second order in $\Delta\boldsymbol{B}$, it is not difficult to show, upon integration by parts, that the second term is of third order in the gradients, hence is negligible. The evaluation of the first term is stated by Brey *et al.* to be

$$\text{term (iv)} \approx -\int_0^\infty d\overline{\tau}\int d\overline{\boldsymbol{x}}\,\langle[e^{i\mathscr{L}\overline{\tau}}\widehat{G}(\mu)]\mathrm{P}\widehat{\boldsymbol{\mathscr{J}}}^{\overline{\beta}}A'^{\overline{\gamma}}(\overline{\mu})\rangle_0 \cdot \boldsymbol{\nabla}B_{\overline{\beta}}(\boldsymbol{r},t)\overline{\boldsymbol{\rho}}\cdot\boldsymbol{\nabla}B_{\overline{\gamma}}(\boldsymbol{r},t). \tag{3.38}$$

Since they do not give the detailed manipulations, I present a proof in appendix F. Note that this term adds to term (iii–c) to change the $\boldsymbol{\mathscr{J}}\boldsymbol{A}'$ in (3.34) to $\mathrm{Q}(\boldsymbol{\mathscr{J}}\boldsymbol{A}')$.

## 3.6. *Summary of the gradient expansion*

One can now collect all of the terms. Brey *et al.* show that the terms involving $(\partial_t\boldsymbol{B})^{(1)}$ can be combined. One ultimately finds

$$\begin{aligned}\langle G\rangle = \langle G\rangle_{\text{Euler}} &- \boldsymbol{k}_{1\boldsymbol{\nabla}}^{\overline{\beta}}[G](\mu,t)\cdot\boldsymbol{\nabla}B_{\overline{\beta}}(\boldsymbol{r},t) - k_{1\Delta}^{\overline{\beta}}[G](\mu,t)B_{\overline{\beta}}(\boldsymbol{r},t)\\ &- \boldsymbol{g}_2^{\overline{\beta}}[G](\mu,t) : \boldsymbol{\nabla}\boldsymbol{\nabla}B_{\overline{\beta}}(\boldsymbol{r},t) - \boldsymbol{h}_2^{\overline{\beta}\overline{\gamma}}[\mathrm{G}](\mu,t) : \boldsymbol{\nabla}B_{\overline{\beta}}(\boldsymbol{r},t)\boldsymbol{\nabla}B_{\overline{\gamma}}(\boldsymbol{r},t)\\ &+ \left[\partial_t\left(\boldsymbol{k}_2^{\overline{\beta}}[G](\mu,t)\cdot\boldsymbol{\nabla}B_{\overline{\beta}}(\boldsymbol{r},t)\right)\right]^{(1)},\end{aligned} \tag{3.39}$$

where

$$\boldsymbol{k}_{1\boldsymbol{\nabla}}^{\overline{\beta}}[G](\mu,t) \doteq \int_0^\infty d\overline{\tau}\int d\overline{\mu}\,\langle\widehat{G}(\mu)e^{-i\mathscr{L}\overline{\tau}}\widehat{\boldsymbol{J}}^{\overline{\beta}}(\overline{\mu})\rangle_0, \tag{3.40a}$$

$$k_{1\Delta}^{\overline{\beta}}[G](\mu,t) \doteq \int_0^\infty d\overline{\tau}\int d\overline{\mu}\,\langle\widehat{G}(\mu)e^{-i\mathrm{Q}\mathscr{L}\overline{\tau}}\dot{A}'^{\overline{\beta}}_{\Delta,\overline{s}}(\overline{\mu})\rangle_0, \tag{3.40b}$$

$$\boldsymbol{g}_2^{\overline{\beta}}[G](\mu,t) \doteq -\int d\overline{\tau}\int d\overline{\mu}\,\langle\widehat{G}(\mu)e^{-i\mathscr{L}\overline{\tau}}\widehat{\boldsymbol{J}}^{\overline{\beta}}(\overline{\mu})\rangle_0\overline{\boldsymbol{\rho}}, \tag{3.40c}$$

$$\boldsymbol{h}_2^{\overline{\beta}\overline{\gamma}}[G](\mu,t) \doteq -\int_0^\infty d\overline{\tau}\int d\overline{\mu}\,\langle\widehat{G}(\mu)e^{-i\mathscr{L}\overline{\tau}}\mathrm{Q}[\widehat{\boldsymbol{\mathscr{J}}}^{\overline{\beta}}A'^{\overline{\gamma}}(\overline{\mu})]\rangle_0\overline{\boldsymbol{\rho}}, \tag{3.40d}$$

$$\boldsymbol{k}_2^{\beta}[G](\mu,t) \doteq \int_0^\infty d\overline{\tau}\int d\overline{\mu}\,\langle\widehat{G}(\mu)e^{-i\mathscr{L}\overline{\tau}}\widehat{\boldsymbol{\mathscr{J}}}^{\beta}\rangle_0\overline{\tau}. \tag{3.40e}$$

We have now found a formula for the nonequilibrium average of any quantity $G$, correct through first order in exchange terms and second order in gradients. That formula can be used to evaluate the right-hand side of the equation for $\partial_t\boldsymbol{a}_s$ [the average of equations (2.26)]. The averages of the conserved fluxes are

$$\begin{aligned}\langle\boldsymbol{J}_s^\alpha\rangle = \langle\boldsymbol{J}_s^\alpha\rangle_{\text{Euler}} &- \underbrace{\boldsymbol{k}_1^{\overline{\beta}}[\boldsymbol{J}_s^\alpha](\mu,t)\cdot\boldsymbol{\nabla}B_{\overline{\beta}}(\boldsymbol{r},t)}_{\boldsymbol{\nabla}\mathrm{NS}^\alpha_{\overline{\beta}}} - \underbrace{\boldsymbol{g}_2^{\overline{\beta}}[\boldsymbol{J}_s^\alpha](\mu,t) : \boldsymbol{\nabla}\boldsymbol{\nabla}B_{\overline{\beta}}(\boldsymbol{r},t)}_{\mathrm{B}^\alpha_{\overline{\beta}}}\\ &- \underbrace{\boldsymbol{h}_2^{\overline{\beta}\overline{\gamma}}[\boldsymbol{J}_s^\alpha](\mu,t) : \boldsymbol{\nabla}B_{\overline{\beta}}(\boldsymbol{r},t)\boldsymbol{\nabla}B_{\overline{\gamma}}(\boldsymbol{r},t)}_{\mathrm{B}^\alpha_{\overline{\beta}\overline{\gamma}}} + \underbrace{\left[\partial_t\left(\boldsymbol{k}_2^{\overline{\beta}}[\boldsymbol{J}_s^\alpha](\mu,t)\cdot\boldsymbol{\nabla}B_{\overline{\beta}}(\boldsymbol{r},t)\right)\right]^{(1)}}_{\mathrm{B}^\alpha\partial_t\boldsymbol{k}_{\overline{\beta}} + \mathrm{B}^\alpha\partial_t B_{\overline{\beta}}},\end{aligned} \tag{3.41}$$

and the averages of the exchange terms are

$$X_s^\alpha \equiv \langle \dot{A}'^{\alpha}_{\Delta,s} \rangle = \underbrace{-\,k_{1\Delta}^{\overline{\beta}}[\dot{A}'^{\alpha}_{\Delta,s}](\mu,t)B_{\overline{\beta}}(\boldsymbol{r},t)}_{{}_\Delta\mathrm{NS}^{\alpha}_{\overline{\beta}}}. \tag{3.42}$$

In the above, the various terms have been concisely identified for future reference. NS and B stand for Navier–Stokes and Burnett, respectively. The field indices can assume the values $n$, $\boldsymbol{p}$, or $e$. For example, in the momentum equation the linear Burnett term generates contributions $\mathrm{B}^{\boldsymbol{p}}_{\boldsymbol{p}}$ and $\mathrm{B}^{\boldsymbol{p}}_{e}$.

## 4. Some specific formulas for transport coefficients

In §4.1 I follow Brey (1983) and write out the general structure of the dissipative fluid equations for the unmagnetised, one-component fluid. Integrals that define the transport coefficients are recorded in §4.2. Then in §4.3 I comment on the first-order exchange effects.

### 4.1. *The Burnett equations for a one-component fluid*

An important reference case is the one-component fluid. Those results were recorded for $\boldsymbol{B}^{\mathrm{ext}} = \boldsymbol{0}$ in appendix A of Brey (1983), and I shall transcribe them here, following Brey's numbering conventions for the various dissipative coefficients. The results will be used in §6 to demonstrate the consistency of the calculations of Catto & Simakov of parallel viscosity with the two-time formalism.

Instead of Brey's $\eta$ and $\lambda$, I shall use $\mu \doteq \eta/nm$ and $\kappa \doteq \lambda/n$, which have the dimensions of a diffusion coefficient. Subscripted $\mu$'s and $\kappa$'s relate to the Burnett corrections and have various dimensions. Some supporting algebra is given by Krommes (2018*c*, §2 and §3).

In the following definitions, $d$ denotes the dimension of space, $e$ denotes the internal-energy density, $h$ denotes the enthalpy, and $s$ denotes the entropy density. For an ideal gas, one has

$$p \to nT, \quad e \to \frac{3}{2}nT, \quad h \to \frac{5}{2}nT, \quad s \to \ln\left[\frac{1}{n}\left(\frac{4\pi m e}{3h_{\mathrm{P}}^2}\right)^{3/2}\right] + \frac{5}{2}. \tag{4.1}$$

(In the last expression, $h_{\mathrm{P}}$ is Planck's constant.) The expansion coefficient $\alpha$ and isothermal compressibility $\kappa_T$ are

$$\alpha \doteq -\frac{1}{n}\left(\frac{\partial n}{\partial T}\right)_p \to \frac{1}{T}, \quad \kappa_T \doteq \frac{1}{n}\left(\frac{\partial n}{\partial p}\right)_T \to \frac{1}{p}. \tag{4.2}$$

The strain rate and vorticity tensors are

$$\boldsymbol{S} \doteq \frac{1}{2}[(\boldsymbol{\nabla u}) + (\boldsymbol{\nabla u})^{\mathrm{T}}], \quad \boldsymbol{\varOmega} \doteq \frac{1}{2}[(\boldsymbol{\nabla u})^{\mathrm{T}} - (\boldsymbol{\nabla u})]; \tag{4.3$a$}$$

the traceless tensor

$$\boldsymbol{W} \doteq (\boldsymbol{\nabla u}) + (\boldsymbol{\nabla u})^{\mathrm{T}} - \frac{2}{d}(\boldsymbol{\nabla}\cdot\boldsymbol{u})\boldsymbol{I} \tag{4.4}$$

is also useful.

4.1.1. The one-component Euler equations

Let us write $\boldsymbol{\tau} = p\mathbf{I} + \boldsymbol{\pi}$. Then the one-component Euler equations are

$$\partial_t n = -\boldsymbol{\nabla}\cdot(n\boldsymbol{u}), \tag{4.5a}$$

$$mn(\partial_t \boldsymbol{u} + \boldsymbol{u}\cdot\boldsymbol{\nabla}\boldsymbol{u}) = -nq(\boldsymbol{E} + c^{-1}\boldsymbol{u}\times\boldsymbol{B}^{\mathrm{ext}}) - \boldsymbol{\nabla}p, \tag{4.5b}$$

$$\partial_t T + \boldsymbol{u}\cdot\boldsymbol{\nabla}T = -T\left(\frac{\partial p}{\partial e}\right)\bigg|_n \boldsymbol{\nabla}\cdot\boldsymbol{u}. \tag{4.5c}$$

For the derivation of (4.5$c$), see the discussion of (II-S:3.126).

4.1.2. Dissipative momentum flux for a one-component fluid

The dissipative momentum flux through Burnett order is

$$\begin{aligned}\boldsymbol{\pi}/nm = &-\mu\mathbf{W} - \zeta(\boldsymbol{\nabla}\cdot\boldsymbol{u})\mathbf{I}\\ &- 2\mu_1(\boldsymbol{\nabla}\boldsymbol{\nabla}p - \kappa_T\boldsymbol{\nabla}p\,\boldsymbol{\nabla}p) + (\mu_7 - \alpha\mu_1)(\boldsymbol{\nabla}p\,\boldsymbol{\nabla}T + \boldsymbol{\nabla}T\,\boldsymbol{\nabla}p)\\ &+ \mu_3\boldsymbol{\nabla}\boldsymbol{\nabla}T + \mu_5\boldsymbol{\nabla}T\,\boldsymbol{\nabla}T\\ &+ \left[\mu_{11} - 2\left(\frac{\partial(n\mu_1)}{\partial n}\right)_s\right](\boldsymbol{\nabla}\cdot\boldsymbol{u})\mathbf{S} + (\mu_{12} - 2\mu_1)\mathbf{S}\cdot\mathbf{S}^{\mathrm{T}} + 2\mu_1\boldsymbol{\Omega}\cdot\boldsymbol{\Omega}^{\mathrm{T}}\\ &+ \mu_{13}(\mathbf{S}^{\mathrm{T}}\cdot\boldsymbol{\Omega} + \boldsymbol{\Omega}^{\mathrm{T}}\cdot\mathbf{S})\\ &+ \Bigg\{-\mu_2[\nabla^2 p - \kappa_T|\boldsymbol{\nabla}p|^2] + \mu_4\nabla^2 T + \mu_6|\boldsymbol{\nabla}T|^2 + (\mu_8 - \alpha\mu_2)\boldsymbol{\nabla}p\cdot\boldsymbol{\nabla}T\\ &\quad + \left[\mu_9 - \left(\frac{\partial(n\mu_2)}{\partial n}\right)_s\right](\boldsymbol{\nabla}\cdot\boldsymbol{u})^2 + (\mu_{10} - \mu_2)\,\mathrm{Tr}(\mathbf{S}\cdot\mathbf{S}^{\mathrm{T}}) + \mu_2\,\mathrm{Tr}(\boldsymbol{\Omega}\cdot\boldsymbol{\Omega}^{\mathrm{T}})\Bigg\}\mathbf{I}.\end{aligned} \tag{4.6}$$

Here the first line gives the Navier–Stokes result; the subsequent lines are the Burnett corrections. The viscosities are defined in the next several paragraphs in terms of certain $K$ quantities that are the integrals of two-time correlation functions and are defined in §4.2. In the following definitions, underbracing indicates the value of an expression for an ideal gas. The origins of the various terms can be traced by noting which $K$ quantity enters the expression for a particular viscosity, then referring to §4.2, where each formula is linked to one or more[37] of the terms in the general result (3.39). $K$'s with wavy underlining, such as $\utilde{K}_3$, are determined by correlation functions with three phase-space points; the others follow from two-point correlations.

The Navier–Stokes viscosity coefficients are

$$\mu \doteq (nmT)^{-1}K^{\mathrm{I}}, \tag{4.7a}$$

$$\zeta \doteq (nmT)^{-1}\left(K^{\mathrm{II}} + \frac{2}{d}K^{\mathrm{I}}\right), \tag{4.7b}$$

where $\mu$ is the kinematic viscosity and $\zeta$ is the bulk viscosity.

The Burnett momentum coefficients are

$$\mu_1 \doteq (nmT)^{-1}K^{\mathrm{IV}}, \tag{4.8a}$$

$$\mu_2 \doteq (nmT)^{-1}K^{\mathrm{V}}, \tag{4.8b}$$

$$\mu_3 \doteq -2(nmT^2)^{-1}K_1, \tag{4.8c}$$

$$\mu_4 \doteq -(nmT^2)^{-1}K_2, \tag{4.8d}$$

[37] For the cases involving projections, multiple terms lead to each of the integrals $K_{20}$–$K_{23}$. For more specific information, please consult the supplementary details in Krommes (2018$c$).

$$\mu_5 \doteq 4(nmT^3)^{-1}K_1 + 2\underbrace{(h/nT)}_{5/2}(nmT^3)^{-1}\underset{\sim}{K_3} - 2(nmT^4)^{-1}\underset{\sim}{K_5} + 2n^{-1}(mT)^{-2}\underbrace{\frac{\partial}{\partial T}\left(\frac{h}{n}\right)_p}_{5/2} K_{20}, \tag{4.8e}$$

$$\mu_6 \doteq 2(nmT^3)^{-1}K_2 + (nmT^3)^{-1}\underbrace{(h/nT)}_{5/2}\underset{\sim}{K_4} - (nmT^4)^{-1}\underset{\sim}{K_6} + n^{-1}(mT)^{-2}\underbrace{\frac{\partial}{\partial T}\left(\frac{h}{n}\right)_p}_{5/2} K_{21}, \tag{4.8f}$$

$$\mu_7 \doteq -(n^2mT^3)^{-1}\underset{\sim}{K_3} + n^{-1}(mT)^{-2}\underbrace{\frac{\partial}{\partial p}\left(\frac{h}{n}\right)_T}_{0} K_{20}, \tag{4.8g}$$

$$\mu_8 \doteq -(n^2mT^3)^{-1}\underset{\sim}{K_4} + n^{-1}(mT)^{-2}\underbrace{\frac{\partial}{\partial p}\left(\frac{h}{n}\right)_T}_{0} K_{21}, \tag{4.8h}$$

$$\mu_9 \doteq -(nmT^2)^{-1}\underset{\sim}{K_7} - (nmT)^{-1}\underbrace{\left(\frac{\partial p}{\partial e}\right)_n}_{2/3} K_{21}, \tag{4.8i}$$

$$\mu_{10} \doteq -2(nmT^2)^{-1}\underset{\sim}{K_8} + 2(nmT)^{-1}K_{21}, \tag{4.8j}$$

$$\mu_{11} \doteq -2(nmT^2)^{-1}(\underset{\sim}{K_9} + \underset{\sim}{K_{10}}) - 2(nmT)^{-1}\underbrace{\left(\frac{\partial p}{\partial e}\right)_n}_{2/3} K_{20}, \tag{4.8k}$$

$$\mu_{12} \doteq -4(nmT^2)^{-1}(\underset{\sim}{K_{11}} + \underset{\sim}{K_{12)}}) + 4(nmT)^{-1}K_{20}, \tag{4.8l}$$

$$\mu_{13} \doteq -2(nmT^2)^{-1}(\underset{\sim}{K_{11}} - \underset{\sim}{K_{12}}) + 2(nmT)^{-1}K_{20}. \tag{4.8m}$$

For a weakly coupled gas, these coefficients can be evaluated in the ideal-gas limit. In that limit, the fact that $\mathrm{Tr}\,\widehat{\boldsymbol{\tau}} = 0$ provides the constraints

$$K^{\mathrm{II}} = -\frac{2}{d}K^{\mathrm{I}}, \quad K^{\mathrm{V}} = -\frac{2}{d}K^{\mathrm{IV}}, \quad K_2 = -\frac{2}{d}K_1, \quad \underset{\sim}{K_4} = -\frac{2}{d}\underset{\sim}{K_3}, \quad \underset{\sim}{K_6} = -\frac{2}{d}\underset{\sim}{K_5}, \quad K_{21} = -\frac{2}{d}K_{20}. \tag{4.9}$$

Thus, for a weakly coupled gas the bulk viscosity $\zeta$ vanishes [see (4.7$b$)], one has

$$\mu_2 = -\frac{2}{d}\mu_1, \quad \mu_4 = -\frac{1}{d}\mu_3, \quad \mu_6 = -\frac{1}{d}\mu_5, \quad \mu_8 = -\frac{2}{d}\mu_7, \quad \mu_9 = -\frac{1}{d}\mu_{11}, \quad \mu_{10} = -\frac{1}{d}\mu_{12}, \tag{4.10}$$

and the dissipative momentum flux reduces to

$$\boldsymbol{\pi}^{\mathrm{wc}}/nm = -\mu\,\boldsymbol{W}$$

$$
\begin{aligned}
&- 2\mu_1 \left[ \left( \boldsymbol{\nabla}\boldsymbol{\nabla} p - \frac{1}{p} \boldsymbol{\nabla} p \, \boldsymbol{\nabla} p \right) - \frac{1}{d} \left( \nabla^2 p - \frac{1}{p} |\boldsymbol{\nabla} p|^2 \right) \boldsymbol{I} \right] \\
&+ (\mu_7 - T^{-1}\mu_1) \left( \boldsymbol{\nabla} p \, \boldsymbol{\nabla} T + \boldsymbol{\nabla} T \, \boldsymbol{\nabla} p - \frac{2}{d} \boldsymbol{\nabla} p \cdot \boldsymbol{\nabla} T \, \boldsymbol{I} \right) \\
&+ \mu_3 \left( \boldsymbol{\nabla}\boldsymbol{\nabla} T - \frac{1}{d} \nabla^2 T \, \boldsymbol{I} \right) + \mu_5 \left( \boldsymbol{\nabla} T \, \boldsymbol{\nabla} T - \frac{1}{d} |\boldsymbol{\nabla} T|^2 \boldsymbol{I} \right) \\
&+ \left[ \mu_{11} - 2 \left( \frac{\partial (n\mu_1)}{\partial n} \right)_s \right] \left( (\boldsymbol{\nabla} \cdot \boldsymbol{u}) \boldsymbol{S} - \frac{1}{d} (\boldsymbol{\nabla} \cdot \boldsymbol{u})^2 \boldsymbol{I} \right) \\
&+ (\mu_{12} - 2\mu_1) \left( (\boldsymbol{S} \cdot \boldsymbol{S}^{\mathrm{T}} - \boldsymbol{\Omega} \cdot \boldsymbol{\Omega}^{\mathrm{T}}) - \frac{1}{d} \operatorname{Tr} \left( \boldsymbol{S} \cdot \boldsymbol{S}^{\mathrm{T}} - \boldsymbol{\Omega} \cdot \boldsymbol{\Omega}^{\mathrm{T}} \right) \boldsymbol{I} \right) \\
&+ \mu_{13} (\boldsymbol{S}^{\mathrm{T}} \cdot \boldsymbol{\Omega} + \boldsymbol{\Omega}^{\mathrm{T}} \cdot \boldsymbol{S}). \qquad (4.11)
\end{aligned}
$$

Each line is separately traceless.[38] The integral expression for the kinematic viscosity $\mu$ will be shown to agree with the one derived in Part I.

#### 4.1.3. Dissipative heat flux for a one-component fluid

The dissipative heat flux $\boldsymbol{j}^e_{\mathrm{diss}} \equiv \boldsymbol{q}$ through Burnett order is

$$
\begin{aligned}
\boldsymbol{q}/n = {} & n^{-1} \boldsymbol{\pi} \cdot \boldsymbol{u} - \kappa \boldsymbol{\nabla} T \\
& + \Bigg[ \kappa_2 - \kappa_1 T \underbrace{\left( \frac{\partial p}{\partial e} \right)_n}_{3/2} \Bigg] \boldsymbol{\nabla} (\boldsymbol{\nabla} \cdot \boldsymbol{u}) + \kappa_3 \nabla^2 \boldsymbol{u} \\
& + \Bigg\{ \kappa_4 - \kappa_1 \underbrace{\left( \frac{\partial p}{\partial e} \right)_n}_{3/2} - \kappa_1 T \Bigg[ \underbrace{\frac{\partial}{\partial T} \left( \frac{\partial p}{\partial e} \right)_n}_{0} \Bigg]_p - \left( \frac{\partial (n\kappa_1)}{\partial n} \right)_s \Bigg\} (\boldsymbol{\nabla} \cdot \boldsymbol{u}) \boldsymbol{\nabla} T \\
& + [(\kappa_5 - \kappa_1) \boldsymbol{S} + (\kappa_6 + \kappa_1) \boldsymbol{\Omega}] \cdot \boldsymbol{\nabla} T \\
& + \Bigg( \kappa_7 \boldsymbol{S} + \Bigg\{ \kappa_8 - \kappa_1 T \Bigg[ \underbrace{\frac{\partial}{\partial p} \left( \frac{\partial p}{\partial e} \right)_n}_{0} \Bigg]_T \Bigg\} (\boldsymbol{\nabla} \cdot \boldsymbol{u}) \boldsymbol{I} \Bigg) \cdot \boldsymbol{\nabla} p, \qquad (4.12)
\end{aligned}
$$

where the Navier–Stokes thermal diffusivity is

$$\kappa \doteq n^{-1} T^{-2} K^{\mathrm{III}} \qquad (4.13)$$

and the Burnett energy coefficients are

$$\kappa_1 \doteq (nT^2)^{-1} K^{\mathrm{VI}}, \qquad (4.14a)$$

$$\kappa_2 \doteq -(nT)^{-1} (K_1 + K_2), \qquad (4.14b)$$

$$\kappa_3 \doteq -(nT)^{-1} K_1, \qquad (4.14c)$$

$$
\begin{aligned}
\kappa_4 \doteq (nT^2)^{-1} \Bigg\{ & K_1 + K_2 - T^{-1} (\utilde{K_{13}} + \utilde{K_{19}}) + (nT)^{-1} h \, \utilde{K_{17}} \\
& + T \Bigg[ \underbrace{\frac{\partial}{\partial T} \left( \frac{\partial p}{\partial n} \right)_e}_{0} \Bigg]_p K_{22} + T \Bigg[ \underbrace{\frac{\partial}{\partial T} \left( \frac{\partial p}{\partial e} \right)_n}_{0} \Bigg]_p K_{23} \Bigg\}, \qquad (4.14d)
\end{aligned}
$$

[38]For the last line of (4.11), recall that the trace of the product of a symmetric and an antisymmetric matrix vanishes.

$$\kappa_5 \doteq (nT^2)^{-1}\Big\{3K_1 + K_2 + 2(nT)^{-1}h\underset{\sim}{K_{16}} - T^{-1}(2\underset{\sim}{K_{18}} + \underset{\sim}{K_{14}} + \underset{\sim}{K_{15}})$$
$$+ \Big[1 + \underbrace{\left(\frac{\partial p}{\partial e}\right)_n}_{2/3}\Big]K_{23} + \Big[\underbrace{\left(\frac{\partial p}{\partial n}\right)_e}_{0} - \underbrace{\frac{h}{n}}_{(5/2)T}\Big]K_{22}\Big\}, \tag{4.14e}$$

$$\kappa_6 \doteq (nT^2)^{-1}\Big\{K_1 - K_2 - T^{-1}(\underset{\sim}{K_{14}} - \underset{\sim}{K_{15}})$$
$$- \Big[1 + \left(\frac{\partial p}{\partial e}\right)_n\Big]K_{23} - \Big[\underbrace{\left(\frac{\partial p}{\partial n}\right)_e}_{0} - \underbrace{\frac{h}{n}}_{(5/2)T}\Big]K_{22}\Big\}, \tag{4.14f}$$

$$\kappa_7 \doteq -2(nT)^{-2}\underset{\sim}{K_{16}} - 2\mu_1, \tag{4.14g}$$

$$\kappa_8 \doteq -(nT)^{-2}\underset{\sim}{K_{17}} - \mu_2 + \frac{1}{nT}\left[\frac{\partial}{\partial p}\underbrace{\left(\frac{\partial p}{\partial n}\right)_e}_{0}\right]_T K_{22} + \frac{1}{nT}\underbrace{\left[\frac{\partial}{\partial p}\left(\frac{\partial p}{\partial e}\right)_n\right]_T}_{0} K_{23}. \tag{4.14h}$$

Note that all of the Burnett energy corrections are negligible when $\boldsymbol{u}$ is ordered small.

### 4.2. *Integrals of correlation functions*

The $\mu$'s and $\kappa$'s that appear in the previous subsections are defined in terms of various integrals of correlation tensors. Symmetry considerations reduce those tensors to a collection of scalar quantities $K^i$ and $K_i$ defined as follows, using Brey's numbering conventions. The origin of each term is indicated by the notation such as $\mathrm{NS}^{\boldsymbol{p}}_{\boldsymbol{p}}$ that was introduced in conjunction with (3.39). The one-component results tabulated here are correct for the unmagnetised case. When $\boldsymbol{B}^{\mathrm{ext}} \neq \mathbf{0}$, additional transport coefficients must be introduced. That was done for linear response in §I:3; I shall eschew that exercise for the Burnett coefficients. For multispecies plasmas, modified propagators must be used instead of $\mathrm{R}_0$.

$$\mathrm{NS}^{\boldsymbol{p}}_{\boldsymbol{p}}: \quad \int_0^\infty d\overline{\tau}\,\langle \widehat{\tau}_{ij}(\mathbf{0})\mathrm{R}_0(\overline{\tau})\widehat{\mathscr{T}_{jk}}\rangle = K^{\mathrm{I}}(\delta_{ik}\delta_{jl} + \delta_{il}\delta_{jk}) + K^{\mathrm{II}}\delta_{ij}\delta_{kl}, \tag{4.15a}$$

$$\mathrm{NS}^{e}_{e}: \quad \int_0^\infty d\overline{\tau}\langle \widehat{J}^E_i(\mathbf{0})\mathrm{R}_0(\overline{\tau})\widehat{\mathscr{J}^E_j}\rangle = K^{\mathrm{III}}\delta_{ij}, \tag{4.15b}$$

$$\mathrm{Bt}^{\boldsymbol{p}}_{\boldsymbol{p}}: \quad \int_0^\infty d\overline{\tau}\,\langle \widehat{\tau}_{ij}(\mathbf{0})\mathrm{R}_0(\overline{\tau})\widehat{\mathscr{T}_{kl}}\rangle\overline{\tau} = K^{\mathrm{IV}}(\delta_{ik}\delta_{jl} + \delta_{il}\delta_{jk}) + K^{\mathrm{V}}\delta_{ij}\delta_{kl}, \tag{4.15c}$$

$$\mathrm{Bt}^{e}_{e}: \quad \int_0^\infty d\overline{\tau}\,\langle \widehat{J}^E_i(\mathbf{0})\mathrm{R}_0(\overline{\tau})\widehat{\mathscr{J}^E_j}\rangle\overline{\tau} = K^{\mathrm{VI}}\delta_{ij}, \tag{4.15d}$$

$$\mathrm{B}^{\boldsymbol{p}}_{e}: \quad \int_0^\infty d\overline{\tau}\int d\overline{\boldsymbol{x}}\,\langle \widehat{\tau}_{ij}(\mathbf{0})\mathrm{R}_0(\overline{\tau})\widehat{J}^E_k(\overline{\boldsymbol{x}})\rangle\overline{x}_l = K_1(\delta_{ik}\delta_{jl} + \delta_{il}\delta_{jk}) + K_2\delta_{ij}\delta_{kl}, \tag{4.15e}$$

$$\mathrm{B}^{\boldsymbol{p}}_{en}: \quad \int d\overline{\tau}\int_0^\infty d\overline{\boldsymbol{x}}\,\langle \widehat{\tau}_{ij}(\mathbf{0})\mathrm{R}_0(\overline{\tau})\widehat{\mathscr{J}^E_k}N'(\overline{\boldsymbol{x}})\rangle\overline{x}_l = \underset{\sim}{K_3}(\delta_{ik}\delta_{jl} + \delta_{il}\delta_{jk}) + \underset{\sim}{K_4}\delta_{ij}\delta_{kl}, \tag{4.15f}$$

$$\mathrm{B}^{\boldsymbol{p}}_{ee}: \quad \int_0^\infty d\overline{\tau}\int d\overline{\boldsymbol{x}}\,\langle \widehat{\tau}_{ij}(\mathbf{0})\mathrm{R}_0(\overline{\tau})\widehat{\mathscr{J}^E_k}E'(\overline{\boldsymbol{x}})\rangle\overline{x}_l = \underset{\sim}{K_5}(\delta_{ik}\delta_{jl} + \delta_{il}\delta_{jk}) + \underset{\sim}{K_6}\delta_{ij}\delta_{kl}, \tag{4.15g}$$

$$\mathrm{B}^{\boldsymbol{p}}_{\boldsymbol{pp}}:\quad \int_0^\infty \mathrm{d}\overline{\tau}\int \mathrm{d}\overline{\boldsymbol{x}}\,\langle\widehat{\tau}_{ij}(\mathbf{0})\mathrm{R}_0(\overline{\tau})\widehat{\mathscr{T}}_{kl}G'_m(\overline{\boldsymbol{x}})\rangle\overline{x}_n$$
$$= \underset{\sim}{K_7}\delta_{ij}\delta_{kl}\delta_{mn} + \underset{\sim}{K_8}\delta_{ij}(\delta_{km}\delta_{ln}+\delta_{kn}\delta_{lm})$$
$$+ \underset{\sim}{K_9}\delta_{kl}(\delta_{im}\delta_{jn}+\delta_{in}\delta_{jm}) + \underset{\sim}{K_{10}}\delta_{mn}(\delta_{ik}\delta_{jl}+\delta_{il}\delta_{jk})$$
$$+ \underset{\sim}{K_{11}}[\delta_{km}(\delta_{il}\delta_{jn}+\delta_{jl}\delta_{in}) + \delta_{lm}(\delta_{ik}\delta_{jn}+\delta_{jk}\delta_{in})]$$
$$+ \underset{\sim}{K_{12}}[\delta_{kn}(\delta_{il}\delta_{jm}+\delta_{jl}\delta_{im}) + \delta_{ln}(\delta_{ik}\delta_{jm}+\delta_{jk}\delta_{im})], \tag{4.15h}$$

$$\mathrm{B}^{e}_{\boldsymbol{ep}}:\quad \int_0^\infty \mathrm{d}\overline{\tau}\int \mathrm{d}\overline{\boldsymbol{x}}\,\langle\widehat{J}^E_i(\mathbf{0})\mathrm{R}_0(\overline{\tau})\widehat{\mathscr{J}}^E_j G'_k(\overline{\boldsymbol{x}})\rangle\overline{x}_l = \underset{\sim}{K_{13}}\delta_{ij}\delta_{kl} + \underset{\sim}{K_{14}}\delta_{ik}\delta_{jl} + \underset{\sim}{K_{15}}\delta_{il}\delta_{jk}, \tag{4.15i}$$

$$\mathrm{B}^{e}_{\boldsymbol{pn}}:\quad \int_0^\infty \mathrm{d}\overline{\tau}\int \mathrm{d}\overline{\boldsymbol{x}}\,\langle\widehat{J}^E_i(\mathbf{0})\mathrm{R}_0(\overline{\tau})\widehat{\mathscr{T}}_{jk}N'(\overline{\boldsymbol{x}})\rangle\overline{x}_l = \underset{\sim}{K_{16}}(\delta_{ij}\delta_{kl}+\delta_{ik}\delta_{jl}) + \underset{\sim}{K_{17}}\delta_{il}\delta_{jk}, \tag{4.15j}$$

$$\mathrm{B}^{e}_{\boldsymbol{pe}}:\quad \int_0^\infty \mathrm{d}\overline{\tau}\int \mathrm{d}\overline{\boldsymbol{x}}\,\langle\widehat{J}^E_i(\mathbf{0})\mathrm{R}_0(\overline{\tau})\widehat{\mathscr{T}}_{jk}E'(\overline{\boldsymbol{x}})\rangle\overline{x}_l = \underset{\sim}{K_{18}}(\delta_{ij}\delta_{kl}+\delta_{ik}\delta_{jl}) + \underset{\sim}{K_{19}}\delta_{il}\delta_{jk}, \tag{4.15k}$$

$$\mathrm{B}^{\boldsymbol{p}}_{-\mathrm{P}(\boldsymbol{pp}+en+ee)}:\quad \int_0^\infty \mathrm{d}\overline{\tau}\int \mathrm{d}\overline{\boldsymbol{x}}\,\langle\widehat{\tau}_{ij}(\mathbf{0})\mathrm{R}_0(\overline{\tau})P_k(\overline{\boldsymbol{x}})\rangle\overline{x}_l$$
$$= K_{20}(\delta_{ik}\delta_{jl}+\delta_{il}\delta_{jk}) + K_{21}\delta_{ij}\delta_{kl}, \tag{4.15l}$$

$$\mathrm{B}^{e}_{-\mathrm{P}(\boldsymbol{p}n+\boldsymbol{p}e+e\boldsymbol{p})}:\quad \int_0^\infty \mathrm{d}\overline{\tau}\int \mathrm{d}\overline{\boldsymbol{x}}\,\langle\widehat{J}^E_i(\mathbf{0})\mathrm{R}_0(\overline{\tau})N(\overline{\boldsymbol{x}})\rangle\overline{x}_j = K_{22}\delta_{ij}, \tag{4.15m}$$

$$\mathrm{B}^{e}_{-\mathrm{P}(\boldsymbol{p}n+\boldsymbol{p}e+e\boldsymbol{p})}:\quad \int_0^\infty \mathrm{d}\overline{\tau}\int \mathrm{d}\overline{\boldsymbol{x}}\,\langle\widehat{J}^E_i(\mathbf{0})\mathrm{R}_0(\overline{\tau})E(\overline{\boldsymbol{x}})\rangle\overline{x}_j = K_{23}\delta_{ij}. \tag{4.15n}$$

These expressions involve two-time correlation functions $C(t,t')$ with $m$ phase-space arguments associated with time $t$ and $n$ phase-space arguments associated with time $t'$; I shall use the notation $C^{(m,n)}(t,t')$. For weakly coupled systems, where the potential parts of the microscopic fluxes may be neglected, one requires only $m=1$ (in later discussion, $m>1$ will be required) and $n=1$ or $n=2$. I shall assume homogeneous, stationary statistics. By spatial homogeneity, $C^{(1,n)}$ is a function of $n$ spatial differences, which by convention I shall refer to $\boldsymbol{x}$ ($\boldsymbol{x}=\mathbf{0}$ in the above formulas). This leads one to introduce $\boldsymbol{\rho}'\doteq\boldsymbol{x}-\boldsymbol{x}'$ and $\boldsymbol{\rho}''\doteq\boldsymbol{x}-\boldsymbol{x}''$. By temporal stationarity, $C^{(m,n)}(t,t')$ depends only on the time difference $\tau\doteq t-t'$; only the one-sided functions $C^{(m;n)}_+(\tau)\doteq H(\tau)C^{(m,n)}(\tau)$ are required. Thus, for example,

$$C^{(1;1)}_+(\underline{1},\tau;\underline{1}') = C^{(1;1)}_+(\boldsymbol{v},\tau;\boldsymbol{\rho}',\boldsymbol{v}') = \int\frac{\mathrm{d}\boldsymbol{k}'}{(2\pi)^3}\,\mathrm{e}^{\mathrm{i}\boldsymbol{k}'\cdot\boldsymbol{\rho}'}\widehat{C}^{(1;1)}_{+;\boldsymbol{k}'}(\boldsymbol{v},\tau;\boldsymbol{v}'), \tag{4.16a}$$
$$C^{(1;2)}_+(\underline{1},\tau;\underline{1}',\underline{1}'') = C^{(1;2)}_+(\boldsymbol{v}_1,\tau;\boldsymbol{\rho}',\boldsymbol{v}',\boldsymbol{\rho}'',\boldsymbol{v}'')$$
$$= \int\frac{\mathrm{d}\boldsymbol{k}'}{(2\pi)^3}\int\frac{\mathrm{d}\boldsymbol{k}''}{(2\pi)^3}\,\mathrm{e}^{\mathrm{i}\boldsymbol{k}'\cdot\boldsymbol{\rho}'+\mathrm{i}\boldsymbol{k}''\cdot\boldsymbol{\rho}'')}\widehat{C}^{(1;2)}_{+;\boldsymbol{k}',\boldsymbol{k}''}(\boldsymbol{v}_1,\tau;\boldsymbol{v}',\boldsymbol{v}''), \tag{4.16b}$$
$$C^{(2;1)}_+(\underline{1},\underline{2};\tau;\underline{1}') = C^{(2;1)}_+(\boldsymbol{\rho},\boldsymbol{v}_1,\boldsymbol{v}_2,\tau;\boldsymbol{\rho}',\boldsymbol{v}')$$
$$= \int\frac{\mathrm{d}\boldsymbol{k}}{(2\pi)^3}\int\frac{\mathrm{d}\boldsymbol{k}'}{(2\pi)^3}\,\mathrm{e}^{\mathrm{i}\boldsymbol{k}\cdot\boldsymbol{\rho}+\mathrm{i}\boldsymbol{k}'\cdot\boldsymbol{\rho}'}\widehat{C}^{(2;1)}_{+,\boldsymbol{k};\boldsymbol{k}'}(\boldsymbol{v}_1,\boldsymbol{v}_2,\tau;\boldsymbol{v}',\boldsymbol{v}''). \tag{4.16c}$$

For weak coupling, formulas that involve the total amounts of a flux (e.g., $\widehat{\mathscr{J}}^E$) involve the spatial integral over $\boldsymbol{x}-\boldsymbol{x}'$ and thus require $\widehat{C}_{\boldsymbol{k}'=\mathbf{0}}$. Formulas that involve an $\overline{\boldsymbol{x}}$ weighting can be expressed in terms of a wavenumber derivative. Specifically, upon replacing $\overline{\boldsymbol{x}}$ by $\boldsymbol{x}''$,

$$\int \mathrm{d}\boldsymbol{x}'\,\mathrm{d}\boldsymbol{x}''\,C_{+}^{(1;2)}(\boldsymbol{x}=\mathbf{0},\boldsymbol{v},\tau;\boldsymbol{x}',\boldsymbol{v}',\boldsymbol{x}'',\boldsymbol{v}'')\boldsymbol{x}''$$

$$= -\int \mathrm{d}\boldsymbol{\rho}'\,\mathrm{d}\boldsymbol{\rho}''\,C_{+}^{(1;2)}(\boldsymbol{v},\tau;\boldsymbol{\rho}',\boldsymbol{v}',\boldsymbol{\rho}'',\boldsymbol{v}'')\boldsymbol{\rho}'' \tag{4.17a}$$

$$= \frac{\partial}{\partial(-\mathrm{i}\boldsymbol{k}'')}\widehat{C}^{(1;2)}_{+;\boldsymbol{k}',\boldsymbol{k}''}(\boldsymbol{v},\tau;\boldsymbol{v}',\boldsymbol{v}'')\bigg|_{\boldsymbol{k}'=\mathbf{0},\,\boldsymbol{k}''=\mathbf{0}}. \tag{4.17b}$$

A discussion of the evaluation of these correlation functions and their time integrals is given in §5.

### 4.3. *Exchange terms*

Equation (3.42), together with the definition (3.40*b*), provides a representation of first-order interspecies momentum and energy exchange in terms of two-time correlations. It is not immediately obvious that those formulas are consistent with the results already known to Braginskii. Therefore, as an example I work out in appendix E the hydrodynamic contribution to momentum exchange and demonstrate complete agreement with the analogous calculation in Part I.

## 5. Theory of two-time correlation functions for weakly coupled plasmas

The formulas in §4.2 express the transport coefficients in terms of various two-time correlation functions. Equations (4.15*f*)–(4.15*k*) require three-point correlations in phase space, while the remainder of equations (4.15) require merely two-point correlations. The exchange terms discussed in appendix E require four-point correlations. Calculating such multipoint correlation functions for the general case of strong coupling is a formidable challenge even though only the low-frequency, long-wavelength behaviour is of interest; however, the task is relatively simple for weak coupling provided that one ignores issues with long-ranged correlations (i.e., does calculations that only retain effects that lead to the Balescu–Lenard or Landau collision operators).

### 5.1. *Heuristic physics of two-time correlations*

Before proceeding to the details, I shall give a qualitative introduction that relies on the fact that cumulants are related to functional derivatives. I have already introduced this topic in §1.3, where I pointed out that the two-time Green's function of a linear equation is the functional derivative of the basic one-time field with respect to an external source $\widehat{\eta}$. The generalization to statistical theory is well known. For example, for continuous classical fields (no particle discreteness effects), Martin *et al.* (1973) have shown that the $n$-time cumulant $C_n$ of the random field $\widetilde{\psi}(t)$ is the functional derivative of $C_{n-1}$ with respect to a source field $\eta(t)$.[39] That is, a *cumulant generating functional*

[39] $\eta$ plays the role of a source in a particular adjoint of the primitive amplitude equation for $\widetilde{\psi}$.

is[40]

$$Z[\eta] \doteq \ln \left\langle \exp\left( \int_0^\infty \mathrm{d}\overline{t}\, \widetilde{\psi}(\overline{t})\eta(\overline{t}) \right) \right\rangle, \tag{5.1}$$

and one has

$$C_1(t) = \frac{\delta Z[\eta]}{\delta\eta(t)}, \quad C_2(t,t') = \frac{\delta^2 Z[\eta]}{\delta\eta(t)\delta\eta(t')} = \frac{\delta C_1(t)}{\delta\eta(t')}, \quad \dots, \tag{5.2}$$

with the physical cumulants following in the limit $\eta \to 0$. For situations in which particle discreteness effects are important, a one-time cumulant generating functional was discussed by Dawson & Nakayama (1967), and the generalization to two-time cumulants was given by Krommes (1975) and Krommes & Oberman (1976*a*). (These topics are discussed in detail in §5.2.) Now suppose that the nonlinear kinetic equation holds (setting $\boldsymbol{B}^{\mathrm{ext}} = \boldsymbol{0}$ for simplicity and assuming a bilinear collision operator such as the Landau operator):

$$\partial_t f + \boldsymbol{v} \cdot \boldsymbol{\nabla} f + (\mathbb{E} f) \cdot \boldsymbol{\partial} f + \mathrm{C}[f, \overline{f}] = 0. \tag{5.3}$$

Here $\mathbb{E}$ is the electric-field operator defined by (I:2.38). Without worrying about details, which will be discussed later, functionally differentiate (5.3) to obtain (Krommes & Oberman 1976*a*)

$$\partial_t C_2(t,t') + \boldsymbol{v} \cdot \boldsymbol{\nabla} C_2 + (\mathbb{E} f) \cdot \boldsymbol{\partial} C + (\boldsymbol{\partial} f) \cdot \mathbb{E} C_2 + \widehat{\mathrm{C}}[f] C_2 = 0, \tag{5.4}$$

where $\widehat{\mathrm{C}}$ is the linearized collision operator. This shows that at long wavelengths the two-time function $C_2(\tau)$ decays on the collisional time scale, and it is only a matter of filling in the details to integrate $C_2$ according to formulas like (4.15*a*) and obtain the same Navier–Stokes transport coefficients that were discussed in Part I. Upon differentiating again, one obtains schematically

$$\frac{\partial C_3(t,t',t'')}{\partial t} + \boldsymbol{v} \cdot \boldsymbol{\nabla} C_3 + \boldsymbol{E} \cdot \boldsymbol{\partial} C_3 + \widehat{\mathrm{C}}[f] C_3 = -\frac{\delta \widehat{\mathrm{C}}[f]}{\delta f} C_2 C_2 - 2(\mathbb{E} C_2) \cdot \boldsymbol{\partial} C_2, \tag{5.5}$$

after which one can set $t'' = t'$. As we shall see later in more detail, the first term on the right-hand side is related to the action of the nonlinear collision operator acting on $C_2$, which then drives the triplet correlation function that determines the non-Gaussian Burnett transport effects. The significance of the second term is discussed in the paragraph before §G.3.1 on page 91.

### 5.2. *The two-time cumulant hierarchy and correlation functions*

I shall now discuss the formal derivation of the previous results. For conciseness, I shall continue to set $\boldsymbol{B}^{\mathrm{ext}} = \boldsymbol{0}$; the way to add the Lorentz force to the final formulas will be clear. Most powerfully, one has available the renormalized theory of Rose (1979), which generalizes continuum statistical dynamics [for example, the formalism of Martin *et al.* (1973)] to include particle discreteness. However, for a weakly coupled, near-equilibrium system it is more expeditious to proceed via a two-time generalization of the BBGKY hierarchy. The one- and two-time hierarchies were discussed in appendix C of Krommes (1975) via a generating-functional approach; the following material is taken directly from that appendix; see also appendix A of Krommes & Oberman (1976*a*). For one-time

[40] Martin *et al.* actually consider a more complicated, time-ordered generating functional that depends on two sources $\eta$ and $\widehat{\eta}$ and is capable of generating both correlation functions and infinitesimal response functions.

physics, define the generating function

$$S[\eta] \doteq \langle \mathrm{e}^{\widetilde{\Phi}_\eta(t)} \rangle, \tag{5.6}$$

where

$$\widetilde{\Phi}_\eta(t) \doteq \int \mathrm{d}\overline{\boldsymbol{q}}\, \widetilde{f}(\overline{\boldsymbol{q}}, t)\eta(\overline{\boldsymbol{q}}) \tag{5.7}$$

is a functional of $\eta(\boldsymbol{q})$. Note that there is no time integration in this formula. The one-time distribution functions are defined in the thermodynamic limit as (Dawson & Nakayama 1967),

$$f_\eta^{(s)}(\underline{1}, \ldots, \underline{s}, t) = \mathrm{D}_s \mathrm{D}_{s-1} \ldots \mathrm{D}_1 \langle S \rangle, \tag{5.8}$$

where

$$\mathrm{D}_s \doteq \frac{\delta}{\delta\eta(\underline{s})} - \sum_{i<s} \overline{n}_s^{-1} \delta(\underline{i} - \underline{s}) \tag{5.9}$$

and an underline signifies that time should be omitted from the implied set of variables. The usual $s$-body distributions are the $\eta \to 0$ limit of $f_\eta^{(s)}$. Similarly, one-time cumulants (denoted here by an overline) follow by differentiating $\ln \langle S \rangle$:

$$\overline{f}_\eta^{(s)} = \mathrm{D}_s \ldots \mathrm{D}_1 \ln \langle S \rangle. \tag{5.10}$$

The properties of the logarithm lead to the well-known cluster expansion

$$f_\eta^{(s)}(\underline{1}, \ldots, \underline{s}) = \sum_{n=1}^{s} \sum_P \overline{f}_\eta^{(N(P_1))}(P_1) \overline{f}_\eta^{(N(P_2))}(P_2) \ldots \overline{f}_\eta^{(N(P_n))}(P_n), \tag{5.11}$$

where $P_i$ is a subset of $\{\underline{1}, \ldots, \underline{s}\}$, $N(P_i)$ is the number of members of $P_i$, $\sum_{i=1}^n N(P_i) = s$, and $\sum_P$ is the sum over all distinct and disjoint subsets of $\{\underline{1}, \underline{2}, \ldots, \underline{s}\}$:

$$\cup_{i=1}^n P_i = \{1, 2, \ldots, s\} \equiv \{s\}. \tag{5.12}$$

To obtain time-evolution equations, consider the time derivative of the generating functional:

$$\frac{\partial S}{\partial t} = \left\langle \frac{\mathrm{d}\widetilde{\Phi}_\eta}{\mathrm{d}t} \mathrm{e}^{\widetilde{\Phi}_\eta} \right\rangle \tag{5.13$a$}$$

$$= \left\langle \left( \int \mathrm{d}\widehat{\boldsymbol{q}}\, \frac{\partial \widetilde{f}(\widehat{\boldsymbol{q}}, t)}{\partial t} \eta(\widehat{\boldsymbol{q}}) \right) \mathrm{e}^{\widetilde{\Phi}_\eta} \right\rangle \tag{5.13$b$}$$

$$= -\int \mathrm{d}\widehat{\boldsymbol{q}} \left\langle \left[ \widehat{\boldsymbol{v}} \cdot \widehat{\boldsymbol{\nabla}} \widetilde{f}(\widehat{\boldsymbol{q}}, t) + \left(\frac{q}{m}\right)_{\widehat{s}} \mathbf{E} \widetilde{f}(\widehat{\boldsymbol{q}}, t) \cdot \frac{\partial \widetilde{f}}{\partial \widehat{\boldsymbol{v}}} \right] \mathrm{e}^{\widetilde{\Phi}_\eta} \right\rangle \eta(\widehat{\boldsymbol{q}}). \tag{5.13$c$}$$

Upon functionally differentiating this equation, one arrives at the usual BBGKY hierarchy

$$0 = \partial_t f_\eta^{(s)} + \sum_{i \in \{s\}} \boldsymbol{v}_i \cdot \boldsymbol{\nabla}_i f_\eta^{(s)} + \sum_{i \neq j \in \{s\}} \boldsymbol{\epsilon}_{ij} \cdot (q_j \boldsymbol{\partial}_i) f_\eta^{(s)} + \sum_{i \in \{s\}} \boldsymbol{\partial}_i \cdot \mathbf{E}(i, s+1) f_\eta^{(s+1)} + O(\eta). \tag{5.14}$$

A similar result holds for the cumulant hierarchy. Upon indicating one-time cumulants with overlines, one has

$$0 = \frac{\mathrm{d}^{(s)}\overline{f}_\eta^{(s)}}{\mathrm{d}t} + \sum_{i\neq j\in\{s\}} \boldsymbol{\epsilon}_{ij}\cdot(q_j\boldsymbol{\partial}_i)\Bigg(\sum_{n=1}^{s-1}\sum_P \overline{f}_\eta^{(s-n)}(\ldots,\underline{i},\ldots)\overline{f}_\eta^{(n)}(\ldots\,\underline{j}\,\ldots) + \overline{f}_\eta^{(s)}(\underline{1},\ldots,\underline{s})\Bigg)$$
$$+ \sum_{i\in\{s\}} \boldsymbol{\partial}_i\cdot\mathbb{E}(i,s+1)\Bigg(\sum_{n=1}^{s}\sum_P \overline{f}_\eta^{(s+1-n)}(\ldots,\underline{i},\ldots)\overline{f}_\eta^{(n)}(\ldots,\underline{s+1},\ldots) + \overline{f}_\eta^{(s+1)}(\underline{1},\ldots,\underline{s+1})\Bigg)$$
$$+ O(\eta). \tag{5.15}$$

Again, $\sum_P$ means to sum over all distinct permutations of disjoint subsets.

I shall use the standard notation $\overline{f}^{(1)} \equiv f$, $\overline{f}^{(2)} \equiv g$, $\overline{f}^{(3)} \equiv h$, and $\overline{f}^{(4)} \equiv k$. Upon noting that $\mathbb{E}(\underline{1},\overline{\underline{1}})f(\overline{\underline{1}},t) = \boldsymbol{E}(1)$ and defining the *Landau operator*[41] for particle 1 as

$$\mathrm{iL}_1 \equiv \mathrm{iL}(1,\overline{1}) \doteq \boldsymbol{v}_1\cdot\boldsymbol{\nabla}_1\delta(\underline{1}-\overline{\underline{1}}) + \boldsymbol{E}(1)\cdot\boldsymbol{\partial}_1\delta(\underline{1}-\overline{\underline{1}}) + \boldsymbol{\partial}_1 f\cdot\mathbb{E}(\underline{1},\overline{\underline{1}}) \tag{5.16}$$

(the last term is responsible for crucial polarization or self-consistent response effects), one can write the first three members of the cumulant hierarchy as

$$0 = \partial_t f(\underline{1},t) + \boldsymbol{v}_1\cdot\boldsymbol{\nabla}_1 + \boldsymbol{E}\cdot\boldsymbol{\partial}_1 f + \boldsymbol{\partial}_1\cdot\mathbb{E}(\underline{1},\overline{\underline{2}})g(\underline{1},\overline{\underline{2}},t), \tag{5.17a}$$
$$0 = (\partial_t + \mathrm{iL}_1 + \mathrm{iL}_2)g(\underline{1},\underline{2},t) + \boldsymbol{\epsilon}_{12}\cdot(q_2\boldsymbol{\partial}_1 - q_1\boldsymbol{\partial}_2)[f(\underline{1},t)f(\underline{2},t) + g(\underline{1},\underline{2},t)]$$
$$+ [\boldsymbol{\partial}_1\cdot\mathbb{E}(\underline{1},\overline{\underline{3}})h(\underline{1},\underline{2},\overline{\underline{3}},t) + (1\leftrightarrow 2)], \tag{5.17b}$$
$$\begin{aligned}
0 = {} & (\partial_t + \mathrm{iL}_1 + \mathrm{iL}_2 + \mathrm{iL}_3)h(\underline{1},\underline{2},\underline{3},t)\\
& + \boldsymbol{\epsilon}_{12}\cdot(q_2\boldsymbol{\partial}_1)[g(\underline{1},\underline{3})f(\underline{2}) + f(\underline{1})g(\underline{2},\underline{3}) + h(\underline{1},\underline{2},\underline{3})]\\
& + \boldsymbol{\epsilon}_{13}\cdot(q_3\boldsymbol{\partial}_1)[g(\underline{1},\underline{2})f(\underline{3}) + f(\underline{1})g(\underline{3},\underline{2}) + h(\underline{1},\underline{2},\underline{3})]\\
& + \boldsymbol{\epsilon}_{21}\cdot(q_1\boldsymbol{\partial}_2)[g(\underline{2},\underline{3})f(\underline{1}) + f(\underline{2})g(\underline{1},\underline{3}) + h(\underline{1},\underline{2},\underline{3})]\\
& + \boldsymbol{\epsilon}_{23}\cdot(q_3\boldsymbol{\partial}_2)[g(\underline{2},\underline{1})f(\underline{3}) + f(\underline{2})g(\underline{3},\underline{1}) + h(\underline{1},\underline{2},\underline{3})]\\
& + \boldsymbol{\epsilon}_{31}\cdot(q_1\boldsymbol{\partial}_3)[g(\underline{3},\underline{2})f(\underline{1}) + f(\underline{3})g(\underline{1},\underline{2}) + h(\underline{1},\underline{2},\underline{3})]\\
& + \boldsymbol{\epsilon}_{32}\cdot(q_2\boldsymbol{\partial}_3)[g(\underline{3},\underline{1})f(\underline{2}) + f(\underline{3})g(\underline{1},\underline{2}) + h(\underline{1},\underline{2},\underline{3})]\\
& + \boldsymbol{\partial}_1\cdot\mathbb{E}(\underline{1},\underline{4})[g(1,2)g(3,4) + g(1,3)g(2,4) + k(\underline{1},\ldots,\underline{4})]\\
& + \boldsymbol{\partial}_2\cdot\mathbb{E}(\underline{2},\underline{4})[g(2,1)g(3,4) + g(2,3)g(1,4) + k(\underline{1},\ldots,\underline{4})]\\
& + \boldsymbol{\partial}_3\cdot\mathbb{E}(\underline{3},\underline{4})[g(3,1)g(2,4) + g(3,2)g(1,4) + k(\underline{1},\ldots,\underline{4})].
\end{aligned} \tag{5.17c}$$

In the above, the terms in $\boldsymbol{\epsilon}_{ij}$ are related to *particle noise*; see Rose (1979, and references therein) for discussion of this concept. If one retains terms only through $O(\epsilon_\mathrm{p})$, the terms in $h$ and $k$ can be ignored.

[41]The Landau operator L, which describes self-consistent linearized Vlasov dynamics, should not be confused with the Landau collision operator $\mathrm{C}^\mathrm{L}$.

### 5.3. *Two-time, two-phase-space-point correlations*

Now consider the extension of these results to two times. Introduce the extended generating functional[42]

$$S_2[\eta,\eta'] \doteq \exp\left(\int \mathrm{d}\overline{\boldsymbol{q}}\,\widetilde{f}(\overline{\boldsymbol{q}},t)\eta(\overline{\boldsymbol{q}}) + \int \mathrm{d}\overline{\boldsymbol{q}}'\,\widetilde{f}(\overline{\boldsymbol{q}}',t')\eta'(\overline{\boldsymbol{q}}')\right), \tag{5.18}$$

which involves two independent functions $\eta(\boldsymbol{q})$ and $\eta'(\boldsymbol{q})$, and define the two-time cumulants

$$C^{(s,1)}(\underline{1},\ldots,\underline{s},t,\underline{1}',t') \doteq \frac{\delta\overline{f}^{(s)}(\underline{1},\ldots,\underline{s},t)}{\delta\eta'(\underline{1}')}. \tag{5.19}$$

The superscript pair $(s,s')$ indicates the number of arguments associated with times $t$ (namely $s$) and $t'$ (namely $s'$). This notation is redundant when the argument list is displayed, but it fosters readability. Importantly, $C^{(1,1)}(1,1') = \langle\delta\widetilde{f}(1)\delta\widetilde{f}(1')\rangle \equiv C(1,1')$ — this is the fundamental two-time Klimontovich correlation function. The key result is that ***the $C^{(s,1)}$ obey the linearization of the one-time BBGKY cumulant hierarchy*** — for example, in the $\partial_t g$ equation replace $g(\underline{1},\underline{2},t) \to g(\underline{1},\underline{2},t) + \epsilon\, C^{(2,1)}(\underline{1},\underline{2},t,\underline{1}',t')$ and collect the terms of first order in $\epsilon$ to find $\partial_t C^{(2,1)} = \cdots$. This follows immediately from the definition of the $C$'s as functional derivatives. Thus, the functional derivative of (5.17*a*) is

$$0 = \partial_t C^{(1,1)}(1,1') + \mathrm{iL}_1[f]C^{(1,1)} + \boldsymbol{\partial}_1\cdot\mathbb{E}(\underline{1},\overline{\underline{2}})C^{(2,1)}(\underline{1},\overline{\underline{2}},t,1'), \tag{5.20}$$

with $C^{(1,1)} \equiv C$. In deriving this result, it was crucial to not assume prematurely that the mean field $\boldsymbol{E}$ vanishes; indeed, $\boldsymbol{E} \neq \boldsymbol{0}$ when $\eta \neq 0$ and the functional derivative of $\boldsymbol{E}[f] = \mathbb{E}(\underline{1},\overline{\underline{1}})f(\overline{\underline{1}},t)$ with respect to $\eta'$ is $\mathbb{E}(\underline{1},\overline{\underline{1}})C^{(1,1)}(\overline{\underline{1}},t,1')$, which produces the last, self-consistent response or polarization term in the Landau operator (5.16). Only after performing all functional derivatives and setting $\eta = 0$ may one assert that $\boldsymbol{E} = \boldsymbol{0}$.

Equation (5.20) is not closed, as it involves the unknown function $C^{(2,1)}$. It is clear that the closure problem cannot be solved merely by performing further functional differentiations. Instead, at some point one needs to express $C^{(n+1,1)}$ in terms of $\{C^{(m,1)} \mid m \leqslant n\}$. When this is done for $n = 1$, the formal equation that results is called the *Dyson equation*. Martin *et al.* (1973) showed how to do this for continuous classical fields,[43] and Rose (1979) provided an elegant generalization that handles particle discreteness as well. Rose's work is very important, and his equations could be used as the basis for the subsequent discussion. Instead, I shall continue with an analysis of the two-time hierarchy, which for weak coupling is somewhat more transparent and leads rather directly to an approximate closure. Thus, the present discussion adds additional perspective to Rose's general results.

Note that near thermal equilibrium one has $C^{(s,s')} = O(\epsilon_{\mathrm{p}}^{s+s'-1})$. In particular, $C^{(1,1)} = O(\epsilon_{\mathrm{p}})$; to find collisional corrections, one needs to work only to $O(\epsilon_{\mathrm{p}}^2)$. To find

---

[42]Krommes (1975) and Krommes & Oberman (1976*a*) introduced the second independent function $\eta'$ to handle the second discrete time $t'$; the integrals are over just the phase-space coordinates, and one applies to $S_2$ functional derivatives with only phase-space arguments. This is to be contrasted with a generating functional of the MSR type, $S[\eta] \doteq \exp(\int \mathrm{d}\overline{1}\,\widetilde{f}(\overline{1})\eta(\overline{1}))$, from which $\widetilde{f}(1')$ can be generated from a functional derivative involving arguments involving both phase space and time. The MSR approach is clearly the more general. Historically, the two-time generating functional was introduced in order to systematize the linearization procedure discussed after (5.19), which was originally noticed by Williams (1973).

[43]The review of Krommes (2002) contains lengthy discussion of the MSR formalism.

a representation for $C^{(2,1)}$, consider its equation, which follows from the linearization of (5.17$b$):

$$\begin{aligned}
&(\partial_t + \mathrm{iL}_1[f] + \mathrm{iL}_2[f])C^{(2,1)}(\underline{1},\underline{2},t,1') \\
&\quad = -\boldsymbol{\epsilon}_{12} \cdot (q_2\boldsymbol{\partial}_1 - q_1\boldsymbol{\partial}_2)[C^{(1,1)}(1,1')f(2) + f(1)C^{(1,1)}(2,1')] \\
&\qquad - \boldsymbol{\partial}_1 g(\underline{1},\underline{2},t) \cdot \mathbf{E}(\underline{1},\underline{\overline{1}})C^{(1,1)}(\underline{\overline{1}},t,1') - \boldsymbol{\partial}_2 g(\underline{1},\underline{2},t) \cdot \mathbf{E}(\underline{2},\underline{\overline{2}})C^{(1,1)}(\underline{\overline{2}},t,1') \\
&\qquad - \boldsymbol{\partial}_1 C^{(1;1)}(1;1') \cdot \mathbf{E}(\underline{1},\underline{\overline{1}})g(\underline{\overline{1}},\underline{2},t) - \boldsymbol{\partial}_2 C^{(1;1)}(2;1') \cdot \mathbf{E}(\underline{2},\underline{\overline{2}})g(\underline{1},\underline{\overline{2}},t) + O(\epsilon_{\mathrm{p}}^3).
\end{aligned} \tag{5.21}$$

Notation such as $f(2)$ means $f(\underline{2},t)$ (i.e., $t_2 = t$). The second and third lines after the equals sign in this equation arise from the linearization of the second and third terms in (5.16).

Because the integrals that determine the transport coefficients are integrated in $\overline{\tau}$ from 0 to $\infty$, we are interested only in the correlations for $t \geqslant t'$. Thus, consider the one-sided version of (5.21):

$$(\partial_t + \mathrm{iL}_1[f] + \mathrm{iL}_2[f])C_+^{(2;1)}(\underline{1},\underline{2},t;1') = \delta(t-t')C^{(0,3)}(\underline{1},\underline{2},\underline{1}',t') + s_+^{(2;1)}(\underline{1},\underline{2},t;1'), \tag{5.22}$$

where

$$\begin{aligned}
s_+^{(2;1)}(\underline{1},\underline{2},t;1') &\doteq -\boldsymbol{\epsilon}_{12} \cdot (q_2\boldsymbol{\partial}_1 - q_1\boldsymbol{\partial}_2)[C_+^{(1;1)}(1;1')f(2) + f(1)C_+^{(1;1)}(2;1')] \\
&- \boldsymbol{\partial}_1 g(\underline{1},\underline{2},t) \cdot \mathbf{E}(\underline{1},\underline{\overline{1}})C^{(1,1)}(\underline{\overline{1}},t,1') - \boldsymbol{\partial}_2 g(\underline{1},\underline{2},t) \cdot \mathbf{E}(\underline{2},\underline{\overline{2}})C^{(1,1)}(\underline{\overline{2}},t,1') \\
&- \boldsymbol{\partial}_1 C_+^{(1;1)}(\underline{1},t;\underline{1}',t') \cdot \mathbf{E}(\underline{1},\underline{\overline{1}})g(\underline{\overline{1}},\underline{2},t) - \boldsymbol{\partial}_2 C_+^{(1;1)}(\underline{2},t;\underline{1}',t') \cdot \mathbf{E}(\underline{2},\underline{\overline{2}})g(\underline{1},\underline{\overline{2}},t).
\end{aligned} \tag{5.23}$$

(The semicolon is used to indicate one-sided functions.) Note that the equal-time function $C^{(0,3)}(t')$ has entered as an initial condition. Let the causal Green's function for the linearized Vlasov equation obey

$$(\partial_t + \mathrm{iL}_1)R(1;1') = \delta(1-1'), \tag{5.24}$$

and represent the solution as

$$R(1;1') = H(t-t')\varXi(1,1'). \tag{5.25}$$

Then the solution of (5.23) is

$$\begin{aligned}
C_+^{(2;1)}(\underline{1},\underline{2},t,1') &= H(t-t')\varXi(\underline{1},t,\underline{\overline{1}},t')\varXi(\underline{2},t,\underline{\overline{2}},t')C^{(0,3)}(\underline{\overline{1}},\underline{\overline{2}};1') \\
&\quad - \int_{t'}^{t} \mathrm{d}\overline{t}\, \varXi(\underline{1},t,\underline{\overline{1}},\overline{t})\varXi(\underline{2},t,\underline{\overline{2}},\overline{t})s_+^{(2;1)}(\underline{\overline{1}},\underline{\overline{2}},\overline{t};1').
\end{aligned} \tag{5.26}$$

If the first line of (5.23) were inserted into (5.26), the last line of (5.26) would have the same form as the solution for the pair correlation function $g$ in standard Balescu–Lenard theory (reviewed in §G.1) except that the product $f(\underline{1},\overline{t})f(\underline{2},\overline{t})$ that appears in the classical derivation is replaced here by the $Cf$ terms in the square brackets in the first line of (5.23). That term emerges as the $O(\epsilon)$ term when $f$ is replaced by $f + \epsilon C_+^{(1;1)}$. This suggests that the $s_+^{(2;1)}$-driven contribution of $C_+^{(2;1)}$ to (5.20) may be related to the linearized collision operator. However, this conclusion is not immediate because in the standard derivation of the Balescu–Lenard operator cancellations occur between the form of $\varXi$, which contains an $f$, and the terms on which the $q_2\boldsymbol{\partial}_1 - q_1\boldsymbol{\partial}_2$ operates; it is unclear how those cancellations occur when it is $C_+^{(1;1)}$ that is differentiated. Furthermore, a formal linearization of the Balescu–Lenard operator ought to contain terms involving fluctuations in the strength of the dielectric shielding. Thus, one must do a serious calculation.

In order to proceed, it is convenient to work with spatial Fourier transforms. With the conventions used in (4.16), the Fourier transform of (5.20) is[44]

$$\begin{aligned}&\partial_\tau C^{(1;1)}_{+;\boldsymbol{k}'}(\underline{1},\tau;\underline{1}') + \mathrm{iL}_{\boldsymbol{k}'}(\underline{1};\underline{\overline{1}})C^{(1;1)}_{+;\boldsymbol{k}''}(\underline{\overline{1}},\tau;\underline{1}')\\&\quad= \delta(\tau)C^{(0,2)}_{\boldsymbol{k}'}(\underline{1},\underline{1}') - \boldsymbol{\partial}\boldsymbol{\cdot}\int\frac{\mathrm{d}\boldsymbol{k}}{(2\pi)^3}\,\mathbb{E}_{\boldsymbol{k}}(\underline{\overline{2}})C^{(2;1)}_{+,\boldsymbol{k};\boldsymbol{k}'}(\underline{1},\underline{2},\tau;\underline{1}'),\end{aligned} \tag{5.27}$$

and the Fourier transform of (5.23) is

$$\begin{aligned}&\partial_\tau C^{(2;1)}_{+,\boldsymbol{k};\boldsymbol{k}'}(\underline{1},\underline{2},\tau;\underline{1}') + \mathrm{iL}_{\boldsymbol{k}+\boldsymbol{k}'}(\underline{1},\underline{\overline{1}})C^{(2;1)}_{+,\boldsymbol{k};\boldsymbol{k}'}(\underline{\overline{1}},\underline{2},\tau;\underline{1}') + \mathrm{iL}_{-\boldsymbol{k}}(\underline{1},\underline{\overline{1}})C^{(2;1)}_{+,\boldsymbol{k};\boldsymbol{k}'}(\underline{1},\underline{\overline{2}},\tau;\underline{1}')\\&\quad= \delta(\tau)C^{(0,3)}_{\boldsymbol{k},\boldsymbol{k}'}(\underline{1},\underline{2},\underline{1}') + s^{(2;1)}_{+,\boldsymbol{k};\boldsymbol{k}'}(\underline{1},\underline{2},\tau;\underline{1}'),\end{aligned} \tag{5.28}$$

where

$$\begin{aligned}&s^{(2;1)}_{+,\boldsymbol{k};\boldsymbol{k}'}(\underline{1},\underline{2},\tau;\underline{1}') \doteq\\&\quad - (q_2\boldsymbol{\partial}_1 - q_1\boldsymbol{\partial}_2)\boldsymbol{\cdot}[\boldsymbol{\epsilon}_{\boldsymbol{k}}C^{(1;1)}_{+;\boldsymbol{k}'}(\underline{1};\underline{1}')f(2) + \epsilon_{\boldsymbol{k}+\boldsymbol{k}'}f(1)C^{(2;1)}_{+;\boldsymbol{k}'}(\underline{2};\underline{1}')\\&\quad - \boldsymbol{\partial}_1 g_{\boldsymbol{k}}(\underline{1},\underline{2},t)\boldsymbol{\cdot}\mathbb{E}_{\boldsymbol{k}'}(\underline{\overline{1}})C^{(1;1)}_{+;\boldsymbol{k}'}(\underline{\overline{1}},\underline{2},\tau;\underline{1}') - \boldsymbol{\partial}_2 g_{\boldsymbol{k}+\boldsymbol{k}'}(\underline{1},\underline{2},t)\boldsymbol{\cdot}\mathbb{E}_{\boldsymbol{k}'}(\underline{\overline{2}})C^{(1;1)}_{+;\boldsymbol{k}'}(\underline{1},\underline{\overline{2}},\tau;\underline{1}')\\&\quad - \boldsymbol{\partial}_1 C^{(1;1)}_{+;\boldsymbol{k}'}(\underline{1},\tau;\underline{1}')\boldsymbol{\cdot}\mathbb{E}_{\boldsymbol{k}}(\underline{\overline{1}})g_{\boldsymbol{k}}(\underline{\overline{1}},\underline{2},t) - \boldsymbol{\partial}_2 C^{(1;1)}_{+;\boldsymbol{k}'}(\underline{1},\tau;\underline{1}')\boldsymbol{\cdot}[\mathbb{E}_{\boldsymbol{k}+\boldsymbol{k}'}(\underline{\overline{1}})g_{\boldsymbol{k}+\boldsymbol{k}'}(\underline{\overline{1}},\underline{1},t)]^*.\end{aligned} \tag{5.29}$$

It was shown in §3.2.1 that the two-time, two-phase-space-point correlation functions required for many of the transport coefficients can be written in terms of weighted velocity integrals of the basic Klimontovich correlation function $C_+(\tau) \equiv C^{(1;1)}_{+;\boldsymbol{k}'=\boldsymbol{0}}(\tau)$. Upon setting $\boldsymbol{k}' = \boldsymbol{0}$, one can solve (5.28) as

$$\begin{aligned}C^{(2;1)}_{+\boldsymbol{k}}(\underline{1},\underline{2},\tau;\underline{1}') &= H(\tau)\varXi_{\boldsymbol{k}}(\underline{1},\underline{\overline{1}},\tau)\varXi^*_{\boldsymbol{k}}(\underline{2},\underline{\overline{2}},\tau)C^{(0,3)}_{\boldsymbol{k},\boldsymbol{0}}(\underline{1},\underline{2},\underline{1}')\\&\quad + \int_0^\tau \mathrm{d}\overline{\tau}\,\varXi_{\boldsymbol{k}}(\underline{1},\underline{\overline{1}},\tau)\varXi^*_{\boldsymbol{k}}(\underline{2},\underline{\overline{2}},\tau)s^{(2;1)}_{+,\boldsymbol{k}}(\underline{\overline{1}},\underline{\overline{2}},\tau-\overline{\tau};\underline{1}'),\end{aligned} \tag{5.30}$$

where

$$\begin{aligned}s^{(2;1)}_{+,\boldsymbol{k}}(\underline{1},\underline{1},\overline{\tau};\underline{1}') &\doteq -\boldsymbol{\epsilon}_{\boldsymbol{k}}\boldsymbol{\cdot}(q_2\boldsymbol{\partial}_1 - q_1\boldsymbol{\partial}_2)[C^{(1;1)}_{+}(\underline{1},\tau;\underline{1}')f(2) + C^{(1;1)}_{+}(\underline{2},\tau;\underline{1}')f(1)]\\&\quad - \boldsymbol{\partial}_1 C^{(1;1)}_{+}(\underline{1},\tau;\underline{1}')\boldsymbol{\cdot}\mathbb{E}_{\boldsymbol{k}}(\underline{\overline{1}})g_{\boldsymbol{k}}(\underline{\overline{1}},\underline{2},t) - \boldsymbol{\partial}_2 C^{(1;1)}_{+}(\underline{2},\tau;\underline{1}')\boldsymbol{\cdot}[\mathbb{E}_{\boldsymbol{k}}(\underline{\overline{1}})g_{\boldsymbol{k}}(\underline{1},\underline{\overline{1}},t)]^*.\end{aligned} \tag{5.31}$$

Here the terms in (5.29) involving $\mathbb{E}_{\boldsymbol{k}'}C^{(1;1)}_{+;\boldsymbol{k}'}|_{\boldsymbol{k}'=\boldsymbol{0}}$ were assumed to vanish.

The $\varXi$'s describe Debye shielding clouds, so their characteristic time scale is the microscopic autocorrelation time $\omega_{\mathrm{p}}^{-1}$. In contrast, all of the $\tau$-dependent terms in (5.31) involve $C_+(\tau)$, which (as will be shown) varies on the collisional timescale. Thus, a Markovian approximation is appropriate and the $s^{(2;1)}_{+}$ contribution to $C^{(2;1)}_{+}$

[44] Here I have redefined the underline notation such that when spatial or wavenumber arguments are displayed explicitly, $\underline{1} \equiv \{\boldsymbol{v}_1, s_1\}$. I have also dropped the hats that signify Fourier transforms. The subscripting convention for a function like $C^{(2;1)}_{+,\boldsymbol{k};\boldsymbol{k}'}$ is as follows. The comma simply separates the + that denotes a one-sided function. The wavenumber arguments before the semicolon are conjugate to the equal-time ($t$) spatial differences referred to $\boldsymbol{x}_1$ (e.g., $\boldsymbol{\rho} \to \boldsymbol{k}$); the ones after the semicolon refer to the arguments at $t'$, again referred to $\boldsymbol{x}_1$ (e.g., $\boldsymbol{\rho}' \to \boldsymbol{k}'$). If there is only one argument at time $t$, there is no equal-time wavenumber and the slot before the semicolon is left blank (e.g., $C^{(1;1)}_{+;\boldsymbol{k}'}$). If all of the arguments are at equal times, the semicolon is omitted [e.g., the initial condition $C^{(0,3)}_{\boldsymbol{k},\boldsymbol{k}'}(\underline{1},\underline{2},\underline{1}')$].

is approximately

$$\left(\int_0^\infty \mathrm{d}\overline{\tau}\, \Xi_{\boldsymbol{k}}(\underline{1},\overline{\underline{1}},\overline{\tau})\Xi^*_{\boldsymbol{k}}(\underline{2},\overline{\underline{2}},\overline{\tau})\right) s^{(2;1)}_{+,\boldsymbol{k}}(\overline{\underline{1}},\overline{\underline{2}},\tau;\underline{1}'). \tag{5.32}$$

The contribution of this term to (5.27) is analysed in §G.2; it is found that it leads to the linearized Balescu–Lenard operator $\widehat{\mathrm{C}}^{\mathrm{BL}}$ acting on $C^{(1;1)}_+$. I shall subsequently drop the BL superscript. Thus, one has

$$\partial_\tau C^{(1;1)}_+(\underline{1},\tau;\underline{1}') + \widehat{\mathrm{C}}C^{(1)}_+ = \delta(t-t')C^{(0,2)}_+(\underline{1},\underline{1}',t') + s^{(1;1)}_+(\underline{1},\tau;\underline{1}'), \tag{5.33}$$

where

$$s^{(1;1)}_+(\underline{1},\tau;\underline{1}') \doteq -\boldsymbol{\partial}_1 \cdot \mathbb{E}(1,2)\Xi(\underline{1},\tau;\overline{\underline{1}})\Xi(\underline{1},\tau;\overline{\underline{2}})C^{(0,3)}(\overline{\underline{1}},\overline{\underline{2}},\underline{1}',t'). \tag{5.34}$$

For the evaluation of transport coefficients, one is required to integrate the solution from 0 to $\infty$ in $\tau$. An equation for $\int_0^\infty \mathrm{d}\overline{\tau}\, C^{(1)}_{+,\boldsymbol{k}\to\mathbf{0}}(\overline{\tau})$ can be obtained by integrating (5.33) from $\overline{\tau} = 0_-$ to $\infty$ and assuming that correlations decay to 0 at long times:

$$\widehat{\mathrm{C}}\int_0^\infty \mathrm{d}\overline{\tau}\, C^{(1;1)}_+(\overline{\tau}) = C^{(0,2)}_{\boldsymbol{k}=\mathbf{0}}(\boldsymbol{v}_1,\boldsymbol{v}_{1'},t') + \int_0^\infty \mathrm{d}\overline{\tau}\, s^{(1;1)}_{+,\boldsymbol{k}=\mathbf{0}}(\boldsymbol{v}_1,\overline{\tau};\boldsymbol{v}'_1), \tag{5.35}$$

where

$$C^{(0,2)}_{\boldsymbol{k}=\mathbf{0}}(\boldsymbol{v}_1,\boldsymbol{v}_{1'},t') = \overline{n}^{-1}_{s_1}\delta_{s_1 s_{1'}}\delta(\boldsymbol{v}_1-\boldsymbol{v}_{1'})f_{s_{1'}}(\boldsymbol{v}_{1'}). \tag{5.36}$$

If the $s_+$ term were negligible, one would find

$$\int \mathrm{d}\overline{\tau}\, C^{(1;1)}_+(\overline{\tau}) = \widehat{\mathrm{C}}^{-1}\overline{n}^{-1}_s\delta_{ss'}\delta(\boldsymbol{v}-\boldsymbol{v}')f_{s'}(\boldsymbol{v}') \tag{5.37}$$

and, with the aid of (3.9*d*), formulas such as (4.15*a*) would reduce to the standard velocity-space matrix elements that emerged in Part I. A discussion that justifies the neglect of $s_+$ is given in §H.3.

## 5.4. *Two-time, three-phase-space-point correlations*

For the integrals (4.15*f*)–(4.15*k*), one requires two-time correlations with three phase-space points, i.e., $C^{(1;2)}_+(\underline{1},t;\underline{1}',t',\underline{1}'',t')$. A natural way of proceeding is to introduce a third source $\eta''$, generate the equation for $C(t,t',t'')$, set $t''=t'$, and evolve forward from $t'$ to $t$. The functional derivative of (5.20) with respect to $\eta''$ is

$$0 = \partial_t C^{(1,1,1)}(1,1',1'') + \mathrm{i}\mathrm{L}_1 C^{(1,1,1)} + \boldsymbol{\partial}_1 \cdot \mathbb{E}(\underline{1},\overline{\underline{2}})C^{(2,1,1)}(\underline{1},\overline{\underline{2}},t,1',1''), \tag{5.38}$$

and the functional derivative of (5.23) with respect to $\eta''$ is

$$(\partial_t + \mathrm{i}\mathrm{L}_1 + \mathrm{i}\mathrm{L}_2)C^{(2;1,1)}_+(\underline{1},\underline{2},t;1',1'') = \delta(t-t')C^{(0,3,1)}(\underline{1},\underline{2},\underline{1}',t',t'') + s^{(2;1,1)}_+(\underline{1},\underline{2},t;1',1''), \tag{5.39}$$

where

$$\begin{aligned} s^{(2;1,1)}_+&(\underline{1},\underline{2},t;1',1'') \doteq \\ &- \boldsymbol{\epsilon}_{12}\cdot(q_2\boldsymbol{\partial}_1 - q_1\boldsymbol{\partial}_2)[C^{(1;1,1)}_+(1;1',1'')f(2) + f(1)C^{(1;1,1)}_+(2;1',1'')] \\ &- \boldsymbol{\epsilon}_{12}\cdot(q_2\boldsymbol{\partial}_1 - q_1\boldsymbol{\partial}_2)[C^{(1;1)}_+(1;1')C^{(1;1)}_+(2;1'') + C^{(1;1)}_+(1;1'')C^{(1;1)}_+(2;1')] \\ &- \boldsymbol{\partial}_1 C^{(1;1)}_+(1;1'')\cdot\mathbb{E}(\underline{1},\overline{\underline{1}})C^{(2;1)}_+(\overline{\underline{1}},\underline{2},t;1') - \boldsymbol{\partial}_2 C^{(1;1)}_+(2;1'')\cdot\mathbb{E}(\underline{2},\overline{\underline{2}})C^{(2;1)}_+(\underline{1},\overline{\underline{2}},t;1'); \end{aligned} \tag{5.40}$$

the last two terms arise by differentiating the $f$ in the last term of the Landau operator (5.16).

One may now set $t'' = t'$. For stationary fluctuations, the one-sided version of (5.38) becomes

$$\partial_t C_+^{(1;2)}(\underline{1}, \tau; \underline{1}', \underline{1}'') + \mathrm{iL}_1(\underline{1}, \overline{\underline{1}}) C_+^{(1;2)}(\overline{\underline{1}}, \tau; \underline{1}', \underline{1}'')$$
$$= \delta(\tau) C^{(0,3)}(\underline{1}, \underline{1}', \underline{1}'', t') - \boldsymbol{\partial}_1 \boldsymbol{\cdot} \mathbf{E}(\underline{1}, \overline{\underline{2}}) C_+^{(2;2)}(\underline{1}, \overline{\underline{2}}, \tau; \underline{1}', \underline{1}''), \tag{5.41}$$

and (5.40) becomes

$$\partial_t C_+^{(2;2)}(\underline{1}, \underline{2}, t; \underline{1}', \underline{1}'', t') + \mathrm{iL}_1(\underline{1}, \overline{\underline{1}}) C_+^{(2;2)}(\overline{\underline{1}}, \underline{2}, \tau; \underline{1}', \underline{1}'') + \mathrm{iL}_2(\underline{2}, \overline{\underline{2}}) C_+^{(2;2)}(\underline{1}, \overline{\underline{2}}, \tau; \underline{1}', \underline{1}'')$$
$$= \delta(\tau) C_+^{(0,4)}(\underline{1}, \underline{2}, \underline{1}', \underline{2}'', t') + s_+^{(2;2)}(\underline{1}, \underline{2}, \tau; \underline{1}', \underline{1}''), \tag{5.42}$$

where

$$s_+^{(2;2)}(\underline{1}, \underline{2}, \tau; \underline{1}', \underline{1}'') \doteq$$
$$- \boldsymbol{\epsilon}_{12} \boldsymbol{\cdot} (q_2 \boldsymbol{\partial}_1 - q_1 \boldsymbol{\partial}_2)[C_+^{(1;2)}(\underline{1}, \tau; \underline{1}', \underline{1}'') f(2) + f(1) C_+^{(1;2)}(\underline{2}, \tau; \underline{1}', \underline{1}'')$$
$$- [\boldsymbol{\partial}_1 C_+^{(1;2)}(\underline{1}, \tau; \underline{1}', \underline{1}'') \boldsymbol{\cdot} \mathbf{E}(\underline{1}, \overline{\underline{1}}) g(\overline{\underline{1}}, \underline{2}) - \boldsymbol{\partial}_2 C_+^{(1;2)}(\underline{2}, \tau; \underline{1}', \underline{1}'') \boldsymbol{\cdot} \mathbf{E}(\underline{2}, \overline{\underline{2}}) g(\underline{1}, \overline{\underline{2}})]$$
$$- \boldsymbol{\epsilon}_{12} \boldsymbol{\cdot} (q_2 \boldsymbol{\partial}_1 - q_1 \boldsymbol{\partial}_2)[C_+^{(1;1)}(\underline{1}, \tau; \underline{1}') C_+^{(1;1)}(\underline{2}, \tau; \underline{1}'') + C_+^{(1;1)}(\underline{1}, \tau; \underline{1}'') C_+^{(1;1)}(\underline{2}, \tau; \underline{1}')]$$
$$- [\boldsymbol{\partial}_1 C_+^{(1;1)}(\underline{1}, \tau; \underline{1}') \boldsymbol{\cdot} \mathbf{E}(\underline{1}, \overline{\underline{1}}) C_+^{(2;1)}(\overline{\underline{1}}, \underline{2}, \tau; \underline{1}'') + (1' \Leftrightarrow 1'')$$
$$+ \boldsymbol{\partial}_2 C_+^{(1;1)}(\underline{2}, \tau; \underline{1}') \boldsymbol{\cdot} \mathbf{E}(\underline{2}, \overline{\underline{2}}) C_+^{(2;1)}(\underline{1}, \overline{\underline{2}}, \tau; \underline{1}'') + (1' \Leftrightarrow 1'')]$$
$$- [\boldsymbol{\partial}_1 C_+^{(2;1)}(\underline{1}, \underline{2}, \tau; 1') \boldsymbol{\cdot} \mathbf{E}(\underline{1}, \overline{\underline{1}}) C_+^{(1;1)}(\overline{\underline{1}}, \tau; \underline{1}'') + (1' \Leftrightarrow 1'')$$
$$+ \boldsymbol{\partial}_2 C_+^{(2;1)}(\underline{1}, \underline{2}, \tau; \underline{1}') \boldsymbol{\cdot} \mathbf{E}(\underline{2}, \overline{\underline{2}}) C_+^{(1;1)}(\overline{\underline{2}}, \tau; \underline{1}'') + (1' \Leftrightarrow 1'')]$$
$$- [\boldsymbol{\partial}_1 g(\underline{1}, \underline{2}) \boldsymbol{\cdot} \mathbf{E}(\underline{1}, \overline{\underline{1}}) C_+^{(1;2)}(\overline{\underline{1}}, \tau; \underline{1}', \underline{1}'') + \boldsymbol{\partial}_2 g(\underline{1}, \underline{2}) \boldsymbol{\cdot} \mathbf{E}(\underline{2}, \overline{\underline{2}}) C_+^{(1;2)}(\overline{\underline{2}}, \tau; \underline{1}', \underline{1}'')]. \tag{5.43}$$

With the additional definition $\boldsymbol{\rho}'' \doteq \boldsymbol{x}_1 - \boldsymbol{x}_1''$, one has, for example,

$$C_+^{(2;2)}(\underline{1}, \underline{2}, t; \underline{1}', \underline{1}'', t') = C_+^{(2;2)}(\boldsymbol{\rho}, \underline{1}, \underline{2}, \tau; \boldsymbol{\rho}', \underline{1}', \boldsymbol{\rho}'', \underline{1}'') \tag{5.44$a$}$$
$$= \int \frac{\mathrm{d}\boldsymbol{k}}{(2\pi)^3} \int \frac{\mathrm{d}\boldsymbol{k}'}{(2\pi)^3} \int \frac{\mathrm{d}\boldsymbol{k}''}{(2\pi)^3} \mathrm{e}^{\mathrm{i}\boldsymbol{k}\boldsymbol{\cdot}\boldsymbol{\rho} + \mathrm{i}\boldsymbol{k}'\boldsymbol{\cdot}\boldsymbol{\rho}' + \mathrm{i}\boldsymbol{k}''\boldsymbol{\cdot}\boldsymbol{\rho}''} C_{+,\boldsymbol{k};\boldsymbol{k}',\boldsymbol{k}''}^{(2;2)}(\underline{1}, \underline{2}, \tau; \underline{1}', \underline{1}''). \tag{5.44$b$}$$

As discussed in §4.2, one requires $\mathrm{i}\boldsymbol{\partial}_{\boldsymbol{k}''} C_{+;\boldsymbol{k}'=\boldsymbol{0},\boldsymbol{k}''}^{(1;2)}(\underline{1}, \tau; \underline{1}', \underline{1}'')|_{\boldsymbol{k}''=\boldsymbol{0}}$ [see (4.17$b$)]. The function $C_{+;\boldsymbol{k}',\boldsymbol{k}''}^{(1;2)}(\tau)$ evolves according to the Fourier transform of (5.41):

$$\partial_\tau C_{+;\boldsymbol{k}',\boldsymbol{k}''}^{(1;2)}(\underline{1}, \tau; \underline{1}', \underline{1}'') + \mathrm{iL}_{\boldsymbol{k}'+\boldsymbol{k}''}(\underline{1}, \overline{\underline{1}}) C_{+;\boldsymbol{k}',\boldsymbol{k}''}^{(1;2)}(\overline{\underline{1}}, \tau; \underline{1}', \underline{1}'')$$
$$= \delta(\tau) C_{\boldsymbol{k}',\boldsymbol{k}''}^{(0,3)}(\underline{1}, \underline{1}', \underline{1}'') - \boldsymbol{\partial}_1 \boldsymbol{\cdot} \int \frac{\mathrm{d}\boldsymbol{k}}{(2\pi)^3} \mathbf{E}_{\boldsymbol{k}}^*(\overline{\underline{2}}) C_{+,\boldsymbol{k};\boldsymbol{k}',\boldsymbol{k}''}^{(2;2)}(\underline{1}, \overline{\underline{2}}, \tau; \underline{1}', \underline{1}''). \tag{5.45}$$

Clearly, $C_+^{(2;2)}$ is crucial for determining the collisional dynamics of $C_+^{(1;2)}$. The Fourier transform of (5.43) is

$$\partial_\tau C_{+,\boldsymbol{k};\boldsymbol{k}',\boldsymbol{k}''}^{(2;2)}(\underline{1}, \underline{2}, \tau; \underline{1}', \underline{1}'')$$
$$+ \mathrm{iL}_{\boldsymbol{k}+\boldsymbol{k}'+\boldsymbol{k}''}(\underline{1}, \overline{\underline{1}}) C_{+,\boldsymbol{k};\boldsymbol{k}',\boldsymbol{k}''}^{(2;2)}(\overline{\underline{1}}, \underline{2}, \tau; \underline{1}', \underline{1}'') + \mathrm{iL}_{\boldsymbol{k}}^*(\underline{2}, \overline{\underline{2}}) C_{+,\boldsymbol{k};\boldsymbol{k}',\boldsymbol{k}''}^{(2;2)}(\boldsymbol{k}, \underline{1}, \overline{\underline{2}}, \tau; \underline{1}', \underline{1}'')$$
$$= \delta(\tau) C_{\boldsymbol{k},\boldsymbol{k}',\boldsymbol{k}''}^{(0,4)}(\underline{1}, \underline{2}, \underline{1}', \underline{1}'') + s_{+,\boldsymbol{k};\boldsymbol{k}',\boldsymbol{k}''}^{(2;2)}(\underline{1}, \underline{2}, \tau; \underline{1}', \underline{1}''), \tag{5.46}$$

where

$$s_{+,\boldsymbol{k};\boldsymbol{k}',\boldsymbol{k}''}^{(2;2)}(\underline{1}, \underline{2}, \tau; \underline{1}', \underline{1}'') \doteq$$
$$- (q_2 \boldsymbol{\partial}_1 - q_1 \boldsymbol{\partial}_2) \boldsymbol{\cdot} [\boldsymbol{\epsilon}_{\boldsymbol{k}} C_{+;\boldsymbol{k}',\boldsymbol{k}''}^{(1;2)}(\underline{1}, \tau; \underline{1}', \underline{1}'') f(\underline{2}) + \boldsymbol{\epsilon}_{\boldsymbol{k}+\boldsymbol{k}'+\boldsymbol{k}''} C_{+;\boldsymbol{k}',\boldsymbol{k}''}^{(1;2)}(\underline{2}, \tau; \underline{1}', \underline{1}'') f(\underline{1})]$$

$$
\begin{aligned}
&- \{\partial_1 C^{(1;2)}_{+;\boldsymbol{k}',\boldsymbol{k}''}(\underline{1},\tau;\underline{1}',\underline{1}'')\cdot \mathbb{E}_{\boldsymbol{k}}(\overline{1}) g_{\boldsymbol{k}}(\overline{1},2) \\
&\qquad + \partial_2 C^{(1;2)}_{+;\boldsymbol{k}',\boldsymbol{k}''}(\underline{2},\tau;\underline{1}',\underline{1}'')\cdot [\mathbb{E}^*_{\boldsymbol{k}+\boldsymbol{k}'+\boldsymbol{k}''}(\overline{2}) g_{\boldsymbol{k}+\boldsymbol{k}'+\boldsymbol{k}''}(\underline{1},\overline{2})]\} \\
&- (q_2\partial_1 - q_1\partial_2)\cdot [\epsilon_{\boldsymbol{k}+\boldsymbol{k}''} C^{(1;1)}_{+;\boldsymbol{k}'}\underline{1},\tau;\underline{1}') C^{(1;1)}_{+;\boldsymbol{k}''}(\underline{2},\tau;\underline{1}'') \\
&\qquad\qquad\qquad + \epsilon_{\boldsymbol{k}+\boldsymbol{k}'} C^{(1;1)}_{+;\boldsymbol{k}''}(\underline{1},\tau;\underline{1}'') C^{(1;1)}_{+;\boldsymbol{k}'}(\underline{2},\tau;\underline{1}')] \\
&- [\partial_1 C^{(1;1)}_{+;\boldsymbol{k}'}(\underline{1},\tau;\underline{1}')\cdot \mathbb{E}_{\boldsymbol{k}}(\overline{1}) C^{(2;1)}_{+,\boldsymbol{k};\boldsymbol{k}''}(\overline{1},\underline{2},\tau;\underline{1}'') \\
&\quad + \partial_1 C^{(1;1)}_{+;\boldsymbol{k}''}(\underline{1},\tau;\underline{1}'')\cdot \mathbb{E}_{\boldsymbol{k}}(\overline{1}) C^{(2;1)}_{+,\boldsymbol{k};\boldsymbol{k}'}(\overline{1},\underline{2},\tau;\underline{1}') \\
&\quad + \partial_2 C^{(1;1)}_{+;\boldsymbol{k}'}(\underline{2},\tau;\underline{1}')\cdot \mathbb{E}^*_{\boldsymbol{k}+\boldsymbol{k}'}(\overline{2}) C^{(2;1)}_{+\boldsymbol{k}+\boldsymbol{k}';\boldsymbol{k}''}(\underline{1},\overline{2},\tau;\underline{1}'') \\
&\quad + \partial_2 C^{(1;1)}_{+;\boldsymbol{k}''}(\underline{2},\tau;\underline{1}'')\cdot \mathbb{E}^*_{\boldsymbol{k}+\boldsymbol{k}'}(\overline{2}) C^{(2;1)}_{+\boldsymbol{k}+\boldsymbol{k}';\boldsymbol{k}'}(\underline{1},\overline{2},\tau;\underline{1}')] \\
&- [\partial_1 C^{(2;1)}_{+,\boldsymbol{k};\boldsymbol{k}'}(\underline{1},\underline{2},\tau;\underline{1}')\cdot \underline{\mathbb{E}_{\boldsymbol{k}''}(\overline{1}) C^{(1;1)}_{+;\boldsymbol{k}''}(\overline{1},\tau;\underline{1}'')} \\
&\quad + \partial_1 C^{(2;1)}_{+,\boldsymbol{k};\boldsymbol{k}''}(\underline{1},\underline{2},\tau;\underline{1}'')\cdot \underline{\mathbb{E}_{\boldsymbol{k}'}(\overline{1}) C^{(1;1)}_{+;\boldsymbol{k}'}(\overline{1},\tau;\underline{1}')} \\
&\quad + \partial_2 C^{(2;1)}_{+\boldsymbol{k}+\boldsymbol{k}'';\boldsymbol{k}'}(\underline{1},\underline{2};\underline{1}')\cdot \underline{\mathbb{E}_{\boldsymbol{k}''}(\overline{2}) C^{(1;1)}_{+;\boldsymbol{k}''}(\overline{2},\tau;\underline{1}'')} \\
&\quad + \partial_2 C^{(2;1)}_{+\boldsymbol{k}+\boldsymbol{k}';\boldsymbol{k}''}(\underline{1},\underline{2},\tau;\underline{1}'')\cdot \underline{\mathbb{E}_{\boldsymbol{k}'}(\overline{2}) C^{(1;1)}_{+;\boldsymbol{k}'}(\overline{2},\tau;\underline{1}')]} \\
&- [\partial_1 g_{\boldsymbol{k}}(\underline{1},\underline{2})\cdot \underline{\mathbb{E}_{\boldsymbol{k}'+\boldsymbol{k}''}(\overline{1}) C^{(1;2)}_{+;\boldsymbol{k}',\boldsymbol{k}''}(\overline{1},\tau;\underline{1}',\underline{1}'')} \\
&\quad + \partial_2 g_{\boldsymbol{k}+\boldsymbol{k}'+\boldsymbol{k}''}(\underline{1},\underline{2})\cdot \underline{\mathbb{E}_{\boldsymbol{k}'+\boldsymbol{k}''}(\overline{2}) C^{(1;2)}_{+\boldsymbol{k}',\boldsymbol{k}''}(\overline{2},\tau;\underline{1}',\underline{1}'')]}.
\end{aligned}
\tag{5.47}
$$

For small $\boldsymbol{k}'$ and $\boldsymbol{k}''$, the underlined terms are negligible. Upon setting $\boldsymbol{k}' = \boldsymbol{0}$, one finds

$$
\begin{aligned}
&s^{(2;2)}_{+,\boldsymbol{k};\boldsymbol{k}''}(\underline{1},\underline{2},\tau;\underline{1}',\underline{1}'') \doteq s^{(2;2)}_{+,\boldsymbol{k};\boldsymbol{k}'=\boldsymbol{0},\boldsymbol{k}''}(\underline{1},\underline{2},\tau;\underline{1}',\underline{1}'') = \\
&\quad - (q_2\partial_1 - q_2\partial_2)\cdot \epsilon_{\boldsymbol{k}}[C^{(1;2)}_{+;\boldsymbol{0},\boldsymbol{k}''}(\underline{1},\tau;\underline{1}',\underline{1}'') f(\underline{2}) + C^{(1;2)}_{+;\boldsymbol{0},\boldsymbol{k}''}(\underline{2},\tau;\underline{1}',\underline{1}'') f(\underline{1})] \\
&\quad - \{\partial_1 C^{(1;2)}_{+;\boldsymbol{0},\boldsymbol{k}''}(\underline{1},\tau;\underline{1}',\underline{1}'')\cdot \mathbb{E}_{\boldsymbol{k}}(\overline{1}) g_{\boldsymbol{k}}(\overline{1},\underline{2}) + \partial_2 C^{(1;2)}_{+;\boldsymbol{0},\boldsymbol{k}''}(\underline{2};\underline{1}',\underline{1}'')\cdot [\mathbb{E}^*_{\boldsymbol{k}}(\overline{2}) g_{\boldsymbol{k}}(\underline{1},\overline{2}]^*\} \\
&\quad - (q_2\partial_1 - q_1\partial_2)\cdot \epsilon_{\boldsymbol{k}}[C^{(1;1)}_{+}(\underline{1},\tau;\underline{1}') C^{(1;1)}_{+;\boldsymbol{k}''}(\underline{2},\tau;\underline{1}'') + C^{(1;1)}_{+;\boldsymbol{k}''}(\underline{1},\tau;\underline{1}'') C^{(1;1)}_{+}(\underline{2},\tau;\underline{1}')] \\
&\quad - [\partial_1 C^{(1;1)}_{+}(\underline{1},\tau;\underline{1}')\cdot \mathbb{E}_{\boldsymbol{k}}(\overline{1}) C^{(2;1)}_{+,\boldsymbol{k};\boldsymbol{k}''}(\overline{1},\underline{2},\tau;\underline{1}'') \\
&\qquad + \partial_1 C^{(1;1)}_{+;\boldsymbol{k}''}(\underline{1},\tau;\underline{1}'')\cdot \mathbb{E}_{\boldsymbol{k}}(\overline{1}) C^{(2;1)}_{+,\boldsymbol{k}}(\overline{1},\underline{2},\tau;\underline{1}') \\
&\qquad + \partial_2 C^{(1;1)}_{+}(\underline{2},\tau;\underline{1}')\cdot \mathbb{E}^*_{\boldsymbol{k}}(\overline{2}) C^{(2;1)}_{+,\boldsymbol{k};\boldsymbol{k}''}(\underline{1},\overline{2},\tau;\underline{1}'') \\
&\qquad + \partial_2 C^{(1;1)}_{+;\boldsymbol{k}''}(\underline{2},\tau;\underline{1}'')\cdot \mathbb{E}^*_{\boldsymbol{k}}(\overline{2}) C^{(2;1)}_{+,\boldsymbol{k}}(\underline{1},\overline{2},\tau;\underline{1}')].
\end{aligned}
\tag{5.48}
$$

Upon following the same arguments as in §5.3, one finds that the second and third lines of (5.40) lead to a term $-\widehat{\mathrm{C}}C^{(1;2)}_{+;\boldsymbol{0},\boldsymbol{k}''}(1;\underline{1}',\underline{1}'',t')$ in (5.45). The remaining terms are analysed in §G.3, where it is shown that they are related to the nonlinear Balescu–Lenard collision operator. One finds

$$
\begin{aligned}
&\partial_\tau C^{(1;2)}_{+;\boldsymbol{0},\boldsymbol{k}''}(\underline{1},\tau;\underline{1}',\underline{1}'') + \mathrm{iL}_{\boldsymbol{k}''}(\underline{1},\overline{1}) C^{(1;2)}_{+;\boldsymbol{0},\boldsymbol{k}''}(\overline{1},\tau;\underline{1}',\underline{1}'') + \widehat{\mathrm{C}}C^{(1;2)}_{+;\boldsymbol{0},\boldsymbol{k}''} \\
&\quad = \delta(\tau) C^{(0;3)}_{\boldsymbol{0},\boldsymbol{k}''}(\underline{1},\underline{1}',\underline{1}'') - \{\mathrm{C}^{\mathrm{BL}}[f;\, C^{(1;1)}_{+;\boldsymbol{0}}(\tau;\underline{1}'),\, \overline{C}^{(1;1)}_{+;\boldsymbol{k}''}(\tau;\underline{1}'')] + (1' \Leftrightarrow 1'')\}.
\end{aligned}
\tag{5.49}
$$

Notice that while the last term of this equation does contain the (generalized, three-argument) nonlinear collision operator [for an explanation of the notation $\mathrm{C}[a;b,\overline{c}]$, see the discussion of (G 23)], that construction is not the $\mathrm{C}[f_1,\overline{f}_1]$ that appears in second-order Chapman–Enskog theory ($f_1$ does not depend on $\tau$, but is rather an integral over all $\tau$). We shall see in the next section how $\mathrm{C}[f_1,\overline{f}_1]$ emerges from the solution of (5.49).

## 6. Comparison to the results of Catto & Simakov

For magnetised plasmas, the most complete calculation that involves some Burnett terms is the one by Catto & Simakov (2004). I shall now discuss the relationship of their calculation to the present results.

### 6.1. *Brief review of the calculation of Catto & Simakov (2004)*

For definiteness, I shall consider the ion version of the calculation of Catto & Simakov. They begin with the Landau kinetic equation

$$\partial_t f(\boldsymbol{x},\boldsymbol{v},t) + \boldsymbol{v}\cdot\boldsymbol{\nabla} f + (\boldsymbol{E} + c^{-1}\boldsymbol{v}\times\boldsymbol{B}^{\text{ext}})\cdot\boldsymbol{\partial} f = -\mathrm{C}[f,\overline{f}], \tag{6.1}$$

where $\mathrm{C}[f,\overline{f}]$ is the bilinear ion Landau collision operator $\mathrm{C}_{ii}+\mathrm{C}_{ie}$. Use of the Landau operator restricts the calculation to weakly coupled plasma,[45] although as is typical for work that extends the class of calculations reviewed by Braginskii (1965), this important point is not stressed. (Weak coupling is a good approximation for magnetically confined fusion plasmas.) With $\boldsymbol{w} \doteq \boldsymbol{v} - \boldsymbol{u}(\boldsymbol{x},t)$, the variable transformation $(\boldsymbol{x},\boldsymbol{v},t)\to(\boldsymbol{x},\boldsymbol{w},t)$ is made; that transforms (6.1) to

$$\begin{aligned}\partial_t f(\boldsymbol{x},\boldsymbol{w},t) + \boldsymbol{u}\cdot\boldsymbol{\nabla} f + \boldsymbol{w}\cdot\boldsymbol{\nabla} f + [(mn)^{-1}(\boldsymbol{\nabla} p + \boldsymbol{\nabla}\cdot\boldsymbol{\pi} - \boldsymbol{R}) - \boldsymbol{w}\cdot(\boldsymbol{\nabla}\boldsymbol{u})]\cdot\partial_{\boldsymbol{w}} f&\\ + \mathrm{i}\widehat{\mathrm{M}} + \mathrm{C}[f,\overline{f}] = 0,&\end{aligned} \tag{6.2}$$

where

$$\mathrm{i}\widehat{\mathrm{M}} \doteq \omega_{\mathrm{c}}\boldsymbol{w}\times\widehat{\boldsymbol{b}}\cdot\partial_{\boldsymbol{w}} \tag{6.3}$$

and the exact form of the momentum equation was used to replace the $\partial_t\boldsymbol{u}$ that arises after the variable transformation. The distribution function is expanded as $f = \sum_{n=0}^{\infty}\epsilon^n f_n$, where $\epsilon$ is an ordering parameter. Catto & Simakov follow the standard procedure of taking $\nabla = O(\epsilon)$,[46] but they do not explicitly describe a multiple-scale procedure as is explicated in appendix I:A; I shall return to this point. Whereas Braginskii implicitly assumes that $\boldsymbol{u} = O(1)$, Catto & Simakov follow Mikhaĭlovskiĭ & Tsypin (1971) in taking $\boldsymbol{u} = O(\epsilon)$.[47] This implies that $\boldsymbol{\nabla}\cdot\boldsymbol{\pi} = O(\epsilon^3)$. The term involving the ion friction force cancels with $\mathrm{C}_{ie}$ to lowest order in the mass ratio. Catto & Simakov are led to the sequence of equations

$$\mathrm{i}\widehat{\mathrm{M}} f_0 + \mathrm{C}[f_0,\overline{f}_0] = 0, \tag{6.4a}$$

$$\mathrm{i}\widehat{\mathrm{M}} f_1 + \widehat{\mathrm{C}} f_1 = \boldsymbol{w}\cdot\boldsymbol{\nabla} f_0 + (mn)^{-1}\boldsymbol{\nabla} p\cdot\partial_{\boldsymbol{w}} f_0, \tag{6.4b}$$

$$\begin{aligned}\mathrm{i}\widehat{\mathrm{M}} f_2 + \widehat{\mathrm{C}} f_2 = \boldsymbol{w}\cdot\boldsymbol{\nabla} f_1 + (mn)^{-1}\boldsymbol{\nabla} p\cdot\partial_{\boldsymbol{w}} f_1&\\ + [\partial_t f_0 + \boldsymbol{u}\cdot\boldsymbol{\nabla} f_0 - \boldsymbol{w}\cdot(\boldsymbol{\nabla}\boldsymbol{u})\cdot\partial_{\boldsymbol{w}} f_0] - \mathrm{C}[f_1,\overline{f}_1].&\end{aligned} \tag{6.4c}$$

Solution of (6.4a) leads to the shifted Maxwellian

$$f_0 = n(2\pi v_{\mathrm{t}}^2)^{-3/2}\mathrm{e}^{-w^2/2v_{\mathrm{t}}^2}, \tag{6.5}$$

---

[45] A further assumption is that the effect of the magnetic field is negligible during a collision (i.e., that $\omega_{\mathrm{c}}/\omega_{\mathrm{p}} \ll 1$).

[46] As noted on page 9, Catto & Simakov define the two ordering parameters $\Delta \doteq \lambda_{\mathrm{mfp}}\nabla_{\parallel}$ and $\delta \doteq \rho\nabla_{\perp}$. They assume that $\delta\sim\Delta\sim\epsilon$.

[47] This is equivalent to assuming that $\boldsymbol{u}_{\perp}$ is of the order of the diamagnetic flow $\boldsymbol{u}_* \doteq \widehat{\boldsymbol{b}}\times\boldsymbol{\nabla} p/(mn\omega_{\mathrm{c}})$, since $u_*/v_{\mathrm{t}} \sim \rho\nabla_{\perp}\ln T = O(\delta)$. It is also equivalent to the ordering $\boldsymbol{u}\sim\boldsymbol{q}/p$, where $\boldsymbol{q} = \boldsymbol{q}_{\parallel} + \boldsymbol{q}_* + \boldsymbol{q}_{\perp}$ is the classical heat flow, since one has $\boldsymbol{q}_{\parallel}/p \sim n(v_{\mathrm{t}}^2/\nu)\nabla_{\parallel}T/(nT) \sim v_{\mathrm{t}}\Delta$ and $\boldsymbol{q}_{\perp}/p \sim n(\rho^2\nu)\nabla_{\perp}T/(nT) \sim v_{\mathrm{t}}\delta(\nu/\omega_{\mathrm{c}}) \sim v_{\mathrm{t}}\delta$.

where $n = n(\boldsymbol{x}_1, t_1, \dots)$ and similarly for the $T$ in $v_{\mathrm{t}}^2 \doteq T/m$. Use of the fluid equations gives

$$\partial_t f_0 + \boldsymbol{u} \cdot \boldsymbol{\nabla} f_0 \approx -\frac{2}{3} x^2 \boldsymbol{\nabla} \cdot \boldsymbol{u} - \left(\frac{2}{3} x^2 - 1\right) \frac{1}{p} \boldsymbol{\nabla} \cdot \boldsymbol{q}, \tag{6.6}$$

where $x^2 \doteq w^2/2v_{\mathrm{t}}^2$.

Because Catto & Simakov do not explicitly describe a multiple-scale procedure, the origin of the terms on the right-hand sides of (6.4*b*) and (6.4*c*) may not be totally clear, so I shall provide a bit more detail. As discussed in appendix I:A, a systematic discussion of Chapman–Enskog theory includes, in addition to expansion of the distribution function, a multiple-scale expansion in both space and time:

$$\frac{\partial}{\partial t} = \frac{\partial}{\partial t_0} + \epsilon \frac{\partial}{\partial t_1} + \epsilon^2 \frac{\partial}{\partial t_2} + \cdots, \tag{6.7}$$

and similarly for $\partial/\partial \boldsymbol{x}$. Here $t_0$ and $\boldsymbol{x}_0$ are the kinetic scales, $t_1$ and $\boldsymbol{x}_1$ are the transit scales, and $t_2$ and $\boldsymbol{x}_2$ are the transport scales; for the physics of those scales, see the discussion after (I:A 2). One then finds [see (I:A 4*a*)–(I:A 4*c*)]

$$\mathrm{i}\widehat{\mathrm{M}} f_0 = -\mathrm{C}[f_0, \overline{f}_0], \tag{6.8a}$$

$$\frac{\mathrm{D} f_0}{\mathrm{D} t_1} = -\widehat{\mathrm{C}} f_1, \tag{6.8b}$$

$$\frac{\mathrm{D} f_0}{\mathrm{D} t_2} + \frac{\mathrm{D} f_1}{\mathrm{D} t_1} = -(\widehat{\mathrm{C}} f_2 + \mathrm{C}[f_1, \overline{f}_1]), \tag{6.8c}$$

which should be compared to the Catto & Simakov equations [(6.4) above]. Clearly, in those equations one should take $\boldsymbol{\nabla} \equiv \boldsymbol{\nabla}_1$. In (6.2) written at first order, one identifies $t \equiv t_1$; the $\partial_{t_1} f_0$ term that is part of the $\mathrm{D} f_0/\mathrm{D} t$ in (6.8*b*) is eliminated by using (6.2), which ultimately leads to the right-hand side of (6.4*b*). The same procedure applied to $\partial_{t_1} f_1$ leads to the right-hand side of (6.4*c*), with the $t$ in $\partial_t f_0$ being identified as $t \equiv t_2$.

### 6.2. *Comparison of results*

Catto & Simakov focus more on obtaining approximate quantitative results (e.g., by means of variational methods) than on the general structure of the transport theory. The presence of a magnetic field complicates the formulas. Magnetic-field-related effects are obviously important for practical applications, as discussed by Catto & Simakov (2004, 2005); some of the effects they calculate are essential for a proper determination of the radial electric field in toroidal devices. In the present paper, my interest is on the general structure of the theory, so I shall not discuss explicit results for magnetic-field corrections at Burnett order. However, a connection to the general unmagnetised formulas may be obtained by examining the result of Catto & Simakov (2004) for the parallel viscosity tensor $\boldsymbol{\pi}_{\parallel}$. It is not hard to see that they obtain only a subset of the terms displayed in the general unmagnetised result (4.6); this is a natural consequence of the subsidiary ordering $\boldsymbol{u} \sim \epsilon$. Terms quadratic in $\boldsymbol{u}$ [the last four lines of (4.11)] are then negligible, being of third order. Furthermore, in this unmagnetised theory one must take the pressure gradient to be $O(\epsilon^2)$, as can be seen from the balance $\partial_t \boldsymbol{u} \approx -(mn)^{-1} \boldsymbol{\nabla} p$ with the ordering $\boldsymbol{u} = O(\epsilon)$. (Time derivatives are at least of first order.) All Burnett terms involving $\boldsymbol{\nabla} p$ are thus negligible, and (4.11) reduces to

$$(nm)^{-1}\boldsymbol{\pi}^{\mathrm{wc}} = -2\mu\left(\frac{1}{2}[\boldsymbol{\nabla}\boldsymbol{u} + (\boldsymbol{\nabla}\boldsymbol{u})^{\mathrm{T}}] - \frac{1}{d}\boldsymbol{\nabla}\boldsymbol{\cdot}\boldsymbol{u}\,\boldsymbol{I}\right), \\ + \mu_3\left(\boldsymbol{\nabla}\boldsymbol{\nabla}T - \frac{1}{d}\nabla^2 T\,\boldsymbol{I}\right) + \mu_5\left(\boldsymbol{\nabla}T\,\boldsymbol{\nabla}T - \frac{1}{d}|\boldsymbol{\nabla}T|^2\boldsymbol{I}\right), \tag{6.9}$$

where $\mu$, $\mu_3$, and $\mu_5$ are given by (4.7$a$), (4.8$c$), and (4.8$e$), respectively. Those coefficients are defined by the integrals $K^{\mathrm{I}}$, $K_1$, and $K_{20}$, which are expressed in terms of two-time correlation functions involving two phase-space points, as well as by $\underset{\sim}{K}_3$ and $\underset{\sim}{K}_5$, which are expressed in terms of two-time correlation functions involving three phase-space points. This unmagnetised result should be compared to the result of Catto & Simakov,

$$(nm)^{-1}\boldsymbol{\pi}_{\parallel} = (nm)^{-1}(p_{\parallel} - p_{\perp})\left(\widehat{\boldsymbol{b}}\widehat{\boldsymbol{b}} - \frac{1}{3}\boldsymbol{I}\right), \tag{6.10}$$

where

$$\begin{aligned}(nm)^{-1}(p_{\parallel} - p_{\perp}) \approx{}& \\ & - c_1\left(\frac{v_{\mathrm{t}}^2}{\nu}\right)\left(\widehat{\boldsymbol{b}}\boldsymbol{\cdot}(\boldsymbol{\nabla}\boldsymbol{u})\boldsymbol{\cdot}\widehat{\boldsymbol{b}} - \frac{1}{3}\boldsymbol{\nabla}\boldsymbol{\cdot}\boldsymbol{u}\right) \\ & - \left(\frac{v_{\mathrm{t}}^2}{\nu}\right)\frac{1}{p}\left[c_2\left(\widehat{\boldsymbol{b}}\boldsymbol{\cdot}(\boldsymbol{\nabla}\boldsymbol{q})\boldsymbol{\cdot}\widehat{\boldsymbol{b}} - \frac{1}{3}\boldsymbol{\nabla}\boldsymbol{\cdot}\boldsymbol{q}\right) + c_3\left(\widehat{\boldsymbol{b}}\boldsymbol{\cdot}(\boldsymbol{\nabla}\boldsymbol{q}_{\parallel})\boldsymbol{\cdot}\widehat{\boldsymbol{b}} - \frac{1}{3}\boldsymbol{\nabla}\boldsymbol{\cdot}\boldsymbol{q}_{\parallel}\right)\right] \\ & + c_4\left(\frac{v_{\mathrm{t}}^2}{\nu}\right)\frac{1}{p}\left(q_{\parallel}\nabla\ln p - \frac{1}{3}\boldsymbol{q}\boldsymbol{\cdot}\boldsymbol{\nabla}\ln p\right) \\ & - c_5\left(\frac{v_{\mathrm{t}}^2}{\nu}\right)\left(\frac{q_{\parallel}}{p}\right)\nabla_{\parallel}\ln T \\ & - c_6\left(\frac{q}{p}\right)^2 + c_7\left(\frac{q_{\parallel}}{p}\right)^2\end{aligned} \tag{6.11}$$

and the $c$'s are positive numerical coefficients — for example,

$$c_1 \doteq 3\boldsymbol{\cdot}\frac{1025}{1068}, \quad c_7 \doteq \frac{1137}{17\,800}. \tag{6.12}$$

(Such rational fractions arise from the evaluation of variational forms.) The last two terms of (6.11) arise from the term $\mathrm{C}[f_1, \overline{f}_1]$ in (6.4$c$). The parallel component of the unmagnetised Navier–Stokes stress [the first line of (6.9)] agrees in form with the first line of (6.11). For the remainder of this discussion, let us examine the parallel physics by replacing in (6.9) $\boldsymbol{\nabla} \to \widehat{\boldsymbol{b}}\nabla_{\parallel}$; then the tensor $\widehat{\boldsymbol{b}}\widehat{\boldsymbol{b}} - \frac{1}{3}\boldsymbol{I}$ emerges, in agreement with (6.10). One anticipates that the $c_6$ and $c_7$ terms in (6.11), which arise from the nonlinear collision operator, are related to the contributions to (6.9) stemming from three-point correlations; I shall discuss this further below.

Clearly, both formalisms will generate the same Navier–Stokes coefficients, so I shall focus on establishing consistency for the Burnett coefficients. That is, one should compare

$$\frac{2}{3}\mu_3\nabla_{\parallel}^2 T + \frac{2}{3}\mu_5(\nabla_{\parallel}T)^2 \quad \text{(Brey)} \tag{6.13}$$

with

$$-\frac{2}{3}\left(\frac{v_{\rm t}^2}{\nu}\right)\frac{1}{p}(c_2+c_3)\nabla_\| q_\| - c_5\left(\frac{v_{\rm t}^2}{\nu}\right)\left(\frac{q_\|}{p}\right)\nabla_\| \ln T$$
$$- c_6\left(\frac{q}{p}\right)^2 + c_7\left(\frac{q_\|}{p}\right)^2 \quad \text{(Catto \& Simakov).} \tag{6.14}$$

Unfortunately, term by term comparison is not possible because Brey expresses his results solely in terms of gradients of the thermodynamic forces whereas Catto & Simakov use a mixed representation in which gradients of the heat flux appear in addition to gradients of pressure and temperature. Since one has $q_\| = -\kappa_\|(T)\nabla_\| T$, a contribution to $(\nabla_\| T)^2$ arises from $\nabla_\| \kappa$ in addition to the explicit $q_\| \nabla_\| T$ term in (6.14). The overall forms of the results, involving terms in $\nabla_\|^2 T$ and $(\nabla_\| T)^2$ (having the proper scaling with the dimensional variables), clearly agree, but this is not surprising since it could have been predicted on the basis of symmetry considerations. Quantitative comparison is also not possible because the two-time formulas are formally exact, whereas the results of Catto & Simakov are approximate. However, one can return to the multiple-scale expansion employed by Catto & Simakov, inquire about the formal expressions (involving, for example, $\widehat{\mathrm{C}}^{-1}$) that it predicts, and compare those to the predictions of the two-time formalism. That is done for a particular example in the next section.

### 6.3. *Example: The Burnett viscosity coefficient of* $\boldsymbol{\nabla} T\,\boldsymbol{\nabla} T$

In (4.11) for the dissipative momentum flux, it is seen that in Brey's notation the coefficient of $\boldsymbol{\nabla} T\,\boldsymbol{\nabla} T$ is called $\mu_5$. Formula (4.8*e*) for $\mu_5$ reduces in the limit of weak coupling to

$$\mu_5 = 4(nmT^3)^{-1}K_1 - 2(nmT^2)^{-1}\left(T^{-2}\underset{\sim}{K}_5 - \frac{5}{2}T^{-1}\underset{\sim}{K}_3 - \frac{5}{2}m^{-1}K_{20}\right). \tag{6.15}$$

By solving the equations for the relevant two- and three-point correlation functions that determine the integrals $K_1$, $\underset{\sim}{K}_3$, $\underset{\sim}{K}_5$, and $K_{20}$, I shall make this formula more explicit and ultimately compare it successfully with the prediction of the Chapman–Enskog formalism.

#### 6.3.1. The $K_1$ integral

The $K_1$ integral is defined by (4.15*e*), repeated here for convenience:

$$I_{1,ijkl} \doteq \int_0^\infty \mathrm{d}\overline{\tau}\int \mathrm{d}\overline{\boldsymbol{x}}\,\langle \widehat{\tau}_{ij}(\mathbf{0})\mathrm{R}_0(\overline{\tau})\widehat{J}^E_k(\overline{\boldsymbol{x}})\rangle_0 \overline{x}_l = K_1(\delta_{ik}\delta_{jl}+\delta_{il}\delta_{jk}) + K_2\delta_{ij}\delta_{kl}. \tag{6.16}$$

It was shown in §3.2.1 how to express the required expectations in terms of two- or three-point correlation functions. With $\overline{\boldsymbol{x}} = -\overline{\boldsymbol{\rho}}$, one has

$$\boldsymbol{I}_1 \doteq -\int \mathrm{d}\overline{\tau}\int \mathrm{d}\overline{\boldsymbol{\rho}}\,\langle \widehat{\boldsymbol{\tau}}(\mathbf{0})\mathrm{R}_0(\overline{\tau})\widehat{\boldsymbol{J}}^E(-\overline{\boldsymbol{\rho}})\rangle_0\overline{\boldsymbol{\rho}} \tag{6.17a}$$

$$= -T\int \mathrm{d}\overline{\tau}\int \mathrm{d}\overline{\boldsymbol{\rho}}\int \mathrm{d}\boldsymbol{v}\int \mathrm{d}\boldsymbol{v}'\,\widehat{\boldsymbol{\tau}}(\boldsymbol{v})C_+^{(1;1)}(\overline{\boldsymbol{\rho}},\boldsymbol{v},\tau,\boldsymbol{v}')\boldsymbol{\beta}(\boldsymbol{v}')\,\overline{\boldsymbol{\rho}} \tag{6.17b}$$

$$= T\int \mathrm{d}\overline{\tau}\int \mathrm{d}\boldsymbol{v}\int \mathrm{d}\boldsymbol{v}'\,\widehat{\boldsymbol{\tau}}(\boldsymbol{v})\boldsymbol{\beta}(\boldsymbol{v}')\left.\frac{\partial C^{(1;1)}_{+;\boldsymbol{k}'}(\boldsymbol{v},\overline{\tau};\boldsymbol{v}')}{\partial(-\mathrm{i}\boldsymbol{k}')}\right|_{\boldsymbol{k}'=\mathbf{0}}, \tag{6.17c}$$

where

$$\boldsymbol{\beta}(\boldsymbol{v}) \equiv \widehat{\boldsymbol{J}}^E(\boldsymbol{v}) \doteq \left(\frac{mv^2/2}{T} - \frac{5}{2}\right)\boldsymbol{v}. \tag{6.18}$$

The two-time correlation function is governed by (5.27), where the last term of that equation is replaced by $-\widehat{\mathrm{C}}C^{(1;1)}_{+;\boldsymbol{k}'}$. The solution is

$$C^{(1;1)}_{+;\boldsymbol{k}'}(\overline{\tau}) = \mathrm{e}^{-(\mathrm{i}\boldsymbol{k}'\cdot\boldsymbol{v}+\widehat{\mathrm{C}})\overline{\tau}}\,C^{(0,2)}_{\boldsymbol{k}'}(\boldsymbol{v},\overline{\boldsymbol{v}}). \tag{6.19}$$

The initial condition contains a singular term, displayed in (5.36), plus a term in $g_{\boldsymbol{k}'}(\boldsymbol{v},\overline{\boldsymbol{v}})$ that can be shown to phase-mix away. The $\overline{\boldsymbol{v}}$ integration can be performed over the singular term. The $\overline{\tau}$ integration can be done, leading to the operator $(\mathrm{i}\boldsymbol{k}'\cdot\boldsymbol{v}+\widehat{\mathrm{C}})^{-1}$. The operator relation $\mathrm{dA}^{-1} = -\mathrm{A}^{-1}(\mathrm{dA})\mathrm{A}^{-1}$ allows the $\boldsymbol{k}'$ derivative to be evaluated. The final result is the velocity-space matrix element

$$I_1 = -T\langle\widehat{\boldsymbol{\tau}}(\boldsymbol{v}) \mid \widehat{\mathrm{C}}^{-1}\boldsymbol{v}\widehat{\mathrm{C}}^{-1} \mid \boldsymbol{\beta}(\boldsymbol{v})\rangle_0. \tag{6.20}$$

6.3.2. Three-point correlations and the nonlinear collision operator

I now turn to the evaluation of the remaining terms in (6.15), namely

$$\underset{\sim}{\mu}_5 \doteq -2(nmT^2)^{-1}\left(T^{-2}\underset{\sim}{K}_5 - \frac{5}{2}T^{-1}\underset{\sim}{K}_3 - \frac{5}{2}m^{-1}K_{20}\right). \tag{6.21}$$

From (4.15*f*), (4.15*g*), and (4.15*l*), one has

$$\begin{aligned}
&\left(T^{-2}\underset{\sim}{K}_5 - \frac{5}{2}T^{-1}\underset{\sim}{K}_3 - \frac{5}{2}m^{-1}K_{20}\right)(\delta_{ik}\delta_{jl}+\delta_{il}\delta_{jk})\\
&\quad + \left(T^{-2}\underset{\sim}{K}_6 - \frac{5}{2}T^{-1}\underset{\sim}{K}_4 - \frac{5}{2}m^{-1}K_{21}\right)\delta_{ij}\delta_{kl}\\
&= \int_0^\infty \mathrm{d}\overline{\tau}\int \mathrm{d}\overline{\boldsymbol{x}}\left\langle \widehat{\tau}_{ij}(\mathbf{0})\mathrm{R}_0(\overline{\tau})(T^{-1}\widehat{\mathscr{J}}^E_k)\left(\frac{1}{T}E'(\overline{\boldsymbol{x}}) - \frac{5}{2}N'(\overline{\boldsymbol{x}})\right)\right\rangle_0 \overline{x}_l\\
&\quad - \frac{5}{2}\int_0^\infty \mathrm{d}\overline{\tau}\int \mathrm{d}\overline{\boldsymbol{x}}\,\langle\widehat{\tau}_{ij}(\mathbf{0})\mathrm{R}_0(\overline{\tau})m^{-1}P_k(\overline{\boldsymbol{x}})\rangle_0\overline{x}_l.
\end{aligned} \tag{6.22}$$

It was shown in §3.2.1 how to express the required expectations in terms of two- or three-point correlation functions. First consider the last integral, which stems from $K_{20}$. The same sequence of operations that led to (6.20) produces

$$I_{20} = -\frac{5}{2}\langle\widehat{\boldsymbol{\tau}}(\boldsymbol{v}) \mid \mathrm{C}^{-1}\boldsymbol{v}\mathrm{C}^{-1} \mid \boldsymbol{v}\rangle_0. \tag{6.23}$$

Unfortunately, this integral is infinite for a one-component plasma since $|\boldsymbol{v}\rangle$ is in the null space of $\widehat{\mathrm{C}}$. Upon inquiring into the origin of this term, one learns that it stems from the expection $\boldsymbol{h}^{ee}_2[\widehat{\boldsymbol{\tau}}](\mu,t)$ [see (3.40*d*)], which contains a Q operator. Brey's integrals $\underset{\sim}{K}_5$ and $K_{20}$ arise by writing $\mathrm{Q} = 1 - \mathrm{P}$; $\underset{\sim}{K}_5$ arises from the 1, and $K_{20}$ arises from the P operation. The role of Q is precisely to prevent such infinities from occurring; thus, one expects that $\underset{\sim}{K}_5$ will contain a singular part that will be exactly cancelled by $K_{20}$, and this will be seen to be true.

Now consider the first line of (6.22),

$$I_{5\text{-}3} \doteq -\int_0^\infty \mathrm{d}\overline{\tau}\int \mathrm{d}\overline{\boldsymbol{\rho}}\left\langle \widehat{\boldsymbol{\tau}}(\mathbf{0})\mathrm{R}_0(\overline{\tau})(T^{-1}\widehat{\mathscr{J}}^E)\left(\frac{1}{T}E'(-\overline{\boldsymbol{\rho}}) - \frac{5}{2}N'(-\overline{\boldsymbol{\rho}})\right)\right\rangle_0\overline{\boldsymbol{\rho}}, \tag{6.24}$$

which can be expressed in terms of the triplet correlation function $C^{(1;2)}_{+;\boldsymbol{k}',\boldsymbol{k}''}$. The presence of $\widehat{\mathscr{J}}^E$ (the total amount of $\widehat{\boldsymbol{J}}^E$) in this expression in (6.24) means that only the $\boldsymbol{k}' = \mathbf{0}$ limit is required, and the $\overline{\boldsymbol{\rho}}$ integral requires only $\partial/\partial(-\mathrm{i}\boldsymbol{k}'')|_{\boldsymbol{k}''=\mathbf{0}}$. It was shown in §G.3 that (5.45) for the triplet correlation function $C^{(1;2)}_+$ becomes approximately (5.49). The solution of (5.49) (which contains an inhomogeneous, $\tau$-dependent source term) is

$$C^{(1;2)}_{+;\mathbf{0},\boldsymbol{k}''}(\underline{1},\tau;\underline{1}',\underline{1}'') = \mathrm{e}^{-(\mathrm{i}\boldsymbol{k}''\cdot\boldsymbol{v}+\widehat{\mathrm{C}})\tau}C^{(0,3)}_{\mathbf{0},\boldsymbol{k}''}(\underline{1},\underline{1}',\underline{1}'') - \int_0^\tau \mathrm{d}\widehat{\tau}\,\mathrm{e}^{-(\mathrm{i}\boldsymbol{k}''\cdot\boldsymbol{v}+\widehat{\mathrm{C}})(\tau-\widehat{\tau})}\{\mathrm{C}[f;\,C^{(1;1)}_{+}(\widehat{\tau};\underline{1}'),\,\overline{C}^{(1;1)}_{+;\boldsymbol{k}''}(\widehat{\tau};\underline{1}'')] + (1'\Leftrightarrow 1'')\}. \tag{6.25}$$

(For brevity, I shall drop the BL label on $\mathrm{C}[f;a,\overline{b}]$.) In addition to the $\boldsymbol{\beta}(\boldsymbol{v})$ defined by (6.18), let us define $\gamma(\boldsymbol{v}) \doteq \frac{1}{2}mv^2/T - \frac{5}{2}$; one has $\boldsymbol{\beta} = \gamma\boldsymbol{v}$. To evaluate $\boldsymbol{I}_{5\text{-}3}$, the solution (6.25) must be multiplied by $\boldsymbol{\beta}(\boldsymbol{v}')$ and $\gamma(\boldsymbol{v}'')$, integrated over $\tau$, $\boldsymbol{v}$, $\boldsymbol{v}'$, and $\boldsymbol{v}''$, and differentiated with respect to $-\mathrm{i}\boldsymbol{k}''$ at $\boldsymbol{k}''=\mathbf{0}$. Performing those operations first on the singular initial-condition term leads[48] to the matrix element

$$-\langle\widehat{\boldsymbol{\tau}}\mid\widehat{\mathrm{C}}^{-1}\boldsymbol{v}\widehat{\mathrm{C}}^{-1}\mid\boldsymbol{\beta}\gamma\rangle_0. \tag{6.26}$$

Write $\boldsymbol{\beta}\gamma = (\mathrm{P}+\mathrm{Q})\boldsymbol{\beta}\gamma$. By symmetry, only the momentum projection contributes. One has

$$\mathrm{P}|\boldsymbol{\beta}\gamma\rangle \to |\boldsymbol{v}\rangle\frac{1}{v_{\mathrm{t}}^2}\cdot\langle\boldsymbol{v}\,\boldsymbol{\beta}\gamma\rangle = \frac{5}{2}|\boldsymbol{v}\rangle. \tag{6.27}$$

This contribution is seen to exactly cancel $\boldsymbol{I}_{20}$, as was predicted.[49] For future use, note that

$$\mathrm{Q}|\boldsymbol{\beta}\gamma\rangle = |\boldsymbol{\beta}\gamma\rangle - \mathrm{P}|\boldsymbol{\beta}\gamma\rangle = \left(\gamma^2-\frac{5}{2}\right)|\boldsymbol{v}\rangle. \tag{6.28}$$

Now consider the processing of the source term (second line) of (6.25). The time integral from 0 to $\infty$ of a one-sided function is its $\omega=0$ Fourier component. Because the source term is in convolution form, its Fourier transform is the product of the individual Fourier transforms of $\exp[-\mathrm{i}(\boldsymbol{k}''\cdot\boldsymbol{v}+\widehat{\mathrm{C}})\tau]$ and $\mathrm{C}[f;C'(\tau),\overline{C}''(\tau)]$. Because $\mathrm{C}[f;a,\overline{b}]$ is multiplicative in its last two slots [involving $a(\boldsymbol{v})b(\overline{\boldsymbol{v}})$, where $\overline{\boldsymbol{v}}$ is the integration variable in the collision operator], that $\tau$ dependence is $\exp[-(\widehat{\mathrm{C}}+\mathrm{i}\boldsymbol{k}''\cdot\overline{\boldsymbol{v}}+\widehat{\overline{\mathrm{C}}})\tau] + (\boldsymbol{v}\Leftrightarrow\overline{\boldsymbol{v}})$. Therefore, the collisional contribution is

$$\boldsymbol{I}^{\mathrm{C}}_{5\text{-}3} = \int\mathrm{d}\boldsymbol{v}\,\widehat{\boldsymbol{\tau}}(\boldsymbol{v})\Big([\mathrm{i}(\boldsymbol{k}''\cdot\boldsymbol{v}+\widehat{\mathrm{C}})]^{-1} \times\{\mathrm{C}[f;(\widehat{\mathrm{C}}+\mathrm{i}\boldsymbol{k}''\cdot\overline{\boldsymbol{v}}+\widehat{\overline{\mathrm{C}}})^{-1}\,|\,\boldsymbol{\beta}f,\overline{\gamma f}] + (\boldsymbol{v}\Leftrightarrow\overline{\boldsymbol{v}})\}\Big)\frac{\overleftarrow{\partial}}{\partial(-\mathrm{i}\boldsymbol{k}'')}\bigg|_{\boldsymbol{k}''=\mathbf{0}}. \tag{6.29}$$

Here I have introduced the notation $\mathrm{C}[f;\lambda\,|\,a,\overline{b}]$ to indicate that the scaling factor $\lambda$ operates on the product $a(\boldsymbol{v})b(\overline{\boldsymbol{v}})$. The $\boldsymbol{k}''$ derivative[50] generates two contributions, one from each of the operator inverses. I shall consider each in turn.

- Differentiating $\mathrm{C}[f;\lambda(\boldsymbol{k}'')\,|\,a,\overline{b}]$: Upon performing the $\boldsymbol{k}''$ derivative on the second line of (6.29), one is led to the integrand

$$I_{kl} \doteq (\widehat{\mathrm{C}}+\widehat{\overline{\mathrm{C}}})^{-1}[\overline{v}_l(\widehat{\mathrm{C}}+\widehat{\overline{\mathrm{C}}})^{-1}\beta_k f\overline{\gamma}\overline{f} + (\boldsymbol{v}\Leftrightarrow\overline{\boldsymbol{v}})]. \tag{6.30}$$

[48] A contribution from the derivative of $C^{(0;3)}_{\mathbf{0},\boldsymbol{k}''}$ with respect to $\boldsymbol{k}''$ can be shown to phase-mix away.

[49] Obviously, this consistency check is being done here in the limit of weak coupling. However, such cancellations must occur in general if the formalism is to be consistent.

[50] The $\boldsymbol{k}''$ integral appears here on the right, operating to the left, in order to retain the proper order of the tensor indices. In the subsequent expressions, I shall not bother with that nicety because the tensor can be shown to be symmetric in its last two indices.

Because $|\overline{\gamma}\rangle$ is a null eigenfunction of $\overline{\widehat{\mathrm{C}}}$, $I_{kl}$ reduces to

$$I_{kl} = (\widehat{\mathrm{C}} + \overline{\widehat{\mathrm{C}}})^{-1}[(\widehat{\mathrm{C}}^{-1}\beta_k f)(\overline{\beta}_l \overline{f}) + (\boldsymbol{v} \Leftrightarrow \overline{\boldsymbol{v}})]. \tag{6.31}$$

Now use the identity

$$(\widehat{\mathrm{C}} + \overline{\widehat{\mathrm{C}}})^{-1} = \overline{\widehat{\mathrm{C}}}^{-1} - (\widehat{\mathrm{C}} + \overline{\widehat{\mathrm{C}}})^{-1}\widehat{\mathrm{C}}\overline{\widehat{\mathrm{C}}}^{-1} \tag{6.32}$$

to find that

$$I_{kl} = (\widehat{\mathrm{C}}^{-1}\beta_k f)(\overline{\widehat{\mathrm{C}}}^{-1}\overline{\beta}_l \overline{f}) - (\widehat{\mathrm{C}} + \overline{\widehat{\mathrm{C}}})^{-1}[\beta_k(\overline{\widehat{\mathrm{C}}}^{-1}\overline{\beta}_l \overline{f}] + (\boldsymbol{v} \Leftrightarrow \overline{\boldsymbol{v}}). \tag{6.33$a$}$$

The construction

$$|\boldsymbol{\zeta}\rangle \doteq -\widehat{\mathrm{C}}^{-1}|\boldsymbol{\beta}\rangle \tag{6.34}$$

can be recognized as the velocity-dependent part of the first-order distribution function driven by temperature gradients; see (I:A 21). Equation (6.33$a$) thus reduces to

$$I_{kl} = 2\zeta_k\zeta_l - I_{lk}, \tag{6.35}$$

which determines the symmetric part of $I_{kl}$ to be

$$S_{kl} \doteq \frac{1}{2}(I_{kl} + I_{lk}) = \zeta_k\zeta_l. \tag{6.36}$$

Only the symmetric part contributes to $\boldsymbol{I}^{\mathrm{C}}_{5\text{-}3}$. Thus, when (6.36) is inserted into (6.29), one obtains

$${}_a\boldsymbol{I}^{\mathrm{C}}_{5\text{-}3} = \int \mathrm{d}\boldsymbol{v}\, \widehat{\boldsymbol{\tau}}(\boldsymbol{v})\widehat{\mathrm{C}}^{-1}\mathrm{C}[f;\boldsymbol{\zeta},\overline{\boldsymbol{\zeta}}]. \tag{6.37}$$

• <u>Differentiating $[\mathrm{i}(\boldsymbol{k}'' \cdot \boldsymbol{v} + \widehat{\mathrm{C}})]^{-1}\mathrm{C}$</u>: Upon differentiating the first term of (6.29), one is led to the contribution

$$\begin{aligned} {}_b\boldsymbol{I}^{\mathrm{C}}_{5\text{-}3} &\doteq \int \mathrm{d}\boldsymbol{v}\, \widehat{\boldsymbol{\tau}}(\boldsymbol{v})\widehat{\mathrm{C}}^{-1}\boldsymbol{v}\widehat{\mathrm{C}}^{-1}\{\mathrm{C}[f;(\widehat{\mathrm{C}} + \overline{\widehat{\mathrm{C}}})^{-1}\,|\,\boldsymbol{\beta}f,\overline{\gamma}\overline{f}] + (\boldsymbol{v} \Leftrightarrow \overline{\boldsymbol{v}})\} & (6.38a)\\ &= -\int \mathrm{d}\boldsymbol{v}\, \widehat{\boldsymbol{\tau}}(\boldsymbol{v})\widehat{\mathrm{C}}^{-1}\boldsymbol{v}\widehat{\mathrm{C}}^{-1}\{\mathrm{C}[f;\boldsymbol{\zeta},\overline{\gamma}\overline{f}] + \mathrm{C}[f;\gamma f,\overline{\boldsymbol{\zeta}}]\}. & (6.38b) \end{aligned}$$

[Recall that $\boldsymbol{\zeta}$ was defined by (6.34).] It is useful to manipulate this result as follows. Define $\alpha(\boldsymbol{v}) \doteq \frac{1}{2}mv^2/T$, then add and subtract a factor of $\frac{1}{2}$ to each of the $\gamma$ factors (note that $\gamma + \frac{1}{2} = \alpha - \frac{3}{2}$):

$$\begin{aligned} \mathrm{C}[\boldsymbol{\zeta},\overline{\gamma}\overline{f}] + \mathrm{C}[f;\gamma f,\overline{\boldsymbol{\zeta}}] &= \mathrm{C}[f;\boldsymbol{\zeta},(\overline{\alpha} - \tfrac{3}{2})\overline{f}] + \mathrm{C}[f;(\alpha - \tfrac{3}{2})f,\overline{\boldsymbol{\zeta}}]\\ &\quad - \frac{1}{2}(\mathrm{C}[f;\widehat{\mathrm{C}}^{-1}\boldsymbol{\beta}f,\overline{f}] + \mathrm{C}[f;f,\widehat{\mathrm{C}}^{-1}\overline{\boldsymbol{\beta}}\overline{f}]). \end{aligned} \tag{6.39}$$

The expression in the second line can be recognized as $-\frac{1}{2}\widehat{\mathrm{C}}(\widehat{\mathrm{C}}^{-1}|\boldsymbol{\beta}\rangle) = -\frac{1}{2}|\boldsymbol{\beta}\rangle$. When this term is inserted into (6.38$b$), it produces a term that cancels half of the matrix element $\boldsymbol{I}_1$ [(6.20)].

I shall now summarize these results in a form that can be directly compared with the predictions of Chapman–Enskog theory, which are obtained in the next section. Upon taking into account the coefficients in (6.15) and the 50% cancellation noted above

(reflected by the factor of $\frac{1}{2}$ in the square brackets in the next equation), one needs

$$\frac{1}{2}\left[4\left(\frac{1}{2}\right)T^{-1}\boldsymbol{l}_1 - (\boldsymbol{l}_{5\text{-}3} + \boldsymbol{l}_{20}\right] = \underbrace{-\langle\widehat{\boldsymbol{\tau}}|\widehat{\mathrm{C}}^{-1}\boldsymbol{v}\widehat{\mathrm{C}}^{-1}|\boldsymbol{\beta}\rangle}_{\text{(a)}} + \underbrace{\langle\widehat{\boldsymbol{\tau}}|\widehat{\mathrm{C}}^{-1}\boldsymbol{v}\widehat{\mathrm{C}}^{-1}|\left(\gamma^2 - \tfrac{5}{2}\right)\boldsymbol{v}\rangle}_{\text{(b)}}$$
$$+\underbrace{\langle\widehat{\boldsymbol{\tau}}|\widehat{\mathrm{C}}^{-1}\boldsymbol{v}\widehat{\mathrm{C}}^{-1}|f^{-1}\{\mathrm{C}[f;\boldsymbol{\zeta},(\overline{\alpha}-\tfrac{3}{2})\overline{f}] + \mathrm{C}[f;(\alpha-\tfrac{3}{2})f,\overline{\boldsymbol{\zeta}}]\}\rangle}_{\text{(c)}}$$
$$\underbrace{-\langle\widehat{\boldsymbol{\tau}}|\widehat{\mathrm{C}}^{-1}|f^{-1}\mathrm{C}[f;\boldsymbol{\zeta},\overline{\boldsymbol{\zeta}}]\rangle}_{\text{(d)}}. \tag{6.40}$$

These tensors are to be contracted with $T^{-2}\boldsymbol{\nabla}T\,\boldsymbol{\nabla}T$. [By symmetry, that introduces a factor of 2, which is accounted for by the first factor of $\frac{1}{2}$ on the left-hand side of (6.40).]

### 6.4. *Explicit formulas from Chapman–Enskog theory*

One may now compare these predictions from the two-time theory to those that follow from standard Chapman–Enskog theory applied to the Landau kinetic equation. One must solve (6.8*c*),

$$\widehat{\mathrm{C}}f_2 = -\frac{\mathrm{D}f_0}{\mathrm{D}t_2} - \frac{\mathrm{D}f_1}{\mathrm{D}t_1} - \mathrm{C}[f_1,\overline{f}_1]. \tag{6.41}$$

The solvability conditions for this equation were discussed in appendix I:A; they amount to requiring that the equations are satisfied at Navier–Stokes order. The term in $\mathrm{D}f_0/\mathrm{D}t_2$ merely leads to second-order multiple-scale corrections to the first-order equation, so I focus on the last two terms. One has

$$f_1 = -\widehat{\mathrm{C}}^{-1}\left[\frac{1}{T}\left(\alpha-\frac{5}{2}\right)\boldsymbol{v}\cdot\boldsymbol{\nabla}T f_0\right]. \tag{6.42}$$

I shall evaluate $-\mathrm{D}f_1/\mathrm{D}t_1$ using the small-flow ordering of Catto & Simakov. Then $\partial_t f_1$ is negligible through second order, as is the term in $\boldsymbol{E}\cdot\boldsymbol{\partial}$. Upon noting that $\partial_T\alpha = -T^{-1}\alpha$ and that $\boldsymbol{\nabla}\ln f_0 = (\alpha-\frac{3}{2})T^{-1}\boldsymbol{\nabla}T$ when only temperature gradients are considered, one has

$$-\boldsymbol{v}\cdot\boldsymbol{\nabla}f_1 = \frac{1}{T^2}\boldsymbol{v}\cdot\widehat{\mathrm{C}}^{-1}\left\{\left[\underbrace{-\left(\alpha-\frac{5}{2}\right)}_{\text{(a)}}f_0\underbrace{-\alpha+\left(\alpha-\frac{5}{2}\right)\left(\alpha-\frac{3}{2}\right)}_{\text{(b)}}\right]\boldsymbol{v}f_0\right\}\cdot\boldsymbol{\nabla}T\,\boldsymbol{\nabla}T$$
$$+\underbrace{\boldsymbol{v}\widehat{\mathrm{C}}^{-1}(\boldsymbol{\nabla}\widehat{\mathrm{C}})f_1\cdot\boldsymbol{\nabla}T}_{\text{(c)}} + O(\nabla^2 T). \tag{6.43}$$

Note that it was necessary to differentiate the collision operator because it is a functional of $f_0$.[51] I shall not calculate the term in $\nabla^2 T$ because in this example I am restricting my attention to just the coefficient of $(\nabla T)^2$.

Upon noting that the previously defined $\boldsymbol{\beta}(\boldsymbol{v}) = [\alpha(\boldsymbol{v})-\frac{5}{2}]\boldsymbol{v}$, one finds that the second-order perturbation driven by term (a) is

$$|{}_a\chi_2\rangle = -T^{-2}\widehat{\mathrm{C}}^{-1}\boldsymbol{v}\widehat{\mathrm{C}}^{-1}|\boldsymbol{\beta}\rangle : \boldsymbol{\nabla}T\,\boldsymbol{\nabla}T. \tag{6.44}$$

[51] Catto & Simakov do not explicitly differentiate a formal collision operator; however, they do differentiate the solution of their first-order Chapman–Enskog equation, which is equivalent.

Term (b) can be rewritten more conveniently as

$$(\mathrm{b}) = \left(\alpha - \frac{5}{2}\right)^2 - \frac{5}{2} = \left(\gamma^2 - \frac{5}{2}\right), \tag{6.45}$$

which gives rise to the contribution

$$|_b\chi_2\rangle = \frac{1}{T^2}\widehat{\mathrm{C}}^{-1}\boldsymbol{v}\widehat{\mathrm{C}}^{-1}\left|\left(\gamma^2 - \frac{5}{2}\right)\boldsymbol{v}\right\rangle : \boldsymbol{\nabla}T\,\boldsymbol{\nabla}T. \tag{6.46}$$

Term (c) can be reduced by explicitly performing the gradient operation on $f_0$ to find

$$(\boldsymbol{\nabla}\widehat{\mathrm{C}})f_1 = T^{-1}\{\mathrm{C}[(\alpha - \tfrac{3}{2})f_0, \overline{\boldsymbol{\zeta}}] + \mathrm{C}[\boldsymbol{\zeta}, (\overline{\alpha} - \tfrac{3}{2})\overline{f}]\} \cdot \boldsymbol{\nabla}T. \tag{6.47}$$

Finally, the nonlinear collision operator drives the correction

$$|_d\chi_2\rangle = -\widehat{\mathrm{C}}^{-1}|f_0^{-1}\mathrm{C}[f_1, \overline{f}_1]\rangle. \tag{6.48}$$

It can now be seen that when one forms the matrix element $\langle\widehat{\boldsymbol{\tau}}|\chi_2\rangle$, each of the contributions (a)–(d) matches with the corresponding ones of (6.40), which summarizes the predictions of the two-time theory for the coefficient of $T^{-2}\boldsymbol{\nabla}T\,\boldsymbol{\nabla}T$. Thus, we have obtained agreement between the two formalisms at least for the coefficients of $\boldsymbol{\nabla}T\,\boldsymbol{\nabla}T$. It should be clear on conceptual grounds that if one is careful — very careful — agreement for the other myriad of Burnett coefficients will follow as well.

It is important to note that agreement has been found only when the Landau collision operator is used. I point out in §G.3.2 that in the more complete Balescu–Lenard theory arises a second-order cross term that involves fluctuations in the dielectric shielding. Formally, that effect is of the same order as that driven by $\mathrm{C}[f_1, \overline{f}_1]$. It could easily be incorporated into the Chapman–Enskog expansion by adding to the right-hand side of (6.41) the last line of (G 46). Calculation of the analogous effect in two-time theory involves a tedious evaluation of the contributions of the last four lines of (G 45), as discussed in §G.3; that is left as an exercise for the future.

## 7. Summary and discussion

The principal contributions and limitations of this paper are as follows:

- An introduction to Burnett effects was given.
  - ∘ A physical picture of one particular mechanism, arising from unbalanced viscous forces in the presence of a temperature gradient, that contributes to the coefficient of $\boldsymbol{\nabla}T\,\boldsymbol{\nabla}T$ in the momentum equation was described.
  - ∘ It was emphasized that the Burnett effects arise from gradient-induced symmetry breaking that lead to non-Gaussian statistics.
  - ∘ Burnett effects arise in contexts more general than many-body theory. In appendix A I describe a simple stochastic model (containing no discrete-particle effects) that exhibits Burnett effects due to a certain kind of non-Gaussian statistics; it demonstrates the role of symmetry breaking. Knowledge of that appendix is not necessary in order to appreciate the body of the paper, but perhaps it adds some additional useful perspectives.
- The time-independent projection-operator formalism of Brey *et al.* (1981) was extended to include multiple species and a magnetic field.
  - ∘ Momentum exchange between species was expressed as a two-time correlation, and an algebraically nontrivial calculation was done to demonstrate the equivalence of the hydrodynamic part of that formula to the corresponding one derived by Braginskii.

◦ Although it was pointed out where Burnett-level exchange effects arise in the formalism, specific formulas for those effects were not displayed.

◦ Also not displayed in detail were the additional Burnett-level formulas for perpendicular transport coefficients in the magnetised limit. (It is clear how to proceed; one must merely relax the symmetry assumptions that hold for $\boldsymbol{B} = \mathbf{0}$, then do perturbation theory for small $\nu/\omega_c$. However, the effects are very small.)

• A formalism appropriate for evaluating the relevant two-time correlation functions, involving either two or three phase-space points, in the weakly coupled limit was described.

• Solutions of the two-time equations were used to evaluate a representative Burnett parallel viscosity, and agreement with one-time Chapman–Enskog theory (applied to the Landau kinetic equation) was obtained.

◦ It was shown where contributions calculated from the Chapman–Enskog expansion employed by Catto & Simakov arise in the two-time formalism.

◦ In particular, nonlinear noise terms are responsible for effects relating to the nonlinear collision operator.

◦ An additional second-order effect involving fluctuations in the dielectric shielding was not evaluated either by Catto & Simakov or in the present paper.

Clearly, setting up and working out the two-time formalism of Brey *et al.* involves a fair amount of effort. In view of the agreement between that formalism and standard Chapman–Enskog theory, which is relatively straightforward, a natural question is, why should one bother? The answer depends on one's goals as well as the physical situation. If one is interested solely in the regular parts of the transport coefficients in the limit of weak coupling, it seems clear that Chapman–Enskog calculations, for example as implemented approximately by Catto & Simakov (2004), are more direct. Indeed, some of the manipulations that were done in the present paper in working out the weakly coupled two-time formulas appear to be almost redundant. For example, the algebra that was done in §G.2 to obtain the linearized Balescu–Lenard operator from the two-time hierarchy and in appendix E to obtain the interspecies momentum transfer involves repeated instances of familiar manipulations involving the Vlasov response function — the same class of manipulations used in deriving the original nonlinear Balescu–Lenard operator (§G.1). This is not surprising because, as I discussed, the two-time equations follow from the one-time ones by functional differentiation. If one already knows the one-time kinetic equation, there is no need to rederive its implications at the two-time level[52] — although the calculations done here provide important confidence-building consistency checks.

However, one knows the one-time kinetic equation only in the limit of weak coupling and with the neglect of long-ranged correlations. For stronger coupling, in order to proceed via Chapman–Enskog expansion, at the very least one would have to derive a collision operator that is more complete than the Balescu–Lenard or Landau operators. If that operator were local in physical space (involving correlation lengths shorter than the mean free path), then the usual Chapman–Enskog formulas would apply; complication would arise only in the quantitative evaluation of certain matrix elements. But in either one- or two-time theory, one must ultimately face up to the concerns expressed in §1 about possible divergences of the transport coefficients due to nonlocal collective effects. Some sort of renormalized formalism is required even at the level of discrete particles. The most complete such description is by Rose (1979), who shows how to handle the effects of both discrete particles and continuum turbulence within a unified framework and presents a so-called particle direct-interaction approximation (PDIA). However, mere possession of a renormalized theory of two-time correlation functions will not solve all issues relating to divergences of the Burnett coefficients. Although the formulas in §4.2 involve correlation functions that could be calculated from a renormalized theory, those formulas follow from the assumption that it is permissible to proceed with a regular perturbation expansion in the gradients. However, various authors[53] have demonstrated that it is necessary to expand in fractional powers of the gradients or to assume more general nonanalytic dependence. Considerable further work is required in order to establish the detailed connections between such research and Rose's formalism.

[52] Obviously, this statement is true only in the hydrodynamic limit and when the focus is merely on the calculation of transport coefficients. A general two-time correlation function contains much more information than can be obtained from the one-time kinetic equation.

It is fortunate that the weak-coupling limit has outsized importance in many plasma-physics applications. Although in that limit it is unnecessary to calculate transport coefficients from two-time formulas, proceeding in that way does have value beyond consistency checking. The calculations in §6.3.2, where I showed how the weakly coupled, nonlinear collision operator emerges in second-order kinetic theory, can be viewed as providing additional perspective on the nonlinear noise terms that have been extensively discussed in standard turbulence theory — generally treated in terms of renormalized approximations such as the direct-interaction approximation (DIA) — and appear as well in Rose's more-encompassing formalism. In turbulence theory, in particular the DIA and related approximations, the noise terms describe the internal stochastic forcing associated with the nonlinear interactions; that forcing is required in order to maintain the fluctuation level against the tendency for it to decay by nonlinear scrambling. More generally, that balance shows up in the consistency that must be enforced between the one- and two-time functions (Rose 1979); see appendix H for further discussion. Contributions from singular initial-condition terms were crucial in establishing in §6.3 the agreement between Chapman–Enskog calculations and two-time theory. Also, as I showed, nonlinear-noise terms involving discrete particles lead directly to contributions to transport coefficients involving the nonlinear collision operator. The key, stochastic-forcing role of nonlinear noise makes it seem inevitable that those nonlinear $\mathrm{C}[f_1, \overline{f}_1]$ effects must be present, although they were missed by plasma physicists prior to Catto & Simakov.

I have done nothing in this paper to address the practical relevance of Burnett-order transport effects. Rather, the discussions and calculations in Part I and the present Part II of this series are intended to raise awareness in the plasma-physics community of the significant utility of projection-operator methods. Applied to many-body theory, they provide a beautiful, compact, and unified representation of dissipative transport by mapping the orthogonal subspace, containing rapid fluctuations, into the slow, hydrodynamic subspace. The formalism evokes poetry:

> They were so amply beautiful, the maps,
> With their blue rivers winding to the sea,
> So calmly beautiful, who could have blamed
> Us for believing, bowed to our drawing boards,
> In a large and ultimate equivalence,
> One map that challenged and replaced the world?
>
> — from Projection, by H. Nemerov (1967)

[53] I have not done a complete literature survey. For some entry points, see the citations in the quotation by Wong *et al.* (1978) reproduced on page 5.

This paper is written in memory of the late Prof. Carl Oberman, who introduced me to plasma kinetic theory. Carl taught me many things, one of which has fostered patience in numerous calculations: He observed that, as one struggles to analytically understand the generic theory research problem, first one incorrectly finds infinity; then (again incorrectly) zero; then the correct, finite answer. Clearly, infinities and zeros abound in projection-operator manipulations. Carl would have enjoyed the present ones.

I am grateful to G. Hammett for useful discussions on the physical and mathematical difficulties with the Burnett equations. P. Catto graciously provided some background to his 2004 paper with Simakov; he also asked insightful questions about the manuscript that led me to include some clarifying remarks. This work was supported by the U. S. Department of Energy Contract DE-AC02-09CH11466.

## Appendix A. A simple stochastic model that exhibits Burnett effects

The only way to obtain a detailed description of Burnett effects for a plasma is to attack the many-body problem head-on with the methods of nonequilibrium statistical mechanics, as was done in the body of the paper. However, the general problem is difficult because of dynamical nonlinearity. Here I shall discuss aspects of a simpler model that, although dynamically linear, is stochastically nonlinear and thus displays under statistical averaging many of the features of the full problem. The model is called the *stochastic oscillator*; variations of it have been frequently used to illustrate aspects of statistical closure.[54] The primitive amplitude equation is taken to be

$$\frac{\mathrm{d}\widetilde{\psi}}{\mathrm{d}t} + \nu\widetilde{\psi} + \mathrm{i}\boldsymbol{k}\cdot\widetilde{\boldsymbol{V}}\widetilde{\psi} = \widetilde{f}(t), \tag{A 1}$$

where $\widetilde{\boldsymbol{V}}$ and $\widetilde{f}$ are random variables whose statistics are prescribed as follows.

The random velocity $\widetilde{\boldsymbol{V}}(t)$ is assumed to be independent of both space and time, with specified (passive) centred, stationary, *non-Gaussian*[55] statistics:

$$\langle\widetilde{\boldsymbol{V}}\rangle = \boldsymbol{0}, \quad \langle\delta\widetilde{\boldsymbol{V}}\delta\widetilde{\boldsymbol{V}}\rangle = U\boldsymbol{I}, \quad \langle\delta\widetilde{\boldsymbol{V}}\delta\widetilde{\boldsymbol{V}}\delta\widetilde{\boldsymbol{V}}\rangle = \boldsymbol{T}, \tag{A 2}$$

where $U$ and $\boldsymbol{T}$ are specified. Because $U$ has the dimensions of velocity squared, it can be written as $U = \overline{u}^2$, where $\overline{u}$ is a characteristic rms velocity fluctuation. In the absence of a preferred direction, the fully symmetric third-rank tensor $\boldsymbol{T}$ would vanish.[56] However, the goal is to emulate features of the hydrodynamic transport problem, in which long-wavelength gradients break the symmetry of the equilibrium state. Therefore, I choose

$$T_{ijk} = W(\delta_{ij}w_k + \delta_{jk}w_i + \delta_{ki}w_j), \tag{A 3}$$

where the coefficient $W$ also has the dimensions of $\overline{u}^2$. The significance of the constant vector $\boldsymbol{w}$ will become clear momentarily.

The random forcing $\widetilde{f}(t)$ is taken to be another independent time series with white-noise statistics:

$$\langle\widetilde{f}(t)\rangle = 0, \quad \langle\delta\widetilde{f}(t+\tau)\delta\widetilde{f}(t)\rangle = 2D_v\delta(\tau). \tag{A 4}$$

[54]Some references in which the stochastic-oscillator model are discussed include Kraichnan (1961), Kraichnan (1964), Krommes (2002), Krommes & Reiman (2009), and Krommes (2015).

[55]In all of the references cited in the previous footnote, the multiplicatively random coefficient is chosen to be Gaussian. In the present discussion, non-Gaussianity is essential.

[56]In the absence of any vector, the only third-rank tensor available from which to construct $T_{ijk}$ is the Levi–Civita tensor $\epsilon_{ijk}$; however, that is antisymmetric. Useful discussions of symmetry considerations are by Robertson (1940) and Mathews & Walker (1970, §3–2).

Finally, the nonrandom relaxation rate $\nu$ (analogous to the collision frequency in kinetic theory) is assumed to vanish in the equation for $\langle\widetilde{\psi}\rangle$. (This emulates the hydrodynamic conservation properties of the collision operator.)

With these assumptions, the mean and fluctuating equations become

$$\frac{\mathrm{d}\langle\widetilde{\psi}\rangle}{\mathrm{d}t} + \mathrm{i}\boldsymbol{k}\cdot\langle\delta\widetilde{\boldsymbol{V}}(t)\delta\widetilde{\psi}(t)\rangle = 0, \tag{A 5a}$$

$$\frac{\mathrm{d}\delta\psi}{\mathrm{d}t} + \nu\delta\psi + \mathrm{i}\boldsymbol{k}\cdot\delta\widetilde{\boldsymbol{V}}(t)\langle\widetilde{\psi}\rangle + \mathrm{i}\boldsymbol{k}\cdot(\delta\widetilde{\boldsymbol{V}}(t)\delta\psi(t) - \langle\ldots\rangle) = \delta\widetilde{f}(t). \tag{A 5b}$$

At $k = 0$, the equation for fluctuations reduces to the well-known classical Langevin equation for the velocity of a Brownian test particle (Wang & Uhlenbeck 1945). Thus, it is clear that a statistically steady state can be reached in which the collisional dissipation balances in mean square against the random forcing, the strength of which is measured by the diffusion coefficient $D_v$.

The $\boldsymbol{k}$-dependent terms drive corrections to that steady state. In terms of the infinitesimal response function (Green's function)

$$R(t;t') \doteq H(t-t')\mathrm{e}^{-\nu(t-t')}, \tag{A 6}$$

where $H(t)$ is the Heaviside unit step function, an exact integral equation is

$$\begin{aligned}\delta\psi(t) = &\int_{-\infty}^{t}\mathrm{d}\overline{t}\,R(t;\overline{t})\delta\widetilde{f}(\overline{t}) - \int_{-\infty}^{t}\mathrm{d}\overline{t}\,R(t;\overline{t})\mathrm{i}\boldsymbol{k}\cdot\delta\widetilde{\boldsymbol{V}}(\overline{t})\langle\psi\rangle(\overline{t})\\ &- \int_{-\infty}^{t}\mathrm{d}\overline{t}\,R(t;\overline{t})\mathrm{i}\boldsymbol{k}\cdot(\delta\widetilde{\boldsymbol{V}}(\overline{t})\delta\psi(\overline{t}) - \langle\ldots\rangle).\end{aligned} \tag{A 7}$$

At first order in $k$ (not first order in the size of the fluctuations!), one has

$$\delta\psi^{(1)}(t) = -\int_{-\infty}^{t}\mathrm{d}\overline{t}\,R(t;\overline{t})\mathrm{i}\boldsymbol{k}\cdot\delta\widetilde{\boldsymbol{V}}(\overline{t})\langle\psi\rangle(\overline{t}), \tag{A 8}$$

which gives an $O(k^2)$ contribution to (A 5a):

$$\mathrm{i}\boldsymbol{k}\cdot\langle\delta\widetilde{\boldsymbol{V}}(t)\delta\psi^{(1)}(t)\rangle = k^2\int_{-\infty}^{t}\mathrm{d}\overline{t}\,R(t;\overline{t})U(t,\overline{t})\langle\psi\rangle(\overline{t}). \tag{A 9a}$$

$$= k^2\int_{0}^{\infty}\mathrm{d}\overline{\tau}\,R(\overline{\tau})U(\overline{\tau})\langle\psi\rangle(t-\overline{\tau}) \tag{A 9b}$$

$$\approx k^2 D^{(1)}\langle\psi\rangle(t), \tag{A 9c}$$

where

$$D^{(1)} \doteq \int_{0}^{\infty}\mathrm{d}\overline{\tau}\,R(\overline{\tau})U(\overline{\tau}). \tag{A 10}$$

Since the velocity correlation function $U(\tau)$ has been chosen to be time-independent, one has $D^{(1)} = \overline{u}^2/\nu$. This agrees with the scaling of the classical Navier–Stokes transport coefficients if $\overline{u}$ is identified with the thermal velocity $v_{\mathrm{t}}$.

At second order in $k$, one has

$$\delta\psi^{(2)}(t) = -\int_{-\infty}^{t} \mathrm{d}\overline{t}\, R(t;\overline{t})\mathrm{i}\boldsymbol{k}\cdot(\delta\widetilde{\boldsymbol{V}}(\overline{t})\delta\psi^{(1)}(\overline{t}) - \langle\ldots\rangle) \tag{A 11a}$$

$$= -\int_{-\infty}^{t} \mathrm{d}\overline{t}\, R(t;\overline{t})\boldsymbol{k}\cdot\left(\delta\widetilde{\boldsymbol{V}}(\overline{t})\int_{-\infty}^{\overline{t}} \mathrm{d}\overline{t}'\, R(\overline{t};\overline{t}')\boldsymbol{k}\cdot\delta\widetilde{\boldsymbol{V}}(\overline{t}')\langle\psi\rangle(\overline{t}') - \langle\ldots\rangle\right); \tag{A 11b}$$

this contributes to (A 5$a$)

$$\mathrm{i}\boldsymbol{k}\cdot\langle\delta\widetilde{\boldsymbol{V}}(t)\delta\psi^{(2)}(t)\rangle$$

$$= -\mathrm{i}\int_{-\infty}^{t} \mathrm{d}\overline{t}\left\langle \boldsymbol{k}\cdot\delta\widetilde{\boldsymbol{V}}(t)R(t;\overline{t})\left(\boldsymbol{k}\cdot\delta\widetilde{\boldsymbol{V}}(\overline{t})\int_{-\infty}^{\overline{t}} \mathrm{d}\overline{t}'\, R(\overline{t};\overline{t}')\boldsymbol{k}\cdot\delta\widetilde{\boldsymbol{V}}(\overline{t}')\langle\psi\rangle(\overline{t}') - \langle\ldots\rangle\right)\right\rangle \tag{A 12a}$$

$$= -3\mathrm{i}k^2 W\boldsymbol{k}\cdot\boldsymbol{w}\int_{-\infty}^{t} \mathrm{d}\overline{t}\, R(t;\overline{t})\int_{-\infty}^{\overline{t}} \mathrm{d}\overline{t}'\, R(\overline{t};\overline{t}')\langle\psi\rangle(\overline{t}'). \tag{A 12b}$$

With the changes of variables $\overline{\tau} \doteq t - \overline{t}$ and $\overline{\tau}' \doteq \overline{t} - \overline{t}'$, one has

$$\int_{-\infty}^{t} \mathrm{d}\overline{t}\, R(t;\overline{t})\int_{-\infty}^{\overline{t}} \mathrm{d}\overline{t}'\, R(\overline{t};\overline{t}')\langle\psi\rangle(\overline{t}') = \int_{0}^{\infty} \mathrm{d}\overline{\tau}\, R(\overline{\tau})\int_{0}^{\infty} \mathrm{d}\overline{\tau}'\, R(\overline{\tau}')\langle\psi\rangle(t - \overline{\tau} - \overline{\tau}') \tag{A 13a}$$

$$\approx \left(\int_{0}^{\infty} \mathrm{d}\overline{\tau}\, R(\overline{\tau})\right)\left(\int_{0}^{\infty} \mathrm{d}\overline{\tau}'\, R(\overline{\tau}')\right)\langle\psi\rangle(t). \tag{A 13b}$$

Therefore, one has

$$\mathrm{i}\boldsymbol{k}\cdot\langle\delta\widetilde{\boldsymbol{V}}(t)\delta\psi^{(2)}(t)\rangle \approx -\mathrm{i}kD^{(2)}(\boldsymbol{k}\cdot\widehat{\boldsymbol{w}})(k\langle\psi\rangle), \tag{A 14}$$

where

$$D^{(2)} \propto (\overline{u}^2/\nu)(\overline{u}/\nu). \tag{A 15}$$

This effect is analogous to the unmagnetised Burnett transport coefficients that I discuss in the body of the paper. They scale with $\nu^{-2}$. (The factor $\overline{u}/\nu$ is analogous to the collisional mean free path $\lambda_{\mathrm{mfp}}$.) The Burnett fluxes, represented here by $-D^{(2)}(\boldsymbol{k}\cdot\widehat{\boldsymbol{w}})(k\langle\psi\rangle)$, are of second order in the gradients and are smaller than the Navier–Stokes ones by a factor of $k\lambda_{\mathrm{mfp}}$. (In many-body theory, the limit $k \to 0$ is taken and the reduction factor becomes $\lambda_{\mathrm{mfp}}/L$, where $L$ is the system size.) In reality, those fluxes involve products of hydrodynamic variables such as $|\boldsymbol{\nabla}T|^2$. Here one of those gradients is represented by the factor $\boldsymbol{k}\cdot\widehat{\boldsymbol{w}}$ because the stochastic model is passive, so the advecting velocity is not linearly proportional to $\psi$.

This calculation demonstrates several important points that generalize to the full many-body problem:

• Transport effects in the equations for mean fields arise from the symmetry breaking of the equilibrium state by the hydrodynamic gradients.

• The Burnett coefficients arise from non-Gaussian statistics. (The many-body theory described in the body of the paper expresses the non-Gaussian effects in terms of two-time Klimontovich correlations involving either two or three phase-space points. It is shown in §6.4 that the contributions calculated by Catto & Simakov (2004) from the nonlinear collision operator are related to three-point Klimontovich correlations.)

## Appendix B. Microscopic and macroscopic forces

To find the time evolution of the microscopic momentum density, differentiate the definition (2.11) to find

$$\frac{\partial \widetilde{\boldsymbol{P}}_s(\boldsymbol{r},t)}{\partial t} = \frac{\partial}{\partial t}\sum_{i\in s} m_s \boldsymbol{v}_i(t)\delta(\boldsymbol{r}-\boldsymbol{x}_i(t)) \tag{B 1a}$$

$$= \sum_{i\in s} m_s \boldsymbol{v}_i[-\boldsymbol{v}_i\boldsymbol{\nabla}\delta(\boldsymbol{r}-\boldsymbol{x}_i)] + \sum_{i\in s}\left(q_s(\boldsymbol{E}^{\rm ext}+c^{-1}\boldsymbol{v}_i\times\boldsymbol{B}^{\rm ext}) - \frac{\partial \widetilde{U}_i}{\partial \boldsymbol{x}_i}\right)\delta(\boldsymbol{r}-\boldsymbol{x}_i), \tag{B 1b}$$

where $\boldsymbol{E}^{\rm ext}$ and $\boldsymbol{B}^{\rm ext}$ are externally imposed electric and magnetic fields (I shall assume that $\boldsymbol{E}^{\rm ext} = \boldsymbol{0}$) and $\widetilde{U}_i \doteq \sum_{j\neq i}^{\mathscr{N}} q_i|\boldsymbol{x}_i-\boldsymbol{x}_j|^{-1}q_j$ is the interparticle potential energy. The first term of (B 1$b$) is $-\boldsymbol{\nabla}\cdot\widetilde{\boldsymbol{\tau}}_{K,s}$, where the kinetic momentum flux is $\widetilde{\boldsymbol{\tau}}_{K,s} \doteq \sum_{i\in s} m_s\boldsymbol{v}_i\boldsymbol{v}_i\delta(\boldsymbol{r}-\boldsymbol{x}_i)$. The Lorentz-force term is easily written as $\omega_{cs}\widetilde{\boldsymbol{P}}_s\times\widehat{\boldsymbol{b}}$, where $\omega_{cs}\doteq(qB/mc)_s$, $B\doteq|\boldsymbol{B}^{\rm ext}|$, and $\widehat{\boldsymbol{b}}\doteq\boldsymbol{B}^{\rm ext}/B$.

While the kinetic term is in the form of a divergence, the potential-energy term is not (yet). A well-known trick is to work with the Fourier transform and to symmetrize the potential term. Upon Fourier transforming, that becomes $-\sum_{i\in s}\sum_{s'}\sum_{j\neq i}^{\mathscr{N}_{s'}}\partial\widetilde{U}_{ij}/\partial\boldsymbol{x}_i\,{\rm e}^{-{\rm i}\boldsymbol{k}\cdot\boldsymbol{x}_i}$. The contributions with $s'=s$ can be symmetrized by interchanging $i$ and $j$; one finds

$$-\frac{1}{2}\sum_{ij}^{\mathscr{N}_s}\left(\frac{\partial\widetilde{U}_{ij}}{\partial\boldsymbol{x}_i}\,{\rm e}^{-{\rm i}\boldsymbol{k}\cdot\boldsymbol{x}_i} + \frac{\partial\widetilde{U}_{ji}}{\partial\boldsymbol{x}_j}\,{\rm e}^{-{\rm i}\boldsymbol{k}\,\boldsymbol{x}_j}\right) = -\frac{1}{2}\sum_{ij}^{\mathscr{N}_s}\left(\frac{\partial\widetilde{U}_{ij}}{\partial\boldsymbol{x}_i}{\rm e}^{-{\rm i}\boldsymbol{k}\cdot\boldsymbol{x}_i} - \frac{\partial\widetilde{U}_{ij}}{\partial\boldsymbol{x}_i}{\rm e}^{-{\rm i}\boldsymbol{k}\cdot\boldsymbol{x}_j}\right) \tag{B 2a}$$

$$= -\frac{1}{2}\sum_{ij}^{\mathscr{N}_s}\frac{\partial\widetilde{U}_{ij}}{\partial\boldsymbol{x}_i}\left(1-{\rm e}^{{\rm i}\boldsymbol{k}\cdot(\boldsymbol{x}_i-\boldsymbol{x}_j)}\right){\rm e}^{-{\rm i}\boldsymbol{k}\cdot\boldsymbol{x}_i} \tag{B 2b}$$

$$= -\frac{1}{2}\sum_{ij}^{\mathscr{N}_s}\frac{\partial\widetilde{U}_{ij}}{\partial\boldsymbol{x}_{ij}}\left(\frac{1-{\rm e}^{{\rm i}\boldsymbol{k}\cdot\boldsymbol{x}_{ij}}}{-{\rm i}\boldsymbol{k}\cdot\boldsymbol{x}_{ij}}\right)(-{\rm i}\boldsymbol{k}\cdot\boldsymbol{x}_{ij}){\rm e}^{-{\rm i}\boldsymbol{k}\cdot\boldsymbol{x}_i} \tag{B 2c}$$

$$= -{\rm i}\boldsymbol{k}\cdot\Delta\widetilde{\boldsymbol{\tau}}_s(\boldsymbol{k}), \tag{B 2d}$$

where

$$\Delta\widetilde{\boldsymbol{\tau}}_s(\boldsymbol{k}) \doteq -\frac{1}{2}\sum_{ij}^{\mathscr{N}_s}\frac{\partial\widetilde{U}_{ij}}{\partial\boldsymbol{x}_{ij}}\boldsymbol{x}_{ij}\left(\frac{1-{\rm e}^{{\rm i}\boldsymbol{k}\cdot\boldsymbol{x}_{ij}}}{-{\rm i}\boldsymbol{k}\cdot\boldsymbol{x}_{ij}}\right){\rm e}^{-{\rm i}\boldsymbol{k}\cdot\boldsymbol{x}_i}. \tag{B 3}$$

The limit of the parenthesized expression[57] as $\boldsymbol{k}\to\boldsymbol{0}$ is 1. The tensor $\Delta\widetilde{\boldsymbol{\tau}}_s(\boldsymbol{k})$ is symmetric for central forces since

$$\frac{\partial U_{ij}(\boldsymbol{r})}{\partial\boldsymbol{r}}\boldsymbol{r} = \widehat{\boldsymbol{r}}\left(\frac{\partial U_{ij}(r)}{\partial r}r\right)\widehat{\boldsymbol{r}}. \tag{B 4}$$

Note that for the Coulomb potential $U_{ij}(r) = q_iq_j/r$, one has $-(\partial U_{ij}/\partial r)r = U_{ij}(r)$. The terms $s'\neq s$ cannot be symmetrized.

From these formulas, one wants to distill a macroscopic electrostatic potential $\phi(\boldsymbol{r})$

---

[57] The form (B 3) agrees with equation (A10) of Brey (1983) since he uses the opposite sign of $\boldsymbol{k}$ in his definition of the Fourier transform; for example, ${\rm i}\boldsymbol{k}\cdot\boldsymbol{x}_{ij}\to-{\rm i}\boldsymbol{k}\cdot\boldsymbol{x}_{ij}=+{\rm i}\boldsymbol{k}\cdot\boldsymbol{x}_{ji}$, which reproduces the exponent in Brey's term.

and microscopic fluctuating stresses for use in (2.26$b$) with $\boldsymbol{E} \doteq -\boldsymbol{\nabla}\phi$ being the macroscopic (collective or internal) electric field. To do so, it is necessary to be more precise about the nature of the statistical ensemble. One possibility is to assume that one works in a large box of volume $\mathscr{V}$ containing $\mathscr{N} = \sum_s \mathscr{N}_s$ particles with zero net charge. However, this obscures the distinction between short-ranged correlations and long-ranged forces. This is crucial for the plasma, which can support nontrivial, spatially varying electric fields $\boldsymbol{E}(\boldsymbol{r}, t)$ over distances much larger than the microscopic correlation length $\lambda_{\mathrm{D}}$. (The paradigmic example are the Langmuir oscillations $\omega^2 \approx \pm\omega_{\mathrm{p}}^2$.) One would like to write the microscopic forces as an average plus a fluctuating piece, $-\boldsymbol{\nabla}\boldsymbol{\cdot}\Delta\widetilde{\boldsymbol{\tau}}_s(\boldsymbol{r}, t) = -\boldsymbol{\nabla}\boldsymbol{\cdot}\langle\Delta\boldsymbol{\tau}_s\rangle(\boldsymbol{r}, t) - \boldsymbol{\nabla}\boldsymbol{\cdot}\delta\boldsymbol{\tau}_s(\boldsymbol{r}, t)$, where notably $\langle\Delta\boldsymbol{\tau}_s\rangle$ depends on $\boldsymbol{r}$. An average with the exact nonequilibrium ensemble accomplishes this; however, one wants to express that average in terms of a local reference distribution. A uniform distribution of particles over the entire box is not satisfactory since symmetry considerations would lead to a spatially constant $\langle\Delta\boldsymbol{\tau}\rangle$. One could imagine dividing the system into boxes whose sides are larger than $\lambda_{\mathrm{D}}$, and local charge imbalances in those boxes would lead to internal electric fields $\boldsymbol{E}^{\mathrm{int}} = -\boldsymbol{\nabla}\phi$. However, such boxes are not necessarily in local thermal equilibrium due to high-frequency oscillations. Instead, the basic coarse-graining procedure discussed in §2.1 uses boxes $\Delta\mathscr{V}$ of side large compared to the collisional mean free path (but small compared to the macroscopic gradient scale length). Now an average over the charge distribution of particles in $\Delta\mathscr{V}$ produces from the terms with $s' = s$ a $\langle\Delta\boldsymbol{\tau}_s\rangle(\boldsymbol{r}, t) = (\overline{n}q)_s\phi_s(\boldsymbol{r}, t)\boldsymbol{I}$, where $\phi_s$ is the contribution to the total low-frequency electrostatic potential due to species $s$. Note that $\sum_{j\in s} = \sum_{j\in s}^{\Delta\mathscr{V}(\boldsymbol{r})} + \sum_{j\in s}^{\Delta\mathscr{V}'\neq\Delta\mathscr{V}}$. The last term gives nonlocal contributions to the mean potential.[58]

The terms $s' \neq s$ can be broken into two kinds of effects: those arising from unlike-species particles all within $\Delta\mathscr{V}$, and contributions from other cells $\Delta\mathscr{V}'$. The former generate local random momentum and energy exchange effects as well as a local contribution to the mean potential; the latter gives the nonlocal unlike-species contribution to the mean potential.

Apparently both coarse-graining procedures mentioned above support collective electric fields. The choice of box size is not irrelevant, however; it affects the approximate procedure (involving Novikov's theorem) that is used in appendix D to establish formulas for the subtracted fluxes in terms of certain thermodynamic derivatives. For a discussion of low-frequency, long-wavelength classical transport, scaling the box side to $\lambda_{\mathrm{mfp}}$ is the correct choide.

Thus, the momentum flux obeys (2.26$b$). With prime denoting the fluctuation from the mean of the potential-energy terms, the portion of the microscopic stress tensor that contributes to local transport coefficients is defined by

$$\widetilde{\boldsymbol{\tau}}_s'(\boldsymbol{r}, t) = \int \frac{\mathrm{d}\boldsymbol{k}}{(2\pi)^3}\,\mathrm{e}^{\mathrm{i}\boldsymbol{k}\boldsymbol{\cdot}\boldsymbol{r}}\widetilde{\boldsymbol{\tau}}_s'(\boldsymbol{k}, t), \quad \widetilde{\boldsymbol{\tau}}_s(\boldsymbol{k}, t) = \sum_{i\in s}[m_s\boldsymbol{v}_i\boldsymbol{v}_i + \Delta\widetilde{\boldsymbol{\tau}}_{s,i}(\boldsymbol{k})]\mathrm{e}^{-\mathrm{i}\boldsymbol{k}\boldsymbol{\cdot}\boldsymbol{x}_i}, \tag{B 5}$$

and

$$\Delta\widetilde{\boldsymbol{\tau}}_{s,i}(\boldsymbol{k}) \doteq -\frac{1}{2}\sum_{j\in s}\frac{\partial\widetilde{U}_{ij}}{\partial\boldsymbol{x}_{ij}}\,\boldsymbol{x}_{ij}\left(\frac{1-\mathrm{e}^{\mathrm{i}\boldsymbol{k}\boldsymbol{\cdot}\boldsymbol{x}_{ij}}}{-\mathrm{i}\boldsymbol{k}\boldsymbol{\cdot}\boldsymbol{x}_{ij}}\right). \tag{B 6}$$

[58] In principle, the densities in each of the boxes can be macroscopically random if the system is turbulent. That effect is ignored in the present work in order to focus on the microscopic fluctuations that give rise to classical transport.

The total amount of this flux is

$$\widetilde{\mathscr{T}}'_s(t) = \int \mathrm{d}\boldsymbol{r}\, \widetilde{\boldsymbol{\tau}}'_s(\boldsymbol{r},t) = \widetilde{\boldsymbol{\tau}}'_s(\boldsymbol{k}=\boldsymbol{0},t) = \sum_{i\in s}[m_s \boldsymbol{v}_i\boldsymbol{v}_i + \Delta\widetilde{\boldsymbol{\tau}}'_{s,i}(\boldsymbol{0})], \tag{B 7}$$

where for the Coulomb interaction $\Delta\widetilde{\boldsymbol{\tau}}_{s,i}(\boldsymbol{0}) = \frac{1}{2}\sum_{j\in s}\widehat{\boldsymbol{x}}_{ij}U_{ij}\widehat{\boldsymbol{x}}_{ij}$.

When the stress tensor is used in formulas for transport coefficients, it invariably appears in its subtracted form (see §D). That subtraction naturally removes any mean potential.

For the exchange terms, one has

$$\dot{\boldsymbol{P}}'_{\Delta,s} = -\sum_{i\in s}\sum_{s'\neq s}\sum_{j\in s'} \boldsymbol{\nabla}_i U_{ij}\delta(\boldsymbol{r}-\boldsymbol{x}_i). \tag{B 8}$$

Since the mean field has already been extracted (denoted by the prime), the $j$ sum should in principle be restricted to particles lying in the same coarse-graining cell. The statistical effects of $\dot{\widetilde{\boldsymbol{P}}}_\Delta$ will be discussed in appendix E.

## Appendix C. The conjugate variables

Here I give some details about the choice (2.34) of conjugate variables $\boldsymbol{B}$. The constraint (2.33) that the mean hydrodynamic variables $\boldsymbol{a}_s$ are given by the local equilibrium average of the microscopic densities leads, according to (2.32*a*), to a set of functional differential equations that determine the $\boldsymbol{B}$'s in terms of the $\boldsymbol{a}$'s:

$$\boldsymbol{a}_s = \begin{pmatrix} n_s \\ \boldsymbol{p}_s \\ e_s \end{pmatrix} = \begin{pmatrix} \delta/\delta \boldsymbol{B}_{n,s} \\ \delta/\delta \boldsymbol{B}_{\boldsymbol{p},s} \\ \delta/\delta \boldsymbol{B}_{e,s} \end{pmatrix} \ln Z_B[\boldsymbol{B}], \tag{C 1}$$

where, upon using the definitions of the microscopic variables, performing the $\overline{\boldsymbol{r}}$ integration implied in the $\star$ operation, and defining $\beta_s \doteq -B_{e,s}$ for convenience, one finds

$$Z_B = \int \mathrm{d}\mathit{\Gamma}\, \exp\left\{\sum_{\overline{s}}\sum_{i\in\overline{s}}\left[B_{n,\overline{s}}(\boldsymbol{x}_i) + m_{\overline{s}}\boldsymbol{v}_i\cdot\boldsymbol{B}_{\boldsymbol{p},\overline{s}}(\boldsymbol{x}_i) - \left(\frac{1}{2}m_{\overline{s}}v_i^2 + U_i\right)\beta_{\overline{s}}(\boldsymbol{x}_i)\right]\right\}. \tag{C 2}$$

Although $U_i$ implicitly depends on all $\boldsymbol{x}_{j\neq i}$, which implies that $Z_B$ contains two-point spatial correlations, all of the dependence on $\boldsymbol{v}_i$ is explicit and uncoupled. The integrations over each of the $\boldsymbol{v}_i$ can therefore be performed by completing the square:

$$Z_B = \int \mathrm{d}X\, \exp\left[\sum_{\overline{s}}\sum_{i\in\overline{s}}\left(\frac{3}{2}\ln[2\pi\sigma^2_{\overline{s}}(\boldsymbol{x}_i)] + B_{n\overline{s}}(\boldsymbol{x}_i) + \frac{1}{2}m_{\overline{s}}|\boldsymbol{B}_{\boldsymbol{p}\overline{s}}|^2(\boldsymbol{x}_i)\beta^{-1}_{\overline{s}}(\boldsymbol{x}_i) - U_i\beta_{\overline{s}}(\boldsymbol{x}_i)\right)\right], \tag{C 3}$$

where $\sigma^2 \doteq (m\beta)^{-1}$. Upon using the basic result

$$\frac{\delta B_{n\overline{s}}(\boldsymbol{x}_i)}{\delta B_{ns}(\boldsymbol{r})} = \delta_{\overline{s}s}\delta(\boldsymbol{x}_i - \boldsymbol{r}), \tag{C 4}$$

one finds from the density component of (C 1) that

$$n_s(\boldsymbol{r},t) = \langle \widetilde{N}_s(\boldsymbol{r},t)\rangle_B, \tag{C 5}$$

where the expectation is taken with the normalized exponential in (C 3). That is,

$$F_B(\Gamma) = \exp\left[-\ln Z_B + \sum_{\overline{s}} \sum_{i\in\overline{s}} \left(\frac{3}{2}\ln[2\pi\sigma_{\overline{s}}^2(\boldsymbol{x}_i)] \right.\right.$$
$$\left.\left. + B_{n\overline{s}}(\boldsymbol{x}_i) + \frac{1}{2}m_{\overline{s}}|\boldsymbol{B}_{\boldsymbol{p}\overline{s}}|^2(\boldsymbol{x}_i)\beta_{\overline{s}}^{-1}(\boldsymbol{x}_i) - U_i\beta_{\overline{s}}(\boldsymbol{x}_i)\right)\right]. \qquad (\text{C } 6)$$

Similarly, upon evaluating the functional derivative of $Z_B$ with respect to $\boldsymbol{B}_{\boldsymbol{p}s}(\boldsymbol{r})$ and making use of (C 5) and the definition (2.13) of the flow velocity, one finds

$$\boldsymbol{B}_{\boldsymbol{p},s}(\boldsymbol{r},t) = (\beta\boldsymbol{u})_s(\boldsymbol{r},t). \qquad (\text{C } 7)$$

Next, the functional derivative with respect to $B_{es} = -\beta_s$ provides contributions to (C 1) from each of the $\sigma^2$, $\beta^{-1}$, and $\beta$ terms. Upon using (2.20) and (2.21), one ultimately finds that the energy component of (C 1) expresses the mean internal energy as

$$u_s(\boldsymbol{r},t) = \langle\widetilde{U}_s\rangle_B \qquad (\text{C } 8)$$

provided that $\beta_s^{-1}(\boldsymbol{r},t)$ is identified with the local temperature $T_s(\boldsymbol{r},t)$. With this crucial result, the analogy to the thermal-equilibrium Gibbs distribution is sufficiently close so that one can identify

$$B_{ns}(\boldsymbol{r},t) = \beta_s(\boldsymbol{r},t)\left(\mu_s(\boldsymbol{r},t) - \frac{1}{2}m_s u_s^2(\boldsymbol{r},t)\right), \qquad (\text{C } 9)$$

where $\mu$ is the chemical potential[59] per particle.

## Appendix D. The subtracted fluxes

In order to calculate the subtracted fluxes, it is necessary to understand the implications of the projection operator P, defined by (2.45). First observe that in the reference distribution $F_0$ there are no cross correlations between the relative velocities $\boldsymbol{w}_i$ and particle positions, and also that $F_0$ is an isotropic function of the $\boldsymbol{w}_i$'s. Thus, the covariance matrix $\boldsymbol{\mathscr{M}}$ decomposes into a $2\times 2$ density–energy submatrix $\boldsymbol{\mathscr{M}}_2$ and

$$\mathscr{M}_{ss'}^{\boldsymbol{pp}}(\boldsymbol{x},\boldsymbol{x}') = \langle\boldsymbol{P}_s'(\boldsymbol{x})\boldsymbol{P}_{s'}'(\boldsymbol{x}')\rangle_0 \qquad (\text{D } 1a)$$
$$= \sum_{i\in s} m_s \sum_{j\in s'} m_{s'}\langle\boldsymbol{w}_i\boldsymbol{w}_j\rangle\langle\delta(\boldsymbol{x}-\boldsymbol{x}_i)\delta(\boldsymbol{x}'-\boldsymbol{x}_j)\rangle_0 \qquad (\text{D } 1b)$$
$$= m_s T_s(\boldsymbol{r},t)\boldsymbol{I}\sum_{i\in s}\langle\delta(\boldsymbol{x}-\boldsymbol{x}_i)\rangle_0\delta(\boldsymbol{x}-\boldsymbol{x}')\delta_{ss'} \qquad (\text{D } 1c)$$
$$= (m\overline{n}T)_s(\boldsymbol{r},t)\delta(\boldsymbol{x}-\boldsymbol{x}')\delta_{ss'}\boldsymbol{I}. \qquad (\text{D } 1d)$$

---

[59]The chemical potential is discussed in virtually every introductory book on statistical physics. Its derivation for a monatomic ideal gas often begins with the Sackur–Tetrode entropy formula

$$\mathscr{S} = k\mathscr{N}\left\{\ln\left[\frac{\mathscr{V}}{\mathscr{N}}\left(\frac{4\pi m\mathscr{E}}{3\mathscr{N}h_{\mathrm{P}}^2}\right)^{3/2}\right] + \frac{5}{2}\right\},$$

then uses the result $\mu = -T\,\partial\mathscr{S}/\partial\mathscr{N}$ to find

$$\beta\mu = \ln\left[n\left(\frac{4\pi me}{3h_{\mathrm{P}}^2}\right)^{3/2}\right]. \qquad (\text{C } 10)$$

In the presence of a mean electrostatic potential $\phi(\boldsymbol{r},t)$, a term $q_s\phi$ must be added to $\mu$.

Here the mean density $\overline{n}$ rather than the true density $n(\boldsymbol{r}, t)$ has entered because the expectation is taken in the reference ensemble, which is translationally invariant; see (2.8).

The specific form of $\boldsymbol{\mathscr{M}}_2$ will not be needed. Note, however, that it is not delta correlated because of spatial correlations arising from internal-energy corrections.

Consider a quantity $\widetilde{\chi}$ that is even in the $\boldsymbol{w}$'s (particularly, $\widetilde{\chi} = \widetilde{N}$ or $\widetilde{E}$) and whose average in the reference ensemble is $\chi \doteq \langle\widetilde{\chi}\rangle_0$. I shall prove that

$$\mathrm{P}\widetilde{\chi}(\boldsymbol{r}, t) \approx \chi(\boldsymbol{r}, t) + \sum_{\overline{s}} \widetilde{N}_{\overline{s}}(\boldsymbol{r}, t)\left(\frac{\partial\chi}{\partial n_{\overline{s}}}\right)_e + \widetilde{E}_{\overline{s}}(\boldsymbol{r}, t)\left(\frac{\partial\chi}{\partial e_{\overline{s}}}\right)_n, \tag{D 2}$$

where the thermodynamic derivatives are taken in the reference ensemble. This result was quoted by Brey (1983) for the one-component fluid. The argument relies on the assumption (used throughout this work) that in a local region the system is close to local thermal equilibrium. Thus, one can imagine dividing the system into cells whose sides are much larger than the spatial correlation length but much smaller than the macroscopic gradient scale length. One expects that all of the cumulants of $\widetilde{N}'$ and $\widetilde{E}'$ will be small when at least two of their spatial arguments lie in different cells. This implies that the coarse-grained $\widetilde{N}'$ and $\widetilde{E}'$ can be considered to be Gaussian random variables to lowest order. Then the expectations required to evaluate $\mathrm{P}\widetilde{\chi}$ can be calculated with the aid of Novikov's theorem (Krommes 2015, appendix B, and references therein), whose basic form for a centred random scalar field $\widetilde{\phi}'$ is

$$\langle\phi'(\boldsymbol{x})\mathscr{F}[\phi']\rangle = \int \mathrm{d}\overline{\boldsymbol{x}}\, C(\boldsymbol{x}, \overline{\boldsymbol{x}}) \left\langle \frac{\delta\mathscr{F}[\phi']}{\delta\phi'(\overline{\boldsymbol{x}})} \right\rangle, \tag{D 3}$$

where $\mathscr{F}$ denotes an arbitrary functional, $C(\boldsymbol{x}, \boldsymbol{x}') \doteq \langle\phi'(\boldsymbol{x})\phi'(\boldsymbol{x}')\rangle$, and $\delta$ denotes a functional derivative. In the present context with several species-dependent fields, this generalizes to

$$\boldsymbol{y}(\mu) \doteq \left\langle \begin{pmatrix} N'(\mu) \\ E'(\mu) \end{pmatrix} \widetilde{\chi} \right\rangle_0 = \boldsymbol{\mathscr{M}}_2(\mu, \overline{\mu}) \star \begin{pmatrix} \langle\delta\widetilde{\chi}/\delta N'(\overline{\mu})\rangle_0 \\ \langle\delta\widetilde{\chi}/\delta E'(\overline{\mu})\rangle_0 \end{pmatrix}. \tag{D 4}$$

Then, upon noting that $\boldsymbol{\mathscr{M}}_2^{-1} \star \boldsymbol{\mathscr{M}}_2 = \boldsymbol{I}$ and omitting dependence on $t$, one finds

$$\mathrm{P}\widetilde{\chi}(\boldsymbol{r}) = \chi(\boldsymbol{r}) + (N',\ E') \star \boldsymbol{\mathscr{M}}_2^{-1} \star \boldsymbol{y} = \sum_{\overline{s}} \int \mathrm{d}\overline{\boldsymbol{x}}\, (N',\ E')_{\overline{s}}(\overline{\boldsymbol{x}}) \boldsymbol{\cdot} \begin{pmatrix} \langle\delta\widetilde{\chi}(\boldsymbol{r})/\delta N'_{\overline{s}}(\overline{\boldsymbol{x}})\rangle_0 \\ \langle\delta\widetilde{\chi}(\boldsymbol{r})/\delta E'_{\overline{s}}(\overline{\boldsymbol{x}})\rangle_0 \end{pmatrix}. \tag{D 5}$$

By definition of the coarse-graining, the support of the functional derivatives lies essentially within the cell centred on $\boldsymbol{r}$, so the first $\overline{\boldsymbol{x}}$ under the integral can be replaced by $\boldsymbol{x}$. The remaining $\overline{\boldsymbol{x}}$ integration changes the functional derivatives to ordinary partial derivatives, and one is led to (D 2).

One can now construct the subtracted fluxes $\widehat{\boldsymbol{J}} \doteq \mathrm{Q}\boldsymbol{J} = (1 - \mathrm{P})\boldsymbol{J}$.

• The subtracted density flux $\widehat{\boldsymbol{J}}^n_s \doteq m_s^{-1}\widehat{\boldsymbol{P}}_s$ vanishes because $\boldsymbol{P}_s$ lies in the hydrodynamic subspace.[60]

• For the subtracted momentum flux, note that the velocities of particles of species $s$ are in the frame moving with $\boldsymbol{u}_s$, so $\langle\widetilde{\boldsymbol{\tau}}'_s\rangle = p_s\boldsymbol{I}$. Therefore, in that frame one has, with

[60]Explicitly, as a simple exercise in the use of the projection operator,

$$\widehat{\boldsymbol{P}}_s = \widetilde{\boldsymbol{P}}_s - (\langle\widetilde{\boldsymbol{P}}_s\rangle_0 + \widetilde{\boldsymbol{P}}' \star \boldsymbol{\mathscr{M}}_{\boldsymbol{pp}}^{-1} \star \langle\widetilde{\boldsymbol{P}}'\, \widetilde{\boldsymbol{P}}_s\rangle_0) = \widetilde{\boldsymbol{P}}_s - (\langle\widetilde{\boldsymbol{P}}\rangle_s + \widetilde{\boldsymbol{P}}'_s) = \boldsymbol{0}.$$

the aid of (D 2), that

$$\widehat{\boldsymbol{\tau}}_s \approx \boldsymbol{\tau}_s - \mathbf{I}\left[p_s + \sum_{\overline{s}} N'_{\overline{s}} \left(\frac{\partial p_s}{\partial n_{\overline{s}}}\right)_e + \sum_{\overline{s}} E'_{\overline{s}} \left(\frac{\partial p_s}{\partial e_{\overline{s}}}\right)_n\right]. \tag{D 6}$$

Without the frame change, terms of $O(u_s^2)$ would appear.

• Now consider the subtracted energy flux. In the local frame, symmetry considerations and the result (D 1*d*) lead to

$$\mathrm{P}\widetilde{\boldsymbol{J}}^e_s(\boldsymbol{r}) = \langle \widetilde{\boldsymbol{J}}^e_s \rangle_0 + \sum_{\overline{s}} \int \mathrm{d}\overline{\boldsymbol{x}}\, \boldsymbol{P}'_{\overline{s}}(\overline{\boldsymbol{x}})(m\overline{n}T)^{-1}_{\overline{s}} \langle \boldsymbol{P}'_{\overline{s}}(\overline{\boldsymbol{x}}) \widetilde{\boldsymbol{J}}^e_s(\boldsymbol{r}) \rangle_0. \tag{D 7}$$

It is not difficult to show that[61]

$$\langle \widetilde{\boldsymbol{J}}^e_s \rangle_0 = h_s \boldsymbol{u}_s = \left(\frac{h}{mn}\right)_s \boldsymbol{p}_s, \tag{D 8}$$

where $h \doteq e + p$ is the enthalpy, and that[62]

$$\boldsymbol{X}_{\overline{s}s}(\boldsymbol{r}, \overline{\boldsymbol{x}}) \doteq \langle \boldsymbol{P}'_{\overline{s}}(\overline{\boldsymbol{x}}) \widetilde{\boldsymbol{J}}^e_s(\boldsymbol{r}) \rangle_0 = T\mathbf{I} \int \frac{\mathrm{d}\boldsymbol{k}}{(2\pi)^3}\, \mathrm{e}^{\mathrm{i}\boldsymbol{k}\cdot(\overline{\boldsymbol{x}}-\boldsymbol{r})} \left(e_0 + \overline{n}T + \frac{1}{3}\overline{n}^{\mathrm{T}} \Delta\boldsymbol{\tau}_0(\boldsymbol{k})\right)_s. \tag{D 9}$$

The $\boldsymbol{k}$-independent terms are proportional to $\delta(\overline{\boldsymbol{x}} - \boldsymbol{r})$. The $\Delta\boldsymbol{\tau}(\boldsymbol{k})$ term decays within one cell; thus,

$$\int \mathrm{d}\overline{\boldsymbol{x}}\, \boldsymbol{P}'(\overline{\boldsymbol{x}}) \cdot \boldsymbol{X}(\boldsymbol{r} - \overline{\boldsymbol{x}}) \approx \boldsymbol{P}'(\boldsymbol{r}) \cdot \int \mathrm{d}\boldsymbol{\rho}\, \boldsymbol{X}(\boldsymbol{\rho}). \tag{D 10}$$

The last integral extracts the $\boldsymbol{k} = \boldsymbol{0}$ component, whereupon $\int \mathrm{d}\boldsymbol{\rho}\, \boldsymbol{X}(\boldsymbol{\rho}) = Th(\overline{n}/n)\mathbf{I}$. Finally,

$$\mathrm{P}\widetilde{\boldsymbol{J}}^e_s(\boldsymbol{r}) = \left(\frac{h}{mn}\right)_s \boldsymbol{p}_s(\boldsymbol{r}) + \left(\frac{h}{mn}\right)_s \boldsymbol{P}'(\boldsymbol{r}) = \left(\frac{h}{mn}\right)_s \boldsymbol{P}_s(\boldsymbol{r}), \tag{D 11}$$

so

$$\widehat{\boldsymbol{J}}^E_s = \widetilde{\boldsymbol{J}}^E_s - \left(\frac{h}{mn}\right)_s \widetilde{\boldsymbol{P}}_s. \tag{D 12}$$

---

[61] Symmetry considerations lead to

$$\begin{aligned}\boldsymbol{J}^e_s(\boldsymbol{r}) &= \int \frac{\mathrm{d}\boldsymbol{k}}{(2\pi)^3}\, \mathrm{e}^{\mathrm{i}\boldsymbol{k}\cdot\boldsymbol{r}} \\ &\times \left\langle \sum_{i\in s} \left(\frac{1}{2}m_s|\boldsymbol{w}_i + \boldsymbol{u}_s|^2(\boldsymbol{w}_i + \boldsymbol{u}_s) + \widetilde{U}_i(\boldsymbol{w}_i + \boldsymbol{u}_s) + \Delta\widetilde{\tau}_i(\boldsymbol{k}) \cdot (\boldsymbol{w}_i + \boldsymbol{u}_s)\right) \mathrm{e}^{-\mathrm{i}\boldsymbol{k}\cdot\boldsymbol{x}_i} \right\rangle_0 \\ &= \Bigg[\underbrace{\left(\frac{1}{2}mnu^2 + \frac{3}{2}nT + u\right)}_{e} + \underbrace{(nT + \Delta p)}_{p}\Bigg]_s \boldsymbol{u}_s = h_s \boldsymbol{u}_s.\end{aligned}$$

[62] One uses the facts that (i) there are no velocity correlations in the reference ensemble, and (ii) $\langle w^{2n} \rangle = (2n+1)!!\, v_{\mathrm{t}}^{2n}$.

## Appendix E. Aspects of the first-order exchange terms

In this section I consider aspects of the first-order exchange term

$$X_s^\alpha \doteq -k_{1\Delta}^{\overline{\beta}}[\dot{A}'^{\alpha}_{\Delta s}](\mu,t)B_{\overline{\beta}}(\boldsymbol{r},t) \quad \text{(E 1}a\text{)}$$

$$= -\int_0^\infty \mathrm{d}\overline{\tau} \int \mathrm{d}\overline{\mu}\, \langle \mathrm{Q}\dot{A}'^{\alpha}_{\Delta s}(\mu)\mathrm{e}^{-\mathrm{iQ}\mathscr{L}\overline{\tau}}\dot{A}'^{\overline{\beta}}_{\Delta}(\overline{\mu})\rangle_0 B_{\overline{\beta},\overline{s}}(\boldsymbol{r},t). \quad \text{(E 1}b\text{)}$$

There is no density contribution to this term since $\dot{A}'^{n}_{\Delta} = 0$. For $\alpha = \boldsymbol{p}$ or $\alpha = e$, one can divide $X^\alpha$ into a hydrodynamic and a nonhydrodynamic part, $X^\alpha = X^\alpha_{\mathrm{h}} + X^\alpha_{\mathrm{nh}}$, corresponding to the two terms associated with the $\mathrm{Q} = 1 - \mathrm{P}$ in the propagator. I shall not give a complete discussion of all of the exchange effects in this appendix, but as an example of the manipulations I shall show that for weakly coupled plasma $X^{\boldsymbol{p}}_{\mathrm{h}}$ reduces to the hydrodynamic contribution to the exchange term calculated in Part I, namely (for two-species plasma)

$$X_s^\alpha = -(nm)_s \nu_{ss'}(\boldsymbol{u}_s - \boldsymbol{u}_{s'}) \quad \text{(E 2)}$$

[see (I:3.28) and (I:3.29$b$)].

Symmetry in velocity space constrains $\overline{\beta}$ to equal $\alpha$. I shall calculate the hydrodynamic momentum exchange ($\alpha = \boldsymbol{p}$). The random exchange force $\dot{\boldsymbol{P}}'_{\Delta}$ is given by (B 8). When only the kinetic parts of the $\boldsymbol{A}$'s are retained, it is easy to see that $\mathrm{P}\dot{\boldsymbol{P}}'_{\Delta,s}$ vanishes by symmetry. Therefore, one must calculate

$$-X^{\boldsymbol{p}}_{\mathrm{h}} \doteq \int_0^\infty \mathrm{d}\overline{\tau} \int \mathrm{d}\overline{\mu}\, \langle \dot{\boldsymbol{P}}'_{\Delta s}(\mu)\mathrm{e}^{-\mathrm{i}\mathscr{L}\overline{\tau}}\dot{\boldsymbol{P}}'_{\Delta\overline{s}}(\overline{\mu})\rangle_0 \boldsymbol{\cdot} \boldsymbol{B}_{\boldsymbol{p},\overline{s}}, \quad \text{(E 3)}$$

where $\boldsymbol{B}_{\boldsymbol{p},\overline{s}} \doteq T_{\overline{s}}^{-1}\boldsymbol{u}_{\overline{s}}$. Upon writing this in terms of the Klimontovich microdensity, one finds

$$-X^{\boldsymbol{p}}_{\mathrm{h}} = (\overline{n}q)_s \sum_{s'\neq s}(\overline{n}q)_{s'} \sum_{\overline{s}}(\overline{n}q)_{\overline{s}} \sum_{\overline{s}'\neq\overline{s}}(\overline{n}q)_{\overline{s}'}$$
$$\times \int_0^\infty \mathrm{d}\overline{\tau}\int \mathrm{d}\boldsymbol{v}\int \mathrm{d}\boldsymbol{r}'\int \mathrm{d}\boldsymbol{v}'\int \mathrm{d}\overline{\boldsymbol{r}}\int \mathrm{d}\overline{\boldsymbol{v}}\int \mathrm{d}\overline{\boldsymbol{r}}'\int \mathrm{d}\overline{\boldsymbol{v}}'\boldsymbol{\epsilon}(\boldsymbol{r}-\boldsymbol{r}')\,\boldsymbol{\epsilon}(\overline{\boldsymbol{r}}-\overline{\boldsymbol{r}}')$$
$$\times \langle \widetilde{f}_s(\boldsymbol{r},\boldsymbol{v},\overline{\tau})\widetilde{f}_{s'}(\boldsymbol{r}',\boldsymbol{v}',\overline{\tau})\widetilde{f}_{\overline{s}}(\overline{\boldsymbol{r}},\overline{\boldsymbol{v}},0)\widetilde{f}_{\overline{s}'}(\overline{\boldsymbol{r}}',\overline{\boldsymbol{v}}',0)\rangle_0 \boldsymbol{\cdot} \boldsymbol{B}_{\boldsymbol{p},\overline{s}}, \quad \text{(E 4)}$$

where

$$\boldsymbol{\epsilon}(\boldsymbol{r}) \doteq -\boldsymbol{\nabla}(r^{-1}) = \int \frac{\mathrm{d}\boldsymbol{k}}{(2\pi)^3}\,\mathrm{e}^{\mathrm{i}\boldsymbol{k}\boldsymbol{\cdot}\boldsymbol{r}}\boldsymbol{\epsilon}_{\boldsymbol{k}}, \quad \boldsymbol{\epsilon}_{\boldsymbol{k}} \doteq -4\pi\mathrm{i}\boldsymbol{k}/k^2 \quad \text{(E 5)}$$

gives the electric field due to a unit point charge at the origin. Cumulant expansion of the four-point correlation function gives schematically

$$\langle \widetilde{f}\widetilde{f}\widetilde{f}\widetilde{f}\rangle = ffff + 6ffC_2 + 4fC_3 + 3C_2C_2 + C_4, \quad \text{(E 6)}$$

where $C_n$ denotes the $n$th-order cumulant. It is assumed that $f$ is a Maxwellian $f_{\mathrm{M}}$ to lowest order; this is equivalent to subtracting out the contributions due to a long-ranged mean potential. Then all of the terms involving $f$ vanish under spatial integration due to the presence of an $\boldsymbol{\epsilon}$ and the isotropy of $f_{\mathrm{M}}$. Of the three terms involving $CC$, where $C \equiv C_2$, one involves the equal-time correlations $C_{ss'}(\boldsymbol{r}-\boldsymbol{r}',\boldsymbol{v},\boldsymbol{v}',0)C_{\overline{s}\overline{s}'}(\overline{\boldsymbol{r}}-\overline{\boldsymbol{r}}',\overline{\boldsymbol{v}},\overline{\boldsymbol{v}}',0)$. That term also does not contribute because to lowest order $C(\boldsymbol{r}-\boldsymbol{r}') = C(|\boldsymbol{r}-\boldsymbol{r}'|)$ and $\int \mathrm{d}\boldsymbol{\rho}\,\boldsymbol{\epsilon}(\boldsymbol{\rho})C(\rho) = \boldsymbol{0}$ due to isotropy. Because for thermal noise $C_n = O(\epsilon_{\mathrm{p}}^{n-1})$, $C_4$ is negligible for weak coupling. This leaves

$$
\begin{aligned}
-X_{\mathrm{h}}^{\boldsymbol{p}} = (\overline{n}q)_s \sum_{s'\neq s} (\overline{n}q)_{s'} \sum_{\overline{s}} (\overline{n}q)_{\overline{s}} \sum_{\overline{s}'\neq\overline{s}} (\overline{n}q)_{\overline{s}'} \\
\times \int_0^\infty \mathrm{d}\overline{\tau} \int \mathrm{d}\boldsymbol{v} \int \mathrm{d}\boldsymbol{r}' \int \mathrm{d}\boldsymbol{v}' \int \mathrm{d}\overline{\boldsymbol{r}} \int \mathrm{d}\overline{\boldsymbol{v}} \int \mathrm{d}\overline{\boldsymbol{r}}' \int \mathrm{d}\overline{\boldsymbol{v}}' \boldsymbol{\epsilon}(\boldsymbol{r}-\boldsymbol{r}')\,\boldsymbol{\epsilon}(\overline{\boldsymbol{r}}-\overline{\boldsymbol{r}}') \\
\times [C_{s\overline{s}}(\boldsymbol{r}-\overline{\boldsymbol{r}},\boldsymbol{v},\overline{\tau};\overline{\boldsymbol{v}})C_{s'\overline{s}'}(\boldsymbol{r}'-\overline{\boldsymbol{r}}',\boldsymbol{v}',\overline{\tau};\overline{\boldsymbol{v}}') \\
+ C_{s\overline{s}'}(\boldsymbol{r}-\overline{\boldsymbol{r}}',\boldsymbol{v},\overline{\tau};\overline{\boldsymbol{v}}')C_{s'\overline{s}}(\boldsymbol{r}'-\overline{\boldsymbol{r}},\boldsymbol{v}',\overline{\tau};\overline{\boldsymbol{v}})] \boldsymbol{\cdot} \boldsymbol{B}_{\boldsymbol{p},\overline{s}}.
\end{aligned}
\tag{E 7}
$$

In the plasma ordering, $\overline{n} = O(\epsilon_{\mathrm{p}}^{-1})$, $q = O(\epsilon_{\mathrm{p}})$, and $T = O(\epsilon_{\mathrm{p}})$; I have already noted that $C = O(\epsilon_{\mathrm{p}})$. The coefficient of $\boldsymbol{u}$ is therefore $O(\epsilon_{\mathrm{p}})$. Thus, it is adequate to calculate $C$ only to lowest order in $\epsilon_{\mathrm{p}}$ (i.e., to use just the collisionless Vlasov response). Since the integration is extended only from 0 to $\infty$, only the one-sided correlation function $C_+(\overline{\tau}) \doteq H(\overline{\tau})C(\overline{\tau})$ enters; its evolution is expressed by the Vlasov response function $R^{(0)}$:

$$
C_+(\overline{\tau}) \approx R^{(0)}(\overline{\tau}) * C(0), \tag{E 8}
$$

where $*$ denotes convolution (i.e., integration/summation over intermediate space, velocity, and species arguments). The initial condition is

$$
C_{ss'}(\boldsymbol{r},\boldsymbol{v},\boldsymbol{r}',\boldsymbol{v}',0) = \delta_{ss'}\delta(\boldsymbol{r}-\boldsymbol{r}')\delta(\boldsymbol{v}-\boldsymbol{v}')\overline{n}_{s'}^{-1} f_{s'}(\boldsymbol{r}',\boldsymbol{v}',0) + g_{ss'}(\boldsymbol{r},\boldsymbol{v},\boldsymbol{r}',\boldsymbol{v}',0), \tag{E 9}
$$

where $g$ is the pair correlation function.[63] Consistent with the assumptions used in deriving the Balescu–Lenard collision operator, I shall use the $g$ appropriate for Debye-length scales; one may use the thermal-equilibrium result since $X^\alpha$ is calculated as a linearization from a Maxwellian. That formula is well known;[64] one finds (for Vlasov-scale wavenumbers)

$$
g_{ss'}(\boldsymbol{k},\boldsymbol{v},\boldsymbol{v}',0) \approx -\left(\frac{q_s q_{s'}}{T}\right)\frac{4\pi/k^2}{\mathscr{D}(\boldsymbol{k},0)}, \tag{E 10}
$$

where the static dielectric function is

$$
\mathscr{D}(\boldsymbol{k},\omega=0) \doteq 1 + \frac{k_{\mathrm{D}}^2}{k^2}. \tag{E 11}
$$

Here the Debye wavenumber $k_{\mathrm{D}}$ obeys $k_{\mathrm{D}}^2 = \sum_s k_{\mathrm{D}s}^2$ with $k_{\mathrm{D}s} \doteq (4\pi\overline{n}q^2/T)_s^{1/2}$. Upon Fourier transformation, one is led to

$$
\begin{aligned}
-X_{\mathrm{h}}^{\boldsymbol{p}} = (\overline{n}q)_s \sum_{s'\neq s} (\overline{n}q)_{s'} \sum_{\overline{s}} (\overline{n}q)_{\overline{s}} \sum_{\overline{s}'\neq\overline{s}} (\overline{n}q)_{\overline{s}'} \int \mathrm{d}\boldsymbol{v} \int \mathrm{d}\boldsymbol{v}' \int \mathrm{d}\overline{\boldsymbol{v}} \int \mathrm{d}\overline{\boldsymbol{v}}' \int \frac{\mathrm{d}\omega}{2\pi} \int \frac{\mathrm{d}\boldsymbol{k}}{(2\pi)^3} \boldsymbol{\epsilon}_{\boldsymbol{k}}\,\boldsymbol{\epsilon}_{\boldsymbol{k}}^* \\
\times [C_{+,s\overline{s}}^*(\boldsymbol{k},\omega,\boldsymbol{v};\overline{\boldsymbol{v}})C_{+,s'\overline{s}'}(\boldsymbol{k},\omega,\boldsymbol{v}';\overline{\boldsymbol{v}}') - C_{+,s\overline{s}'}^*(\boldsymbol{k},\omega,\boldsymbol{v};\overline{\boldsymbol{v}}')C_{+,s'\overline{s}}(\boldsymbol{k},\omega,\boldsymbol{v}';\overline{\boldsymbol{v}})] \boldsymbol{\cdot} \boldsymbol{B}_{\boldsymbol{p},\overline{s}}.
\end{aligned}
\tag{E 12}
$$

To be specific, consider $s = e$ and a single species of ions. The expression (E 12) then reduces to

$$
\begin{aligned}
-{}_aX_{\mathrm{h}}^{\boldsymbol{p}} = (\overline{n}q)_e^2(\overline{n}q)_i^2 \int \mathrm{d}\boldsymbol{v} \int \mathrm{d}\boldsymbol{v}' \int \mathrm{d}\overline{\boldsymbol{v}} \int \mathrm{d}\overline{\boldsymbol{v}}' \int \frac{\mathrm{d}\omega}{2\pi} \int \frac{\mathrm{d}\boldsymbol{k}}{(2\pi)^3} \boldsymbol{\epsilon}_{\boldsymbol{k}}\,\boldsymbol{\epsilon}_{\boldsymbol{k}}^* \\
\times [C_{+,ee}^*(\boldsymbol{k},\boldsymbol{v};\overline{\boldsymbol{v}})C_{+,ii}(\boldsymbol{k},\omega,\boldsymbol{v}';\overline{\boldsymbol{v}}') - C_{+,ei}^*(\boldsymbol{k},\omega,\boldsymbol{v};\overline{\boldsymbol{v}}')C_{+,ie}(\boldsymbol{v};\overline{\boldsymbol{v}})] \boldsymbol{\cdot} \Delta\boldsymbol{B},
\end{aligned}
\tag{E 13}
$$

[63] Both $f$ and $g$ are to be calculated in the reference ensemble, so they should be adorned with the subscript 0; however, I shall drop that in the remainder of this appendix.

[64] Several different approaches are described by Montgomery & Tidman (1964). One route is worked out in §G.1.3.

where

$$\Delta\boldsymbol{B} \doteq \boldsymbol{B}_{\boldsymbol{p},e} - \boldsymbol{B}_{\boldsymbol{p},i} = (\beta\boldsymbol{u})_e - (\beta\boldsymbol{u})_i \approx T^{-1}(\boldsymbol{u}_e - \boldsymbol{u}_i). \tag{E 14}$$

To obtain (E 13), I interchanged $\overline{\boldsymbol{v}}$ and $\overline{\boldsymbol{v}}'$ after expanding out the species dependence of the $\boldsymbol{B}_i$ term. The approximation of a common temperature in the last form of (E 14) is justified because I am ignoring second-order exchange effects.

To complete the calculation, one needs according to (E 8) an expression for the response function $R^{(0)}$. It is well known (Krommes 2002) that operational methods lead to the general expression for the fully renormalized electrostatic response function

$$R = r - r\,\boldsymbol{\partial} f\,\mathscr{D}^{-1}\boldsymbol{\cdot}\mathbb{E}r, \tag{E 15}$$

where $r$ is the single-particle response function, $\boldsymbol{\partial} \doteq (q/m)\partial/\partial_{\boldsymbol{v}}$, $\mathscr{D}$ is the dielectric function

$$\mathscr{D} = 1 + \mathbb{E}r\boldsymbol{\cdot}\boldsymbol{\partial} f, \tag{E 16}$$

and $\mathbb{E}$ is the electric-field operator whose kernel is $\mathbb{E}_{s,\overline{s}}(\boldsymbol{k},\omega,\boldsymbol{v};\overline{\boldsymbol{v}}) = \boldsymbol{\epsilon}_{\boldsymbol{k}}(\overline{n}q)_{\overline{s}}$. In the presence of fluctuations, the actual evaluation of $r$ and $\mathscr{D}$ is entirely nontrivial and occupies a good portion of the formal discussion of plasma turbulence theory (Krommes 2002). But, as noted above, one only requires the collisionless approximation $R^{(0)}$. In that case

$$r^{(0)}_{s\overline{s}}(\boldsymbol{k},\omega,\boldsymbol{v};\overline{\boldsymbol{v}}) = \frac{\delta_{s\overline{s}}\delta(\boldsymbol{v}-\overline{\boldsymbol{v}})}{-\mathrm{i}(\omega - \boldsymbol{k}\boldsymbol{\cdot}\boldsymbol{v} + \mathrm{i}\epsilon)} \tag{E 17}$$

(the positive infinitesimal $\epsilon$ enforces causality) and

$$\mathscr{D}(\boldsymbol{k},\omega) = 1 + \sum_{\overline{s}}\chi_{\overline{s}}(\boldsymbol{k},\omega), \tag{E 18}$$

where the zeroth-order susceptibility is

$$\chi^{(0)}_{\overline{s}}(\boldsymbol{k},\omega) = \frac{\omega^2_{\mathrm{p}\overline{s}}}{k^2}\int \mathrm{d}\overline{\boldsymbol{v}}\,\frac{\boldsymbol{k}\boldsymbol{\cdot}\partial_{\overline{\boldsymbol{v}}}f_{\overline{s}}(\overline{\boldsymbol{v}})}{\omega - \boldsymbol{k}\boldsymbol{\cdot}\overline{\boldsymbol{v}} + \mathrm{i}\epsilon}. \tag{E 19}$$

The delta functions in (E 17) then allow (E 15) to be simplified to

$$R^{(0)}_{s\overline{s}}(\boldsymbol{k},\omega,\boldsymbol{v};\overline{\boldsymbol{v}}) = \frac{\delta_{s\overline{s}}\delta(\boldsymbol{v}-\overline{\boldsymbol{v}})}{-\mathrm{i}(\omega - \boldsymbol{k}\boldsymbol{\cdot}\boldsymbol{v} + \mathrm{i}\epsilon)} - \left(\frac{\boldsymbol{\epsilon}_{\boldsymbol{k}}\boldsymbol{\cdot}\boldsymbol{\partial} f_s}{-\mathrm{i}(\omega - \boldsymbol{k}\boldsymbol{\cdot}\boldsymbol{v} + \mathrm{i}\epsilon)}\right)\frac{1}{\mathscr{D}^{(0)}(\boldsymbol{k},\omega)}\left(\frac{(\overline{n}q)_{\overline{s}}}{-\mathrm{i}(\omega - \boldsymbol{k}\boldsymbol{\cdot}\overline{\boldsymbol{v}} + \mathrm{i}\epsilon)}\right). \tag{E 20}$$

This expression can now be used in conjunction with (E 8) to evaluate (E 13). After somewhat lengthy but straightforward algebra that includes a number of cancellations, one finds

$$-\,{}_aX^{\boldsymbol{p}}_{\mathrm{h}} = (\overline{n}q^2)_e(\overline{n}q^2)_i\int \mathrm{d}\boldsymbol{v}\int \mathrm{d}\boldsymbol{v}'\int\frac{\mathrm{d}\omega}{2\pi}\int\frac{\mathrm{d}\boldsymbol{k}}{(2\pi)^3}\,\frac{1}{\mathscr{D}(\boldsymbol{k},0)}\,\boldsymbol{\epsilon}_{\boldsymbol{k}}\,\boldsymbol{\epsilon}^*_{\boldsymbol{k}} \times\left(\frac{1 + \chi^{(0)*}_i(\boldsymbol{k},\omega) + \chi^{(0)}_e(\boldsymbol{k},\omega)}{|\mathscr{D}^{(0)}(\boldsymbol{k},\omega)|^2}\right)\left(\frac{f_e(\boldsymbol{v})f_i(\boldsymbol{v}')}{(\omega - \boldsymbol{k}\boldsymbol{\cdot}\boldsymbol{v} - \mathrm{i}\epsilon)(\omega - \boldsymbol{k}\boldsymbol{\cdot}\boldsymbol{v}' + \mathrm{i}\epsilon)}\right). \tag{E 21}$$

The static shielding factor $\mathscr{D}(\boldsymbol{k},0)^{-1}$ that appears in this expression is built from two contributions corresponding to the two terms of (E 9):

$$\frac{1}{\mathscr{D}(\boldsymbol{k},0)} = 1 - \frac{k^2_{\mathrm{D}}/k^2}{\mathscr{D}(\boldsymbol{k},0)}. \tag{E 22}$$

It reflects the shielding of a test particle [the first term of (E 9)] by polarization [the second term of (E 9)].

The following remarks are relevant to the goal of obtaining expression (E 2): (i) (E 21) cannot be immediately reduced by residue methods applied to the $\omega$ integration because of the factor $|\mathscr{D}^{(0)}(\boldsymbol{k},\omega)|^{-2}$, which contains structure in both of the top and bottom halves of the complex $\omega$ plane; (ii) the expression (E 2) was derived from the Landau collision operator linearized around a Maxwellian; (iii) the Landau operator is an approximation to the more fundamental Balescu–Lenard operator. It is rather clear that from the present approach one should ultimately obtain the linearized-Balescu–Lenard generalization of (E 2). However, (E 21) holds for arbitrary $f$; to obtain agreement, one needs to specialize that to $f = f_{\mathrm{M}}$ and is free to use special properties of the Maxwellian.

One algebraic route is as follows. Because all susceptibilities from this point forward will be evaluated at zeroth order, I shall drop the (0) superscripts in order to unclutter the notation. Note that

$$1+\chi_i^*+\chi_e = 1+\chi_i^*+\chi_e^*+(\chi_e-\chi_e^*) = \mathscr{D}^*+2\mathrm{i}\,\mathrm{Im}\,\chi_e. \tag{E 23}$$

The $\mathscr{D}^*$ term cancels one factor of the $|\mathscr{D}|^{-2}$, leaving the integral

$$I_1 \doteq \int \frac{\mathrm{d}\omega}{2\pi}\frac{1}{\mathscr{D}(\boldsymbol{k},\omega)(\omega-\boldsymbol{k}\cdot\boldsymbol{v}-\mathrm{i}\epsilon)(\omega-\boldsymbol{k}\cdot\boldsymbol{v}'+\mathrm{i}\epsilon)} \tag{E 24$a$}$$

$$= \frac{\mathrm{i}}{\mathscr{D}(\boldsymbol{k},\boldsymbol{k}.\boldsymbol{v})[\boldsymbol{k}\cdot(\boldsymbol{v}-\boldsymbol{v}')+\mathrm{i}\epsilon]} \tag{E 24$b$}$$

$$= \mathrm{i}\left(\frac{\mathscr{D}'(\boldsymbol{k},\boldsymbol{k}\cdot\boldsymbol{v})-\mathrm{i}\mathscr{D}''(\boldsymbol{k},\boldsymbol{k}\cdot\boldsymbol{v})}{|\mathscr{D}(\boldsymbol{k},\boldsymbol{k}\cdot\boldsymbol{v})|^2}\right)\left[\mathrm{Pr}\left(\frac{1}{\boldsymbol{k}\cdot\Delta\boldsymbol{v}}\right)-\mathrm{i}\pi\,\delta(\boldsymbol{k}\cdot\Delta\boldsymbol{v})\right], \tag{E 24$c$}$$

where prime and double prime denote the real and imaginary parts. Only the real part of this expression is required:

$$\mathrm{Re}\,I_1 = \frac{1}{|\mathscr{D}(\boldsymbol{k},\boldsymbol{k}\cdot\boldsymbol{v})|^2}\Bigg[[\underbrace{1}_{(\mathrm{a}_1)}+\underbrace{\chi_e'(\boldsymbol{k},\boldsymbol{k}\cdot\boldsymbol{v})}_{(\mathrm{b}_1)}+\underbrace{\chi_i'(\boldsymbol{k},\boldsymbol{k}\cdot\boldsymbol{v})}_{(\mathrm{c}_1)}]\pi\,\delta(\boldsymbol{k}\cdot\Delta\boldsymbol{v})$$
$$+[\underbrace{\chi_e''(\boldsymbol{k},\boldsymbol{k}\cdot\boldsymbol{v})}_{(\mathrm{d}_1)}+\underbrace{\chi_i''(\boldsymbol{k},\boldsymbol{k}\cdot\boldsymbol{v})}_{(\mathrm{e}_1)}]\,\mathrm{Pr}\left(\frac{1}{\boldsymbol{k}\cdot\Delta\boldsymbol{v}}\right)\Bigg]. \tag{E 25}$$

The last term of (E 23) contributes

$$I_2 \doteq 2\mathrm{i}\int\frac{\mathrm{d}\omega}{2\pi}\left(\frac{\chi_e''(\boldsymbol{k},\omega)}{|\mathscr{D}(\boldsymbol{k},\omega)|^2}\right)\left(\frac{1}{(\omega-\boldsymbol{k}\cdot\boldsymbol{v}-\mathrm{i}\epsilon)(\omega-\boldsymbol{k}\cdot\boldsymbol{v}'+\mathrm{i}\epsilon)}\right), \tag{E 26}$$

the real part of which is

$$\mathrm{Re}\,I_2 = -\,\mathrm{Pr}\left(\frac{1}{\boldsymbol{k}\cdot\Delta\boldsymbol{v}}\right)\left(\underbrace{\frac{\chi_e''(\boldsymbol{k},\boldsymbol{k}\cdot\boldsymbol{v})}{|\mathscr{D}(\boldsymbol{k},\boldsymbol{k}\cdot\boldsymbol{v})|^2}}_{(\mathrm{d}_2)}+\underbrace{\frac{\chi_e''(\boldsymbol{k},\boldsymbol{k}\cdot\boldsymbol{v}')}{|\mathscr{D}(\boldsymbol{k},\boldsymbol{k}\cdot\boldsymbol{v}')|^2}}_{(\mathrm{f}_2)}\right). \tag{E 27}$$

Term ($\mathrm{d}_2$) cancels with term ($\mathrm{d}_1$).

The remaining terms must be averaged over $f_e(\boldsymbol{v})$ and $f_i(\boldsymbol{v}')$ according to (E 21). Progress can be made upon specializing to $f = f_{\mathrm{M}}$ and using the property $\partial_{\boldsymbol{v}} f_{\mathrm{M}} = -(\boldsymbol{v}/v_{\mathrm{t}}^2)f_{\mathrm{M}}$. It is then easy to show that

$$\chi_s(\boldsymbol{k},\omega) = \frac{k_{\mathrm{D}s}^2}{k^2} - \omega\left(\frac{k_{\mathrm{D}s}^2}{k^2}\right)\int \mathrm{d}\overline{\boldsymbol{v}}\,\frac{f_s(\overline{\boldsymbol{v}})}{\omega-\boldsymbol{k}\cdot\overline{\boldsymbol{v}}+\mathrm{i}\epsilon}, \tag{E 28}$$

or

$$\chi_s'(\boldsymbol{k},\omega) = \frac{k_{\mathrm{D}s}^2}{k^2} - \omega\left(\frac{k_{\mathrm{D}s}^2}{k^2}\right)\int \mathrm{d}\overline{\boldsymbol{v}}\ \mathrm{Pr}\left(\frac{1}{\omega - \boldsymbol{k}\cdot\overline{\boldsymbol{v}}}\right) f_s(\overline{\boldsymbol{v}}), \tag{E 29a}$$

$$\chi_s''(\boldsymbol{k},\omega) = \pi\omega\left(\frac{k_{\mathrm{D}s}^2}{k^2}\right)\int \mathrm{d}\overline{\boldsymbol{v}}\,\delta(\omega - \boldsymbol{k}\cdot\overline{\boldsymbol{v}}) f_s(\overline{\boldsymbol{v}}). \tag{E 29b}$$

Upon using (E 29*b*) to evaluate term ($\mathrm{f}_2$), then interchanging $\boldsymbol{v}$ and $\overline{\boldsymbol{v}}$, one can show that

$$-\int \mathrm{d}\boldsymbol{v}\,\mathrm{d}\boldsymbol{v}'\ \mathrm{Pr}\left(\frac{1}{\boldsymbol{k}\cdot(\boldsymbol{v}-\boldsymbol{v}')}\right)\underbrace{\frac{\chi_e''(\boldsymbol{k},\boldsymbol{k}\cdot\boldsymbol{v}')}{|\mathscr{D}(\boldsymbol{k},\boldsymbol{k}\cdot\boldsymbol{v}')|^2}}_{(\mathrm{f}_2)} f_e(\boldsymbol{v})f_i(\boldsymbol{v}')$$
$$= \pi\int \mathrm{d}\boldsymbol{v}\,\mathrm{d}\boldsymbol{v}'\left(\underbrace{\frac{k_{\mathrm{De}}^2}{k^2}}_{(\mathrm{a}_e)} - \underbrace{\chi_e'(\boldsymbol{k},\boldsymbol{k}\cdot\boldsymbol{v})}_{(\mathrm{b}_2)}\right)\frac{\delta(\boldsymbol{k}\cdot(\boldsymbol{v}-\boldsymbol{v}'))}{|\mathscr{D}(\boldsymbol{k},\boldsymbol{k}\cdot\boldsymbol{v})|^2} f_e(\boldsymbol{v})f_i(\boldsymbol{v}'). \tag{E 30}$$

Term ($\mathrm{a}_e$) adds to term ($\mathrm{a}_1$), while term ($\mathrm{b}_2$) cancels with term ($\mathrm{b}_1$). Similarly, the contribution of term ($\mathrm{e}_1$) can be evaluated by interchanging $\boldsymbol{v}'$ and $\overline{\boldsymbol{v}}$; one finds

$$\int \mathrm{d}\boldsymbol{v}\,\mathrm{d}\boldsymbol{v}'\ \mathrm{Pr}\left(\frac{1}{\boldsymbol{k}\cdot(\boldsymbol{v}-\boldsymbol{v}')}\right)\underbrace{\frac{\chi_e''(\boldsymbol{k},\boldsymbol{k}\cdot\boldsymbol{v}')}{|\mathscr{D}(\boldsymbol{k},\boldsymbol{k}\cdot\boldsymbol{v}')|^2}}_{(\mathrm{e}_1)} f_e(\boldsymbol{v})f_i(\boldsymbol{v}')$$
$$= \pi\int \mathrm{d}\boldsymbol{v}\,\mathrm{d}\boldsymbol{v}'\left(\underbrace{\frac{k_{\mathrm{D}i}^2}{k^2}}_{(\mathrm{a}_i)} - \underbrace{\chi_i'(\boldsymbol{k},\boldsymbol{k}\cdot\boldsymbol{v})}_{(\mathrm{g}_1)}\right)\frac{\delta(\boldsymbol{k}\cdot(\boldsymbol{v}-\boldsymbol{v}'))}{|\mathscr{D}(\boldsymbol{k},\boldsymbol{k}\cdot\boldsymbol{v})|^2} f_e(\boldsymbol{v})f_i(\boldsymbol{v}'). \tag{E 31}$$

[Note the overall sign difference between (E 30) and (E 31).] Term ($\mathrm{a}_i$) adds to term ($\mathrm{a}_1$), while term ($\mathrm{g}_1$) cancels with term ($\mathrm{c}_1$). One obtains

$$X_e^{\boldsymbol{p}} = -\left(\pi(\overline{n}q^2)_e\overline{n}q^2)_i\int \mathrm{d}\boldsymbol{v}\int \mathrm{d}\overline{\boldsymbol{v}}\int\frac{\mathrm{d}\boldsymbol{k}}{(2\pi)^3}\frac{\boldsymbol{\epsilon_k}\,\boldsymbol{\epsilon_k}^*}{|\mathscr{D}(\boldsymbol{k},\boldsymbol{k}\cdot\boldsymbol{v})|^2}\delta(\boldsymbol{k}\cdot(\boldsymbol{v}-\overline{\boldsymbol{v}}))f_e(\boldsymbol{v})f_i(\overline{\boldsymbol{v}})\right)\cdot\Delta\boldsymbol{B}, \tag{E 32}$$

where $\Delta\boldsymbol{B}$ is given by (E 14). This is the Balescu–Lenard generalization of the matrix element that defines the hydrodynamic part of the momentum exchange term. Specifically, the positive-definite term in large parentheses is (to within a normalization factor) $-\mathrm{i}(\boldsymbol{\Omega}_\mathrm{C})^{\boldsymbol{p}}_{\boldsymbol{p}}$, where $\boldsymbol{\Omega}_\mathrm{C}$ is the collisional contribution to the frequency matrix defined by (I:2.54*b*). The Landau form of this term (used by Braginskii) is obtained in the standard way by setting $\mathscr{D}(\boldsymbol{k},\boldsymbol{k}\cdot\boldsymbol{v})$ to one and integrating in wavenumber magnitude $k$ between $k_\mathrm{D}$ and $k_\mathrm{max}$. (Classically, $k_\mathrm{max} = b_0^{-1}$, where $b_0$ is the impact parameter for 90° scattering. See footnote 70 on page 86 for more discussion about the cutoff.) The final result is in complete agreement with the calculations of Part I.

## Appendix F. Evaluation of term (iv)

Here I provide some details that were omitted by Brey *et al.* (1981). Upon inserting the first term of (2.74) into (3.37), one has

$$\text{term (iv)} \approx \int_0^t \mathrm{d}\overline{s}\, \langle G'\mathrm{U}(\overline{s})\mathrm{i}\mathscr{L}\boldsymbol{A}'\rangle_0 \star \boldsymbol{M}^{-1} \star \langle \boldsymbol{A}'\psi_{\Delta\boldsymbol{B}}(t-\overline{s})\boldsymbol{A}'^{\mathrm{T}}\rangle_0 \star \Delta\boldsymbol{B}(t-\overline{s}) \tag{F 1$a$}$$

$$= -\int_0^t \mathrm{d}\overline{s} \int_0^{t-\overline{s}} \mathrm{d}\overline{s}' \int \mathrm{d}\overline{\boldsymbol{x}} \int \mathrm{d}\overline{\boldsymbol{x}}'\, \langle G'\mathrm{U}(\overline{s})\mathrm{i}\mathscr{L}\boldsymbol{A}'\rangle_0 \star \boldsymbol{M}^{-1}$$
$$\star \langle \boldsymbol{A}'[U(\overline{s}')\mathrm{i}\mathscr{L}\mathrm{A}'^{\beta}(\overline{\mu}')]A'^{\gamma}(\overline{\mu}')\rangle_0 \Delta B_\beta(\overline{\mu}', t-\overline{s}-\overline{s}')\Delta B_\gamma(\overline{\mu}, t-\overline{s}). \tag{F 1$b$}$$

Upon replacing $\overline{s}'$ by $\overline{\tau} \doteq \overline{s} + \overline{s}'$, one is led to

$$\text{term (iv)} = -\int_0^t \mathrm{d}\overline{s} \int_s^t \mathrm{d}\overline{\tau} \int \mathrm{d}\overline{\boldsymbol{x}} \int \mathrm{d}\overline{\boldsymbol{x}}'\, \langle \widehat{G}(\mu)\mathrm{U}(\overline{s})\mathrm{i}\mathscr{L}\boldsymbol{A}'\rangle_0 \star \boldsymbol{M}^{-1}$$
$$\star \langle \boldsymbol{A}'[\mathrm{U}(\overline{\tau}-\overline{s})\mathrm{i}\mathscr{L}A'^{\beta}(\overline{\mu}')]A'^{\gamma}(\overline{\mu})\rangle_0 \Delta B_\beta(\boldsymbol{r}-\overline{\boldsymbol{\rho}}', t-\overline{\tau})\Delta B_\gamma(\boldsymbol{r}-\overline{\boldsymbol{\rho}}, t-\overline{s}) \tag{F 2$a$}$$

$$\approx \int_0^\infty \mathrm{d}\overline{\tau} \int_0^{\overline{\tau}} \mathrm{d}\overline{s} \int \mathrm{d}\overline{\boldsymbol{x}}\, \langle \widehat{G}(\mu)\mathrm{U}(\overline{s})\mathrm{i}\mathscr{L}\boldsymbol{A}'\rangle_0 \star \boldsymbol{M}^{-1}$$
$$\star \langle \boldsymbol{A}'[\mathrm{U}(\overline{\tau}-\overline{s})\widehat{\mathscr{J}}^{\beta}]A'^{\gamma}(\overline{\mu})\rangle_0 \cdot \boldsymbol{\nabla}B_\beta(\boldsymbol{r},t)\overline{\boldsymbol{\rho}} \cdot \boldsymbol{\nabla}B_\gamma(\boldsymbol{r},t). \tag{F 2$b$}$$

This requires evaluation of

$$\boldsymbol{I} \doteq \int_0^{\overline{\tau}} \mathrm{d}\overline{s}\, \mathrm{U}_1(\overline{s})\mathrm{i}\mathscr{L}_1\boldsymbol{A}_1'\mathrm{U}_2(\overline{\tau}-\overline{s}). \tag{F 3}$$

(The subscripts distinguish the propagators, which appear under different expectations and thus operate on different variables.) Use the identity (3.25) and integrate the $\mathrm{U}_1$ factor by parts:

$$\boldsymbol{I} = (\boldsymbol{A}_1' - \mathrm{e}^{-\mathrm{Q}_1\mathrm{i}\mathscr{L}_1\overline{\tau}}\boldsymbol{A}_1')\underbrace{\mathrm{U}_2(0)}_{1} - \int_0^{\overline{\tau}} \mathrm{d}\overline{s}\, (\boldsymbol{A}_1' - \mathrm{e}^{-\mathrm{Q}_1\mathrm{i}\mathscr{L}_1\overline{s}}\boldsymbol{A}_1')\mathrm{U}_2(\overline{\tau}-\overline{s})\mathrm{Q}_2\mathrm{i}\mathscr{L}_2\mathrm{Q}_2. \tag{F 4}$$

The first $\boldsymbol{A}_1'$ factor in each parenthesis does not contribute because $\langle \widehat{G}(\mu)\boldsymbol{A}'\rangle = 0$. Use of the identity (3.30) shows that to lowest order in the gradients one can replace $\exp(-\mathrm{Q}\mathrm{i}\mathscr{L}\overline{\tau})$ by $\exp(-\mathrm{i}\mathscr{L}\overline{\tau})$. From the last term in (F 4) arises the quantity

$$\boldsymbol{K} \doteq \langle \boldsymbol{A}'[\mathrm{e}^{-\mathrm{Q}\mathrm{i}\mathscr{L}\mathrm{Q}(\overline{\tau}-\overline{s})}\mathrm{Q}\mathrm{i}\mathscr{L}\mathrm{Q}\widehat{\mathscr{J}}^{\beta}]A'^{\gamma}\rangle_0 = -\langle [\mathrm{Q}\mathrm{i}\mathscr{L}\mathrm{Q}\mathrm{e}^{\mathrm{Q}\mathrm{i}\mathscr{L}\mathrm{Q}(\overline{\tau}-\overline{s})}\boldsymbol{A}'A'^{\gamma}]\widehat{\mathscr{J}}^{\beta}\rangle_0 \tag{F 5$a$}$$

$$= -\langle [\mathrm{Q}\mathrm{i}\mathscr{L}\mathrm{e}^{\mathrm{Q}\mathrm{i}\mathscr{L}(\overline{\tau}-\overline{s})}\mathrm{Q}(\boldsymbol{A}'A'^{\gamma})]\widehat{\mathscr{J}}^{\beta}\rangle_0. \tag{F 5$b$}$$

Because $\mathrm{Q}(\boldsymbol{A}'\boldsymbol{A}') = \boldsymbol{A}'\boldsymbol{A}' - \boldsymbol{A}' \star \boldsymbol{M}^{-1} \star \langle \boldsymbol{A}'\boldsymbol{A}'\boldsymbol{A}'\rangle$ and $\mathscr{L}$ is a linear operator, it can be seen that $\boldsymbol{K}$ contains at least one power of $\mathrm{i}\mathscr{L}\boldsymbol{A}'$ and thus is at least of first order in the gradients. The contribution of the integral term in (F 4) is thus negligible, and one obtains

$$\text{term (iv)} \approx -\int_0^\infty \mathrm{d}\overline{\tau} \int \mathrm{d}\overline{\boldsymbol{x}}\, \langle [\mathrm{e}^{\mathrm{i}\mathscr{L}\overline{\tau}}\widehat{G}(\mu)]\mathrm{P}\widehat{\mathscr{J}}^{\beta}A'^{\gamma}(\overline{\mu})\rangle_0 \cdot \boldsymbol{\nabla}B_\beta(\boldsymbol{r},t)\overline{\boldsymbol{\rho}} \cdot \boldsymbol{\nabla}B_\gamma(\boldsymbol{r},t). \tag{F 6}$$

This is the result quoted by Brey *et al.*

## Appendix G. Balescu–Lenard theory

Various of the calculations in the main text based on multiple-time hierarchies produce equations that after appropriate Markovian approximations lead to some variant of the Balescu–Lenard collision operator. Here I shall review the basic manipulations.

### G.1. *The nonlinear Balescu–Lenard operator*

Although the simplest derivation of the Balescu–Lenard operator is accomplished with the Klimontovich formalism (Klimontovich 1967), I shall proceed instead from the BBGKY hierarchy since that was used in the main text. [A historically important monograph on plasma kinetic theory that contains closely related manipulations is by Montgomery & Tidman (1964). See also the modern introduction to that subject by Swanson (2008).] The first member of the BBGKY hierarchy has the form

$$\partial_t f(1) + \cdots = -\boldsymbol{\partial}\cdot\mathbb{E}(\underline{\bar{2}})g(\underline{1},\underline{\bar{2}},t) \to -\mathrm{C}[f]. \tag{G 1}$$

The last result, which introduces the nonlinear, irreversible collision operator $\mathrm{C}[f]$ as a functional of the one-particle distribution function $f$, holds for space-time scales much longer than the microscopic correlation scales, which for weakly coupled, unmagnetised plasmas are the Debye length $\lambda_{\mathrm{D}}$ and inverse plasma frequency $\omega_{\mathrm{p}}^{-1}$. Thus, one must calculate the pair correlation function $g$ in that limit.

#### G.1.1. Representation of the collision operator in terms of the pair correlation function

The pair correlation function $g$ obeys (5.17*b*), which for weakly coupled plasma reduces to

$$(\partial_t + \mathrm{iL}_1 + \mathrm{iL}_2)g(\underline{1},\underline{2},t) = -\boldsymbol{\epsilon}_{12}\cdot(q_2\boldsymbol{\partial}_1 - q_1\boldsymbol{\partial}_2)f(\underline{1},t)f(\underline{2},t). \tag{G 2}$$

With Green's function for the linearized Vlasov equation being defined as in (5.24) and (5.25), the inhomogeneous solution[65] is

$$g(\underline{1},\underline{2},t) = -\int_{-\infty}^{t} \mathrm{d}\bar{t}\, \Xi(\underline{1},t;\underline{\bar{1}},\bar{t})\Xi(\underline{2},t;\underline{\bar{2}},\bar{t})\boldsymbol{\epsilon}_{\underline{\bar{1}}\,\underline{\bar{2}}}\cdot(q_{\bar{2}}\boldsymbol{\partial}_{\underline{\bar{1}}} - q_{\bar{1}}\boldsymbol{\partial}_{\underline{\bar{2}}})f(\underline{\bar{1}},\bar{t})f(\underline{\bar{2}},\bar{t}). \tag{G 3}$$

(Summations/integrations over repeated barred indices are understood.) It is assumed that on the timescale for the formation of a Debye shielding cloud $f$ is essentially stationary; this justifies the Markovian approximation $f(\bar{t}) \to f(t)$. The time integral can then be extended to $\infty$. It is also assumed that $f$ is essentially spatially constant on the scale of the Debye length. Use of Parseval's theorem leads one to the representation

$$\mathrm{C}[f] = \boldsymbol{\partial}\cdot\int\frac{\mathrm{d}\boldsymbol{k}}{(2\pi)^3}\,\mathbb{E}^{*}_{\boldsymbol{k}}(\underline{\bar{2}})g_{\boldsymbol{k}}(\underline{1},\underline{\bar{2}},\infty), \tag{G 4}$$

where

$$g_{\boldsymbol{k}}(\underline{1},\underline{2},\infty) = \int\frac{\mathrm{d}\omega}{2\pi}\,g_{\boldsymbol{k},\omega}(\underline{1},\underline{2}) \tag{G 5}$$

with

$$g_{\boldsymbol{k},\omega}(\underline{1},\underline{2}) \doteq -\Xi_{\boldsymbol{k},\omega}(\underline{1};\underline{\bar{1}})\Xi^{*}_{\boldsymbol{k},\omega}(\underline{2};\underline{\bar{2}})\boldsymbol{\epsilon}_{\boldsymbol{k}}\cdot(q_{\bar{2}}\boldsymbol{\partial}_{\bar{1}} - q_{\bar{1}}\boldsymbol{\partial}_{\bar{2}})f(\underline{\bar{1}})f(\underline{\bar{2}}). \tag{G 6}$$

The result for the Vlasov response function $\Xi$ is given by (E 15)–(E 19). Because one requires only the electric-field operator applied to $g$, it is possible to immediately simplify (G 4) by using the fundamental shielding identity

$$\mathbb{E}\Xi = \mathscr{D}^{-1}\mathbb{E}r. \tag{G 7}$$

[65] The contribution from the initial condition can be shown to phase-mix away.

However, for various purposes it is useful to first find a formula for $g_{\boldsymbol{k}}$ itself.

G.1.2. Calculation of the pair correlation function $g$

To simplify (G 6), use the result (E 15) to find (upon omitting underlines as well as frequency and wavenumber arguments for brevity)

$$g_{\boldsymbol{k},\omega}(1,2) = \left(r(1) - r(1)(\boldsymbol{\partial} f)_1 \cdot \frac{1}{\mathscr{D}}\mathbb{E}(\overline{1})r(\overline{1})\right)\left(r^*(2) - r^*(2)(\boldsymbol{\partial} f)_2 \cdot \frac{1}{\mathscr{D}^*}\mathbb{E}^*(\overline{2})r^*(\overline{2})\right) \times (q_2\boldsymbol{\partial}_1 - q_1\boldsymbol{\partial}_2)\cdot\boldsymbol{\epsilon}^*_{\boldsymbol{k}} f(\overline{1})f(\overline{2}). \tag{G 8}$$

[The minus sign in (G 6) was absorbed by using $-\boldsymbol{\epsilon}_{\boldsymbol{k}} = \boldsymbol{\epsilon}^*_{\boldsymbol{k}}$.] Here

$$r(1) \doteq [-\mathrm{i}(\omega - \boldsymbol{k}\cdot\boldsymbol{v}_1 + \mathrm{i}\epsilon)]^{-1}, \quad \mathbb{E}(\overline{1}) \doteq \boldsymbol{\epsilon}_{\boldsymbol{k}}(\overline{n}q)_{\overline{1}}, \quad \mathscr{D} \doteq 1 + \mathbb{E}r\cdot\boldsymbol{\partial} f. \tag{G 9}$$

Only derivatives in the direction of $\boldsymbol{k}$ enter, so introduce the scalar quantities

$$\mathbb{E}_{\boldsymbol{k}} \doteq \widehat{\boldsymbol{k}}\cdot\mathbb{E}_{\boldsymbol{k}}, \quad \epsilon_k \doteq \widehat{\boldsymbol{k}}\cdot\boldsymbol{\epsilon}_{\boldsymbol{k}} = -4\pi\mathrm{i}/k, \quad \partial_{\boldsymbol{k}} \doteq \widehat{\boldsymbol{k}}\cdot\boldsymbol{\partial}, \tag{G 10}$$

so that $\mathscr{D} = 1 + \mathbb{E}r\partial f$. Also define

$$F \doteq \mathbb{E}rqf. \tag{G 11}$$

Then, upon multiplying out the right-hand side of (G 8) and using the above definitions, one finds

$$\begin{aligned} g_{\boldsymbol{k},\omega}(1,2) = r(1)r^*(2)\Bigg( & \underbrace{(\partial f)_1(qf)_2}_{(\mathrm{a}_1)} - \underbrace{(qf)_1(\partial f)_2}_{(\mathrm{b}_1)} \\ & - (\partial f)_1\mathscr{D}^{-1}(\underbrace{\mathscr{D}}_{(\mathrm{a}_2)} - 1)(qf)_2 + \underbrace{(\partial f)_1\mathscr{D}^{-1}F(\partial f)_2}_{(\mathrm{c}_1)} \\ & + (\partial f)_2\mathscr{D}^{-1*}(\underbrace{\mathscr{D}}_{(\mathrm{b}_2)} - 1)^*(qf)_1 - \underbrace{(\partial f)_2\mathscr{D}^{-1*}F^*(\partial f)_1}_{(\mathrm{d}_1)} \\ & + \frac{(\partial f)_1(\partial f)_2}{|\mathscr{D}|^2}[(\underbrace{\mathscr{D}}_{(\mathrm{d}_2)} - 1)F^* - (\underbrace{\mathscr{D}}_{(\mathrm{c}_2)} - 1)^*F]\Bigg)\epsilon^*_k. \end{aligned} \tag{G 12}$$

Due to the cancellations indicated by the underbraces, this reduces to[66]

$$g_{\boldsymbol{k},\omega}(1,2) = r(1)r^*(2)\left(\frac{(\partial f)_1(qf)_2}{\mathscr{D}} - \frac{(\partial f)_2(qf)_1}{\mathscr{D}^*} + \frac{(\partial f)_1(\partial f)_2}{|\mathscr{D}|^2}2\mathrm{i}F''\right)\epsilon^*_k, \tag{G 13}$$

where $r(1)r^*(2) = [(\omega - \boldsymbol{k}\cdot\boldsymbol{v}_1 + \mathrm{i}\epsilon)(\omega - \boldsymbol{k}\cdot\boldsymbol{v}_2 - \mathrm{i}\epsilon)]^{-1}$, double prime denotes imaginary part, and

$$F'' = -\mathrm{i}\pi\epsilon_k \sum_{\overline{s}} (\overline{n}q^2)_{\overline{s}} \int \mathrm{d}\overline{\boldsymbol{v}}\,\delta(\omega - \boldsymbol{k}\cdot\overline{\boldsymbol{v}})f_{\overline{s}}(\overline{\boldsymbol{v}}). \tag{G 14}$$

The formula (G 14) must be integrated over frequency according to (G 5). Upon noting that $r_\omega(1)$ and $\mathscr{D}(\omega)$ are analytic in the upper half-plane, the first term of (G 13) can be integrated by residues by closing the contour in the upper half-plane, where $r^*_\omega(2)$ has a simple pole at $\boldsymbol{k}\cdot\boldsymbol{v}_2 + \mathrm{i}\epsilon$. Similarly, the second term can be integrated by closing in the

---

[66] As a partial sign check, note that in general one has $g_{\boldsymbol{k}}(1,2) = g^*_{\boldsymbol{k}}(2,1)$. That symmetry is obeyed by (G 14). (Note that $\boldsymbol{\epsilon}^*_{\boldsymbol{k}} = -\boldsymbol{\epsilon}_{\boldsymbol{k}}$.)

lower half-plane. One finds

$$g_{\boldsymbol{k}}(1,2) = \frac{1}{(\boldsymbol{k}\cdot\boldsymbol{v}_1 - \boldsymbol{k}\cdot\boldsymbol{v}_2 - 2\mathrm{i}\epsilon)}\left(\frac{(\partial f)_1 (qf)_2}{\mathscr{D}(\boldsymbol{k}\cdot\boldsymbol{v}_2)} + \frac{(\partial f)_2 (qf)_1}{\mathscr{D}^*(\boldsymbol{k}\cdot\boldsymbol{v}_1)}\right)\epsilon_k^* + 2\mathrm{i}\int\frac{\mathrm{d}\omega}{2\pi}\, r_\omega(1) r_\omega^*(2) (\partial f)_1\,(\partial f)_2 \frac{F''(\omega)}{|\mathscr{D}(\omega)|^2}\epsilon_k^*. \tag{G 15}$$

The presence of the delta function in (G 14) for Im $F$ allows one to perform the frequency integration in the last term of (G 15), as I shall do later. However, a different approach is useful for the reduction to thermal equilibrium, which I shall discuss in the next section.

### G.1.3. Reduction of the pair correlation function to thermal equilibrium

A physically relevant limit that is also useful for checking signs is the case of thermal equilibrium. Here one has $f = f_{\mathrm{M}}$ and the result $\boldsymbol{\partial} f_{\mathrm{M}} = -(q/T)\boldsymbol{v} f_{\mathrm{M}}$. A simple manipulation of the formula for $\mathscr{D}(\omega)$ then leads to

$$F(\omega) = -(k/\omega)T[\mathscr{D}(\omega) - \mathscr{D}_0], \tag{G 16}$$

where $\mathscr{D}_0 \doteq 1 + k_{\mathrm{D}}^2/k^2$ is the static dielectric function. Upon taking the imaginary part of (G 16) to obtain $F'' = -(k/\omega)T\mathscr{D}''$ (this relation expresses the equilibrium balance between emission and absorption of fluctuations by the discrete particles), one finds that the last term of (G 15) can be written as

$$I \doteq 4\pi\mathrm{i}T\int\frac{\mathrm{d}\omega}{2\pi\omega}\, r_\omega(1) r_\omega^*(2)\left(\frac{1}{\mathscr{D}(\omega)} - \frac{1}{\mathscr{D}^*(\omega)}\right)(\partial f)_1\,(\partial f)_2. \tag{G 17}$$

Note that the integrand of this integral has no singularity at $\omega = 0$. Thus, one may employ the standard trick [which amounts to an application of the Kramers–Kronig relations (Ichimaru 1973)] of deforming the contour to include a semicircular arc around the origin and closing at $\infty$.[67] For the term in $\mathscr{D}^{-1}$, the simplest contour is shown in figure 2. For the term in $(\mathscr{D}^*)^{-1}$, the reflection of that contour into the lower half-plane is appropriate. The contributions from the residues at $\omega = \boldsymbol{k}\cdot\boldsymbol{v}_1$ and $\omega = \boldsymbol{k}\cdot\boldsymbol{v}_2$ cancel the first two terms of (G 15). The contributions from the arc around the origin then lead[68] to the final, well-known result[69]

$$g_{\boldsymbol{k}}^{(\mathrm{eq})}(1,2) = -\frac{q_1 q_2}{T}\frac{4\pi}{k^2\mathscr{D}_0(\boldsymbol{k})}. \tag{G 18}$$

### G.1.4. Final formula for the Balescu–Lenard operator

Formula (G 13) can be used to obtain the Balescu–Lenard collision operator according to (G 4). One has

$$\mathbb{E}^*(\overline{2})g(1,\overline{2}) = r(1)\left(\frac{(\partial f)_1}{\mathscr{D}}F^* - \frac{(\mathscr{D}^*-1)(qf)_1}{\mathscr{D}^*} + \frac{(\partial f)_1(\mathscr{D}^*-1)}{|\mathscr{D}|^2}(F - F^*)\right)\epsilon_k^*. \tag{G 19}$$

---

[67] This trick is commonly used in discussions of the fluctuation–dissipation theorem [see, for example, Birdsall & Langdon (1985, §12–3)], where one evaluates the integral Im $\int \mathrm{d}\omega\,[\omega\mathscr{D}(\omega)]^{-1}$. In that case, there is a contribution from the arc at $\infty$ because $\mathscr{D}(\infty) = 1$. There is no such contribution in (G 17) because of the additional factors of $r_\omega r_\omega^* \sim \omega^{-2}$.

[68] One has $r_0(1) r_0^*(2) = [k^2(\widehat{\boldsymbol{k}}\cdot\boldsymbol{v}_1)(\widehat{\boldsymbol{k}}\cdot\boldsymbol{v}_2)]^{-1}$. The velocities cancel with those arising from $(\partial f_{\mathrm{M}})_1(\partial f_{\mathrm{M}})_2$.

[69] As a sign check, note that $g_k^{(\mathrm{eq})} < 0$ for like-signed particles, signifying repulsion.

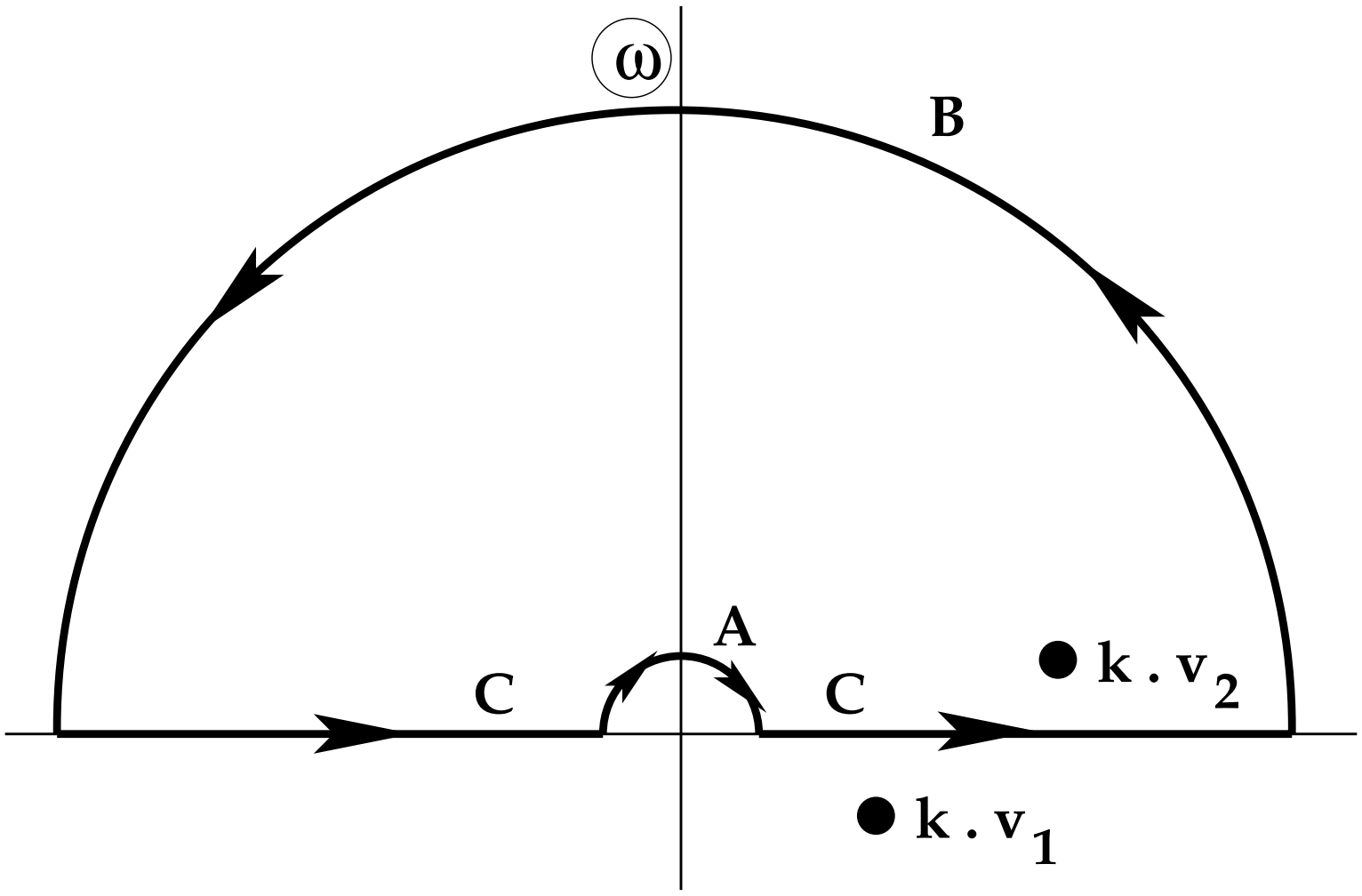

FIGURE 2. Contour of integration used for evaluating the term involving $\mathscr{D}^{-1}$ in (G 17).

Upon rearranging this expression and integrating over $\omega$, one finds

$$\int \frac{\mathrm{d}\omega}{2\pi}\,\mathbb{E}^*(\overline{2})g_{\boldsymbol{k},\omega}(1,\overline{2}) = \int \frac{\mathrm{d}\omega}{2\pi}\,r_\omega(1)\Bigg[(\partial f)_1\Bigg(\underbrace{\frac{F^*(\omega)}{\mathscr{D}(\omega)}}_{(\mathrm{a}_1)} + \frac{\mathscr{D}^*(\omega)}{|\mathscr{D}(\omega)|^2}[\underbrace{F(\omega)}_{0} - \underbrace{F^*(\omega)}_{(\mathrm{a}_2)}]\Bigg)$$
$$- \underbrace{\left(1 - \frac{1}{\mathscr{D}^*(\omega)}\right)(qf)_1}_{(\mathrm{c})} - \underbrace{\frac{(\partial f)_1}{|\mathscr{D}(\omega)|^2}2\mathrm{i}F''(\omega)}_{(\mathrm{d})}\Bigg]\epsilon_k^*. \tag{G 20}$$

Terms $(\mathrm{a}_1)$ and $(\mathrm{a}_2)$ cancel. The underbraced term that vanishes does so by Cauchy's theorem because its sole frequency dependence arises from the product $r_\omega(1)\mathscr{D}^{-1}(\omega)F(\omega)$, which is analytic in the upper half-plane and falls off sufficiently rapidly at $\infty$. Term (c) can be evaluated by residues, and term (d) can be reduced by integrating over the delta function in (G 14). Only the real part survives integration over $\boldsymbol{k}$. With $\boldsymbol{v}_1 \to \boldsymbol{v}$, the final result is $\mathrm{C}_s[f] = \sum_{\overline{s}} \mathrm{C}^{\mathrm{BL}}_{s\overline{s}}[f]$, where

$$\mathrm{C}^{\mathrm{BL}}_{s\overline{s}}[f] \doteq \pi\,(\overline{n}m)_s^{-1}(\overline{n}q^2)_s(\overline{n}q^2)_{\overline{s}}\frac{\partial}{\partial \boldsymbol{v}}\cdot\int \mathrm{d}\overline{\boldsymbol{v}}\int\frac{\mathrm{d}\boldsymbol{k}}{(2\pi)^3}\,\frac{\boldsymbol{\epsilon}_{\boldsymbol{k}}\,\boldsymbol{\epsilon}^*_{\boldsymbol{k}}}{|\mathscr{D}(\boldsymbol{k},\boldsymbol{k}\cdot\boldsymbol{v})|^2}\delta(\boldsymbol{k}\cdot\boldsymbol{v}-\boldsymbol{k}\cdot\overline{\boldsymbol{v}})$$
$$\cdot\left(\frac{1}{m_{\overline{s}}}\frac{\partial}{\partial\overline{\boldsymbol{v}}} - \frac{1}{m_s}\frac{\partial}{\partial\boldsymbol{v}}\right)f_s(\boldsymbol{v})f_{\overline{s}}(\overline{\boldsymbol{v}}), \tag{G 21}$$

which is the Balescu–Lenard operator. Its most important feature is that the natural interacting entities are shielded test particles $[\boldsymbol{\epsilon}_{\boldsymbol{k}}/\mathscr{D}(\boldsymbol{k},\boldsymbol{k}\cdot\boldsymbol{v})]$ rather than bare particles. The first term describes polarization drag, while the second term describes velocity-space diffusion. For future use, note that with my convention for the sign of C polarization drag enters with a plus sign while diffusion enters with a minus sign.

For practical calculations, it is customary to approximate the effect of dielectric shielding by setting $\mathscr{D} \to 1$ and inserting a cutoff at a small wavenumber magnitude of the order of the Debye wavenumber $k_{\mathrm{D}}$. This results in the Landau collision operator

$$\begin{aligned}\mathrm{C}^{\mathrm{L}}_{s\overline{s}}[f] &\doteq 2\pi\,(\overline{n}m)_s^{-1}(\overline{n}q^2)_s(\overline{n}q^2)_{\overline{s}}\ln\Lambda_{s\overline{s}}\\ &\times\frac{\partial}{\partial\boldsymbol{v}}\boldsymbol{\cdot}\int\mathrm{d}\overline{\boldsymbol{v}}\,\boldsymbol{U}(\boldsymbol{v}-\overline{\boldsymbol{v}})\boldsymbol{\cdot}\left(\frac{1}{m_{\overline{s}}}\frac{\partial}{\partial\overline{\boldsymbol{v}}}-\frac{1}{m_s}\frac{\partial}{\partial\boldsymbol{v}}\right)f_s(\boldsymbol{v})f_{\overline{s}}(\overline{\boldsymbol{v}}),\end{aligned} \tag{G 22}$$

where $\Lambda_{s\overline{s}} \doteq \lambda_{\mathrm{D}}/b_{\mathrm{min},s\overline{s}}$ ($b_{\mathrm{min}}$ being classically the impact parameter for 90° scattering[70]) and $\boldsymbol{U}(\boldsymbol{v}) \doteq (\boldsymbol{I}-\widehat{\boldsymbol{v}}\,\widehat{\boldsymbol{v}})/|\boldsymbol{v}|$. Because this operator is bilinear in form, it is often represented as $\mathrm{C}^{\mathrm{L}}[f] = \mathrm{C}^{\mathrm{L}}[f,\,\overline{f}]$. The Balescu–Lenard operator has a more complicated functional dependence, since $\mathscr{D} = \mathscr{D}[f]$. A nonstandard notation is to write

$$\mathrm{C}^{\mathrm{BL}}[f] \equiv \mathrm{C}^{\mathrm{BL}}[f;\,f,\,\overline{f}], \tag{G 23}$$

thus defining an operator $\mathrm{C}[a;\,b,\,\overline{c}]$ that is bilinear in its last two slots; the first slot handles the functional dependence of the dielectric properties. The linearization of that operator, discussed in the next section, is then

$$\Delta\mathrm{C}[f,\Delta f] = \int\mathrm{d}\underline{1}\,\left.\frac{\delta\mathrm{C}[a;\,f,\,\overline{f}]}{\delta a(\underline{1})}\right|_{a=f}\Delta f(\underline{1}) + \mathrm{C}[f;\,f,\,\overline{\Delta f}] + \mathrm{C}[f;\,\Delta f,\,\overline{f}] \tag{G 24}$$

or explicitly

$$\begin{aligned}\Delta\mathrm{C}^{\mathrm{BL}}[f,\Gamma] &= \pi\,(\overline{n}m)_s^{-1}(\overline{n}q^2)_s\frac{\partial}{\partial\boldsymbol{v}}\boldsymbol{\cdot}\sum_{\overline{s}}(\overline{n}q^2)_{\overline{s}}\int\mathrm{d}\overline{\boldsymbol{v}}\int\frac{\mathrm{d}\boldsymbol{k}}{(2\pi)^3}\frac{\boldsymbol{\epsilon}_{\boldsymbol{k}}\,\boldsymbol{\epsilon}^*_{\boldsymbol{k}}}{|\mathscr{D}(\boldsymbol{k},\boldsymbol{k\cdot v})|^2}\delta(\boldsymbol{k\cdot v}-\boldsymbol{k\cdot}\overline{\boldsymbol{v}})\\ &\times\left[-2\,\mathrm{Re}\left(\frac{\Delta\mathscr{D}(\boldsymbol{k},\boldsymbol{k\cdot v})}{\mathscr{D}(\boldsymbol{k},\boldsymbol{k\cdot v})}\right)\left(\frac{1}{m_{\overline{s}}}\frac{\partial}{\partial\overline{\boldsymbol{v}}}-\frac{1}{m_s}\frac{\partial}{\partial\boldsymbol{v}}\right)f_s(\boldsymbol{v})f_{\overline{s}}(\overline{\boldsymbol{v}})\right.\\ &\left.+\left(\frac{1}{m_{\overline{s}}}\frac{\partial}{\partial\overline{\boldsymbol{v}}}-\frac{1}{m_s}\frac{\partial}{\partial\boldsymbol{v}}\right)[f_s(\boldsymbol{v})\Gamma_{\overline{s}}(\overline{\boldsymbol{v}})+\Gamma_s(\boldsymbol{v})f_{\overline{s}}(\overline{\boldsymbol{v}})]\right],\end{aligned} \tag{G 25}$$

where

$$\Delta\mathscr{D}[\Gamma] \doteq \mathbb{E}r\partial\Gamma. \tag{G 26}$$

For perturbations around a Maxwellian equilibrium, the $\Delta\mathscr{D}/\mathscr{D}$ term vanishes. In §G.3, we shall encounter the second-order operator $\mathrm{C}^{\mathrm{BL}}[f;\,\Gamma',\,\overline{\Gamma}''] + \mathrm{C}^{\mathrm{BL}}[f;\,\Gamma'',\,\overline{\Gamma}']$ for certain correlation functions $\Gamma'$ and $\Gamma''$; that operator plays a crucial role in second-order Chapman–Enskog theory and the Burnett transport coefficients.

[70] The wavenumber integration is logarithmically divergent at large $k$ (small impact parameter) because large-angle scattering is not treated correctly. At low temperatures, that can be rectified by asymptotically matching to the Boltzmann operator (Frieman & Book 1963) or, more commonly, by inserting a cutoff at $k_{\max} = b_0^{-1}$, where $b_{0,s\overline{s}} \doteq q_s q_{\overline{s}}/T$. ($b_0$ is the impact parameter for 90° scattering; the distance of closest approach is $2b_0$. In a more precise definition, $T$ is replaced by $\mu\, v_{\mathrm{rel}}^2$, where $\mu$ is the reduced mass and $v_{\mathrm{rel}}$ is the relative velocity.) The Landau operator is accurate only to terms of $O(1)$ relative to $\ln\Lambda$.

When the de Broglie wavelength $\lambda_{\mathrm{B},s\overline{s}} \doteq h_{\mathrm{P}}/(\mu_{s\overline{s}}|\boldsymbol{v}-\overline{\boldsymbol{v}}|)$ is greater than $b_0$ (which is true for sufficiently high temperatures), $b_0$ must be replaced by $\lambda_{\mathrm{B}}$. The argument is summarized, and original references are given, by Krommes (2018*a*).

The two-time hierarchy equations for both two- and three-point correlation functions lead to terms of the form

$$\widehat{\mathrm{D}}[f,\Gamma] \doteq -\boldsymbol{\partial}_1 \cdot \int \frac{\mathrm{d}\boldsymbol{k}}{(2\pi)^3} \int \frac{\mathrm{d}\omega}{2\pi} \mathbb{E}^*_{\boldsymbol{k}}(\overline{2}) \Xi_{\boldsymbol{k},\omega}(1;\overline{1}) \Xi^*_{\boldsymbol{k},\omega}(\overline{2};\overline{2}')$$
$$\times \{\boldsymbol{\epsilon_k} \cdot (q_{\overline{2}'} \boldsymbol{\partial}_{\overline{1}} - q_{\overline{1}} \boldsymbol{\partial}_{\overline{2}'})[\Gamma(\overline{1}) f(\overline{2}') + \Gamma(\overline{2}') f(\overline{1})]$$
$$+ \boldsymbol{\partial}_{\overline{1}} \Gamma(\overline{1}) \cdot [\mathbb{E}^*_{\boldsymbol{k}}(\widehat{1}) g_{\boldsymbol{k}}(\overline{2}', \widehat{1})]^* + \boldsymbol{\partial}_{\overline{2}'} \Gamma(\overline{2}') \cdot [\mathbb{E}^*_{\boldsymbol{k}}(\widehat{2}) g_{\boldsymbol{k}}(\overline{1}, \widehat{2})]\} \qquad (\mathrm{G}\,27)$$

for some $\Gamma$ [e.g., $\Gamma(1) = C^{(1;1)}_{+;\boldsymbol{k}'=\boldsymbol{0}}(\underline{1},\tau;\underline{1}')$ or $\Gamma(1) = C^{(1;2)}_{+;\boldsymbol{k}'=\boldsymbol{0},\boldsymbol{k}''}(\underline{1},\tau;\underline{1}',\underline{1}'')$]. I shall show that $\widehat{\mathrm{D}}$ is the linearized Balescu–Lenard operator: $\widehat{\mathrm{D}}[f,\Gamma] = \Delta\mathrm{C}^{\mathrm{BL}}[f,\Gamma]$ [see formula (G 25)].

To begin, note that the construction $\mathbb{E}^*_{\boldsymbol{k}}(\widehat{2}) g_{\boldsymbol{k}}(\overline{1},\widehat{2})$ (the coefficient of the last $\boldsymbol{\partial}_{\overline{2}'}\Gamma$ term) is the same one [the last line of formula (G 20)] used in the calculation of the Balescu–Lenard operator in the last section; the coefficient of $\boldsymbol{\partial}_{\overline{1}}\Gamma$ can be obtained from that result by replacing $\overline{1} \to \overline{2}'$ and complex conjugating. Cancellations occur between the unshielded part of the $qf$ terms in the last line of (G 27) and the $qf$ terms in the second line of (G 27). Upon omitting most $\boldsymbol{k}$ arguments for brevity, using the notation $\omega_1 \equiv \boldsymbol{k} \cdot \boldsymbol{v}_1$ [e.g., $\mathscr{D}(1) \equiv \mathscr{D}(\boldsymbol{k}, \boldsymbol{k}\cdot\boldsymbol{v}_1)$] and $r_\omega(1) \equiv r_\omega(\boldsymbol{v}_1) \equiv [-\mathrm{i}(\omega - \boldsymbol{k}\cdot\boldsymbol{v}_1 + \mathrm{i}\epsilon)]^{-1}$, and using $-\boldsymbol{\epsilon_k} = \boldsymbol{\epsilon}^*_{\boldsymbol{k}}$, one thus has

$$\widehat{\mathrm{D}}[f,\Gamma] = \partial_1 \cdot \int \frac{\mathrm{d}\boldsymbol{k}}{(2\pi)^3} \int \frac{\mathrm{d}\omega}{2\pi} \Xi_\omega(1;\overline{1}) \frac{1}{\mathscr{D}^*(\omega)} \mathbb{E}^*(\overline{2}) r^*_\omega(\overline{2})$$
$$\times \Bigg[ -(q\Gamma)_{\overline{1}} (\partial f)_{\overline{2}} + (q\Gamma)_{\overline{2}} (\partial f)_{\overline{1}}$$
$$+ (\partial\Gamma_{\overline{1}}) \left( \frac{(qf)_{\overline{2}}}{\mathscr{D}(\overline{2})} + 2\mathrm{i} \int \frac{\mathrm{d}\varpi}{2\pi} r^*_\varpi(\overline{2}) (\partial f)_{\overline{2}} \frac{F''(\varpi)}{|\mathscr{D}(\varpi)|^2} \right)$$
$$- (\partial\Gamma)_{\overline{2}} \left( \frac{(qf)_{\overline{1}}}{\mathscr{D}^*(\overline{1})} - 2\mathrm{i} \int \frac{\mathrm{d}\varpi}{2\pi} r_\varpi(\overline{1}) (\partial f)_{\overline{1}} \frac{F''(\varpi)}{|\mathscr{D}(\varpi)|^2} \right) \Bigg] \epsilon^*_k, \qquad (\mathrm{G}\,28a)$$

$$= \partial_1 \int \frac{\mathrm{d}\boldsymbol{k}}{(2\pi)^3} \int \frac{\mathrm{d}\omega}{2\pi} \Xi_\omega(1;\overline{1}) \frac{1}{\mathscr{D}^*(\omega)}$$
$$\times \Bigg[ -(q\Gamma)_{\overline{1}} [\mathscr{D}^*(\omega) - 1] + (\mathbb{E} r_\omega q\Gamma)^* (\partial f)_{\overline{1}}$$
$$+ (\partial\Gamma)_{\overline{1}} \left( \mathbb{E}^*(\overline{2}) r^*_\omega(\overline{2}) \frac{(qf)_{\overline{2}}}{\mathscr{D}(\overline{2})} + 2\mathrm{i} \int \frac{\mathrm{d}\varpi}{2\pi} [\mathbb{E}(\overline{2}) r_\omega(\overline{2}) r_\varpi(\overline{2}) (\partial f)_{\overline{2}}]^* \frac{F''(\varpi)}{|\mathscr{D}(\varpi)|^2} \right)$$
$$- \Delta\mathscr{D}^*(\omega) \left( \frac{(qf)_{\overline{1}}}{\mathscr{D}^*(\overline{1})} - 2\mathrm{i} \int \frac{\mathrm{d}\varpi}{2\pi} r_\varpi(\overline{1}) (\partial f)_{\overline{1}} \frac{F''(\varpi)}{|\mathscr{D}(\varpi)|^2} \right) \Bigg] \epsilon^*_k, \qquad (\mathrm{G}\,28b)$$

where the $\omega$ in term (d) of (G 20) has been replaced by $\varpi$ and

$$\Delta\mathscr{D}(\omega) \doteq \mathbb{E}(\overline{2}) r_\omega(\overline{2}) (\partial\Gamma)_{\overline{2}}. \qquad (\mathrm{G}\,29)$$

As I noted previously, the $\varpi$ integral can be performed because

$$F''(\varpi) = -\mathrm{i}\pi\, \mathbb{E}_{\overline{s}}(\overline{\boldsymbol{v}}) \delta(\varpi - \boldsymbol{k}\cdot\overline{\boldsymbol{v}}) (qf)_{\overline{s}}(\overline{\boldsymbol{v}}). \qquad (\mathrm{G}\,30)$$

However, I choose to defer that for notational clarity. Upon inserting the formula [cf.

(E 20)] $\Xi_\omega(1;\overline{1})A(\overline{1}) = r_\omega(1)A(1) - r_\omega(1)\partial_1 f \mathscr{D}^{-1}(\omega)\mathbb{E}(\overline{1})r_\omega(\overline{1})A(\overline{1})$, one finds

$$\widehat{\mathrm{D}}[f,\Gamma] = \boldsymbol{\partial}_1 \cdot \int \frac{\mathrm{d}\boldsymbol{k}}{(2\pi)^3}\int\frac{\mathrm{d}\omega}{2\pi}\Bigg\{ r_\omega(1)\Bigg[-(q\Gamma)_1\left(1-\frac{1}{\mathscr{D}^*(\omega)}\right) + \underbrace{(\mathbb{E}r_\omega q\Gamma)^*\frac{(\partial f)_1}{\mathscr{D}^*(\omega)}}_{(\mathrm{a}_1)}\Bigg]$$
$$+ r_\omega(1)\frac{(\partial\Gamma)_1}{\mathscr{D}^*(\omega)}\left(\mathbb{E}^*(\overline{2})r^*_\omega(\overline{2})\frac{(qf)_{\overline{2}}}{\mathscr{D}(\overline{2})} + 2\mathrm{i}\int\frac{\mathrm{d}\varpi}{2\pi}[\mathbb{E}(\overline{2})r_\omega(\overline{2})r_{\overline{\varpi}}(\overline{2})(\partial f)_{\overline{2}}]^*\frac{F''(\varpi)}{|\mathscr{D}(\varpi)|^2}\right)$$
$$+ r_\omega(1)(-)\left(\frac{\Delta\mathscr{D}(\omega)}{\mathscr{D}(\omega)}\right)^*\left(\frac{(qf)_1}{\mathscr{D}^*(1)} - 2\mathrm{i}\int\frac{\mathrm{d}\varpi}{2\pi}r_{\overline{\varpi}}(1)(\partial f)_1\frac{F''(\varpi)}{|\mathscr{D}(\varpi)|^2}\right)$$
$$- r_\omega(1)\frac{(\partial f)_1}{\mathscr{D}(\omega)}\Bigg[-(\mathbb{E}r_\omega q\Gamma)\Bigg(\underbrace{1}_{(\mathrm{b})} - \frac{1}{\mathscr{D}^*(\omega)}\Bigg) + \left(\frac{\mathbb{E}r_\omega q\Gamma}{\mathscr{D}(\omega)}\right)^*[\underbrace{\mathscr{D}(\omega)}_{(\mathrm{a}_2)} - 1]\Bigg]$$
$$- r_\omega(1)\frac{(\partial f)_1}{|\mathscr{D}(\omega)|^2}\Bigg[\Delta\mathscr{D}(\omega)\Bigg(\mathbb{E}^*(\overline{2})r^*_\omega(\overline{2})\frac{(qf)_{\overline{2}}}{\mathscr{D}(\overline{2})}$$
$$+ 2\mathrm{i}\int\frac{\mathrm{d}\varpi}{2\pi}[\mathbb{E}(\overline{2})r_\omega(\overline{2})r_{\overline{\varpi}}(\overline{2})(\partial f)_{\overline{2}}]^*\frac{F''(\varpi)}{|\mathscr{D}(\varpi)|^2}\Bigg)\Bigg]$$
$$- r_\omega(1)\frac{(\partial f)_1}{|\mathscr{D}(\omega)|^2}\Bigg[-\Delta\mathscr{D}^*(\omega)\Bigg(\mathbb{E}(\overline{2})r_\omega(\overline{2})\frac{(qf)_{\overline{2}}}{\mathscr{D}^*(\overline{2})}$$
$$- 2\mathrm{i}\int\frac{\mathrm{d}\varpi}{2\pi}\mathbb{E}r_\omega(\overline{1})r_{\overline{\varpi}}(\overline{1})(\partial f)_{\overline{1}}\frac{F''(\varpi)}{|\mathscr{D}(\varpi)|^2}\Bigg)\Bigg]\Bigg\}\epsilon^*_k. \quad (\mathrm{G\,31}a)$$

Terms $(\mathrm{a}_1)$ and $(\mathrm{a}_2)$ cancel. Term (b) vanishes by analyticity in the upper half of the $\omega$ plane. The sum of the square-bracketed terms (times $\epsilon^*_k$) in the last two lines can be recognized as twice the imaginary part of the term in $\Delta\mathscr{D}^*(\omega)$. I shall now reduce each of the terms. In the algebra, I shall omit the $\partial_1 \cdot (2\pi)^{-3}\int\mathrm{d}\boldsymbol{k}$ for brevity; however, it is important to remember the $\boldsymbol{k}$ integration, which enforces the reality of the final expression.

### G.2.1. The term in $q\Gamma$

The $\omega$ integration can be performed by closing the contour in the lower half-plane, giving rise to $-(q\Gamma)_1\{1 - [\mathscr{D}^*(1)]^{-1}\}\epsilon^*_k$. The first term vanishes by reality or upon interchanging $\boldsymbol{k}$ and $-\boldsymbol{k}$. The remaining expression can be written as

$$\frac{(q\Gamma)_1}{|\mathscr{D}(1)|^2}[1 + \mathbb{E}(\overline{1})r_1(\overline{1})(\partial f)_{\overline{1}}]\epsilon^*_k. \quad (\mathrm{G\,32})$$

The first term again vanishes by reality, and only $\mathrm{Re}\, r_1(\overline{1}) = \pi\delta(\boldsymbol{k}\cdot\boldsymbol{v}_1 - \boldsymbol{k}\cdot\boldsymbol{v}_{\overline{1}})$ survives. One is led to the $+\Gamma\,\overline{\partial f}$ term[71] in the linearized Balescu–Lenard operator (G 25).

[71] When quoting this and similar results, I include an explicit plus or minus sign so that one can quickly check against the expected terms in the linearized Balescu–Lenard operator. Those are

$$-2\,\mathrm{Re}\left(\frac{\Delta\mathscr{D}}{\mathscr{D}}\right)(f\,\overline{\partial f} - \partial f\,\overline{f}) + f\,\overline{\partial\Gamma} - \partial f\,\overline{\Gamma} + \overline{\partial f}\,\Gamma - \overline{f}\,\partial\Gamma.$$

When eyeballing signs, it is helpful to remember that $\epsilon\epsilon^* > 0$ and $\mathrm{Re}\, r > 0$.

### G.2.2. The term in $\partial\Gamma$

The $\omega$ integration can be performed by closing the contour in the lower half-plane, giving rise to

$$-\frac{(\partial\Gamma)_1}{\mathscr{D}^*(1)}\Bigg(\underbrace{\mathbb{E}^*(\bar{2})r_1^*(\bar{2})\frac{(qf)_{\bar{2}}}{\mathscr{D}(\bar{2})}}_{\text{(c)}}+\underbrace{2\mathrm{i}\int\frac{\mathrm{d}\varpi}{2\pi}\,[\mathbb{E}(\bar{2})r_{\boldsymbol{k}\cdot\boldsymbol{v}_1-\mathrm{i}\epsilon}(\bar{2})r_{\varpi}(\bar{2})(\partial f)_{\bar{2}}]^*\frac{F''(\varpi)}{|\mathscr{D}(\varpi)|^2}}_{\text{(d)}}\Bigg)\epsilon_k^*. \tag{G 33}$$

Now invoke the identity

$$r_\omega(\bar{2})r_{\varpi}(\bar{2})=\frac{r_\omega(\bar{2})-r_{\varpi}(\bar{2})}{\mathrm{i}(\omega-\varpi)} \tag{G 34}$$

and perform the $\varpi$ integration using (G 26) to find that term (d) becomes

$$-\frac{(\partial\Gamma)_1}{\mathscr{D}^*(1)}\mathbb{E}(\bar{1})r_1(\bar{1})[\underbrace{\mathscr{D}^*(1)}_{(\mathrm{d}_1)}-\underbrace{\mathscr{D}^*(\bar{1})}_{(\mathrm{d}_2)}]\Big)\frac{(qf)_{\bar{1}}}{|\mathscr{D}(\bar{1})|^2}\epsilon_k^*. \tag{G 35}$$

Term $(\mathrm{d}_2)$ cancels term (c). In term $(\mathrm{d}_1)$, the $\mathscr{D}^*$'s cancel and under the $\boldsymbol{k}$ integration only $\mathrm{Re}\,r_1(\bar{1})$ contributes. One is led to the $-(\partial\Gamma)\overline{f}$ term of (G 25).

### G.2.3. The term in $(\Delta\mathscr{D}/\mathscr{D})^*$

The $\omega$ integration can be performed by closing the contour in the lower half-plane, giving rise to

$$-\left(\frac{\Delta\mathscr{D}(1)}{\mathscr{D}(1)}\right)^*\Bigg(\underbrace{\frac{(qf)_1}{\mathscr{D}^*(1)}}_{\text{(e)}}-\underbrace{2\mathrm{i}\int\frac{\mathrm{d}\varpi}{2\pi}\,r_{\varpi}(1)(\partial f)_1\frac{F''(\varpi)}{|\mathscr{D}(\varpi)|^2}}_{(\mathrm{f}_1)}\Bigg)\epsilon_k^*. \tag{G 36}$$

Term $(\mathrm{f}_1)$ will combine with a later term. (Only its real part contributes.) Term (e) can be manipulated as follows:

$$-\left(\frac{\Delta\mathscr{D}(1)}{\mathscr{D}(1)}\right)^*\frac{(qf)_1}{\mathscr{D}^*(1)}\epsilon_k^*=-\frac{(qf)_1}{|\mathscr{D}(1)|^2}\left[\left(\frac{\Delta\mathscr{D}(1)}{\mathscr{D}(1)}\right)^*\mathscr{D}(1)\right]\epsilon_k^*. \tag{G 37}$$

The term in square brackets can be arranged as

$$\left(\frac{\Delta\mathscr{D}}{\mathscr{D}}\right)^*\mathscr{D}=\left[\left(\frac{\Delta\mathscr{D}}{\mathscr{D}}\right)^*+\frac{\Delta\mathscr{D}}{\mathscr{D}}\right]\mathscr{D}-\Delta\mathscr{D} \tag{G 38a}$$

$$=2\,\mathrm{Re}\left(\frac{\Delta\mathscr{D}}{\mathscr{D}}\right)(\underbrace{1}_{(\mathrm{e}_1)}+\underbrace{\mathbb{E}r\partial f}_{(\mathrm{e}_2)})-\underbrace{\Delta\mathscr{D}}_{(\mathrm{e}_3)}. \tag{G 38b}$$

Term $(\mathrm{e}_1)$ vanishes by reality under the $\boldsymbol{k}$ integration. Only $\mathrm{Re}\,r$ contributes to term $(\mathrm{e}_2)$; one is led to the $-f\,\overline{\partial f}$ term in (G 25). The contribution of term $(\mathrm{e}_3)$ is easily seen to produce the $+f\,\overline{\partial\Gamma}$ term in (G 25).

### G.2.4. The terms in $(\partial f)(\mathbb{E}\Gamma)$

These terms combine to

$$-\int\frac{\mathrm{d}\omega}{2\pi}\,r_\omega(1)\frac{(\partial f)_1}{|\mathscr{D}(\omega)|^2}2\,\mathrm{Re}[(\mathbb{E}r_\omega q\Gamma)\epsilon_k^*]. \tag{G 39}$$

This reduces to the $-(\partial f)\overline{\Gamma}$ term of (G 25).

G.2.5. The remaining terms in $\partial f$

The last two terms in $(\partial f)_1$ reduce to

$$\int \frac{\mathrm{d}\omega}{2\pi} r_\omega(1) \frac{(\partial f)_1}{|\mathscr{D}(\omega)|^2} \times 2\,\mathrm{Re}\left[\Delta\mathscr{D}^*(\omega)\Bigg(\underbrace{\mathbb{E}(\bar{2}) r_\omega(\bar{2}) \frac{(qf)_{\bar{2}}}{\mathscr{D}^*(\bar{2})}}_{(\mathrm{g}_1)} - \underbrace{2\mathrm{i}\int \frac{\mathrm{d}\overline{\omega}}{2\pi}\, \mathbb{E} r_\omega(\bar{1}) r_{\overline{\omega}}(\bar{1})(\partial f)_{\bar{1}} \frac{F''(\overline{\omega})}{|\mathscr{D}(\overline{\omega})|^2}}_{(\mathrm{h})}\Bigg)\epsilon_k^*\right]. \tag{G 40}$$

Thus, only $\mathrm{Re}\, r_\omega(1) = \pi\delta(\omega - \boldsymbol{k}\cdot\boldsymbol{v}_1)$ contributes. Use of the identity (G 34) and the result (G 30) leads to the expression of term (h) as

$$2\mathrm{i}\int \frac{\mathrm{d}\overline{\omega}}{2\pi}\, \mathbb{E} r_w(\bar{1}) r_{\overline{\omega}}(\bar{1})(\partial f)_1 \frac{F''(\overline{\omega})}{|\mathscr{D}(\overline{\omega})|^2} = \mathbb{E}(\overline{\boldsymbol{v}})\left(\frac{\mathscr{D}(\omega) - \mathscr{D}(\boldsymbol{k}\cdot\overline{\boldsymbol{v}})}{\mathrm{i}(\omega - \boldsymbol{k}\cdot\overline{\boldsymbol{v}})}\right)\left(\frac{(qf)_{\overline{\boldsymbol{v}}}}{|\mathscr{D}(\boldsymbol{k}\cdot\overline{\boldsymbol{v}})|^2}\right). \tag{G 41}$$

In the first term in large parentheses, there is no singularity at $\omega = \boldsymbol{k}\cdot\overline{\boldsymbol{v}}$. Thus, the value of that term is inessentially changed by replacing $[\mathrm{i}(\omega - \boldsymbol{k}\cdot\overline{\boldsymbol{v}})]^{-1} \to [\mathrm{i}(\omega - \boldsymbol{k}\cdot\overline{\boldsymbol{v}} + \mathrm{i}\epsilon]^{-1} = -r_\omega(\boldsymbol{k}\cdot\overline{\boldsymbol{v}})$. Expression (G 41) thus becomes

$$\underbrace{\mathscr{D}(\omega)\mathbb{E}(\overline{\boldsymbol{v}}) r_\omega(\overline{\boldsymbol{v}}) \frac{(qf)_{\overline{\boldsymbol{v}}}}{|\mathscr{D}(\boldsymbol{k}\cdot\overline{\boldsymbol{v}})|^2}}_{(\mathrm{f}_2)} - \underbrace{\mathbb{E}(\overline{\boldsymbol{v}}) r_\omega(\overline{\boldsymbol{v}}) \frac{(qf)_{\overline{\boldsymbol{v}}}}{\mathscr{D}^*(\boldsymbol{k}\cdot\overline{\boldsymbol{v}})}}_{(\mathrm{g}_2)}. \tag{G 42}$$

Term $(\mathrm{g}_2)$ cancels term $(\mathrm{g}_1)$. Upon performing the $\omega$ integration, one finds that the entire value of term $(\mathrm{f}_2)$ becomes

$$(\mathrm{f}_2) = \mathrm{Re}\left[\left(\frac{\Delta\mathscr{D}(1)}{\mathscr{D}(1)}\right)\mathbb{E}(\overline{\boldsymbol{v}}) r_1(\overline{\boldsymbol{v}}) \frac{(qf)_{\overline{\boldsymbol{v}}}}{|\mathscr{D}(\boldsymbol{k}\cdot\overline{\boldsymbol{v}})|^2}\epsilon_k^*\right]. \tag{G 43}$$

This can be seen to combine with term $(\mathrm{f}_1)$ in (G 36). Upon using $r_1(\overline{\boldsymbol{v}}) + r_{\overline{\boldsymbol{v}}}(1) = 2\,\mathrm{Re}\, r_1(\overline{\boldsymbol{v}})$, one readily finds that the sum of terms $(\mathrm{f}_1)$ and $(\mathrm{f}_2)$ reduces to the $+\partial f\,\overline{f}$ term in (G 25).

G.2.6. Summary of the reduction of $\widehat{\mathrm{D}}$

This completes the reduction of the operator $\widehat{\mathrm{D}}$. I have shown that $\widehat{\mathrm{D}}[f, \varGamma]$ is indeed the linearized Balescu–Lenard operator $\Delta\mathrm{C}^{\mathrm{BL}}[f, \varGamma]$. As I discuss in the main text, this operator appears in various of the equations for two-time correlation functions.

## G.3. *Nonlinear noise terms and the Balescu–Lenard operator*

I now turn to the analysis of the contributions from the terms in the last five lines of (5.48) for the source term $s_+^{(2;2)}$ in (5.45) for $C_+^{(1;2)}$. Equation (5.46) for $C_+^{(2;2)}$ can be solved in terms of the unperturbed response function as usual, so one must evaluate

$$\mathrm{D}[\varGamma_{\boldsymbol{k}''}] \doteq -\boldsymbol{\partial}_1 \cdot \int_0^\tau \mathrm{d}\overline{\tau} \int \frac{\mathrm{d}\boldsymbol{k}}{(2\pi)^3}\, \mathbf{E}_{\boldsymbol{k}}^*(\underline{2})\varXi_{\boldsymbol{k}}(\underline{1}, \overline{\tau}; \bar{1})\varXi_{\boldsymbol{k}}^*(\underline{2}, \overline{\tau}; \bar{2})\Delta s_{+,\boldsymbol{k};\boldsymbol{k}''}^{(2;2)}(\bar{1}, \bar{2}, \tau - \overline{\tau}; \underline{1}', \underline{1}''), \tag{G 44}$$

where

$$
\begin{aligned}
&\Delta s^{(2;2)}_{+,\boldsymbol{k};\boldsymbol{k}''}(\underline{1},\underline{2},\tau;\underline{1}',\underline{1}'') \doteq \\
&\quad -(q_2\boldsymbol{\partial}_1 - q_1\boldsymbol{\partial}_2)\boldsymbol{\cdot}\boldsymbol{\epsilon}_{\boldsymbol{k}}[C^{(1;1)}_{+}(\underline{1},\tau;\underline{1}')C^{(1;1)}_{+;\boldsymbol{k}''}(\underline{2},\tau;\underline{1}'') + C^{(1;1)}_{+;\boldsymbol{k}''}(\underline{1},\tau;\underline{1}'')C^{(1;1)}_{+}(\underline{2},\tau;\underline{1}')] \\
&\quad -[\boldsymbol{\partial}_1 C^{(1;1)}_{+}(\underline{1},\tau;\underline{1}')\boldsymbol{\cdot}\mathbb{E}_{\boldsymbol{k}}(\overline{\underline{1}})C^{(2;1)}_{+,\boldsymbol{k};\boldsymbol{k}''}(\overline{\underline{1}},\underline{2},\tau;\underline{1}'') \\
&\qquad + \boldsymbol{\partial}_1 C^{(1;1)}_{+;\boldsymbol{k}''}(\underline{1},\tau;\underline{1}'')\boldsymbol{\cdot}\mathbb{E}_{\boldsymbol{k}}(\overline{\underline{1}})C^{(2;1)}_{+,\boldsymbol{k}}(\overline{\underline{1}},\underline{2},\tau;\underline{1}')] \\
&\quad -[\boldsymbol{\partial}_2 C^{(1;1)}_{+}(\underline{2},\tau;\underline{1}')\boldsymbol{\cdot}\mathbb{E}^{*}_{\boldsymbol{k}}(\overline{\underline{2}})C^{(2;1)}_{+,\boldsymbol{k};\boldsymbol{k}''}(\underline{1},\overline{\underline{2}},\tau;\underline{1}'') \\
&\qquad + \boldsymbol{\partial}_2 C^{(1;1)}_{+;\boldsymbol{k}''}(\underline{2},\tau;\underline{1}'')\boldsymbol{\cdot}\mathbb{E}^{*}_{\boldsymbol{k}}(\overline{\underline{2}})C^{(2;1)}_{+,\boldsymbol{k}}(\underline{1},\overline{\underline{2}},\tau;\underline{1}')].
\end{aligned}
\tag{G 45}
$$

Consider the evaluation of the $\overline{\tau}$ integral in (G 44), and notice that the $\tau$ dependence of the first line of (G 45) differs in character from that of the subsequent four lines. The $C^{(1;1)}_{+}$'s in the first line vary on the collisional timescale, as they are all evaluated at $\boldsymbol{k}'$ or $\boldsymbol{k}''$, both of which are small. That is true as well of the $\boldsymbol{\partial}C^{(1;1)}_{+}$'s in the last four lines. However, the $C^{(2;1)}_{+}$ factors in those lines are evaluated at $\boldsymbol{k}$, which appears under an integral and is a characteristic Debye-scale wavenumber. Thus, they vary on the microscopic autocorrelation timescale, which is the same as that on which the $\varXi$'s vary. The consequence is that while a Markovian approximation is immediately valid for the contribution of the first line, the remaining terms require further processing.

Before proceeding with the details, it is helpful to gain an intuitive understanding of what to expect. For a weakly coupled plasma, collisional effects on the one-body distribution function $f$ are captured by the nonlinear Balescu–Lenard operator $\mathrm{C}^{\mathrm{BL}}[f;f,\overline{f}]$. Upon expanding $f = f_0 + \epsilon f_1 + \epsilon^2 f_2$, one is led at first order to the linearized operator $\widehat{\mathrm{C}}f_1 \doteq \mathrm{C}[f_0;f_0,\overline{f}_1] + \mathrm{C}[f_0;f_1,\overline{f}_0]$; the first-order variation with respect to the first argument (which describes the functional dependence of the dielectric function on the distribution function) vanishes. At second order, one finds

$$
\begin{aligned}
\mathrm{C}[f;f,\overline{f}]_2 &= \widehat{\mathrm{C}}f_2 + \mathrm{C}[f_0;f_1,\overline{f}_1] \\
&\quad + \int \mathrm{d}\boldsymbol{v}' \left( \left.\frac{\delta \mathrm{C}[a;f_0,\overline{f}_1]}{\delta a(\boldsymbol{v}')}\right|_{a=f_0} + \left.\frac{\delta \mathrm{C}[a;f_1,\overline{f}_0]}{\delta a(\boldsymbol{v}')}\right|_{a=f_0} \right) f_1(\boldsymbol{v}').
\end{aligned}
\tag{G 46}
$$

(The second variation with respect to the first argument vanishes.) These terms are mirrored in the theory of two-time correlations. The term analogous to $\widehat{\mathrm{C}}f_2$ was derived in §G.2. The term analogous to $\mathrm{C}[f_0;f_1,\overline{f}_1]$ will be shown in the next section to stem from the $C^{(1;1)}C^{(1;1)}$ terms [the second line of (G 45)]. This leaves the last four lines of (G 45) to capture the effect described by the last line of (G 46), which arises from the first-order variation of $\widehat{\mathrm{C}}[f_0]$ (i.e., fluctuations in the dielectric shielding).

### G.3.1. The $C^{(1;1)}_{+}C^{(1;1)}_{+}$ contribution

The initial manipulations of the $C^{(1;1)}_{+}C^{(1;1)}_{+}$ terms proceed as in the derivation of the original Balescu–Lenard operator discussed in §G.1. Because the second line of (G 45) is identical in form to the driving term for the pair correlation function $g$ [see (G 3)], it is easy to see that the contribution from the $\varGamma\varGamma$ terms to $D[\varGamma]$ is

$$
\mathrm{C}^{\mathrm{BL}}[f;\,C^{(1;1)}_{+}(\tau;\underline{1}'),\,\overline{C}^{(1;1)}_{+}(\tau;\underline{1}'')] + (1' \Leftrightarrow 1''). \tag{G 47}
$$

The crucial role of this term in the theory of triplet correlations will be analysed in §6.3.2.

G.3.2. The $\mathbb{E}C_{+}^{(2;1)}$ terms

I now briefly discuss the last four lines of (G 45). A term like $C_{+,\boldsymbol{k}}^{(2;1)}(\underline{\overline{1}}, \underline{2}, \tau; 1')$ is evaluated by time-convolving $[\varXi_{\boldsymbol{k}}(\underline{\overline{1}}, \tau)\varXi_{\boldsymbol{k}}^{*}(\underline{2}, \tau)]$ with a certain, slowly varying source $S$. That result must be convolved with another pair of $\varXi$'s according to (G 44). The resulting contribution to $\mathrm{D}[\varGamma_{\boldsymbol{k}''}]$ drives $C_{+}^{(1;2)}$ according to (5.45). Ultimately, contributions to transport coefficients are determined from $\int_0^\infty \mathrm{d}\overline{\tau}\, C_{+}^{(1;2)}(\overline{\tau})$. Thus, with $\varXi_2 \doteq \varXi\varXi$, one must consider (schematically)

$$I \doteq \int_0^\infty \mathrm{d}\overline{\tau} \int_0^{\overline{\tau}} \mathrm{d}\widehat{\tau}\, \varXi_2(\widehat{\tau}) \int_0^{\overline{\tau}-\widehat{\tau}} \mathrm{d}\tau'\, \varXi_2(\tau') S(\overline{\tau} - \widehat{\tau} - \tau'). \tag{G 48}$$

Upon interchanging the order of integration of the first and second integrals, then of the second and third ones, one finds that

$$I = \int_0^\infty \mathrm{d}\widehat{\tau}\, \varXi_2(\widehat{\tau}) \int_0^\infty \mathrm{d}\tau'\, \varXi_2(\tau') \int_0^\infty \mathrm{d}\overline{\tau}\, S(\overline{\tau}). \tag{G 49}$$

Each $\varXi_2$ integral can be expressed in frequency space as $(2\pi)^{-1} \int_{-\infty}^{\infty} \mathrm{d}\omega\, \varXi(\omega)\varXi^{*}(\omega)$. The second of each pair of $\varXi$'s is operated upon by $\mathbb{E}$, which allows simplifications according to (G 7).

Thus, evaluation of the contributions of the $\mathbb{E}C_{+}^{(2;1)}$ terms is feasible in principle, as it involves nothing more than multiple instances of the manipulations done in §G.2 to obtain the linearized Balescu–Lenard operator. The algebra is tedious, however, and I shall not pursue it here. Note that effects related to fluctuations in the dielectric properties of the plasma disappear when dielectric shielding is taken into account by a cutoff at the Debye wavenumber $k_{\mathrm{D}}$ and if fluctuations in $k_{\mathrm{D}}$ are subsequently ignored. I shall return to this point in §6.4.

## Appendix H. One-sided correlations and nonlinear noise

Equation (5.33) and similar equations contain (i) a left-hand side that includes collisional damping, (ii) an initial condition, and (iii) a source term. These are special cases of Dyson equations, the general form of which is

$$(\partial_t + \boldsymbol{v} \cdot \boldsymbol{\nabla})\mathrm{C}(\boldsymbol{q}, t, \boldsymbol{q}', t') + \int_0^t \mathrm{d}\overline{t}\, \mathrm{d}\overline{\boldsymbol{q}}\, \varSigma(\boldsymbol{q}, t; \overline{\boldsymbol{q}}, \overline{t})\mathrm{C}(\overline{\boldsymbol{q}}, \overline{t}, q', t')$$
$$= \int_0^{t'} \mathrm{d}\overline{t}\, \mathrm{d}\overline{\boldsymbol{q}}\, F(\boldsymbol{q}, t, \overline{\boldsymbol{q}}, \overline{t}) R(\boldsymbol{q}', t'; \overline{\boldsymbol{q}}, \overline{t}), \tag{H 1}$$

where in the choice of signs and symbols for $\varSigma$ and $F$ I have adopted the conventions used in Krommes (2002). For example, Rose (1979) has proposed his PDIA, which has this form.

I shall discuss some elementary implications of equations with the Dyson form. Note that in (H 1) there is no ordering of $t$ and $t'$; $C(t, t')$ is two-sided, not one-sided. It is useful to understand the relationship between this equation and the one-sided equations discussed earlier in the paper. To develop intuition, assume that $\varSigma$ and $F$ are both positive.[72] Then if $F$ were omitted, $C$ would decay to zero. That is incompatible with the fact that $C(t, t)$ has a nonzero value in thermal equilibrium (or in steady-state turbulence). Therefore, an $F$ term must always be present. Another way of saying this is that $F$ is necessary in order that conservation laws associated with the nonlinearity are satisfied.

[72]For turbulence, that is indeed true for $F$; the nonlinear contribution to $\varSigma$ is not always positive as a function of wavenumber, but it is typically so.

### H.1. *Equations for one-sided functions*

In order to make explicit the nonzero value of $C(t,t)$, it is useful to introduce functions that are one-sided in time. Define

$$C_+(t,t') \doteq H(t-t')C(t,t'), \quad C_-(t,t') \doteq H(t'-t)C(t,t'), \tag{H 2}$$

such that $C(t,t') = C_+(t,t') + C_-(t,t')$. Consider an equation of the general form

$$\partial_t C(t,t') + \nu C = s(t,t'). \tag{H 3}$$

This has the same form as the general Dyson equation (with nondissipative terms ignored), with a positive damping coefficient on the left-hand side of the equation for $C$ (not $C_+$ or $C_-$). The equation for $C_+$ is readily found by time-differentiating its definition:

$$\partial_t C_+(t,t') = \delta(t-t')C(t,t) + H(t-t')\partial_t C \tag{H 4a}$$

$$= \delta(t-t')C(t,t) + H(t-t')[-\nu C(t,t') + s(t,t')]; \tag{H 4b}$$

thus,

$$\partial_t C_+(t,t') + \nu C_+ = \delta(t-t')C(t,t) + s_+(t,t'). \tag{H 5}$$

This has the same form as (5.33). Green's function for the operator on the left-hand side is

$$R(t;t') = H(t-t')\Xi(t,t'), \quad \Xi(t,t') \doteq \mathrm{e}^{-\nu(t-t')}. \tag{H 6}$$

Thus, the solution of (H 5) is, with $C_0 \doteq C(t',t')$

$$C_+(t,t') = \int_{-\infty}^{\infty} \mathrm{d}\bar{t}\, R(t;\bar{t})[\delta(\bar{t}-t')C_0 + s_+(\bar{t},t')] \tag{H 7a}$$

$$= H(t-t')\left(\Xi(t;t')C_0 + \int_{t'}^{t} \mathrm{d}\bar{t}\, \Xi(t,\bar{t})s_+(\bar{t},t')\right). \tag{H 7b}$$

Notably, the contribution from the initial condition decays as $t \to \infty$.

In a similar fashion, one finds

$$\partial_t C_-(t,t') + \nu C_- = -\delta(t-t')C_0 + s_-(t,t'). \tag{H 8}$$

(Note the minus sign in the initial-condition term.) With the ansatz

$$C_-(\tau) = H(-\tau)\mathrm{e}^{-\nu\tau}D(\tau), \tag{H 9}$$

one has

$$\partial_\tau C_- = -\nu C_- - \delta(\tau)D(0) + H(-\tau)\mathrm{e}^{-\nu\tau}\partial_\tau D \tag{H 10a}$$

$$= -\nu C_- - \delta(\tau)C_0 + H(-\tau)\widehat{s}(\tau), \tag{H 10b}$$

where $s_-(\tau) = H(-\tau)\widehat{s}(\tau)$ and (H 10$b$) is a rewrite of (H 8). For consistency, one must therefore satisfy $D(0) = C_0$ and

$$D(\tau) = C_0 + \int_0^\tau \mathrm{d}\overline{\tau}\, \mathrm{e}^{\nu\overline{\tau}}\widehat{s}(\overline{\tau}). \tag{H 11}$$

Now if $C(t,t')$ is to describe a physical correlation function, it should decay as $\tau \to -\infty$. However, the contribution to (H 9) from the initial-condition term in (H 11) does not do that; rather, it explodes. Therefore, it is clear that $s_-$ cannot vanish and must be related in a particular way to $C_0$.

### H.2. *Example: The classical Langevin equation*

To see how this works for a simple example, consider the classical Langevin equation (Wang & Uhlenbeck 1945)

$$\dot{v} + \nu v = \delta a(t), \tag{H 12}$$

where $\delta a$ is centred Gaussian white noise such that $\langle \delta a(t)\delta a(t')\rangle = 2D_v \delta(t-t')$. The two-time correlation function obeys

$$\partial_t C(t,t') + \nu C = \langle \delta a(t)\delta v(t')\rangle \tag{H 13a}$$

$$= \int_{-\infty}^{t'} \mathrm{d}\bar{t}\, \langle \delta a(t) \mathrm{e}^{-\nu(t-\bar{t})}\delta a(\bar{t})\rangle \tag{H 13b}$$

$$= 2D_v \int_{-\infty}^{t'} \mathrm{d}\bar{t}\, \mathrm{e}^{-\nu(t'-\bar{t})}\delta(t-\bar{t}) \tag{H 13c}$$

$$= \begin{cases} 0 & \text{for } t > t', \\ D_v & \text{for } t = t', \\ 2D_v \mathrm{e}^{-\nu(t'-t)} & \text{for } t < t'. \end{cases} \tag{H 13d}$$

One therefore concludes that for this model

$$s_+(t,t') = 0, \quad s_-(t,t') = 2D_v H(t'-t)\Xi(t',t). \tag{H 14}$$

The integral in (H 11) can now be done; it is $2D_v \int_0^\tau \mathrm{d}\overline{\tau}\, \mathrm{e}^{2\nu\overline{\tau}} = (D_v/\nu)\left(\mathrm{e}^{2\nu\tau} - 1\right)$. From (H 9), the final solution is therefore

$$C_-(\tau) = H(-\tau)\left[\mathrm{e}^{-\nu\tau}\left(C_0 - \frac{D_v}{\nu}\right) + \frac{D_v}{\nu}\mathrm{e}^{\nu\tau}\right]. \tag{H 15}$$

To ensure convergent behaviour as $\tau \to -\infty$, one must require $C_0 = D_v/\nu$; then $C_-(\tau) = H(-\tau)\mathrm{e}^{\nu\tau}C_0$ and the two-sided solution is

$$C(\tau) = \mathrm{e}^{-\nu|\tau|}C_0. \tag{H 16}$$

This is a well-known result for the classical Langevin problem.

This example shows that the two-time source and the one-time 'initial' value are not independent for a physical correlation function; one concludes that the noise term $F$ cannot be neglected. This establishes the conceptual connection between the treatment in the main text, which used the multiple-time BBGKY hierarchy and one-sided functions, and the Dyson equations used by Rose, which do not involve one-sided correlation functions explicitly. When confronted with the complicated hierarchy equations, one might have naively attempted to ignore the initial-condition terms. As hopefully clarified by the above example, that is incorrect, as it would deal inconsistently with the nonlinear noise. Indeed, one sees explicitly in the calculations of §6.3.2 that a contribution from an initial condition plays a key role in establishing the correct correspondence between the two-time formalism and Chapman–Enskog theory.

### H.3. *The $s_+^{(1;1)}$ term*

Now it is possible to discuss the significance of the $s_+^{(1;1)}$ term on the right-hand side of (5.33) or (5.35). Its size is set by $C^{(0,3)} = O(\epsilon_\mathrm{p}^2)$. It is thus nominally of the same order as the $\widehat{\mathrm{C}}C_+^{(1)}$ term on the left-hand side of (5.33); it describes a contribution to nonlinear noise. However, the correlation time of $s_+^{(1;1)}$, being set by $\Xi(\tau)$, is the short Debye-cloud autocorrelation time $\omega_\mathrm{p}^{-1}$; therefore, the $\int_0^\infty \mathrm{d}\overline{\tau}\, s_+^{(1;1)}$ term in (5.35) is negligible, being $O(\epsilon_\mathrm{p}^2)$, relative to the initial-condition term $C^{(0,2)} = O(\epsilon_\mathrm{p})$. The small $s_+^{(1;1)}$ term is analogous to the $s_+$ term in the above Langevin calculation. There the term vanishes altogether as a consequence of the assumption of zero autocorrelation time for the random acceleration. The $s_+^{(1;1)}$ term would play a role for short times, but it can be neglected in a one-sided equation coarse-grained on the kinetic timescale.

## Appendix I. Notation

The following list of notation merges symbols used in both Part I and Part II.

### I.1. *Basic variables and physics symbols*

**– a –**

$\boldsymbol{A}(\boldsymbol{v})$ — The fundamental vector in the $\mu$-space description: $\boldsymbol{A} \doteq (1,\ \boldsymbol{P}',\ K')^\mathrm{T}$.

$\widetilde{\boldsymbol{A}}(\boldsymbol{x},t)$ — The fundamental random variables in the $\Gamma$-space description: $\widetilde{\boldsymbol{A}} \doteq (\widetilde{N},\ \widetilde{\boldsymbol{P}},\ \widetilde{E})^\mathrm{T}$.

$\boldsymbol{A}_\Delta$ — Momentum and energy exchange terms

$\boldsymbol{a}(\boldsymbol{x},t)$ — The fundamental mean densities: $\boldsymbol{a} \doteq \langle \widetilde{\boldsymbol{A}} \rangle$.

$\alpha$ — Thermal expansion coefficient: $\alpha \doteq -n^{-1}(\partial n/\partial T)_p$.

$\alpha(\boldsymbol{v})$ — The normalized kinetic energy: $\alpha(\boldsymbol{v}) \doteq \frac{1}{2}mv^2/T$.

**– b –**

$\boldsymbol{B}$ — Variables conjugate to $\widetilde{\boldsymbol{A}}$

$\boldsymbol{B}^\mathrm{ext}$ — External magnetic field

$B$ — $B \doteq |\boldsymbol{B}^\mathrm{ext}|$.

$\boldsymbol{b}$ — Auxiliary vector approximately equal to $\Delta\boldsymbol{B}$, the deviation of the conjugate variables from their values at the reference point; see (2.62).

$\widehat{\boldsymbol{b}}$ — Unit vector in the direction of the magnetic field

$b_0$ — Impact parameter for 90° scattering: $b_0 \doteq q_1 q_2/T$.

$\beta$ — Inverse temperature: $\beta \doteq T^{-1}$.

$\boldsymbol{\beta}(\boldsymbol{v})$ — Subtracted kinetic energy flux: $\boldsymbol{\beta}(\boldsymbol{v}) \doteq [a(\boldsymbol{v}) - \frac{5}{2}]\boldsymbol{v} = \gamma(\boldsymbol{v})\boldsymbol{v}$.

**– c –**

$C(t,t')$ — Correlation function

$\mathrm{C}[f]$ — Nonlinear collision operator

$\widehat{\mathrm{C}}$ — Linearized collision operator: $\partial_t|\chi\rangle + \cdots = -\widehat{\mathrm{C}}|\chi\rangle$, where $f \doteq (1+\chi)f_\mathrm{M}$.

$\mathrm{C}[f,\overline{f}]$ — The bilinear Landau operator. The first slot refers to test particles, the second to field particles.

$\mathrm{C}^\mathrm{BL}[f;a,\overline{b}]$ — The nonlinear Balescu–Lenard operator. The first slot describes functional dependence of the dielectric function, the second slot operates on test particles, and the third slot operates on field particles.

$c_\mathrm{p}$, $c_\mathrm{v}$ — Specific heats at constant pressure and volume

$c_\mathrm{s}$ — Sound speed: $c_\mathrm{s} \doteq (ZT_e/m_i)^{1/2}$.

$\chi_{\boldsymbol{k},\omega}$ — Dielectric susceptibility: $\mathscr{D} = 1 + \chi$.

$\chi(\mu)$ — Correction to the lowest-order one-particle distribution function: $f = (1+\chi)f_0$.

**– d –**

$D$ — Diffusion coefficient

$\mathscr{D}(\boldsymbol{k},\omega)$ — Dielectric function

$\mathscr{D}_0(\boldsymbol{k})$ — Static dielectric function: $\mathscr{D}_0(\boldsymbol{k}) \doteq 1 + k_\mathrm{D}^2/k^2$.

$\mathscr{D}_\perp$ — Dielectric constant for strongly magnetised plasma: $\mathscr{D}_\perp \doteq \omega_{\mathrm{p}i}^2/\omega_{\mathrm{c}i}^2$.

$\Delta$ — Ordering parameter: $\Delta \doteq k_{\parallel}\lambda_{\text{mfp}}$.

$\Delta\boldsymbol{B}_{\overline{s}s}(\overline{\boldsymbol{r}}, \boldsymbol{r}, t)$ — $\boldsymbol{B}_{\overline{s}}(\overline{\boldsymbol{r}}, t) - \boldsymbol{B}_s(\boldsymbol{r}, t)$

$\Delta\boldsymbol{B}$ — $\Delta\boldsymbol{B}(\overline{\mu}) \equiv \Delta\boldsymbol{B}_{\overline{s}}(\overline{\boldsymbol{r}}, t) \equiv \Delta\boldsymbol{B}_{\overline{s}\,\overline{s}}(\overline{\boldsymbol{r}}, \boldsymbol{r}, t)$.

$\Delta T$ — Interspecies temperature difference: $\Delta T \doteq T_e - T_i$.

$\Delta\boldsymbol{u}$ — Interspecies flow-velocity difference: $\Delta\boldsymbol{u} \doteq \boldsymbol{u}_e - \boldsymbol{u}_i$.

$d$ — Number of spatial dimensions

$\delta$ — Ordering parameter: $\delta \doteq k_{\perp}\rho$.

$\delta(x)$ — Dirac delta function

$\delta_{ij}$ — Kronecker delta function

**– e –**

$E$ — Energy

$\widetilde{E}(\boldsymbol{x}, t)$ — Microscopic energy density

$\boldsymbol{E}(\boldsymbol{x}, t)$ — Electric field

$\mathbb{E}$ — Electric-field operator

$e$ — Electronic charge (positive)

$e(\boldsymbol{x}, t)$ — Macroscopic energy density. For an ideal gas, $e = \frac{3}{2}nT$.

$\epsilon$ — Ordering parameter; positive infinitesimal

$\epsilon_{\text{p}}$ — Plasma parameter: $\epsilon_{\text{p}} \doteq 1/n\lambda_{\text{D}}^3$.

$\boldsymbol{\epsilon}(\boldsymbol{x}_1, \boldsymbol{x}_2)$ — Interparticle electric field for unit charges: $\boldsymbol{\epsilon}_{12} \doteq -\boldsymbol{\nabla}_1|\boldsymbol{x}_1 - \boldsymbol{x}_2|^{-1}$; $\boldsymbol{\epsilon}(\boldsymbol{x}) = (2\pi)^{-3}\int \mathrm{d}\boldsymbol{k}\, \mathrm{e}^{\mathrm{i}\boldsymbol{k}\cdot\boldsymbol{x}}\boldsymbol{\epsilon}_{\boldsymbol{k}}$.

$\boldsymbol{\epsilon}_{\boldsymbol{k}}$ — $\boldsymbol{\epsilon}_{\boldsymbol{k}} \doteq -4\pi\,\mathrm{i}\boldsymbol{k}/k^2$.

**– f –**

$F(\Gamma, t)$ — Liouville distribution: $F \equiv P_{\mathcal{N}}$.

$F_B(\Gamma; t)$ — Local-equilibrium distribution

$F_0(\Gamma; \boldsymbol{r}, t)$ — Reference distribution

$f(\mu, t)$ — One-particle distribution function: $f \doteq \langle \widetilde{f} \rangle$.

$\widetilde{f}(\mu, t)$ — Klimontovich phase-space distribution function: $\widetilde{f} \doteq \overline{n}^{-1}\sum_i \delta(\mu - \widetilde{\mu}(t))$.

$f_{\text{M}}$ — Maxwellian distribution function

$f_{\text{lM}}$ — Local Maxwellian

$\boldsymbol{f}$ — Random force

$\phi$ — Electrostatic potential

**– g –**

$\Gamma$ — Phase-space coordinates of the many-body system

$g(\underline{1}, \underline{2}, t)$ — Pair correlation function

$\gamma(\boldsymbol{v})$ — $\alpha(\boldsymbol{v}) - \frac{5}{2}$

**– h –**

$H(\tau)$ — Heaviside unit step function

$\mathrm{He}_n$ — Probabilistic Hermite polynomial

$h_{\text{P}}$ — Planck's constant

$h(\boldsymbol{x}, t)$ — Enthalpy density. For an ideal gas, $h = \frac{5}{2}nT$.

$h(\underline{1}, \underline{2}, \underline{3}, t)$ — Triplet correlation function

$\boldsymbol{\eta}$ — Matrix of transport coefficients

$\eta$, $\widehat{\eta}$ — Source functions

**– j –**

$\boldsymbol{J}$ — Generic flux or current

$\widehat{\boldsymbol{J}}$ — Subtracted flux: $\widehat{\boldsymbol{J}} \doteq \mathrm{Q}\boldsymbol{J}$.

**– k –**

$K(\boldsymbol{v})$ — Kinetic energy: $K \doteq \frac{1}{2}mv^2$.

$\boldsymbol{k}$ — Wavevector

$k_{\text{D}}$ — Debye wavenumber: $k_{\text{D}}^2 \doteq \sum_s k_{\text{D}s}^2$, where $k_{\text{D}s} \doteq (4\pi nq^2/T)_s^{1/2}$.

$\kappa$ — Thermal conductivity

$\kappa_T$ — Isothermal compressibility: $\kappa_T \doteq n^{-1}(\partial n/\partial p)_T$.

**– l –**

$L$ — Box size or gradient scale length

$\mathrm{L}_1$ — Landau operator for particle 1

$\vec{\mathrm{L}}^2$ — Square of the angular momentum operator

$\mathscr{L}$ — Liouville operator

$\Lambda$ — Argument of the Coulomb logarithm: $\Lambda \doteq \lambda_{\text{D}}/b_0$ in the classical limit.

$\lambda_{\text{mfp}}$ — Mean free path: $\lambda_{\text{mfp}} \doteq v_{\text{t}}/\nu$.

$\lambda_{\text{B}}$ — de Broglie wavelength: $\lambda_{\text{B},s\overline{s}} \doteq \eta_{\text{P}}/(\mu_{s\overline{s}}|\boldsymbol{v} - \overline{\boldsymbol{v}}|)$.

$\lambda_{\text{D}}$ — Debye length: $\lambda_{\text{D}} \doteq k_{\text{D}}^{-1}$.

**– m –**

$\widehat{\mathrm{M}}$ — Magnetic-field operator: $\widehat{\mathrm{M}} \doteq -\mathrm{i}\omega_{\text{c}}\partial/\partial\zeta$.

$M_{ss'}$ — Total mass: $M_{ss'} \doteq m_s + m_{s'}$.

$\mathscr{M}$ — Covariance matrix in the reference ensemble: $\mathscr{M} \doteq \langle \boldsymbol{A}' \boldsymbol{A}'^{\mathrm{T}} \rangle_0$.

$m$ — Mass

$\mathbf{m}$ — Fourth-rank viscosity tensor

$\mu$ — Viscosity; generic observer arguments $\mu \doteq \{\boldsymbol{x}, \boldsymbol{v}, s\}$ or $\{\boldsymbol{x}, s\}$; field index; mass ratio $m_e/m_i$

$\mu_{ss'}$ — Reduced mass: $\mu_{ss'}^{-1} = m_s^{-1} + m_{s'}^{-1}$.

**– n –**

$\widetilde{N}(\boldsymbol{x}, t)$ — Microscopic number density

$\mathscr{N}$, $N$ — Total number of particles

$n(\boldsymbol{x}, t)$ — Macroscopic (averaged) number density

$\overline{n}$ — Mean density: $\overline{n} \doteq \mathscr{N}/\mathscr{V}$.

$\nu$ — Collision frequency

**– ξ –**

$\Xi(\tau)$ — Response function for the linearized Vlasov equation, sans $H(\tau)$: $R(\tau) = H(\tau)\Xi(\tau)$.

**– ω –**

$\boldsymbol{\Omega}$ — Frequency matrix; vorticity tensor: $\boldsymbol{\Omega} \doteq \frac{1}{2}[(\boldsymbol{\nabla u})^{\mathrm{T}} - (\boldsymbol{\nabla u})]$.

$\omega$ — Fourier transform variable conjugate to $t$

$\omega_{\mathrm{p}}$ — Plasma frequency: $\omega_{\mathrm{p}}^2 = \sum_s \omega_{\mathrm{p}s}^2$, where $\omega_{\mathrm{p}s} \doteq (4\pi n q^2/m)_s^{1/2}$.

**– p –**

$\mathrm{P}$ — Projection operator

$P_m^l$ — Associated Legendre function of the first kind

$\boldsymbol{P}(\boldsymbol{v})$ — $\mu$-space kinetic momentum flux: $\boldsymbol{P} \doteq m\boldsymbol{v}$.

$\widetilde{\boldsymbol{P}}(\boldsymbol{x}, t)$ — Microscopic momentum density

$\boldsymbol{p}(\boldsymbol{x}, t)$ — Macroscopic momentum density

$p(\boldsymbol{x}, t)$ — Macroscopic pressure. For an ideal gas, $p = nT$.

$\boldsymbol{\pi}$ — Pressureless part of the stress tensor: $\boldsymbol{\pi} \doteq \boldsymbol{\tau} - p\mathbf{I}$.

**– q –**

$\mathrm{Q}$ — Orthogonal projection operator: $\mathrm{Q} \doteq 1 - \mathrm{P}$.

$Q$ — Heat generation

$q$ — Signed charge

$\boldsymbol{q}$ — Heat-flow vector

**– r –**

$R_{\{0,1,2\}}$ — $\mathrm{R}_0(\tau) \doteq \mathrm{e}^{-\mathrm{i}\mathscr{L}\tau}$, $\mathrm{R}_1(\tau) \doteq \mathrm{e}^{-\mathrm{Qi}\mathscr{L}\tau}$, $\mathrm{R}_2(\tau) \doteq \mathrm{e}^{-\mathrm{Qi}\mathscr{L}\mathrm{Q}\tau}$.

$R(t; t')$ — Causal infinitesimal response function

$\boldsymbol{R}$ — Friction force

$r(t; t')$ — Single-particle causal response function

$\boldsymbol{r}$ — Reference position at which the fluid equations are evaluated

$\rho(\boldsymbol{r}, t)$ — Charge density. Poisson's equation is $-\nabla^2\phi = 4\pi\rho$.

$\rho_s$ — Gyroradius of species $s$.

$\rho_{\mathrm{s}}$ — Sound radius: $\rho_{\mathrm{s}} \doteq c_{\mathrm{s}}/\omega_{\mathrm{c}i}$.

$\boldsymbol{\rho}$ — Spatial difference: $\boldsymbol{\rho} \doteq \boldsymbol{x} - \boldsymbol{x}'$.

**– s –**

$S$ — Number of species

$S_{ss'}$ — Coefficient in the Landau operator: $S_{ss'} \doteq (nq^2)_s(nq^2)_{s'} \ln \Lambda_{ss'}$.

$\mathbf{S}$ — Rate-of-strain tensor: $\mathbf{S} \doteq \frac{1}{2}[(\boldsymbol{\nabla u}) + (\boldsymbol{\nabla u})^{\mathrm{T}}]$.

$s$ — Species index (e.g., $s \in \{e, i\}$); also an arbitrary time for the evolution of orthogonal perturbations

$s(\boldsymbol{x}, t)$ — Macroscopic entropy density

**– t –**

$T$ — Temperature

$t$ — Time at which the fluid equations are evaluated (cf. $s$)

$\tau$ — Time lag: $\tau \doteq t - t'$.

$\tau_e$, $\tau_i$ — Collision times

$\boldsymbol{\tau}$ — Stress tensor: $\boldsymbol{\tau} = p\mathbf{I} + \boldsymbol{\pi}$.

**– u –**

$\mathrm{U}$ — Modified propagator: $\mathrm{U}(\tau) \doteq \mathrm{Q}\mathrm{e}^{-\mathrm{iQ}\mathscr{L}\mathrm{Q}\tau}\mathrm{Q}$.

$U$ — Potential energy

$\boldsymbol{\mathsf{U}}$ — Projection operator in the Landau operator: $\boldsymbol{\mathsf{U}}(\boldsymbol{v}) \doteq (\boldsymbol{\mathsf{I}} - \widehat{\boldsymbol{v}}\,\widehat{\boldsymbol{v}})/v$.

$\boldsymbol{u}(\boldsymbol{x}, t)$ — Fluid velocity

**– v –**

$\mathscr{V}$, $V$ — System volume

$v_{\mathrm{t}}$ — Thermal velocity: $v_{\mathrm{t}} \doteq (T/m)^{1/2}$.

$\boldsymbol{v}$ — Velocity

**– w –**

$W(\varGamma, s; \boldsymbol{r}, t)$ — $W \doteq \ln\boldsymbol{(}F(\varGamma, s)/F_0(\varGamma; \boldsymbol{r}, t)\boldsymbol{)}$

$\boldsymbol{\mathsf{W}}$ — $\boldsymbol{\mathsf{W}} \doteq \boldsymbol{\nabla u} + (\boldsymbol{\nabla u})^{\mathrm{T}} - \frac{2}{3}(\boldsymbol{\nabla} \cdot \boldsymbol{u})\boldsymbol{\mathsf{I}}$.

$\boldsymbol{w}(\boldsymbol{x}, \boldsymbol{v}, t)$ — Peculiar velocity: $\boldsymbol{w}(\boldsymbol{x}, \boldsymbol{v}, t) \doteq \boldsymbol{v} - \boldsymbol{u}(\boldsymbol{x}, t)$.

**– x –**

$X^{\alpha}$ — Mean exchange terms

$\boldsymbol{x}$ — Generic spatial position (cf. $\boldsymbol{r}$)

$\varXi$ — Vlasov response function sans $H(\tau)$

**– y –**

$Y_m^l$ — Spherical harmonic

$\varUpsilon_{\Delta}$ — Orthogonal source term for exchange effects

**– z –**

$Z$ — Partition function; atomic number

$\zeta$ — Bulk viscosity; gyroangle

## I.2. *Miscellaneous notation*

$\widetilde{A}$ — A tilde indicates a random quantity.

$A'$ — Fluctuation in the reference ensemble: $A' \doteq \widetilde{A} - \langle A \rangle_0$.

$A^*$ — Complex conjugate of $A$

$\widehat{A}$ — On most symbols, a hat indicates the orthogonal projection: $\widehat{A} \doteq \mathrm{Q}A$. The symbol $\widehat{\mathrm{C}}$ is an exception that denotes the linearized collision operator.

$A_+$ — One-sided function

$\boldsymbol{\mathsf{A}}^{\mathrm{T}}$ — Transpose of $\boldsymbol{\mathsf{A}}$

$\underset{\approx}{A}$ — A quantity dependent on a correlation function involving three phase-space points

$A^{\mu}$, $A_{\mu}$ — Contravariant and covariant components of $\boldsymbol{A}$

$A_{(s)}$ — Parentheses around an index defeat the Einstein summation convention.

$\Delta A$ — First-order linearization of $A$

$\mathscr{A}$ — Total amount of $A$: $\mathscr{A} = \int \mathrm{d}\boldsymbol{r}\, A(\boldsymbol{r})$.

$A[\psi]$ — Square brackets indicate functional dependence

$\boldsymbol{\partial}$ — $(q/m)\partial/\partial_{\boldsymbol{v}}$

$\underline{1}$ — A set of coordinates sans time

$\overline{x}$ — Typically, an overline denotes an integration variable or summation index.

$\doteq$ — Definition

$\star$ — Nonlocal integral in the local equilibrium distribution: $A \star B \doteq \sum_{\overline{s}} \int \mathrm{d}\overline{\boldsymbol{r}}\, \boldsymbol{A}_{\overline{s}}(\overline{\boldsymbol{r}}, t) \cdot \boldsymbol{B}_{\overline{s}}(\overline{\boldsymbol{r}}, t)$.

$*$ — Convolution

## I.3. *Acronyms and abbreviations*

B — Burnett

BBGKY — Bogoliubov, Born, Green, Kirkwood, and Yvon

NS — Navier–Stokes

ODE — ordinary differential equation

PDF — probability density function

DIA — direct-interaction approximation

PDIA — particle direct-interaction approximation

# Projection-operator methods for classical transport in magnetized plasmas. II. Nonlinear response and the Burnett equations — Supplement

**John A. Krommes†**

Princeton Plasma Physics Laboratory, P. O. Box 451, MS 28, Princeton, New Jersey 08543–0451 USA

Some technical details of the reduction of the general Burnett equations of J. J. Brey *et al.* [Physica **109A**, 425–444 (1981)] to a one-component neutral fluid are given in order to support the results quoted by J. J. Brey [J. Chem. Phys. **79**, 4585–4598 (1983)]. The material is intended to supplement the paper of J. A. Krommes, 'Projection-operator methods for classical transport in magnetized plasmas. II. Nonlinear response and the Burnett equations', `http://arxiv.org/abs/1711.02202` (101 pages).

CONTENTS

## 1. Introduction

The following calculations are to be read in conjunction with the paper of Brey *et al.* (1981), which describes a general projection-operator formalism for obtaining the Burnett equations for an unmagnetized one-component fluid, and Appendix A of Brey (1983), where the formulas are written out more explicitly. The purpose of this Supplement is to provide a reference for the details of that reduction, which were not given by Brey. Those details provide necessary background for the discussion by Krommes (2017) (Part II for short). The exposition will be somewhat more informal than the published Part II.

† Email address for correspondence: `krommes@princeton.edu`

### 1.1. *Thermodynamic relations*

Let us review various thermodynamic relations. With $E$ being the total (mean) amount of energy in the system and $S$ being the total entropy, the fundamental expression is

$$\mathrm{d}E = T\,\mathrm{d}S - p\,\mathrm{d}V + \mu\,\mathrm{d}N. \tag{1.1}$$

Here $T$ is the temperature, $p$ is the pressure, $\mu$ is the chemical potential, and $N$ is the total number of particles. One also has the Euler expression

$$E = TS - pV + \mu N, \tag{1.2}$$

which leads to the Gibbs–Duhem relation

$$0 = S\,\mathrm{d}T - V\,\mathrm{d}p + N\,\mathrm{d}\mu. \tag{1.3}$$

It is convenient to recast these expressions in terms of the densities $n \doteq N/V$, $e \doteq E/V$, and $s \doteq S/V$. Thus, (1.1) becomes

$$\mathrm{d}(Ve) = T\,\mathrm{d}(Vs) - p\,\mathrm{d}V + \mu\,\mathrm{d}(Vn), \tag{1.4}$$

or

$$\mathrm{d}e = T\,\mathrm{d}s + \mu\,\mathrm{d}n + (-e + Ts - p + n\mu)\mathrm{d}V. \tag{1.5}$$

But dividing (1.2) by $V$ gives

$$e = Ts - p + \mu\,n, \tag{1.6}$$

so (1.5) simplifies to

$$\mathrm{d}e = T\,\mathrm{d}s + \mu\,\mathrm{d}n, \tag{1.7}$$

and the Gibbs–Duhem relation is

$$0 = s\,\mathrm{d}T - \mathrm{d}p + n\,\mathrm{d}\mu. \tag{1.8}$$

From (1.7), one has

$$T = \left(\frac{\partial e}{\partial s}\right)_n, \quad \mu = \left(\frac{\partial e}{\partial n}\right)_s. \tag{1.9}$$

This implies the Maxwell relation

$$\left(\frac{\partial T}{\partial n}\right)_s = \left(\frac{\partial \mu}{\partial s}\right)_n. \tag{1.10}$$

The Gibbs–Duhem relation (1.8) also leads to

$$\left(\frac{\partial \mu}{\partial T}\right)_p = -\frac{s}{n}, \quad \left(\frac{\partial \mu}{\partial p}\right)_T = \frac{1}{n}. \tag{1.11}$$

### 1.2. *The microscopic fluxes*

For a one-component system, the time derivatives of the microscopic densities can be written as the divergence of microscopic fluxes or currents according to

$$\partial_t \widetilde{\boldsymbol{A}}(\boldsymbol{r},t) = -\boldsymbol{\nabla}\cdot\widetilde{\boldsymbol{J}}(\boldsymbol{r},t). \tag{1.12}$$

Those currents were discussed in §II:2.3. As a summary of the principle notation and results, one has

$$\partial_t \widetilde{N} = -\boldsymbol{\nabla} \cdot (m^{-1}\widetilde{\boldsymbol{P}}), \tag{1.13$a$}$$

$$\partial_t \widetilde{\boldsymbol{P}} = -\boldsymbol{\nabla} \cdot \widetilde{\boldsymbol{\tau}}, \tag{1.13$b$}$$

$$\partial_t \widetilde{E} = -\boldsymbol{\nabla} \cdot \widetilde{\boldsymbol{J}}^E, \tag{1.13$c$}$$

where

$$\widetilde{\boldsymbol{\tau}}(\boldsymbol{k}) \doteq \sum_{i=1}^{\mathscr{N}} [m\boldsymbol{v}_i\boldsymbol{v}_i + \Delta\widetilde{\boldsymbol{\tau}}_i(\boldsymbol{k})]\mathrm{e}^{-\mathrm{i}\boldsymbol{k}\cdot\boldsymbol{x}_i}, \tag{1.14$a$}$$

$$\widetilde{\boldsymbol{J}}^E(\boldsymbol{k}) \doteq \sum_{i=1}^{\mathscr{N}} [E_i\boldsymbol{v}_i + \Delta\widetilde{\boldsymbol{\tau}}_i(\boldsymbol{k})\cdot\boldsymbol{v}_i]\mathrm{e}^{-i\boldsymbol{k}\cdot\boldsymbol{x}_i}, \tag{1.14$b$}$$

with $\Delta\widetilde{\boldsymbol{\tau}}_i(\boldsymbol{k})$ being defined by (II:B 6).

### 1.3. *The general formula for the dissipative fluxes*

Brey *et al.* (1981) show that the general expression for the fluxes through second order is

$$\begin{aligned}\langle \boldsymbol{J}^\alpha \rangle = \langle \boldsymbol{J}^\alpha \rangle_{\text{Euler}} - \underbrace{\boldsymbol{k}_1^\beta[\boldsymbol{J}^\alpha]\cdot\boldsymbol{\nabla}B_\beta(\boldsymbol{r},t)}_{\mathrm{NS}^\alpha_\beta} - \underbrace{\boldsymbol{g}_2^\beta[\boldsymbol{J}^\alpha] : \boldsymbol{\nabla}\boldsymbol{\nabla}B_\beta(\boldsymbol{r},t)}_{\mathrm{B}^\alpha_\beta} \\ - \underbrace{\boldsymbol{h}_2^{\beta\gamma}[\boldsymbol{J}^\alpha] : \boldsymbol{\nabla}B_\beta(\boldsymbol{r},t)\boldsymbol{\nabla}B_\gamma(\boldsymbol{r},t)}_{\mathrm{B}^\alpha_{\beta\gamma}} + \underbrace{\left[\frac{\partial}{\partial t}\left(\boldsymbol{k}_2^\beta[\boldsymbol{J}^\alpha]\cdot\boldsymbol{\nabla}B_\beta(\boldsymbol{r},t)\right)\right]^{(1)}}_{\partial\mathrm{B}^\alpha_\beta}.\end{aligned} \tag{1.15}$$

Here the various terms are tersely identified for future reference; NS and B stand for Navier–Stokes and Burnett, respectively. The indices $\alpha$ and $\beta$ refer to[1] $N$, $\boldsymbol{P}$, or $E$. For example, in the momentum equation the linear Burnett term generates contributions $\mathrm{B}^{\boldsymbol{P}}_{\boldsymbol{P}}$ and $\mathrm{B}^{\boldsymbol{P}}_{E}$. The various coefficients are defined as

$$\boldsymbol{k}_1^\beta[\widetilde{\boldsymbol{J}}^\alpha](\mu,t) \doteq \int_0^\infty \mathrm{d}s \int \mathrm{d}\boldsymbol{r}'\, \langle \widehat{\boldsymbol{J}}^\alpha(\boldsymbol{r})\mathrm{e}^{-\mathrm{i}\mathscr{L}s}\widehat{\boldsymbol{J}}^\beta(\boldsymbol{r}')\rangle_0 = \int_0^\infty \mathrm{d}s\, \langle \widehat{\boldsymbol{J}}^\alpha(\boldsymbol{r})\mathrm{e}^{-\mathrm{i}\mathscr{L}s}\widehat{\boldsymbol{\mathscr{J}}}^\beta\rangle_0, \tag{1.16$a$}$$

$$\boldsymbol{k}_2^\beta[\widetilde{\boldsymbol{J}}^\alpha](\mu,t) \doteq \int_0^\infty \mathrm{d}s\, s\langle \widehat{\boldsymbol{J}}^\alpha(\boldsymbol{r})\mathrm{e}^{-\mathrm{i}\mathscr{L}s}\widehat{\boldsymbol{\mathscr{J}}}^\beta\rangle_0, \tag{1.16$b$}$$

$$\boldsymbol{g}_2^\beta[\widetilde{\boldsymbol{J}}^\alpha](\mu,t) \doteq \int_0^\infty \mathrm{d}s \int \mathrm{d}\boldsymbol{r}'\, \langle \widehat{\boldsymbol{J}}^\alpha(\boldsymbol{r})\mathrm{e}^{-\mathrm{i}\mathscr{L}s}\widehat{\boldsymbol{J}}^\beta(\boldsymbol{r}')\rangle_0(\boldsymbol{r}'-\boldsymbol{r}), \tag{1.16$c$}$$

$$\boldsymbol{h}_2^{\beta\gamma}[\widetilde{\boldsymbol{J}}^\alpha](\mu,t) \doteq \int_0^\infty \mathrm{d}s \int \mathrm{d}\boldsymbol{r}'\, \langle \widehat{\boldsymbol{J}}^\alpha(\boldsymbol{r})\mathrm{e}^{-\mathrm{i}\mathscr{L}s}\mathrm{Q}[\widehat{\boldsymbol{\mathscr{J}}}^\beta A'^\gamma(\boldsymbol{r}')]\rangle_0. \tag{1.16$d$}$$

(Brey writes these formulas with $\mathrm{e}^{\mathrm{i}\mathscr{L}s}$ on the left, which is a permissible manipulation with the Liouville operator.) As discussed in Part II, the hats denote subtracted quanties. With previous notation, one has specifically $\widetilde{\boldsymbol{J}}^\alpha = (m^{-1}\widetilde{\boldsymbol{P}},\, \widetilde{\boldsymbol{\tau}},\, \widetilde{\boldsymbol{J}}^E)^{\mathrm{T}}$. I shall now work out all of the contributions one by one. Note that all of the terms with $\alpha = N$ vanish because $\widehat{\boldsymbol{J}}^N = \boldsymbol{0}$. Also, one can make use of the isotropy of the reference state to conclude that various integrals vanish *to lowest order in the gradients* — e.g., $\langle \widehat{\boldsymbol{\tau}}\, \mathrm{e}^{-\mathrm{i}\mathscr{L}s}\widehat{\boldsymbol{J}}^E\rangle = O(\boldsymbol{\nabla})$.

[1]In Part II, lower case is used for these indices.

Finally, one has $B_\beta = \big(\beta(\mu - \frac{1}{2}mu^2),\ \beta\boldsymbol{u},\ -\beta\big)^{\mathrm{T}}$. (Here $\beta$ is being used as both an index and as the inverse temperature.)

In the subsequent reductions, one makes use of the various integrals of correlation functions defined in Eq. (A7) of Brey (1983) and also tabulated in §II:4.2.

## 2. Navier–Stokes terms ($\mathrm{NS}^\alpha_\beta$)

### 2.1. *Navier–Stokes momentum flux* ($\mathrm{NS}^{\boldsymbol{P}}_{\boldsymbol{P}}$)

$$\langle \boldsymbol{\tau}\rangle^{\mathrm{NS}}_{\mathrm{diss}} = -\int_0^\infty \mathrm{d}s\, \langle \widehat{\boldsymbol{\tau}}(\boldsymbol{r})\mathrm{e}^{-\mathrm{i}\mathscr{L}s}\widehat{\boldsymbol{\mathscr{J}}}^{\boldsymbol{P}}\rangle \cdot \boldsymbol{\nabla} \boldsymbol{B}_{\boldsymbol{P}} - \int_0^\infty \mathrm{d}s\, \underbrace{\langle \widehat{\boldsymbol{\tau}}(\boldsymbol{r})\mathrm{e}^{-\mathrm{i}\mathscr{L}s}\widehat{\boldsymbol{\mathscr{J}}}^{E}\rangle}_{0} \cdot \boldsymbol{\nabla} B_E \tag{2.1$a$}$$

$$= -\int_0^\infty \mathrm{d}s\, \langle \widehat{\tau}_{ij}(\boldsymbol{r})\mathrm{e}^{-\mathrm{i}\mathscr{L}s}\widehat{\mathscr{T}}_{kl}\rangle \nabla_l(\beta u_k) \tag{2.1$b$}$$

$$= -[K^{\mathrm{I}}(\delta_{ik}\delta_{jl} + \delta_{il}\delta_{jk}) + K^{\mathrm{II}}\delta_{ij}\delta_{kl}]\frac{1}{T}\nabla_l u_k \tag{2.1$c$}$$

$$= -\frac{1}{T}\{K^{\mathrm{I}}[(\boldsymbol{\nabla u})^{\mathrm{T}} + (\boldsymbol{\nabla u})] + K^{\mathrm{II}}(\boldsymbol{\nabla}\cdot\boldsymbol{u})\,\boldsymbol{I}\} \tag{2.1$d$}$$

$$= -\frac{1}{T}\left[K^{\mathrm{I}}\left((\boldsymbol{\nabla u})^{\mathrm{T}} + (\boldsymbol{\nabla u}) - \frac{2}{d}(\boldsymbol{\nabla}\cdot\boldsymbol{u})\,\boldsymbol{I}\right) + \left(K^{\mathrm{II}} + \frac{2}{d}K^{\mathrm{I}}\right)(\boldsymbol{\nabla}\cdot\boldsymbol{u})\boldsymbol{I}\right] \tag{2.1$e$}$$

$$= -\eta\left((\boldsymbol{\nabla u}) + (\boldsymbol{\nabla u})^{\mathrm{T}} - \frac{2}{d}(\boldsymbol{\nabla}\cdot\boldsymbol{u})\boldsymbol{I}\right) - \zeta(\boldsymbol{\nabla}\cdot\boldsymbol{u})\,\boldsymbol{I}, \tag{2.1$f$}$$

where the kinematic and bulk viscosities are

$$\boxed{\eta \doteq \frac{1}{T}K^{\mathrm{I}},} \quad \boxed{\zeta \doteq \frac{1}{T}\left(K^{\mathrm{II}} + \frac{2}{d}K^{\mathrm{I}}\right).} \tag{2.2}$$

Notice that the kinematic, $\eta$ contribution has been constructed to be traceless.

In the limit of weak coupling, it is well known that the bulk viscosity $\zeta$ vanishes. To show this, one proceeds as follows. One has

$$\zeta\,\delta_{ij} = \frac{1}{dT}\int_0^\infty \mathrm{d}s\, \langle \widehat{\tau}_{ij}(\boldsymbol{r})\mathrm{e}^{-\mathrm{i}\mathscr{L}s}\,\mathrm{Tr}\,\widehat{\mathscr{T}}\rangle. \tag{2.3}$$

Now

$$\widehat{\boldsymbol{\tau}}(\boldsymbol{r}') = \widetilde{\boldsymbol{\tau}}(\boldsymbol{r}') - \boldsymbol{I}[p(\boldsymbol{r}') + N'(\boldsymbol{r}')p_n + E'(\boldsymbol{r}')p_e]. \tag{2.4}$$

For the weakly coupled gas, one has $p \approx nT$ and $e \approx (d/2)nT$, so $(p_n)|_e = 0$ and $(p_e)|_n = 2/d$. Since those thermodynamic derivatives are constants, one has

$$\widehat{\mathscr{T}} = \int \mathrm{d}\boldsymbol{r}'\,[\widetilde{\boldsymbol{\tau}}(\boldsymbol{r}') - p(\boldsymbol{r}')\boldsymbol{I}], \tag{2.5}$$

since $\int \mathrm{d}\boldsymbol{r}'\,N'(\boldsymbol{r}') = 0$ and similarly $\int E' = 0$. Also, one may ignore the internal-energy contributions to $\widetilde{\boldsymbol{\tau}}$ and $p$. Thus,

$$\widehat{\mathscr{T}} \to \sum_{i=1}^{\mathscr{N}} m(\boldsymbol{w}_i\boldsymbol{w}_i - d^{-1}\langle w_i^2\rangle\boldsymbol{I}) \tag{2.6}$$

and

$$\mathrm{Tr}\,\widehat{\mathscr{T}} \to \sum_{i=1}^{\mathscr{N}} m(w_i^2 - \langle w_i^2\rangle). \tag{2.7}$$

In the absence of particle correlations, all particles will be equivalent. The term $mw^2 - (d/2)T$ is the orthogonalized single-particle kinetic energy — a null eigenfunction of the weakly coupled collision operator that lies in the hydrodynamic subspace. The remainder of the expression for $\zeta$ will involve a weakly coupled Q; thus, $\zeta$ vanishes in the limit of weak coupling.

### 2.2. *Navier–Stokes energy flux* ($\mathrm{NS}_E^E$)

$$\langle \boldsymbol{J}^E \rangle_{\text{diss}}^{\text{NS}} = -\int_0^\infty \mathrm{d}s \, \underbrace{\langle \widehat{\boldsymbol{J}}^E(\boldsymbol{r}) \mathrm{e}^{-\mathrm{i}\mathscr{L}s} \widehat{\boldsymbol{\mathscr{J}}}^{\boldsymbol{P}} \rangle_0}_{0} \cdot \boldsymbol{\nabla} \boldsymbol{B}_{\boldsymbol{P}} - \int_0^\infty \mathrm{d}s \, \langle \widehat{\boldsymbol{J}}^E(\boldsymbol{r}) \mathrm{e}^{-\mathrm{i}\mathscr{L}s} \widehat{\boldsymbol{\mathscr{J}}}^E \rangle_0 \cdot \boldsymbol{\nabla} B_E \tag{2.8a}$$

$$= -(K^{\text{III}} \boldsymbol{I}) \cdot \boldsymbol{\nabla}(-T^{-1}) \tag{2.8b}$$

$$= -\lambda \boldsymbol{\nabla} T, \tag{2.8c}$$

where

$$\boxed{\lambda \doteq \frac{1}{T^2} K^{\text{III}}.} \tag{2.9}$$

## 3. Burnett terms

### 3.1. *Linear Burnett terms* ($\mathrm{B}_\beta^\alpha$)

#### 3.1.1. Linear Burnett — momentum ($\mathrm{B}_E^{\boldsymbol{P}}$)

$$\langle \boldsymbol{\tau} \rangle_{\text{diss}}^{\mathrm{B}_1} = -\int_0^\infty \mathrm{d}s \int \mathrm{d}\boldsymbol{r}' \, \underbrace{\langle \widehat{\boldsymbol{\tau}}(\boldsymbol{r}) \mathrm{e}^{-\mathrm{i}\mathscr{L}s} \widehat{\boldsymbol{J}}^{\boldsymbol{P}}(\boldsymbol{r}') \rangle_0 (\boldsymbol{r}' - \boldsymbol{r})}_{0} : \boldsymbol{\nabla\nabla} \boldsymbol{B}_{\boldsymbol{P}}$$

$$- \int_0^\infty \mathrm{d}s \int \mathrm{d}\boldsymbol{r}' \, \langle \widehat{\boldsymbol{\tau}}(\boldsymbol{r}) \mathrm{e}^{-\mathrm{i}\mathscr{L}s} \widehat{\boldsymbol{J}}^E(\boldsymbol{r}') \rangle (\boldsymbol{r}' - \boldsymbol{r}) : \boldsymbol{\nabla\nabla} B_E \tag{3.1a}$$

$$= -[K_1(\delta_{ik}\delta_{jl} + \delta_{il}\delta_{jk}) + K_2 \delta_{ij}\delta_{kl}] \nabla_k \nabla_l (-T^{-1}). \tag{3.1b}$$

Now

$$\boldsymbol{\nabla\nabla}(-T^{-1}) = \boldsymbol{\nabla}(T^{-2}\boldsymbol{\nabla}T) = T^{-2}\boldsymbol{\nabla\nabla}T - 2T^{-3}(\boldsymbol{\nabla}T)(\boldsymbol{\nabla}T). \tag{3.2}$$

Thus,

$$\langle \boldsymbol{\tau} \rangle_{\text{diss}}^{\mathrm{B}_1} = -[T^{-2}(K_1 \boldsymbol{\nabla\nabla} T + K_2 \nabla^2 T \, \boldsymbol{I})] + 2T^{-3}[2K_1(\boldsymbol{\nabla}T)(\boldsymbol{\nabla}T) + K_2 |\boldsymbol{\nabla}T|^2 \boldsymbol{I}] \tag{3.3a}$$

$$= \eta_3 \boldsymbol{\nabla\nabla} T + \eta_4 \nabla^2 T \boldsymbol{I} + \eta_5^{\mathrm{B}_1} (\boldsymbol{\nabla}T)(\boldsymbol{\nabla}T) + \eta_6^{\mathrm{B}_1} |\boldsymbol{\nabla}T|^2 \boldsymbol{I}, \tag{3.3b}$$

where

$$\boxed{\eta_3 \doteq -2T^{-2}K_1,} \quad \boxed{\eta_4 \doteq -T^{-2}K_2,} \quad \boxed{\eta_5^{\mathrm{B}_1} \doteq 4T^{-3}K_1,} \quad \boxed{\eta_6^{\mathrm{B}_1} \doteq 2T^{-3}K_2.} \tag{3.4}$$

3.1.2. Linear Burnett — energy ($\mathrm{B}^{E}_{\boldsymbol{P}}$)

$$\begin{aligned}\langle \boldsymbol{J}^E\rangle^{\mathrm{B}_1}_{\mathrm{diss}} &= -\int_0^\infty \mathrm{d}s \int \mathrm{d}\boldsymbol{r}'\, \langle \widehat{\boldsymbol{J}}^E(\boldsymbol{r})\mathrm{e}^{-\mathrm{i}\mathscr{L}s}\widehat{\boldsymbol{J}}^{\boldsymbol{P}}(\boldsymbol{r}')\rangle(\boldsymbol{r}'-\boldsymbol{r}) : \boldsymbol{\nabla\nabla B_P} \\ &\quad - \int_0^\infty \mathrm{d}s \int \mathrm{d}\boldsymbol{r}'\, \underbrace{\langle \widehat{\boldsymbol{J}}^E \mathrm{e}^{-\mathrm{i}\mathscr{L}s}\widehat{\boldsymbol{J}}^E\rangle_0(\boldsymbol{r}'-\boldsymbol{r})}_{0} : \boldsymbol{\nabla\nabla} B_E \qquad (3.5a) \\ &= -\int_0^\infty \mathrm{d}s \int \mathrm{d}\boldsymbol{r}'\, \langle \widehat{J}^E_i(\boldsymbol{r})\mathrm{e}^{-\mathrm{i}\mathscr{L}s}\widehat{\tau}_{jk}(\boldsymbol{r}')\rangle_0(\boldsymbol{r}'-\boldsymbol{r})_l \nabla_k\nabla_l(T^{-1}u_j). \qquad (3.5b)\end{aligned}$$

Now

$$\begin{aligned}\nabla_k\nabla_l(T^{-1}u_j) &= \nabla_k[T^{-1}\nabla_l u_j - T^{-2}(\nabla_l T)u_j] \qquad (3.6a)\\ &= T^{-1}\nabla_k\nabla_l u_j - T^{-2}[(\nabla_k T)(\nabla_l u_j) + (\nabla_l T)(\nabla_k u_j)]. \qquad (3.6b)\end{aligned}$$

Thus,

$$\begin{aligned}\langle \boldsymbol{J}^E\rangle^{\mathrm{B}_1}_{\mathrm{diss}} &= -[K_1(\delta_{ji}\delta_{kl} + \delta_{jl}\delta_{ki}) + K_2\delta_{jk}\delta_{il}] \\ &\qquad \times [T^{-1}\nabla_l\nabla_k u_j - T^{-2}[(\nabla_l T)(\nabla_k u_j) + (\nabla_k T)(\nabla_l u_j)] \qquad (3.7a)\\ &= -\{T^{-1}K_1[\nabla^2\boldsymbol{u} + \boldsymbol{\nabla}(\boldsymbol{\nabla}\cdot\boldsymbol{u})] + T^{-1}K_2\boldsymbol{\nabla}(\boldsymbol{\nabla}\cdot\boldsymbol{u})\} \\ &\quad + T^{-2}K_1\{2\boldsymbol{\nabla}T\cdot(\boldsymbol{\nabla u}) + [(\boldsymbol{\nabla u})\cdot\boldsymbol{\nabla}T) + \boldsymbol{\nabla}T(\boldsymbol{\nabla}\cdot\boldsymbol{u})]\} \\ &\quad + T^{-2}K_2[\boldsymbol{\nabla}T(\boldsymbol{\nabla}\cdot\boldsymbol{u}) + (\boldsymbol{\nabla u})\cdot\boldsymbol{\nabla}T]\} \qquad (3.7b)\\ &= -T^{-1}K_1\nabla^2\boldsymbol{u} - T^{-1}(K_1+K_2)\boldsymbol{\nabla}(\boldsymbol{\nabla}\cdot\boldsymbol{u}) \\ &\quad + T^{-2}\{K_1[2(\boldsymbol{\nabla u})^{\mathrm{T}} + (\boldsymbol{\nabla u})] + K_2(\boldsymbol{\nabla u})\}\cdot\boldsymbol{\nabla}T \\ &\quad + T^{-2}(K_1+K_2)\boldsymbol{\nabla}T(\boldsymbol{\nabla}\cdot\boldsymbol{u}). \qquad (3.7c)\end{aligned}$$

Now introduce

$$\boldsymbol{S} \doteq \frac{1}{2}[(\boldsymbol{\nabla u})^{\mathrm{T}} + (\boldsymbol{\nabla u})], \quad \boldsymbol{\Omega} \doteq \frac{1}{2}[(\boldsymbol{\nabla u})^{\mathrm{T}} - (\boldsymbol{\nabla u})] \qquad (3.8)$$

such that

$$(\boldsymbol{\nabla u})^{\mathrm{T}} = \boldsymbol{S} + \boldsymbol{\Omega}, \quad (\boldsymbol{\nabla u}) = \boldsymbol{S} - \boldsymbol{\Omega}. \qquad (3.9)$$

The second line of the last equation then becomes

$$T^{-2}[(3K_1+K_2)\boldsymbol{S} + (K_1-K_2)\boldsymbol{\Omega}] \qquad (3.10)$$

and one finds

$$\langle \boldsymbol{J}^E\rangle^{\mathrm{B}_1}_{\mathrm{diss}} = \lambda_2\boldsymbol{\nabla}(\boldsymbol{\nabla}\cdot\boldsymbol{u}) + \lambda_3\nabla^2\boldsymbol{u} + \lambda_4^{\mathrm{B}_1}(\boldsymbol{\nabla}\cdot\boldsymbol{u})\boldsymbol{\nabla}T + \lambda_5^{\mathrm{B}_1}\boldsymbol{S}\cdot\boldsymbol{\nabla}T + \lambda_6^{\mathrm{B}_1}\boldsymbol{\Omega}\cdot\boldsymbol{\nabla}T, \qquad (3.11)$$

where

$$\boxed{\lambda_2 \doteq -T^{-1}(K_1+K_2),} \quad \boxed{\lambda_3 \doteq -T^{-1}K_1,} \qquad (3.12)$$

and

$$\boxed{\lambda_4^{\mathrm{B}_1} \doteq T^{-2}(K_1+K_2),} \quad \boxed{\lambda_5^{\mathrm{B}_1} \doteq T^{-2}(3K_1+K_2),} \quad \boxed{\lambda_6^{\mathrm{B}_1} \doteq T^{-2}(K_1-K_2).} \qquad (3.13)$$

### 3.2. *Nonlinear Burnett terms* ($\mathrm{B}^{\alpha}_{\beta\gamma}$)

The terms involving $(\boldsymbol{\nabla}B)(\boldsymbol{\nabla}B)$ stem from $\boldsymbol{h}_2^{\beta\gamma}[\boldsymbol{J}^\alpha](\mu,t)$, which contains $\mathrm{Q} = 1 - \mathrm{P}$. I shall work out the terms coming from the 1 and coming from the P separately.

I shall signify the various terms by $\mathrm{B}^{\alpha}_{\beta\gamma} \equiv (1-\mathrm{P})(\alpha,\beta,\gamma)$. (Of course, the P does not multiply the entire term; this is just a notation.) Note that the $\beta$ component involves a total flux $\mathscr{J}^{\beta}$; thus, there is no $\beta = N$ contribution.

3.2.1. Nonlinear Burnett — momentum

Since $\boldsymbol{P}$ is odd in velocity, the $(\beta,\gamma)$ components must be either both even or both odd. Therefore,

$$\langle \boldsymbol{\tau} \rangle^{\mathrm{B}^{(2)}}_{\mathrm{diss}} = (1-\mathrm{P})[(\boldsymbol{P},E,N) + (\boldsymbol{P},E,E) + (\boldsymbol{P},\boldsymbol{P},\boldsymbol{P})]. \tag{3.14}$$

---

<u>$(\boldsymbol{P},E,N)$</u>:

$$(\boldsymbol{P},E,N) = -\int_0^\infty \mathrm{d}s \int \mathrm{d}\boldsymbol{r}' \, \langle \widehat{\tau}_{ij}(\boldsymbol{r})\mathrm{e}^{-\mathrm{i}\mathscr{L}s} \widehat{\mathscr{J}}^E_k N'(\boldsymbol{r}')\rangle (\boldsymbol{r}'-\boldsymbol{r})_l (\nabla_k B_E)(\nabla_l B_N) \tag{3.15a}$$

$$= -[K_3(\delta_{ik}\delta_{jl} + \delta_{il}\delta_{jk}) + K_4\delta_{ij}\delta_{kl}](T^{-2}\nabla_k T)(\nabla_l B_N) \tag{3.15b}$$

$$= -T^{-2}[K_3(\boldsymbol{\nabla} T\,\boldsymbol{\nabla} B_N + \boldsymbol{\nabla} B_N\,\boldsymbol{\nabla} T) + K_4(\boldsymbol{\nabla} T \cdot \boldsymbol{\nabla} B_N)\boldsymbol{I}]. \tag{3.15c}$$

Now

$$\boldsymbol{\nabla} B_N = \boldsymbol{\nabla}\left[\beta\left(\mu - \frac{1}{2}mu^2\right)\right] \to \boldsymbol{\nabla}(\beta\mu) \tag{3.16a}$$

$$= -T^{-2}\mu\boldsymbol{\nabla} T + T^{-1}\boldsymbol{\nabla}\mu. \tag{3.16b}$$

One can think of $\mu$ as depending on $T$ and $P$. Thus,

$$\boldsymbol{\nabla} B_N = -\frac{1}{T}\left[\frac{\mu}{T} - \left(\frac{\partial\mu}{\partial T}\right)_p\right]\boldsymbol{\nabla} T + \frac{1}{T}\left[\left(\frac{\partial\mu}{\partial p}\right)_T\right]\boldsymbol{\nabla} p. \tag{3.17}$$

One only needs to evaluate the thermodynamic derivatives in the reference ensemble, so one can use the local thermodynamic relations discussed in §1.1. From (1.11), (3.17) becomes

$$\boldsymbol{\nabla} B_N = -T^{-2}(\mu + Tsn^{-1})\boldsymbol{\nabla} T + T^{-1}n^{-1}\boldsymbol{\nabla} p \tag{3.18a}$$

$$= (nT)^{-1}(-hT^{-1}\boldsymbol{\nabla} T + \boldsymbol{\nabla} p), \tag{3.18b}$$

where again $h \doteq e + p$ and the Euler equation (1.6) was used.

Thus,

$$(\boldsymbol{P},E,N) = -(\overline{n}T^3)^{-1}\{K_3[\boldsymbol{\nabla} T(-T^{-1}h\boldsymbol{\nabla} T + \boldsymbol{\nabla} p) + (-T^{-1}h\boldsymbol{\nabla} T + \boldsymbol{\nabla} p)\boldsymbol{\nabla} T] + K_4\boldsymbol{\nabla} T \cdot (-T^{-1}h\boldsymbol{\nabla} T + \boldsymbol{\nabla} p)\,\boldsymbol{I}] \tag{3.19a}$$

$$= (\overline{n}T^4)^{-1}[2K_3h\boldsymbol{\nabla} T\,\boldsymbol{\nabla} T + K_4h|\boldsymbol{\nabla} T|^2\,\boldsymbol{I}] - (\overline{n}T^3)^{-1}[K_3(\boldsymbol{\nabla} T\,\boldsymbol{\nabla} p + \boldsymbol{\nabla} p\,\boldsymbol{\nabla} T) + K_4\boldsymbol{\nabla} T \cdot \boldsymbol{\nabla} p\,\boldsymbol{I}] \tag{3.19b}$$

$$= \eta_5^{\mathrm{B}^{(2)}_{EN}}\boldsymbol{\nabla} T\,\boldsymbol{\nabla} T + \eta_6^{\mathrm{B}^{(2)}_{EN}}|\boldsymbol{\nabla} T|^2 + \eta_7^{\mathrm{B}^{(2)}_{EN}}(\boldsymbol{\nabla} T\,\boldsymbol{\nabla} p + \boldsymbol{\nabla} p\,\boldsymbol{\nabla} T) + \eta_8^{\mathrm{B}^{(2)}_{EN}}\boldsymbol{\nabla} T \cdot \boldsymbol{\nabla} p\,\boldsymbol{I}, \tag{3.19c}$$

where

$$\boxed{\eta_5^{\mathrm{B}^{(2)}_{EN}} \doteq 2(\overline{n}T^4)^{-1}hK_3,} \quad \boxed{\eta_6^{\mathrm{B}^{(2)}_{EN}} \doteq (\overline{n}T^4)^{-1}hK_4,} \tag{3.20}$$

and

$$\boxed{\eta_7^{\mathrm{B}^{(2)}_{EN}} \doteq -(\overline{n}T^3)^{-1}K_3,} \quad \boxed{\eta_8^{\mathrm{B}^{(2)}_{EN}} \doteq -(\overline{n}T^3)^{-1}K_4.} \tag{3.21}$$

$\underline{-\mathrm{P}(\boldsymbol{P}, E, N)}$:

$$- \mathrm{P}(\boldsymbol{P}, E, N) = \int_0^\infty \mathrm{d}s \int \mathrm{d}\boldsymbol{r}' \, \langle \widehat{\tau}_{ij}(\boldsymbol{r}) \mathrm{e}^{-\mathrm{i}\mathscr{L}s} \mathrm{P}[\widehat{\mathscr{J}}^E_k N'(\boldsymbol{r}')] \rangle (\boldsymbol{r}' - \boldsymbol{r})_l (\nabla_k B_E)(\nabla_l B_N). \tag{3.22}$$

Now

$$\mathrm{P}[\widehat{\mathscr{J}}^E_k N'(\boldsymbol{r}')] = \underbrace{\langle \widehat{\mathscr{J}}^E_k N'(\boldsymbol{r}') \rangle}_{0} + (N' \, E') * \boldsymbol{\mathscr{M}}_2^{-1} * \underbrace{\langle (N' \, E')^{\mathrm{T}} \widehat{\mathscr{J}}^E_k N'(\boldsymbol{r}') \rangle}_{0}$$
$$+ \boldsymbol{P}' * \boldsymbol{\mathscr{M}}^{-1}_{\boldsymbol{PP}} * \langle \boldsymbol{P}' \widehat{\mathscr{J}}^E_k N'(\boldsymbol{r}') \rangle. \tag{3.23}$$

The averages vanish because they are vectors, so they must be proportional to $\boldsymbol{u}$, which vanishes in the local frame. Because $\boldsymbol{\mathscr{M}}_{\boldsymbol{PP}}(\overline{\boldsymbol{x}}, \overline{\boldsymbol{x}}') = m\overline{n}T\delta(\overline{\boldsymbol{x}} - \overline{\boldsymbol{x}}')$, the last term is

$$\int \mathrm{d}\overline{\boldsymbol{x}} \, \mathrm{d}\overline{\boldsymbol{x}}' \, \boldsymbol{P}'(\overline{\boldsymbol{x}}) \cdot \boldsymbol{\mathscr{M}}^{-1}_{\boldsymbol{PP}}(\overline{\boldsymbol{x}}, \overline{\boldsymbol{x}}') \cdot \langle \boldsymbol{P}'(\overline{\boldsymbol{x}}') \widehat{\boldsymbol{\mathscr{J}}}^E N'(\boldsymbol{r}') \rangle_0$$
$$= (m\overline{n}T)^{-1} \int \mathrm{d}\overline{\boldsymbol{x}} \, \boldsymbol{P}'(\overline{\boldsymbol{x}}) \cdot \langle \boldsymbol{P}'(\overline{\boldsymbol{x}}) \widehat{\boldsymbol{\mathscr{J}}}^E N'(\boldsymbol{r}') \rangle_0. \tag{3.24}$$

Now

$$\langle \boldsymbol{P}'(\overline{\boldsymbol{x}}) \widehat{\boldsymbol{\mathscr{J}}}^E N'(\boldsymbol{r}') \rangle = \Bigg\langle \Bigg( \sum_{i=1}^{\mathscr{N}} m\boldsymbol{w}_i \delta(\overline{\boldsymbol{x}} - \boldsymbol{x}_i) \Bigg)$$
$$\times \left[ \sum_{j=1}^{\mathscr{N}} E_j \boldsymbol{w}_j + \Delta\widetilde{\boldsymbol{\tau}}_j \cdot \boldsymbol{w}_j - \left(\frac{h}{n}\right)_i \boldsymbol{w}_j \right] N'(\boldsymbol{r}') \Bigg\rangle_0. \tag{3.25}$$

Averaging over velocity restricts $j$ to $i$. Thus,

$$\langle \boldsymbol{P}'(\overline{\boldsymbol{x}}) \widehat{\boldsymbol{\mathscr{J}}}^E N'(\boldsymbol{r}') \rangle = \sum_{i=1}^{\mathscr{N}} \left\langle \delta(\overline{\boldsymbol{x}} - \boldsymbol{x}_i) m\boldsymbol{w}_i \left[ E_i \boldsymbol{I} + \Delta\widetilde{\boldsymbol{\tau}}_i - \left(\frac{h}{n}\right)_i \boldsymbol{I} \right] \cdot \boldsymbol{w}_i N'(\boldsymbol{r}') \right\rangle_0 \tag{3.26a}$$

$$= T\boldsymbol{I} \sum_{i=1}^{\mathscr{N}} \left\langle \delta(\overline{\boldsymbol{x}} - \boldsymbol{x}_i) \left[ \frac{5}{2}T + U_i + \frac{1}{3} \operatorname{Tr} \Delta\widetilde{\boldsymbol{\tau}}_i - \left(\frac{h}{n}\right)_i \right] N'(\boldsymbol{r}') \right\rangle_0 \tag{3.26b}$$

$$= T\boldsymbol{I} \left\langle \left(\frac{h}{n}\right)' (\overline{\boldsymbol{x}}) \, N'(\boldsymbol{r}') \right\rangle_0. \tag{3.26c}$$

One may replace $\langle \ldots \rangle_0 \approx \langle \ldots \rangle_{\boldsymbol{B}}$. Then

$$\langle \boldsymbol{P}'(\overline{\boldsymbol{x}}) \widehat{\boldsymbol{\mathscr{J}}}^E N'(\boldsymbol{r}') \rangle = \overline{n}T\boldsymbol{I} \frac{\delta h(\overline{\boldsymbol{x}})}{\delta[(\beta\mu)(\boldsymbol{r}')]}. \tag{3.27}$$

The functional derivative essentially gives back a $\delta(\overline{\boldsymbol{x}} - \boldsymbol{r}')$, so performing the integral in the projection brings one to

$$-\mathrm{P}(\boldsymbol{P}, E, E) = [K_{20}(\delta_{ik}\delta_{jl} + \delta_{il}\delta_{jk}) + K_{21}\delta_{ij}\delta_{kl}]$$
$$\times \left[ \frac{\partial}{\partial(\beta\mu)} \left(\frac{h}{mn}\right) \right]_\beta (T^{-2}\nabla_k T)[(nT)^{-1}(-hT^{-1}\nabla_l T + \nabla_l p]. \tag{3.28}$$

From the Gibbs–Duhem relation

$$s \, \mathrm{d}T - \mathrm{d}p + n \, \mathrm{d}\mu = 0 \tag{3.29}$$

at constant $T$, one finds

$$\left(\frac{\partial\chi}{\partial(\beta\mu)}\right)_\beta = nT\left(\frac{\partial\chi}{\partial p}\right)_T. \tag{3.30}$$

Thus, the pressure term gives a contribution

$$-\,\mathrm{P}(\boldsymbol{P},E,N)_{\boldsymbol{\nabla}p} = \eta_7^{\mathrm{B}^{(2)}_{\mathrm{P}EN}}(\boldsymbol{\nabla}T\,\boldsymbol{\nabla}p + \boldsymbol{\nabla}p\,\boldsymbol{\nabla}T) + \eta_8^{\mathrm{B}^{(2)}_{\mathrm{P}EN}}(\boldsymbol{\nabla}T\cdot\boldsymbol{\nabla}p), \tag{3.31}$$

where

$$\boxed{\eta_7^{\mathrm{B}^{(2)}_{\mathrm{P}EN}} \doteq T^{-2}\frac{\partial}{\partial p}\left(\frac{h}{mn}\right)K_{20},} \quad \boxed{\eta_8^{\mathrm{B}^{(2)}_{\mathrm{P}EN}} \doteq T^{-2}\frac{\partial}{\partial p}\left(\frac{h}{mn}\right)K_{21}.} \tag{3.32}$$

The remaining term in $-h\boldsymbol{\nabla}T$ will be canceled by a contribution from $-\mathrm{P}(\boldsymbol{P},E,E)$.

---

<u>$(\boldsymbol{P},E,E)$</u>:

$$(\boldsymbol{P},E,E) = -\int_0^\infty \mathrm{d}s\int \mathrm{d}\boldsymbol{r}'\,\langle\widehat{\tau}_{ij}(\boldsymbol{r})\mathrm{e}^{-\mathrm{i}\mathscr{L}s}\,\widehat{\mathscr{J}}^E_k E'(\boldsymbol{r}')\rangle(\boldsymbol{r}'-\boldsymbol{r})_l(\nabla_k B_E)(\nabla_l B_E) \tag{3.33a}$$

$$= -[K_5(\delta_{ik}\delta_{jl}+\delta_{il}\delta_{jk}) + K_6\delta_{ij}\delta_{kl}]T^{-4}\nabla_k T\nabla_l T \tag{3.33b}$$

$$= -T^{-4}[2K_5\boldsymbol{\nabla}T\,\boldsymbol{\nabla}T + K_6|\boldsymbol{\nabla}T|^2\,\boldsymbol{I}] \tag{3.33c}$$

$$= \eta_5^{\mathrm{B}^{(2)}_{EE}}\boldsymbol{\nabla}T\,\boldsymbol{\nabla}T + \eta_6^{\mathrm{B}^{(2)}_{EE}}|\boldsymbol{\nabla}T|^2\boldsymbol{I}, \tag{3.33d}$$

where

$$\boxed{\eta_5^{\mathrm{B}^{(2)}_{EE}} \doteq -2T^{-4}K_5,} \quad \boxed{\eta_6^{\mathrm{B}^{(2)}_{EE}} \doteq -T^{-4}K_6.} \tag{3.34}$$

---

<u>$-\mathrm{P}(\boldsymbol{P},E,E)$</u>:

The evaluation of this term proceeds along the same lines as for $-\mathrm{P}(\boldsymbol{P},E,N)$, and leads to a contribution proportional to

$$\frac{\delta}{\delta[-\beta(\boldsymbol{r}')]}\left(\frac{h(\overline{\boldsymbol{x}})}{mn}\right)T^{-4}\boldsymbol{\nabla}T\,\boldsymbol{\nabla}T. \tag{3.35}$$

Now

$$\left(\frac{\partial}{\partial(-\beta)}\right)_{\beta\mu} = \left(\frac{\partial}{\partial(-\beta)}\right)_p + \left(\frac{\partial p}{\partial(-\beta)}\right)_{\beta\mu}\left(\frac{\partial}{\partial p}\right)_T \tag{3.36a}$$

$$= T^2\left(\frac{\partial h}{\partial T}\right)_p + T^2\left(\frac{\partial p}{\partial T}\right)_{\beta\mu}\left(\frac{\partial h}{\partial p}\right)_T. \tag{3.36b}$$

To evaluate $\partial p/\partial T$, rewrite the Gibbs–Duhem relation

$$s\,\mathrm{d}T - \,\mathrm{d}p + n\,\mathrm{d}\mu = 0 \tag{3.37}$$

as

$$s\,\mathrm{d}T - \,\mathrm{d}p + n\beta^{-1}\mathrm{d}(\beta\mu) - n\mu\underbrace{\beta\,\mathrm{d}\beta}_{dT/T} = 0 \tag{3.38}$$

or at constant $\beta\mu$,

$$\left(\frac{\partial p}{\partial T}\right)_{\beta\mu} = \beta(Ts+n\mu) = \beta h. \tag{3.39}$$

The $\partial/\partial p$ term cancels with the $\boldsymbol{\nabla} T$ term in $-\mathrm{P}(\boldsymbol{P}, E, N)$. Thus, one gets a contribution

$$-\mathrm{P}(\boldsymbol{P}, E, E)_{\boldsymbol{\nabla} T} = \eta_5^{\mathrm{B}^{(2)}_{\mathrm{P}EE}} \boldsymbol{\nabla} T\, \boldsymbol{\nabla} T + \eta_6^{\mathrm{B}^{(2)}_{\mathrm{P}EE}} |\boldsymbol{\nabla} T|^2, \tag{3.40}$$

where

$$\boxed{\eta_5^{\mathrm{B}^{(2)}_{\mathrm{P}EE}} \doteq T^{-2}\left[\frac{\partial}{\partial T}\left(\frac{h}{mn}\right)\right]_p K_{20},} \quad \boxed{\eta_6^{\mathrm{B}^{(2)}_{\mathrm{P}EE}} \doteq T^{-2}\left[\frac{\partial}{\partial T}\left(\frac{h}{mn}\right)\right]_p K_{21}.} \tag{3.41}$$

---

<u>$(\boldsymbol{P}, \boldsymbol{P}, \boldsymbol{P})$:</u>

$$(\boldsymbol{P}, \boldsymbol{P}, \boldsymbol{P}) = -\int_0^\infty \mathrm{d}s \int \mathrm{d}\boldsymbol{r}' \,\langle \widehat{\tau}_{ij}(\boldsymbol{r}) \mathrm{e}^{-\mathrm{i}\mathscr{L}s} \widehat{\mathscr{T}}_{kl} G'_m(\boldsymbol{r}')\rangle_0 (\boldsymbol{r}' - \boldsymbol{r})_n \nabla_l (T^{-1} u_k) \nabla_n (T^{-1} u_m) \tag{3.42a}$$

$$\begin{aligned}
&= -\{K_7 \delta_{ij}\delta_{kl}\delta_{mn} + K_8 \delta_{ij}(\delta_{km}\delta_{ln} + \delta_{kn}\delta_{lm}) \\
&\quad + K_9 \delta_{kl}(\delta_{im}\delta_{jn} + \delta_{in}\delta_{jm}) + K_{10}\delta_{mn}(\delta_{ik}\delta_{jl} + \delta_{il}\delta_{jk}) \\
&\quad + K_{11}[\delta_{km}(\delta_{il}\delta_{jn} + \delta_{jl}\delta_{in}) + \delta_{lm}(\delta_{ik}\delta_{jn} + \delta_{jk}\delta_{in})] \\
&\quad + K_{12}[\delta_{kn}(\delta_{il}\delta_{jm} + \delta_{jl}\delta_{im}) + \delta_{ln}(\delta_{ik}\delta_{jm} + \delta_{jk}\delta_{im})]\} \\
&\quad \times T^{-2}(\nabla_l u_k)(\nabla_n u_m)
\end{aligned} \tag{3.42b}$$

$$\begin{aligned}
&= -T^{-2}K_7(\boldsymbol{\nabla}\cdot\boldsymbol{u})^2 \boldsymbol{I} - T^{-2}K_8(\nabla_l u_k)(\nabla_l u_k + \nabla_k u_l)\,\boldsymbol{I} \\
&\quad - T^{-2}\{K_9(\boldsymbol{\nabla}\cdot\boldsymbol{u})[(\boldsymbol{\nabla}\boldsymbol{u})^{\mathrm{T}} + (\boldsymbol{\nabla}\boldsymbol{u})] + K_{10}(\boldsymbol{\nabla}\cdot\boldsymbol{u})[(\boldsymbol{\nabla}\boldsymbol{u})^{\mathrm{T}} + (\boldsymbol{\nabla}\boldsymbol{u})]\} \\
&\quad - T^{-2}K_{11}[(\nabla_i u_k)(\nabla_j u_k) + (\nabla_j u_k)(\nabla_i u_k) \\
&\qquad\qquad + (\nabla_l u_i)(\nabla_j u_l) + (\nabla_l u_j)(\nabla_i u_l)] \\
&\quad - T^{-2}K_{12}[(\nabla_i u_k)(\nabla_k u_j) + (\nabla_j u_k)(\nabla_k u_i) \\
&\qquad\qquad + (\nabla_l u_i)(\nabla_l u_j) + (\nabla_l u_j)(\nabla_l u_i)].
\end{aligned} \tag{3.42c}$$

$$\begin{aligned}
&= -T^{-2}K_7(\boldsymbol{\nabla}\cdot\boldsymbol{u})^2\boldsymbol{I} - T^{-2}K_8(\boldsymbol{S} + \not{\boldsymbol{\Omega}}) : (2\boldsymbol{S})\boldsymbol{I} - T^{-2}(K_9 + K_{10})(\boldsymbol{\nabla}\cdot\boldsymbol{u})(2\boldsymbol{S}) \\
&\quad - T^{-2}\{K_{11}[2(\boldsymbol{\nabla}\boldsymbol{u})\cdot(\boldsymbol{\nabla}\boldsymbol{u})^{\mathrm{T}} + (\boldsymbol{\nabla}\boldsymbol{u})^{\mathrm{T}}\cdot(\boldsymbol{\nabla}\boldsymbol{u})^{\mathrm{T}} + (\boldsymbol{\nabla}\boldsymbol{u})\cdot(\boldsymbol{\nabla}\boldsymbol{u})] \\
&\quad + K_{12}[(\boldsymbol{\nabla}\boldsymbol{u})\cdot(\boldsymbol{\nabla}\boldsymbol{u}) + (\boldsymbol{\nabla}\boldsymbol{u})^{\mathrm{T}}\cdot(\boldsymbol{\nabla}\boldsymbol{u})^{\mathrm{T}} + 2(\boldsymbol{\nabla}\boldsymbol{u})^{\mathrm{T}}\cdot(\boldsymbol{\nabla}\boldsymbol{u})]\}.
\end{aligned} \tag{3.42d}$$

The last two lines are

$$\begin{aligned}
&-T^{-2}\{K_{11}[2(\boldsymbol{S} - \boldsymbol{\Omega})\cdot(\boldsymbol{S} + \boldsymbol{\Omega}) + (\boldsymbol{S} + \boldsymbol{\Omega})\cdot(\boldsymbol{S} + \boldsymbol{\Omega}) + (\boldsymbol{S} - \boldsymbol{\Omega})\cdot(\boldsymbol{S} - \boldsymbol{\Omega})] \\
&\quad + K_{12}[(\boldsymbol{S} - \boldsymbol{\Omega})\cdot(\boldsymbol{S} - \boldsymbol{\Omega}) + (\boldsymbol{S} + \boldsymbol{\Omega})\cdot(\boldsymbol{S} + \boldsymbol{\Omega}) + 2(\boldsymbol{S} + \boldsymbol{\Omega})\cdot(\boldsymbol{S} - \boldsymbol{\Omega})
\end{aligned} \tag{3.43a}$$

$$= -T^{-2}[4(K_{11} + K_{12})\boldsymbol{S}\cdot\boldsymbol{S} + 2(K_{11} - K_{12})(\boldsymbol{S}\cdot\boldsymbol{\Omega} - \boldsymbol{\Omega}\cdot\boldsymbol{S}). \tag{3.43b}$$

Thus,

$$\begin{aligned}
(\boldsymbol{P}, \boldsymbol{P}, \boldsymbol{P}) &= \eta_9^{\mathrm{B}^{(2)}_{\boldsymbol{PP}}}(\boldsymbol{\nabla}\cdot\boldsymbol{u})\boldsymbol{I} + \eta_{10}^{\mathrm{B}^{(2)}_{\boldsymbol{PP}}}\boldsymbol{S} : \boldsymbol{S} + \eta_{11}^{\mathrm{B}^{(2)}_{\boldsymbol{PP}}}(\boldsymbol{\nabla}\cdot\boldsymbol{u})\boldsymbol{S} \\
&\quad + \eta_{12}^{\mathrm{B}^{(2)}_{\boldsymbol{PP}}}\boldsymbol{S}\cdot\boldsymbol{S} + \eta_{13}^{\mathrm{B}^{(2)}_{\boldsymbol{PP}}}(\boldsymbol{S}\cdot\boldsymbol{\Omega} - \boldsymbol{\Omega}\cdot\boldsymbol{S}),
\end{aligned} \tag{3.44a}$$

where

$$\boxed{\eta_9^{\mathrm{B}^{(2)}_{\boldsymbol{PP}}} \doteq -T^{-2}K_7,} \quad \boxed{\eta_{10}^{\mathrm{B}^{(2)}_{\boldsymbol{PP}}} \doteq -2T^{-2}K_8,} \quad \boxed{\eta_{11}^{\mathrm{B}^{(2)}_{\boldsymbol{PP}}} \doteq -2T^{-2}(K_9 + K_{10}),} \tag{3.45}$$

and

$$\boxed{\eta_{12}^{\mathrm{B}^{(2)}_{\boldsymbol{PP}}} \doteq -4T^{-2}(K_{11} + K_{12}),} \quad \boxed{\eta_{13}^{\mathrm{B}^{(2)}_{\boldsymbol{PP}}} \doteq -2T^{-2}(K_{11} - K_{12}).} \tag{3.46}$$

$\underline{-\mathrm{P}(\boldsymbol{P},\boldsymbol{P},\boldsymbol{P})}$:

$$-\mathrm{P}(\boldsymbol{P},\boldsymbol{P},\boldsymbol{P}) = \int_0^\infty \mathrm{d}s \int \mathrm{d}\boldsymbol{r}' \, \langle \widehat{\tau}_{ij}(\boldsymbol{r}) \mathrm{e}^{-\mathrm{i}\mathscr{L}s} \mathrm{P} \widehat{\mathscr{T}}_{mn} G_s'(\boldsymbol{r}') \rangle (\boldsymbol{r}'-\boldsymbol{r})_l T^{-2} (\nabla_n u_m)(\nabla_l u_s). \tag{3.47}$$

Symmetry implies that only $\boldsymbol{P}'$ enters the projection. Thus, one needs to calculate $\langle \boldsymbol{P}'(\overline{\boldsymbol{x}}) \widehat{\boldsymbol{\mathscr{T}}} \boldsymbol{P}'(\boldsymbol{r}') \rangle_{kmns}$. Recall that

$$\widehat{\boldsymbol{\tau}} = \boldsymbol{\tau} - \boldsymbol{I}(p + N'p_n + E'p_e). \tag{3.48}$$

Therefore,

$$\widehat{\boldsymbol{\mathscr{T}}} = \boldsymbol{\mathscr{T}} - \boldsymbol{I}\biggl[ \int \mathrm{d}\boldsymbol{y}\, p(\boldsymbol{y}) + \left( \widetilde{\mathscr{E}} - \int \mathrm{d}\boldsymbol{y}\, e(\boldsymbol{y}) \right) \underbrace{\left( \frac{\partial p}{\partial e} \right)_n}_{p_e} \biggr]. \tag{3.49}$$

[Note $\int \mathrm{d}\boldsymbol{y}\, N'(\boldsymbol{y}) = 0$.] One thus has

$$\begin{aligned} \langle \boldsymbol{P}'(\overline{\boldsymbol{x}}) \widehat{\boldsymbol{\mathscr{T}}} \boldsymbol{P}'(\boldsymbol{r}') \rangle = m \sum_{ijk} \Biggl\langle & \delta(\overline{\boldsymbol{x}} - \boldsymbol{x}_i) \delta(\boldsymbol{r}' - \boldsymbol{x}_k) \\ & \times \left[ m\boldsymbol{w}_i (m\boldsymbol{w}_j\boldsymbol{w}_j + \Delta\boldsymbol{\tau}_{\boldsymbol{j}}) - \boldsymbol{I}N^{-1} \left( \int p + \sum_l (E_l - \langle E_l \rangle) \right) \right] \boldsymbol{w}_k \Biggr\rangle. \end{aligned} \tag{3.50$a$}$$

Consider the integration over velocity directions. If $k \neq i$, this vanishes. There are then the possibilities $j \neq i$ (with $l = j$ and $\lambda_{\neq} j$) and $j = i$. For $j \neq i$, one has

$$\left\langle m\boldsymbol{w}_j\boldsymbol{w}_j + \Delta\boldsymbol{\tau}_j - \boldsymbol{I}N - 1 \int p \right\rangle = 0. \tag{3.51}$$

The energy terms contribute

$$-\boldsymbol{I}_{mn} \langle m\boldsymbol{w}_i\boldsymbol{w}_i (E_i - \langle E_i \rangle) \rangle = -a\delta_{ks}\delta_{mn}, \tag{3.52}$$

where

$$a = \frac{1}{3} \left\langle mw^2 \left( \frac{1}{2} m (w^2 - \langle w^2 \rangle) \right) \right\rangle = \frac{1}{6} T^2 (15 - 9) = T^2. \tag{3.53}$$

Remaining is

$$m^2 \left\langle \boldsymbol{w} \left( \boldsymbol{w}\boldsymbol{w} - \frac{1}{3} \langle w^2 \rangle \boldsymbol{I} \right) \boldsymbol{w} \right\rangle_{kmns} = a\delta_{ks}\delta_{mn} + b(\delta_{km}\delta_{ns} + \delta_{kn}\delta_{ms}). \tag{3.54}$$

Taking traces with $(ks)$ and $(mn)$ gives

$$3(3a + 2b) = m^2 \langle w^2 (w^2 - \langle w^2 \rangle) \rangle = 6. \tag{3.55}$$

Taking traces with $(km)$ and $(ns)$ gives

$$3a + 12b = \langle w^2 w^2 - \tfrac{1}{3} \langle w^2 \rangle w^2 \rangle = 15 - 3 = 12. \tag{3.56}$$

The solution of the system

$$3a + 2b = 2, \tag{3.57$a$}$$

$$3a + 12b = 12 \tag{3.57$b$}$$

is $a = 0$ and $b = 1$. Therefore,

$$\langle \boldsymbol{P}'(\overline{\boldsymbol{x}})\widehat{\mathscr{T}}\boldsymbol{P}'(\boldsymbol{r}')\rangle_{kmns} = mnT^2\delta(\overline{\boldsymbol{x}} - \boldsymbol{r}')(\delta_{km}\delta_{ns} + \delta_{kn}\delta_{ms} - p_e\,\delta_{ks}\delta_{mn}). \tag{3.58}$$

Upon dividing by $m\overline{n}T$ (from $\mathscr{M}_{\boldsymbol{PP}}$) and performing the integral over $\overline{\boldsymbol{x}}$, one has

$$\begin{aligned}
-\,&\mathrm{P}(\boldsymbol{P},\boldsymbol{P},\boldsymbol{P})\\
&= T^{-1}[K_{20}(\delta_{ik}\delta_{jl} + \delta_{il}\delta_{jk}) + K_{21}\delta_{ij}\delta_{kl}]\\
&\quad\times[\delta_{km}\delta_{ns} + \delta_{kn}\delta_{ms} - p_e\,\delta_{ks}\delta_{mn}](\nabla_n u_m)(\nabla_l u_s) &(3.59a)\\
&= T^{-1}[K_{20}(\delta_{ik}\delta_{jl} + \delta_{il}\delta_{jk}) + K_{21}\delta_{ij}\delta_{kl}]\\
&\quad\times[(\nabla_n u_k)(\nabla_l u_n) + (\nabla_k u_m)(\nabla_l u_m) - p_e\,(\boldsymbol{\nabla}\cdot\boldsymbol{u})(\nabla_l u_k)] &(3.59b)\\
&= T^{-1}K_{20}\{(\nabla_n u_i)(\nabla_j u_n) + (\nabla_n u_j)(\nabla_i u_n) + (\nabla_i u_m)(\nabla_j u_m) + (\nabla_j u_m)(\nabla_i u_m)\\
&\qquad - p_e(\boldsymbol{\nabla}\cdot\boldsymbol{u}))[(\nabla_j u_i) + (\nabla_i u_j)]\}\\
&\quad + T^{-1}K_{21}\delta_{ij}[(\nabla_n u_l)(\nabla_l u_n) + (\nabla_k u_m)(\nabla_k u_m) - p_e(\boldsymbol{\nabla}\cdot\boldsymbol{u})^2] &(3.59c)\\
&= T^{-1}K_{20}[(\boldsymbol{\nabla}\boldsymbol{u})^{\mathrm{T}}\cdot(\boldsymbol{\nabla}\boldsymbol{u})^{\mathrm{T}} + (\boldsymbol{\nabla}\boldsymbol{u})\cdot(\boldsymbol{\nabla}\boldsymbol{u}) + (\boldsymbol{\nabla}\boldsymbol{u})\cdot(\boldsymbol{\nabla}\boldsymbol{u})^{\mathrm{T}} + (\boldsymbol{\nabla}\boldsymbol{u})\cdot(\boldsymbol{\nabla}\boldsymbol{u})^{\mathrm{T}}\\
&\quad - p_e(\boldsymbol{\nabla}\cdot\boldsymbol{u})((\boldsymbol{\nabla}\boldsymbol{u})^{\mathrm{T}} + (\boldsymbol{\nabla}\boldsymbol{u}))]\\
&\quad + T^{-1}K_{21}\boldsymbol{I}\{\mathrm{Tr}[(\boldsymbol{\nabla}\boldsymbol{u})^{\mathrm{T}} + (\boldsymbol{\nabla}\boldsymbol{u})]\cdot(\boldsymbol{\nabla}\boldsymbol{u})] - p_e(\boldsymbol{\nabla}\cdot\boldsymbol{u})^2\} &(3.59d)\\
&= T^{-1}K_{20}[4\boldsymbol{S}\cdot\boldsymbol{S} + 2(\boldsymbol{S}\cdot\boldsymbol{\Omega} - \boldsymbol{\Omega}\cdot\boldsymbol{S}) - 2p_e(\boldsymbol{\nabla}\cdot\boldsymbol{u})\boldsymbol{S}] + T^{-1}K_{21}\boldsymbol{I}[2\boldsymbol{S}:\boldsymbol{S} - p_e(\boldsymbol{\nabla}\cdot\boldsymbol{u})^2] &(3.59e)\\
&= \eta_{12}^{\mathrm{B}^{(2)}_{\mathrm{P}\boldsymbol{PP}}}\boldsymbol{S}\cdot\boldsymbol{S} + \eta_{13}^{\mathrm{B}^{(2)}_{\mathrm{P}\boldsymbol{PP}}}(\boldsymbol{S}\cdot\boldsymbol{\Omega} - \boldsymbol{\Omega}\cdot\boldsymbol{S}) + \eta_{11}^{\mathrm{B}^{(2)}_{\mathrm{P}\boldsymbol{PP}}}(\boldsymbol{\nabla}\cdot\boldsymbol{u})\boldsymbol{S}\\
&\quad + \eta_{10}^{\mathrm{B}^{(2)}_{\mathrm{P}\boldsymbol{PP}}}\boldsymbol{S}:\boldsymbol{S} + \eta_{9}^{\mathrm{B}^{(2)}_{\mathrm{P}\boldsymbol{PP}}}(\boldsymbol{\nabla}\cdot\boldsymbol{u})^2\boldsymbol{I}, &(3.59f)
\end{aligned}$$

where

$$\boxed{\eta_{9}^{\mathrm{B}^{(2)}_{\mathrm{P}\boldsymbol{PP}}} \doteq -\mathrm{T}^{-1}p_e K_{21},}\quad \boxed{\eta_{10}^{\mathrm{B}^{(2)}_{\mathrm{P}\boldsymbol{PP}}} \doteq 2T^{-1}K_{21},} \tag{3.60}$$

and

$$\boxed{\eta_{11}^{\mathrm{B}^{(2)}_{\mathrm{P}\boldsymbol{PP}}} \doteq -2T^{-1}p_e K_{20},}\quad \boxed{\eta_{12}^{\mathrm{B}^{(2)}_{\mathrm{P}\boldsymbol{PP}}} \doteq 4T^{-1}K_{20},}\quad \boxed{\eta_{13}^{\mathrm{B}^{(2)}_{\mathrm{P}\boldsymbol{PP}}} \doteq 2T^{-1}K_{20}.} \tag{3.61}$$

3.2.2. Nonlinear Burnett — energy

Again there are no $\beta = N$ terms, so

$$\langle \boldsymbol{J}^E\rangle_{\mathrm{diss}}^{\mathrm{B}^{(2)}} = (E,\boldsymbol{P},N) + (E,\boldsymbol{P},E) + (E,E,\boldsymbol{P}). \tag{3.62}$$

---

<u>$(E,E,\boldsymbol{P})$</u>:

$$\begin{aligned}
(E,E,\boldsymbol{P}) &= -\int_0^\infty \mathrm{d}s\int \mathrm{d}\boldsymbol{r}'\,\langle \widehat{J}_i^E(\boldsymbol{r})\mathrm{e}^{-\mathrm{i}\mathscr{L}s}\,\widehat{\mathscr{J}}_j^E\mathrm{G}_k'\rangle(\boldsymbol{r}' - \boldsymbol{r})_l[\nabla_j(-T^{-1})](\nabla_l(T^{-1}u_k)) &(3.63a)\\
&= -(K_{13}\delta_{ij}\delta_{kl} + K_{14}\delta_{ik}\delta_{jl} + K_{15}\delta_{il}\delta_{jk})T^{-3}(\nabla_j T)(\nabla_l u_k) &(3.63b)\\
&= -T^{-3}[K_{13}(\boldsymbol{\nabla}\cdot\boldsymbol{u})\boldsymbol{\nabla}T + K_{14}(\boldsymbol{\nabla}\boldsymbol{u})^{\mathrm{T}}\cdot\boldsymbol{\nabla}T + K_{15}(\boldsymbol{\nabla}\boldsymbol{u})\cdot\boldsymbol{\nabla}T] &(3.63c)\\
&= \lambda_4^{\mathrm{B}^{(2)}_{\boldsymbol{EP}}}(\boldsymbol{\nabla}\cdot\boldsymbol{u})\boldsymbol{\nabla}T + \lambda_5^{\mathrm{B}^{(2)}_{\boldsymbol{EP}}}\boldsymbol{S}\cdot\boldsymbol{\nabla}T + \lambda_6^{\mathrm{B}^{(2)}_{\boldsymbol{EP}}}\boldsymbol{\Omega}\cdot\boldsymbol{\nabla}T, &(3.63d)
\end{aligned}$$

where

$$\boxed{\lambda_4^{\mathrm{B}^{(2)}_{EP}} \doteq -T^{-3}K_{13},} \quad \boxed{\lambda_5^{\mathrm{B}^{(2)}_{EP}} \doteq -T^{-3}(K_{14}+K_{15}),} \quad \boxed{\lambda_6^{\mathrm{B}^{(2)}_{EP}} \doteq -T^{-3}(K_{14}-K_{15}).} \tag{3.64}$$

---

$-\mathrm{P}(E,E,\boldsymbol{P})$:

$$-\mathrm{P}(E,E,\boldsymbol{P}) = T^{-3}\int_0^\infty \mathrm{d}s \int \mathrm{d}\boldsymbol{r}'\, \langle \widehat{J}_i^E(\boldsymbol{r})\mathrm{e}^{-\mathrm{i}\mathscr{L}s}\mathrm{P}[\widehat{\mathscr{J}}_j^E G_k'(\boldsymbol{r}')]\rangle(\boldsymbol{r}'-\boldsymbol{r})_l(\nabla_j T)(\nabla_l u_k). \tag{3.65a}$$

By symmetry, only the scalar quantities enter the projection. One thus needs to work out $\langle N'(\overline{\boldsymbol{x}})\widehat{\boldsymbol{\mathscr{J}}}^E\boldsymbol{P}'(\boldsymbol{r}')\rangle$ and $\langle E'(\overline{\boldsymbol{x}})\widehat{\boldsymbol{\mathscr{J}}}^E\boldsymbol{P}'(\boldsymbol{r}')\rangle$. Then

$$-\mathrm{P}(E,E,\boldsymbol{P}) = T^{-3}\begin{pmatrix}\int_0^\infty \mathrm{d}s\int\mathrm{d}\boldsymbol{r}'\,\langle \widehat{J}_i^E\mathrm{e}^{-\mathrm{i}\mathscr{L}s}N'(\overline{\boldsymbol{x}})\rangle(\boldsymbol{r}'-\boldsymbol{r})_l\\ \int_0^\infty \mathrm{d}s\int\mathrm{d}\boldsymbol{r}'\,\langle \widehat{J}_i^E\mathrm{e}^{-\mathrm{i}\mathscr{L}s}E'(\overline{\boldsymbol{x}})\rangle(\boldsymbol{r}'-\boldsymbol{r})_l\end{pmatrix}^{\mathrm{T}}$$
$$\cdot\,\boldsymbol{\mathscr{M}}_2^{-1}(\overline{\boldsymbol{x}},\overline{\boldsymbol{x}}')\cdot\begin{pmatrix}\langle N'(\overline{\boldsymbol{x}}')\widehat{\mathscr{J}}_j^E G_k'(\boldsymbol{r}')\rangle\\ \langle E'(\overline{\boldsymbol{x}}')\widehat{\mathscr{J}}_j^E G_k'(\boldsymbol{r}')\rangle\end{pmatrix}(\nabla_j T)(\nabla_l u_k). \tag{3.66a}$$

From Novikov's theorem for arbitrary $A$ ($= N'$ or $E'$),

$$\boldsymbol{\mathscr{M}}_2^{-1}(\overline{\boldsymbol{x}},\overline{\boldsymbol{x}}')\cdot\begin{pmatrix}\langle N'(\overline{\boldsymbol{x}}')\widehat{\mathscr{J}}_j^E G_k'(\boldsymbol{r}')\rangle\\ \langle E'(\overline{\boldsymbol{x}}')\widehat{\mathscr{J}}_j^E G_k'(\boldsymbol{r}')\rangle\end{pmatrix} = \begin{pmatrix}\left\langle \dfrac{\delta[\widehat{\boldsymbol{\mathscr{J}}}^E\boldsymbol{P}'(\boldsymbol{r}')]}{\delta N'(\overline{\boldsymbol{x}})}\right\rangle\\ \left\langle \dfrac{\delta[\widehat{\boldsymbol{\mathscr{J}}}^E\boldsymbol{P}'(\boldsymbol{r}')]}{\delta E'(\overline{\boldsymbol{x}})}\right\rangle\end{pmatrix}. \tag{3.67}$$

Now

$$\widehat{\boldsymbol{\mathscr{J}}}^E\boldsymbol{P}'(\boldsymbol{r}') = \sum_{i=1}^{\mathscr{N}}[E_i\boldsymbol{v}_i + \Delta\boldsymbol{\tau}_i\cdot\boldsymbol{v}_i - (h/n)\boldsymbol{v}_i]\sum_{j=1}^{\mathscr{N}} m\boldsymbol{v}_j\delta(\boldsymbol{r}'-\boldsymbol{x}_j). \tag{3.68}$$

Because $N'$ and $E'$ are scalars, one can average over the velocity angles. Individual particles are uncorrelated, so $j=i$. Isotropy makes the result proportional to the unit tensor. Thus,

$$\langle\widehat{\boldsymbol{\mathscr{J}}}^E\boldsymbol{P}'(\boldsymbol{r}')\rangle_\Omega = \sum_{i=1}^{\mathscr{N}}\left[\left(\frac{1}{2}mv_i^2+U_i\right)\frac{1}{3}mv_i^2\boldsymbol{I} + \Delta\boldsymbol{\tau}\left(\frac{1}{3}mv_i^2\right) - \left(\frac{h}{n}\right)\left(\frac{1}{3}mv_i^2\right)\boldsymbol{I}\right]\delta(\boldsymbol{r}'-\boldsymbol{x}_i). \tag{3.69}$$

Now average over velocity magnitudes:

$$\langle\widehat{\boldsymbol{\mathscr{J}}}^E\boldsymbol{P}'(\boldsymbol{r}')\rangle_{\boldsymbol{v}} = \sum_{i=1}^{\mathscr{N}}\left[\left(\frac{5}{2}T^2+TU_i\right)\boldsymbol{I} + \Delta\boldsymbol{\tau}_i - \left(\frac{h}{n}\right)T\boldsymbol{I}\right]\delta(\boldsymbol{r}'-\boldsymbol{x}_i). \tag{3.70}$$

After the remaining average, the entire result will be proportional to $\boldsymbol{I}$, so $\Delta\boldsymbol{\tau}_i \to \frac{1}{3}\,\mathrm{Tr}\,\Delta\boldsymbol{\tau}$. Note that the last, $h$ term is proportional to $\widetilde{N}(\boldsymbol{r}')$. One has

$$-\left(\frac{h}{n}\right)\widetilde{N} = -\left(\frac{h}{n}\right)N' - h, \tag{3.71}$$

so

$$\langle \widehat{\mathscr{J}}^{E} \boldsymbol{P}'(\boldsymbol{r}') \rangle_{\boldsymbol{v}} = T \Bigg[ \underbrace{\frac{3}{2} N'(\boldsymbol{r}') T + U'(\boldsymbol{r}')}_{E'} + P'(\boldsymbol{r}') - \left(\frac{h}{n}\right) N'(\boldsymbol{r}') \Bigg], \tag{3.72}$$

where the identification with $E'$ is made at constant $T$. Thus,

$$\mathscr{M}_2^{-1}(\overline{\boldsymbol{x}}, \overline{\boldsymbol{x}}') \cdot \begin{pmatrix} \langle N'(\overline{\boldsymbol{x}}') \widehat{\mathscr{J}}_j^E G_k'(\boldsymbol{r}') \rangle \\ \langle E'(\overline{\boldsymbol{x}}') \widehat{\mathscr{J}}_j^E G_k'(\boldsymbol{r}') \rangle \end{pmatrix} = \begin{pmatrix} \left\langle \dfrac{\delta[\widehat{\mathscr{J}}^E \boldsymbol{P}'(\boldsymbol{r}')]}{\delta N'(\overline{\boldsymbol{x}})} \right\rangle \\ \left\langle \dfrac{\delta[\widehat{\mathscr{J}}^E \boldsymbol{P}'(\boldsymbol{r}')]}{\delta E'(\overline{\boldsymbol{x}})} \right\rangle \end{pmatrix} \tag{3.73a}$$

$$= T \begin{pmatrix} \left\langle \dfrac{\delta}{\delta N'(\overline{\boldsymbol{x}})} \left[ E'(\boldsymbol{r}') + P'(\boldsymbol{r}') - \left(\dfrac{h}{n}\right) N'(\boldsymbol{r}') \right] \right\rangle \\ \left\langle \dfrac{\delta}{\delta E'(\overline{\boldsymbol{x}})} \left[ [E'(\boldsymbol{r}') + P'(\boldsymbol{r}') - \left(\dfrac{h}{n}\right) N'(\boldsymbol{r}') \right] \right\rangle \end{pmatrix} \tag{3.73b}$$

$$= T \begin{pmatrix} \left\langle \dfrac{\delta P'(\boldsymbol{r}')}{\delta N'(\overline{\boldsymbol{x}})} - \left(\dfrac{h}{n}\right) \delta(\overline{\boldsymbol{x}} - \boldsymbol{r}') \right\rangle \\ \delta(\overline{\boldsymbol{x}} - \boldsymbol{r}') + \left\langle \dfrac{\delta P'(\boldsymbol{r}')}{\delta E'(\overline{\boldsymbol{x}})} \right\rangle \end{pmatrix} \tag{3.73c}$$

and

$$-\mathrm{P}(E, E, \boldsymbol{P}) = T^{-2} \left[ K_{22} \left( \frac{\partial p}{\partial n} - \frac{h}{n} \right) + K_{23} \left( 1 + \frac{\partial p}{\partial e} \right) \right] (\nabla_j T) \underbrace{(\nabla_i u_j)}_{\boldsymbol{S} - \boldsymbol{\Omega}}. \tag{3.74a}$$

$$= \lambda_5^{\mathrm{B}^{(2)}_{\mathrm{P}E\boldsymbol{P}}} \boldsymbol{S} \cdot \boldsymbol{\nabla} T + \lambda_6^{\mathrm{B}^{(2)}_{\mathrm{P}E\boldsymbol{P}}} \boldsymbol{\Omega} \cdot \boldsymbol{\nabla} T, \tag{3.74b}$$

where

$$\boxed{\lambda_5^{\mathrm{B}^{(2)}_{\mathrm{P}E\boldsymbol{P}}} \doteq T^{-2} K_{22} \left( \frac{\partial p}{\partial n} - \frac{h}{n} \right) + K_{23} \left( 1 + \frac{\partial p}{\partial e} \right),} \tag{3.75a}$$

$$\boxed{\lambda_6^{\mathrm{B}^{(2)}_{\mathrm{P}E\boldsymbol{P}}} \doteq -T^{-2} \left[ K_{22} \left( \frac{\partial p}{\partial n} - \frac{h}{n} \right) + K_{23} \left( 1 + \frac{\partial p}{\partial e} \right) \right].} \tag{3.75b}$$

$\underline{(E, \boldsymbol{P}, N)}$:

$$(E, \boldsymbol{P}, N) = -\int_0^\infty \mathrm{d}s \int \mathrm{d}\boldsymbol{r}' \, \langle \widehat{J}_i^E(\boldsymbol{r}) \mathrm{e}^{-\mathrm{i}\mathscr{L}s} \widehat{\mathscr{T}}_{jk} N' \rangle (\boldsymbol{r}' - \boldsymbol{r})_l [\nabla_k (T^{-1} u_j)] (\nabla_l B_N) \quad (3.76a)$$

$$= -T^{-1}[K_{16}(\delta_{ij}\delta_{kl} + \delta_{ik}\delta_{jl}) + K_{17}\delta_{il}\delta_{jk}](\nabla_k u_j)(\nabla_l B_N) \quad (3.76b)$$

$$= -T^{-1}[2K_{16}\boldsymbol{\mathsf{S}} \cdot \boldsymbol{\nabla} B_N + K_{17}(\boldsymbol{\nabla} \cdot \boldsymbol{u}) \boldsymbol{\nabla} B_N \quad (3.76c)$$

$$= -(\overline{n}T^2)^{-1}[2K_{16}\boldsymbol{\mathsf{S}} \cdot (-hT^{-1}\boldsymbol{\nabla} T + \boldsymbol{\nabla} p) + K_{17}(\boldsymbol{\nabla} \cdot \boldsymbol{u})(-hT^{-1}\boldsymbol{\nabla} T + \boldsymbol{\nabla} p)] \quad (3.76d)$$

$$= \lambda_4^{\mathrm{B}^{(2)}_{\boldsymbol{P}N}} (\boldsymbol{\nabla} \cdot \boldsymbol{u}) \boldsymbol{\nabla} T + \lambda_5^{\mathrm{B}^{(2)}_{\boldsymbol{P}N}} \boldsymbol{\mathsf{S}} \cdot \boldsymbol{\nabla} T + \lambda_7^{\mathrm{B}^{(2)}_{\boldsymbol{P}N}} \boldsymbol{\mathsf{S}} \cdot \boldsymbol{\nabla} p + \lambda_8^{\mathrm{B}^{(2)}_{\boldsymbol{P}N}} (\boldsymbol{\nabla} \cdot \boldsymbol{u}) \boldsymbol{\nabla} p, \quad (3.76e)$$

where

$$\boxed{\lambda_4^{\mathrm{B}^{(2)}_{\boldsymbol{P}N}} \doteq (\overline{n}T^3)^{-1} h K_{17},} \quad \boxed{\lambda_5^{\mathrm{B}^{(2)}_{\boldsymbol{P}N}} \doteq 2(\overline{n}T^3)^{-1} h K_{16},} \quad (3.77)$$

and

$$\boxed{\lambda_7^{\mathrm{B}^{(2)}_{\boldsymbol{P}N}} \doteq -2(\overline{n}T^2)^{-1} K_{16},} \quad \boxed{\lambda_8^{\mathrm{B}^{(2)}_{\boldsymbol{P}N}} \doteq -(\overline{n}T^2)^{-1} K_{17}.} \quad (3.78)$$

---

$\underline{-\mathrm{P}(E, \boldsymbol{P}, N)}$:

$$-\mathrm{P}(E, \boldsymbol{P}, N) = T^{-1} \int_0^\infty \mathrm{d}s \int \mathrm{d}\boldsymbol{r}' \, \langle \widehat{J}_i^E(\boldsymbol{r}) \mathrm{e}^{-\mathrm{i}\mathscr{L}s} \mathrm{P} \widehat{\mathscr{T}}_{jk} N' \rangle (\boldsymbol{r}' - \boldsymbol{r})_l (\nabla_k u_j)(\nabla_l B_N) \quad (3.79a)$$

$$= (\overline{n}T^2)^{-1} \begin{pmatrix} \int_0^\infty \mathrm{d}s \int \mathrm{d}\boldsymbol{r}' \, \langle \widehat{J}_i^E \mathrm{e}^{-\mathrm{i}\mathscr{L}s} N'(\overline{\boldsymbol{x}}) \rangle (\boldsymbol{r}' - \boldsymbol{r})_l \\ \int_0^\infty \mathrm{d}s \int \mathrm{d}\boldsymbol{r}' \, \langle \widehat{J}_i^E \mathrm{e}^{-\mathrm{i}\mathscr{L}s} E'(\overline{\boldsymbol{x}}) \rangle (\boldsymbol{r}' - \boldsymbol{r})_l \end{pmatrix}^{\mathrm{T}} \cdot \boldsymbol{\mathscr{M}}_2^{-1}(\overline{\boldsymbol{x}}, \overline{\boldsymbol{x}}') \cdot \begin{pmatrix} \langle N'(\overline{\boldsymbol{x}}') \widehat{\mathscr{T}}_{jk} N'(\boldsymbol{r}') \rangle \\ \langle E'(\overline{\boldsymbol{x}}') \widehat{\mathscr{T}}_{jk} N'(\boldsymbol{r}') \rangle \end{pmatrix} (\nabla_k u_j)(-hT^{-1}\nabla_l T + \nabla_l p). \quad (3.79b)$$

By symmetry,

$$\langle N'(\overline{\boldsymbol{x}}) \widehat{\mathscr{T}}_{jk} N'(\boldsymbol{r}') \rangle = \frac{1}{3} \langle N'(\overline{\boldsymbol{x}}') \operatorname{Tr} \widehat{\boldsymbol{\mathscr{T}}} N'(\boldsymbol{r}') \rangle \delta_{jk} \quad (3.80a)$$

$$= -\left\langle N'(\overline{\boldsymbol{x}}') \left( \int d\boldsymbol{y} \, [N'(\boldsymbol{y}) p_n(\boldsymbol{y}) + E'(\boldsymbol{y}) p_e(\boldsymbol{y})] \right) N'(\boldsymbol{r}') \right\rangle \delta_{jk}. \quad (3.80b)$$

By Novikov's theorem for arbitrary $A$,

$$V \doteq \int d\overline{\boldsymbol{x}} \, A(\overline{\boldsymbol{x}}) \boldsymbol{\mathscr{M}}_2^{-1}(\overline{\boldsymbol{x}}, \overline{\boldsymbol{x}}') \cdot \int d\overline{\boldsymbol{x}} \, A(\overline{\boldsymbol{x}}) \begin{pmatrix} \langle N'(\overline{\boldsymbol{x}}') \widehat{\mathscr{T}}_{jk} N'(\boldsymbol{r}') \rangle \\ \langle E'(\overline{\boldsymbol{x}}') \widehat{\mathscr{T}}_{jk} N'(\boldsymbol{r}') \rangle \end{pmatrix}$$

$$= \begin{pmatrix} \left\langle \dfrac{\delta}{\delta N'(\overline{\boldsymbol{x}})} \left[ \left( \int d\boldsymbol{y} \, [N'(\boldsymbol{y}) p_n(\boldsymbol{y}) + E'(\boldsymbol{y}) p_e(\boldsymbol{y})] \right) N'(\boldsymbol{r}') \right] \right\rangle \\ \left\langle \dfrac{\delta}{\delta E'(\overline{\boldsymbol{x}})} \left[ \left( \int d\boldsymbol{y} \, [N'(\boldsymbol{y}) p_n(\boldsymbol{y}) + E'(\boldsymbol{y}) p_e(\boldsymbol{y})] \right) N'(\boldsymbol{r}') \right] \right\rangle \end{pmatrix}. \quad (3.81a)$$

In the first line, the expectation vanishes if the $\delta/\delta N'$ operates on either of the $N'$'s. Thus,

$$V = \int d\overline{\boldsymbol{x}}\, A(\overline{\boldsymbol{x}}) \begin{pmatrix} \left\langle N'(\boldsymbol{r}') \int d\boldsymbol{y} \left( N'(\boldsymbol{y}) \dfrac{\delta p_n(\boldsymbol{y})}{\delta N'(\overline{\boldsymbol{x}})} + E'(\boldsymbol{y}) \dfrac{\delta p_e(\boldsymbol{y})}{\delta N'(\overline{\boldsymbol{x}})} \right) \right\rangle \\ \left\langle N'(\boldsymbol{r}') \int d\boldsymbol{y} \left( N'(\boldsymbol{y}) \dfrac{\delta p_n(\boldsymbol{y})}{\delta E'(\overline{\boldsymbol{x}})} + E'(\boldsymbol{y}) \dfrac{\delta p_e(\boldsymbol{y})}{\delta E'(\overline{\boldsymbol{x}})} \right) \right\rangle \end{pmatrix} \tag{3.82a}$$

$$\approx A(\boldsymbol{r}') \begin{pmatrix} \dfrac{\partial}{\partial(\beta\mu)} \left( \dfrac{\partial p}{\partial n} \right) \\ \dfrac{\partial}{\partial(\beta\mu)} \left( \dfrac{\partial p}{\partial e} \right) \end{pmatrix}. \tag{3.82b}$$

Thus,

$$-\mathrm{P}(E, \boldsymbol{P}, N) = (\overline{n}T^2)^{-1} \left[ K_{22} \frac{\partial}{\partial(\beta\mu)} \left( \frac{\partial p}{\partial n} \right) + K_{23} \frac{\partial}{\partial(\beta\mu)} \left( \frac{\partial p}{\partial e} \right) \right] \times (\boldsymbol{\nabla} \cdot \boldsymbol{u})(-hT^{-1}\boldsymbol{\nabla} T + \boldsymbol{\nabla} P) \tag{3.83a}$$

$$= T^{-1} \left[ K_{22} \frac{\partial}{\partial p} \left( \frac{\partial p}{\partial n} \right) + K_{23} \frac{\partial}{\partial p} \left( \frac{\partial p}{\partial e} \right) \right] (\boldsymbol{\nabla} \cdot \boldsymbol{u})(-hT^{-1}\boldsymbol{\nabla} T + \boldsymbol{\nabla} p). \tag{3.83b}$$

In obtaining the last line, I used the result

$$\left( \frac{\partial}{\partial(\beta\mu)} \right)_\beta = nT \frac{\partial}{\partial p}, \tag{3.84}$$

which was derived in (3.30). The term in $(\boldsymbol{\nabla} \cdot \boldsymbol{u})\boldsymbol{\nabla} p$ is thus

$$\lambda_8^{\mathrm{B}^{(2)}_{\mathrm{P}\boldsymbol{P}N}} (\boldsymbol{\nabla} \cdot \boldsymbol{u}) \boldsymbol{\nabla} p, \tag{3.85}$$

where

$$\boxed{\lambda_8^{\mathrm{B}^{(2)}_{\mathrm{P}\boldsymbol{P}N}} \doteq \frac{\partial}{\partial p} \left( \frac{\partial p}{\partial n} \right) K_{22} + \frac{\partial}{\partial p} \left( \frac{\partial p}{\partial e} \right) K_{23}.} \tag{3.86}$$

The term in $\boldsymbol{\nabla} T$ will cancel against a contribution from $-\mathrm{P}(E, \boldsymbol{P}, E)$.

---

<u>$(E, \boldsymbol{P}, E)$</u>:

$$(E, \boldsymbol{P}, E) = -\int_0^\infty \mathrm{d}s \int \mathrm{d}\boldsymbol{r}' \, \langle \widehat{J}_i^E(\boldsymbol{r}) \mathrm{e}^{-\mathrm{i}\mathscr{L}s} \widehat{\mathscr{T}}_{jk} E' \rangle (\boldsymbol{r}' - \boldsymbol{r})_l [\nabla_k (T^{-1} u_j)](\nabla_l B_E) \tag{3.87a}$$

$$= -T^{-3} [K_{18} (\delta_{ij}\delta_{kl} + \delta_{ik}\delta_{jl}) + K_{19} \delta_{il}\delta_{jk}] (\nabla_k u_j)(\nabla_l T) \tag{3.87b}$$

$$= -T^{-3} [2K_{18} \mathsf{S} \cdot \boldsymbol{\nabla} T + K_{19} (\boldsymbol{\nabla} \cdot \boldsymbol{u}) \boldsymbol{\nabla} T] \tag{3.87c}$$

$$= \lambda_4^{\mathrm{B}^{(2)}_{\boldsymbol{P}E}} (\boldsymbol{\nabla} \cdot \boldsymbol{u}) \boldsymbol{\nabla} T + \lambda_5^{\mathrm{B}^{(2)}_{\boldsymbol{P}E}} \mathsf{S} \cdot \boldsymbol{\nabla} T, \tag{3.87d}$$

where

$$\boxed{\lambda_4^{\mathrm{B}^{(2)}_{\boldsymbol{P}E}} \doteq -T^{-3} K_{19}} \quad \boxed{\lambda_5^{\mathrm{B}^{(2)}_{\boldsymbol{P}E}} \doteq -2T^{-3} K_{18}.} \tag{3.88}$$

$-\mathrm{P}(E, \boldsymbol{P}, E)$:

The evaluation of $\mathrm{P}(E, \boldsymbol{P}, E)$ proceeds similarly to that for $\mathrm{P}(E, \boldsymbol{P}, N)$, and leads to

$$-\mathrm{P}(E, \boldsymbol{P}, E) = T^{-3}\left[K_{22}\frac{\partial}{\partial(-\beta)}\left(\frac{\partial p}{\partial n}\right) + K_{23}\frac{\partial}{\partial(-\beta)}\left(\frac{\partial p}{\partial e}\right)\right](\boldsymbol{\nabla}\cdot\boldsymbol{u})\boldsymbol{\nabla}T. \tag{3.89}$$

Now

$$\left(\frac{\partial}{\partial(-\beta)}\right)_{(\beta\mu)} = T^2\left(\frac{\partial}{\partial T}\right)_{(\beta\mu)}. \tag{3.90}$$

We have

$$\left(\frac{\partial}{\partial T}\right)_{(\beta\mu)} = \left(\frac{\partial}{\partial T}\right)_p + \left(\frac{\partial p}{\partial T}\right)_{(\beta\mu)}\left(\frac{\partial}{\partial p}\right)_T, \tag{3.91}$$

where the result

$$\left(\frac{\partial p}{\partial T}\right)_{(\beta\mu)} = \beta h \tag{3.92}$$

was derived in (3.39). The $\partial/\partial p$ term cancels with the $\boldsymbol{\nabla}T$ term in $-\mathrm{P}(E, \boldsymbol{P}, N)$. Thus, the $\boldsymbol{\nabla}T$ contributions from the $E$ projections are

$$-\left[\mathrm{P}(E, \boldsymbol{P}, N) + \mathrm{P}(E, \boldsymbol{P}, E)\right]_{\boldsymbol{\nabla}T} = \lambda_4^{\mathrm{B}^{(2)}_{\mathrm{P}\boldsymbol{P}(N+E)}}(\boldsymbol{\nabla}\cdot\boldsymbol{u})\boldsymbol{\nabla}T, \tag{3.93}$$

where

$$\boxed{\lambda_4^{\mathrm{B}^{(2)}_{\mathrm{P}\boldsymbol{P}(N+E)}} \doteq T^{-1}\frac{\partial}{\partial T}\left(\frac{\partial p}{\partial n}\right)K_{22} + T^{-1}\frac{\partial}{\partial T}\left(\frac{\partial p}{\partial e}\right)K_{23}.} \tag{3.94}$$

### 3.3. *Nonlinear Burnett terms — time derivatives*

The nonlinear Burnett terms relating to first-order time derivatives are

$$\underbrace{\left[\frac{\partial}{\partial t}\left(\boldsymbol{k}_2^{\beta}[\boldsymbol{J}^{\alpha}](\mu,t)\cdot\boldsymbol{\nabla}B_{\beta}(\boldsymbol{r},t)\right)\right]^{(1)}}_{\mathrm{B}^{\alpha}\partial_t\boldsymbol{k}_{\beta} + \mathrm{B}^{\alpha}\partial_t B_{\beta}}, \tag{3.95}$$

where

$$\boldsymbol{k}_2^{\beta}[\widetilde{\boldsymbol{J}}^{\alpha}](\mu,t) \doteq \int_0^{\infty}\mathrm{d}s\int\mathrm{d}\boldsymbol{r}'\,s\,\langle\widehat{\boldsymbol{J}}^{\alpha}(\boldsymbol{r})\mathrm{e}^{-\mathrm{i}\mathscr{L}s}\widehat{\boldsymbol{J}}^{\beta}(\boldsymbol{r}')\rangle = \int_0^{\infty}\mathrm{d}s\,s\langle\widehat{\boldsymbol{J}}^{\alpha}(\boldsymbol{r})\mathrm{e}^{-\mathrm{i}\mathscr{L}s}\widehat{\mathscr{J}}^{\beta}\rangle_0. \tag{3.96}$$

#### 3.3.1. Nonlinear Burnett — momentum ($\partial_t$)

Note that $\widehat{\mathscr{N}} = 0$, so $(\boldsymbol{P}, N)_{\partial_t} = 0$.

$(\boldsymbol{P}, \boldsymbol{P})_{\partial_t}$:

$$(\boldsymbol{P}, \boldsymbol{P})_{\partial_t} = \frac{\partial}{\partial t}\Bigg(\underbrace{\int_0^{\infty}\mathrm{d}s\,s\,\langle\widehat{\boldsymbol{\tau}}(\boldsymbol{r})\mathrm{e}^{-\mathrm{i}\mathscr{L}s}\widehat{\mathscr{T}}\rangle_{ijkl}}_{K^{\mathrm{IV}}(\delta_{ik}\delta_{jl}+\delta_{il}\delta_{jk}) + K^{\mathrm{V}}\delta_{ij}\delta_{kl}} : [\boldsymbol{\nabla}(\beta\boldsymbol{u})]_{kl}\Bigg)^{(1)}. \tag{3.97a}$$

One needs

$$\nabla \partial_t(\beta \boldsymbol{u}) = \nabla(-T^{-2}\partial_t T\, \boldsymbol{u} + T^{-1}\partial_t \boldsymbol{u}) \tag{3.98a}$$

$$= -T^{-2}(\partial_t T)(\nabla \boldsymbol{u}) - T^{-2}(\nabla T)(\partial_t \boldsymbol{u}) + T^{-1}\nabla(\partial_t \boldsymbol{u}). \tag{3.98b}$$

The momentum equation is through first order

$$\partial_t \boldsymbol{u} = -\boldsymbol{u}\cdot\nabla\boldsymbol{u} - (mn)^{-1}\nabla p. \tag{3.99}$$

Hence

$$\nabla \partial_t(\beta\boldsymbol{u}) = -T^{-2}(\partial_t T)(\nabla\boldsymbol{u}) + [\ldots], \tag{3.100}$$

where

$$[\ldots] \doteq (mnT^2)^{-1}(\nabla T)(\nabla p) - T^{-1}[(\nabla\boldsymbol{u})\cdot(\nabla\boldsymbol{u}) + (mn)^{-1}\nabla\nabla p - (mn)^{-1}(n^{-1}\nabla n)(\nabla p)]. \tag{3.101}$$

Define

$$\alpha \doteq -\frac{1}{n}\left(\frac{\partial n}{\partial T}\right)_p \quad \text{(expansion coefficient)}, \tag{3.102a}$$

$$\kappa_T \doteq \frac{1}{n}\left(\frac{\partial n}{\partial p}\right)_T \quad \text{(isothermal compressibility)} \tag{3.102b}$$

Then the contribution from $\partial_t\nabla(\beta\boldsymbol{u})$ is

$$\begin{aligned}&[K^{\mathrm{IV}}(\delta_{ik}\delta_{jl} + \delta_{il}\delta_{jk}) + K^{\mathrm{V}}\delta_{ij}\delta_{kl}]\\ &\quad\times\{-T^{-2}(\partial_t T)(\nabla u)\\ &\qquad + [(mnT)^{-1}(T^{-1}-\alpha)(\nabla T)(\nabla p)\\ &\qquad - T^{-1}(\nabla\boldsymbol{u})\cdot(\nabla\boldsymbol{u}) - (mnT)^{-1}\nabla\nabla p + (mnT)^{-1}\kappa_T(\nabla p)(\nabla p)]\},\end{aligned} \tag{3.103}$$

where the $[\ldots]$ piece is

$$\begin{aligned}&(mn)^{-1}\eta_1(T^{-1}-\alpha)(\nabla T\,\nabla p + \nabla p\,\nabla T) + (mn)^{-1}\eta_2(T^{-1}-\alpha)\nabla T\cdot\nabla p\\ &\quad - 2\eta_1[\boldsymbol{S}\cdot\boldsymbol{S} + \boldsymbol{\Omega}\cdot\boldsymbol{\Omega}] - \eta_2(\boldsymbol{S}:\boldsymbol{S} + \boldsymbol{\Omega}:\boldsymbol{\Omega})\\ &\quad - 2(mn)^{-1}\eta_1(\nabla\nabla p - \kappa_T\nabla p\nabla p) - (mn)^{-1}\eta_2(\nabla^2 p - \kappa_T|\nabla p|^2)\end{aligned} \tag{3.104}$$

with

$$\boxed{\eta_1 \doteq T^{-1}K^{\mathrm{IV}},} \quad \boxed{\eta_2 \doteq T^{-1}K^{\mathrm{V}}.} \tag{3.105}$$

The first term combines with the time derivative of the $K$'s to give

$$2(\partial_t\eta_1)\boldsymbol{S} + (\partial_t\eta_2)(\nabla\cdot\boldsymbol{u})\boldsymbol{I}. \tag{3.106}$$

If one uses $n$ and $s$ as variables, the time derivatives can be worked out in terms of $\partial_t n = -\nabla\cdot(n\boldsymbol{u}) \to -n\nabla\cdot\boldsymbol{u}$ and $\partial_t s$. To find the entropy equation, begin with

$$\mathrm{d}s = \beta(\mathrm{d}e - \mu\,\mathrm{d}n). \tag{3.107}$$

Thus, to first order,

$$\partial_t s = \beta(\partial_t e - \mu\,\partial_t n) \tag{3.108a}$$

$$= \beta(-h\nabla\cdot\boldsymbol{u} + n\mu\nabla\cdot\boldsymbol{u}) = -\beta(e + p - n\mu)(\nabla\cdot\boldsymbol{u}) \tag{3.108b}$$

$$= -s\nabla\cdot\boldsymbol{u}. \tag{3.108c}$$

In a homogeneous ensemble, one can choose a constant of integration that makes

$s(n,T)=0$. Assume that this can still be done to lowest order in the gradients in the reference ensemble. Then the derivatives of the $\eta$'s do not enter, and one obtains

$$\boxed{(\boldsymbol{P},\boldsymbol{P})_{\partial_t} = -2n\left(\frac{\partial\eta_1}{\partial n}\right)_s(\boldsymbol{\nabla}\cdot\boldsymbol{u})\mathsf{S} - n\left(\frac{\partial\eta_2}{\partial n}\right)_s(\boldsymbol{\nabla}\cdot\boldsymbol{u})^2\mathsf{I}.} \tag{3.109}$$

---

$(\boldsymbol{P},E)_{\partial_t}$:

$$(\boldsymbol{P},E)_{\partial_t} = \frac{\partial}{\partial t}\left(\int_0^\infty \mathrm{d}s\, s\,\langle\widehat{\boldsymbol{\tau}}(\boldsymbol{r})\mathrm{e}^{-\mathrm{i}\mathscr{L}s}\widehat{\mathscr{J}}^E\rangle\cdot(T^{-2}\boldsymbol{\nabla}T)\right)^{(1)}. \tag{3.110}$$

By symmetry, the expection (a third-rank tensor) vanishes to lowest order, but in general it can be built from a gradient such as $\boldsymbol{\nabla}T$. Since it vanishes to lowest order, the time derivatives of the $T$ factors do not enter through second order. Thus, one only needs to evaluate $\partial_t\langle\widehat{\boldsymbol{\tau}}\mathrm{e}^{-\mathrm{i}\mathscr{L}s}\widehat{\mathscr{J}}^E\rangle$. It can be shown that the contribution from this term cancels the $T^{-1}$ term in the $(T^{-1}-\alpha)$ factors in (3.103). For more detailed discussion, see the closely related calculation in §3.3.2.

3.3.2. Nonlinear Burnett — energy ($\partial_t$)

Note that $\widehat{\mathscr{N}}=0$, so $(\boldsymbol{P},N)_{\partial_t}=0$.

---

$(E,\boldsymbol{P})_{\partial_t}$:

$$(E,\boldsymbol{P})_{\partial_t} = \frac{\partial}{\partial t}\left(\int_0^\infty \mathrm{d}s\, s\,\langle\widehat{\boldsymbol{J}}^E(\boldsymbol{r})\mathrm{e}^{-\mathrm{i}\mathscr{L}s}\widehat{\mathscr{T}}\rangle : [\boldsymbol{\nabla}(T^{-1}\boldsymbol{u})]\right)^{(1)}. \tag{3.111}$$

The expectation is a third-rank tensor. By symmetry in velocity space, it vanishes to lowest order; however, in general it does not vanish, as it can be built from a vector such as $\boldsymbol{u}$.

Since the expectation vanishes to lowest order, the term in $\partial_t(\boldsymbol{\nabla}\boldsymbol{u})$ will not contribute through second order in the gradients. Thus, one need to calculate

$$\mathsf{T} \doteq \frac{\partial}{\partial t}\left(\int_0^\infty \mathrm{d}s\, s\,\langle\widehat{\boldsymbol{J}}^E(\boldsymbol{r})\mathrm{e}^{-\mathrm{i}\mathscr{L}s}\widehat{\mathscr{T}}\rangle\right). \tag{3.112}$$

One has

$$\widehat{\boldsymbol{J}}^E = \int\frac{\mathrm{d}\boldsymbol{k}}{(2\pi)^3}\,\mathrm{e}^{\mathrm{i}\boldsymbol{k}\cdot(\boldsymbol{r}-\boldsymbol{x}_i)}\sum_{i=1}^{\mathscr{N}}\left[\widetilde{E}_i\boldsymbol{v}_i + \Delta\widetilde{\boldsymbol{\tau}}_i(\boldsymbol{k})\cdot\boldsymbol{v}_i - \left(\frac{h}{n}\right)\boldsymbol{v}_i\right], \tag{3.113$a$}$$

$$\widehat{\mathscr{T}} = \sum_{i=1}^{\mathscr{N}} m[\boldsymbol{v}_i\boldsymbol{v}_i + \Delta\widetilde{\boldsymbol{\tau}}_i(0)] - \mathsf{I}\int\mathrm{d}\boldsymbol{y}\,[p + N'p_n + E'p_e)(\boldsymbol{y})]. \tag{3.113$b$}$$

Contributions to the time derivative come from the projected parts of $\widehat{\boldsymbol{J}}^E$ and $\widehat{\mathscr{T}}$, as well as from the time derivative of $f_0$. The integrals of the projections leave leave expectations that are odd in velocity, so those will be proportional to $\boldsymbol{u}$, which one can take to vanish. Thus, we focus on $\partial_t f_0$. One has

$$f_0 = \mathrm{e}^{-\beta\widetilde{\mathscr{E}}_0}/Z_0, \tag{3.114}$$

where

$$\widetilde{\mathscr{E}}_0 \doteq \sum_{i=1}^{\mathscr{N}} \left( \frac{1}{2} m(\boldsymbol{v}_i - \boldsymbol{u})^2 + \widetilde{U}_i \right), \quad Z_0 \doteq \int \mathrm{d}\varGamma\, \mathrm{e}^{-\beta \widetilde{\mathscr{E}}}. \tag{3.115}$$

Thus,

$$\begin{aligned} \partial_t f_0 &= (-\beta)(-\partial_t \boldsymbol{u}) \boldsymbol{\cdot} \sum_{i=1}^{\mathscr{N}} m(\boldsymbol{v}_i - \boldsymbol{u}) f_0 - \partial_t \beta\, \widetilde{\mathscr{E}}_0 f_0 \\ &\quad - f_0 \left\langle \left( (-\beta)(-\partial_t \boldsymbol{u}) \boldsymbol{\cdot} \sum_{i=1}^{\mathscr{N}} m(\boldsymbol{v}_i - \boldsymbol{u}) - \partial_t \beta\, \widetilde{\mathscr{E}}_0 \right) \right\rangle \qquad (3.116a) \\ &= \beta(\partial_t \boldsymbol{u}) \boldsymbol{\cdot} \left( \sum_{i=1}^{\mathscr{N}} m \boldsymbol{w}_i \right) f_0 - (\partial_t \beta) \widetilde{\mathscr{E}}_0' f_0. \qquad (3.116b) \end{aligned}$$

The $\partial_t \beta$ term again leads to an expectation that is odd in velocity, which one can ignore. Thus, one finds that

$$\begin{aligned} \boldsymbol{T} = \beta(\partial_t \boldsymbol{u}) \boldsymbol{\cdot} \int_0^\infty \mathrm{d}s\, s \int \frac{\mathrm{d}\boldsymbol{k}}{(2\pi)^3} \Big\langle \mathrm{e}^{\mathrm{i}\boldsymbol{k}\boldsymbol{\cdot}(\boldsymbol{r}-\boldsymbol{x}_i)} \sum_{i=1}^{\mathscr{N}} \\ \times \left[ \widetilde{E}_i m \boldsymbol{w}_i \boldsymbol{w}_i + m \boldsymbol{w}_i \Delta \widetilde{\boldsymbol{\tau}}_i(\boldsymbol{k}) \boldsymbol{\cdot} \boldsymbol{w}_i - \left(\frac{h}{n}\right) m \boldsymbol{w}_i \boldsymbol{w}_i \right] \mathrm{e}^{-\mathrm{i}\mathscr{L}s} \widehat{\boldsymbol{\mathscr{T}}} \Big\rangle. \end{aligned} \tag{3.117}$$

From the Euler equation for $\boldsymbol{u}$, one can replace $\partial_t \boldsymbol{u} \to -(mn)^{-1} \boldsymbol{\nabla} p$. Now the integral $\beta \int_0^\infty \mathrm{d}s\, s\, \ldots$ has the same dimensions as $\int_0^\infty \mathrm{d}s\, s\, \langle \widehat{\boldsymbol{\tau}} \mathrm{e}^{-\mathrm{i}\mathscr{L}s} \widehat{\boldsymbol{\mathscr{T}}} \rangle$, so it can be written as that term plus a correction:

$$\begin{aligned} [\ldots] &= (\widetilde{E}_i - T) m \boldsymbol{w}_i \boldsymbol{w}_i + T m \boldsymbol{w}_i \boldsymbol{w}_i + m \left( \boldsymbol{w}_i \boldsymbol{w}_i - \frac{1}{3} \langle w^2 \rangle \boldsymbol{I} \right) : \Delta \widetilde{\boldsymbol{\tau}}_i(\boldsymbol{k}) + T \Delta \widetilde{\boldsymbol{\tau}}_i(\boldsymbol{k}) \\ &\quad - \left(\frac{h}{n}\right) m \left( \boldsymbol{w}_i \boldsymbol{w}_i - \frac{1}{3} \langle w^2 \rangle \boldsymbol{I} \right) - \left(\frac{h}{n}\right) \boldsymbol{I} T \qquad (3.118a) \\ &= T[m \boldsymbol{w}_i \boldsymbol{w}_i + \Delta \widetilde{\boldsymbol{\tau}}_i(\boldsymbol{k}) - n^{-1} \boldsymbol{I}(p + N' p_n + E' p_e)] \\ &\quad + (\widetilde{\boldsymbol{E}}_i - T) m \left( \boldsymbol{w}_i \boldsymbol{w}_i - \frac{1}{3} \langle w^2 \rangle \boldsymbol{I} \right) + (\widetilde{\boldsymbol{E}}_i - T) T \boldsymbol{I} + m \left( \boldsymbol{w}_i \boldsymbol{w}_i - \frac{1}{3} \langle w^2 \rangle \boldsymbol{I} \right) : \Delta \widetilde{\boldsymbol{\tau}}_i(\boldsymbol{k}) \\ &\quad - \left(\frac{h}{n}\right) m \left( \boldsymbol{w}_i \boldsymbol{w}_i - \frac{1}{3} \langle w^2 \rangle \boldsymbol{I} \right) - n^{-1} e T \boldsymbol{I} + n^{-1} \boldsymbol{I}(N' p_n + E' p_e). \qquad (3.118b) \end{aligned}$$

The first line leads to contributions

$$\begin{aligned} &- (mn)^{-1} \boldsymbol{\nabla} p \boldsymbol{\cdot} [K^{\mathrm{IV}}(\delta_{ik}\delta_{jl} + \delta_{il}\delta_{jk}) + K^{\mathrm{V}} \delta_{ij}\delta_{kl}](T^{-1} \nabla_l u_k) \\ &\quad = \lambda_7^{\partial \mathrm{B}_{\boldsymbol{P}}^E} \boldsymbol{S} \boldsymbol{\cdot} \boldsymbol{\nabla} p + \lambda_8^{\partial \mathrm{B}_{\boldsymbol{P}}^E} (\boldsymbol{\nabla} \boldsymbol{\cdot} \boldsymbol{u}) \boldsymbol{\nabla} p, \qquad (3.119a) \end{aligned}$$

where

$$\boxed{\lambda_7^{\partial \mathrm{B}_{\boldsymbol{P}}^E} \doteq -2(mn)^{-1} \eta_1, \quad \lambda_8^{\partial \mathrm{B}_{\boldsymbol{P}}^E} \doteq -(mn)^{-1} \eta_2,} \tag{3.120}$$

$\eta_1$ and $\eta_2$ having already been defined in (3.105).

$\underline{(E,E)_{\partial_t}}$:

$$(E,E)_{\partial_t} = \frac{\partial}{\partial t}\Bigg(\underbrace{\int_0^\infty \mathrm{d}s\, s\, \langle \widehat{\boldsymbol{J}}^E \mathrm{e}^{-\mathrm{i}\mathscr{L}s} \widehat{\boldsymbol{\mathscr{J}}}^E \rangle}_{K^{\mathrm{VI}}\boldsymbol{I}} \boldsymbol{\cdot}\boldsymbol{\nabla} B_E\Bigg)^{(1)}. \tag{3.121$a$}$$

First let us calculate $\partial_t \boldsymbol{\nabla} B_E$. One has

$$\boldsymbol{\nabla}\partial_t B_E = \boldsymbol{\nabla}\partial_t(-T^{-1}) \tag{3.122$a$}$$
$$= \boldsymbol{\nabla}(T^{-2}\partial_t T) \tag{3.122$b$}$$
$$= -2T^{-3}(\partial_t T)\boldsymbol{\nabla}T + T^{-2}\boldsymbol{\nabla}(\partial_t T). \tag{3.122$c$}$$

It is convenient to express $T$ in terms of $n$ and $e$, which have simple Euler equations:

$$\partial_t T(n,e) = T_n(\partial_t n) + T_e(\partial_t e) \tag{3.123$a$}$$
$$= -[T_n(n\boldsymbol{\nabla}\boldsymbol{\cdot}\boldsymbol{u} + \boldsymbol{u}\boldsymbol{\cdot}\boldsymbol{\nabla} n) + T_e(h\boldsymbol{\nabla}\boldsymbol{\cdot}\boldsymbol{u} + \boldsymbol{u}\boldsymbol{\cdot}\boldsymbol{\nabla} e)]. \tag{3.123$b$}$$

For $\partial_t T$ itself, after setting $\boldsymbol{u} = \boldsymbol{0}$,

$$\partial_t T \to -(nT_n + hT_e)(\boldsymbol{\nabla}\boldsymbol{\cdot}\boldsymbol{u}). \tag{3.124}$$

Now one can show that

$$n\,T_n + h\,T_e = Tp_e, \tag{3.125}$$

so the Euler temperature equation is[2]

$$\partial_t T = -Tp_e\boldsymbol{\nabla}\boldsymbol{\cdot}\boldsymbol{u}. \tag{3.126}$$

To prove (3.125), note that from

$$\mathrm{d}e = T\,\mathrm{d}s + \mu\,\mathrm{d}n \tag{3.127}$$

one has

$$0 = T\,\mathrm{d}s + \mu\,\mathrm{d}n \quad (e = \text{const}), \tag{3.128$a$}$$
$$\mathrm{d}e = T\,\mathrm{d}s \quad (n = \text{const}). \tag{3.128$b$}$$

Thus, upon expressing $p = p(n,s)$,

$$\left(\frac{\partial p}{\partial e}\right)_n = \left(\frac{\partial s}{\partial e}\right)_n \left(\frac{\partial p}{\partial s}\right)_n = \frac{1}{T}\left(\frac{\partial p}{\partial s}\right)_n, \tag{3.129$a$}$$
$$\left(\frac{\partial T}{\partial n}\right)_e = \left(\frac{\partial T}{\partial n}\right)_s + \left(\frac{\partial s}{\partial n}\right)_e \left(\frac{\partial T}{\partial s}\right)_n = \left(\frac{\partial T}{\partial n}\right)_s - \left(\frac{\mu}{T}\right)\left(\frac{\partial T}{\partial s}\right)_n, \tag{3.129$b$}$$
$$\left(\frac{\partial T}{\partial e}\right)_n = \left(\frac{\partial s}{\partial e}\right)_n \left(\frac{\partial T}{\partial s}\right)_n = \frac{1}{T}\left(\frac{\partial T}{\partial s}\right)_n. \tag{3.129$c$}$$

Then

$$n\,(T_n)|_e + h\,(T_e)|_n = n\,(T_n)|_s + T^{-1}(\underbrace{-n\mu + e + p}_{\mathrm{s}})\,(T_s)|_n. \tag{3.130}$$

From the Gibbs–Duhem relation,

$$0 = s\,\mathrm{d}T - \mathrm{d}p + n\,\mathrm{d}\mu, \tag{3.131}$$

[2]When the ideal-gas values of $p$ and $e$ are used, (3.126) reduces to $\frac{3}{2}n\,\partial_t T = -nT\boldsymbol{\nabla}\boldsymbol{\cdot}\boldsymbol{u}$, which is the Euler part of the familiar Braginskii temperature equation.

so

$$(p_s)|_n = s\,(T_s)|_n + n\,(\mu_s)|_n. \tag{3.132}$$

The relationship is then proven if

$$(\mu_s)|_n = (T_n)|_s. \tag{3.133}$$

But this is the Maxwell relation that follows from

$$\mathrm{d}e = T\,\mathrm{d}s + \mu\,\mathrm{d}n. \tag{3.134}$$

Thus,

$$\boldsymbol{\nabla}\partial_t B_E = 2T^{-2}\left(\frac{\partial p}{\partial e}\right)_n (\boldsymbol{\nabla}\cdot\boldsymbol{u})\boldsymbol{\nabla}T - T^{-2}\boldsymbol{\nabla}[T_n(n\boldsymbol{\nabla}\cdot\boldsymbol{u}+\boldsymbol{u}\cdot\boldsymbol{\nabla}n)+T_e(h\boldsymbol{\nabla}\cdot\boldsymbol{u}+\boldsymbol{u}\cdot\boldsymbol{\nabla}e)]. \tag{3.135}$$

The last term is $-T^{-2}[\dots]$, where

$$\begin{aligned}
[\dots] &= (\boldsymbol{\nabla}T_n)n(\boldsymbol{\nabla}\cdot\boldsymbol{u}) + (\boldsymbol{\nabla}T_e)h(\boldsymbol{\nabla}\cdot\boldsymbol{u}) \\
&\quad + T_n[(\boldsymbol{\nabla}n)(\boldsymbol{\nabla}\cdot\boldsymbol{u}) + n\boldsymbol{\nabla}(\boldsymbol{\nabla}\cdot\boldsymbol{u}) + (\boldsymbol{\nabla}\boldsymbol{u})\cdot\boldsymbol{\nabla}n] \\
&\quad + T_e[(\boldsymbol{\nabla}h)(\boldsymbol{\nabla}\cdot\boldsymbol{u}) + h\boldsymbol{\nabla}(\boldsymbol{\nabla}\cdot\boldsymbol{u}) + (\boldsymbol{\nabla}\boldsymbol{u})\cdot\boldsymbol{\nabla}e]. && (3.136a) \\
&= (\boldsymbol{\nabla}T_n)n(\boldsymbol{\nabla}\cdot\boldsymbol{u}) + (\boldsymbol{\nabla}T_e)h(\boldsymbol{\nabla}\cdot\boldsymbol{u}) \\
&\quad + (n\,T_n + h\,T_e)\boldsymbol{\nabla}(\boldsymbol{\nabla}\cdot\boldsymbol{u}) + (\boldsymbol{\nabla}\cdot\boldsymbol{u})\boldsymbol{I}(T_n\boldsymbol{\nabla}n + T_e\boldsymbol{\nabla}h) \\
&\qquad + (\boldsymbol{\nabla}\boldsymbol{u})\cdot(T_n\boldsymbol{\nabla}n + T_e\boldsymbol{\nabla}e) && (3.136b) \\
&= [\boldsymbol{\nabla}(nT_n + hT_e)](\boldsymbol{\nabla}\cdot\boldsymbol{u}) + (n\,T_n + h\,T_e)\boldsymbol{\nabla}(\boldsymbol{\nabla}\cdot\boldsymbol{u}) + (\boldsymbol{\nabla}u)\cdot\boldsymbol{\nabla}T && (3.136c) \\
&= \left[\frac{\partial(Tp_e)}{\partial T}\boldsymbol{\nabla}T + \frac{\partial(Tp_e)}{\partial p}\boldsymbol{\nabla}p\right](\boldsymbol{\nabla}\cdot\boldsymbol{u}) + Tp_e\boldsymbol{\nabla}(\boldsymbol{\nabla}\cdot\boldsymbol{u}) + (\boldsymbol{\nabla}u)\cdot\boldsymbol{\nabla}T. && (3.136d)
\end{aligned}$$

The full set of terms one needs to calculate is

$$(\partial_t K^{\mathrm{VI}})T^{-2}\boldsymbol{\nabla}T + K^{\mathrm{VI}}\boldsymbol{\nabla}\partial_t B_E. \tag{3.137}$$

Now $K^{\mathrm{VI}}$ depends on time only through its dependence on the state variables. It is convenient to view it as a function of $n$ and $s$, since the Euler equation for $s$ is trivial. Thus,

$$\partial_t K^{\mathrm{VI}} = \left(\frac{\partial K^{\mathrm{VI}}}{\partial n}\right)_s (\partial_t n) = -\left[n\left(\frac{\partial K^{\mathrm{VI}}}{\partial n}\right)_s + s\left(\frac{\partial K^{\mathrm{VI}}}{\partial s}\right)_n\right](\boldsymbol{\nabla}\cdot\boldsymbol{u}). \tag{3.138}$$

One therefore has

$$\begin{aligned}
(\partial_t K^{\mathrm{VI}})T^{-2}\boldsymbol{\nabla}T &= -\left[n\left(\frac{\partial K^{\mathrm{VI}}}{\partial n}\right)_s + s\left(\frac{\partial K^{\mathrm{VI}}}{\partial s}\right)_n\right](\boldsymbol{\nabla}\cdot\boldsymbol{u})T^{-2}\boldsymbol{\nabla}T && (3.139a) \\
&= \left[-n\frac{\partial}{\partial n}(T^{-2}K^{\mathrm{VI}}) - 2T^{-3}n\left(\frac{\partial T}{\partial n}\right)_s K^{\mathrm{VI}} - s\left(\frac{\partial K^{\mathrm{VI}}}{\partial s}\right)_n\right](\boldsymbol{\nabla}\cdot\boldsymbol{u})\boldsymbol{\nabla}T. && (3.139b)
\end{aligned}$$

The second term adds to the first term of (3.135) to give

$$2T^{-3}(\boldsymbol{\nabla}\cdot\boldsymbol{u})\boldsymbol{\nabla}T\left[T\left(\frac{\partial p}{\partial e}\right)_n - n\left(\frac{\partial T}{\partial n}\right)_s\right]K^{\mathrm{VI}} = 2T^{-3}(\boldsymbol{\nabla}\cdot\boldsymbol{u})\boldsymbol{\nabla}T\left[s\left(\frac{\partial T}{\partial s}\right)_n\right]K^{\mathrm{VI}}; \tag{3.140}$$

adding that to the third term gives a correction

$$-\,s\frac{\partial}{\partial s}(T^{-2}K^{\mathrm{VI}})(\boldsymbol{\nabla}\cdot\boldsymbol{u})\boldsymbol{\nabla}T \tag{3.141}$$

which vanishes with $s = 0$. (Compare the analogous situation with $\eta_1$.)

If one defines

$$\boxed{\lambda_1 \doteq T^{-2} K^{\mathrm{VI}},} \tag{3.142}$$

then

$$(E, E)_{\partial_t} = -n \left(\frac{\partial \lambda_1}{\partial n}\right)_s - \lambda_1 \frac{\partial}{\partial T}\left[\left(\frac{\partial p}{\partial e}\right)_n\right](\boldsymbol{\nabla}\cdot\boldsymbol{u})\boldsymbol{\nabla}T - \lambda_1 T \frac{\partial}{\partial p}\left[\left(\frac{\partial p}{\partial e}\right)_n\right](\boldsymbol{\nabla}\cdot\boldsymbol{u})\boldsymbol{\nabla}P$$
$$- \lambda_1 T \left(\frac{\partial p}{\partial e}\right)_n \boldsymbol{\nabla}(\boldsymbol{\nabla}\cdot\boldsymbol{u}) - \lambda_1(\boldsymbol{S} - \boldsymbol{\Omega})\cdot\boldsymbol{\nabla}T, \tag{3.143}$$

which generates all of the $\lambda_1$ terms in Eq. (A2) of Brey (1983).

## 4. Summary of the dissipative fluxes

Here I summarize the dissipative fluxes that have been derived. The results agree with those of Brey (1983), although I have slightly reordered some of the terms.

The dissipative part of the momentum flux is

$$\begin{aligned}\boldsymbol{\tau}_{\mathrm{diss}} = {}& -\eta\left((\boldsymbol{\nabla}\boldsymbol{u}) + (\boldsymbol{\nabla}\boldsymbol{u})^{\mathrm{T}} - \frac{2}{d}(\boldsymbol{\nabla}\cdot\boldsymbol{u})\boldsymbol{I}\right) - \zeta(\boldsymbol{\nabla}\cdot\boldsymbol{u})\,\boldsymbol{I} \\
& + \eta_3 \boldsymbol{\nabla}\boldsymbol{\nabla}T + \eta_5 \boldsymbol{\nabla}T\,\boldsymbol{\nabla}T \\
& + [\eta_7 - (mn)^{-1}\alpha\eta_1](\boldsymbol{\nabla}T\,\boldsymbol{\nabla}p + \boldsymbol{\nabla}p\,\boldsymbol{\nabla}T) - 2(mn)^{-1}\eta_1(\boldsymbol{\nabla}\boldsymbol{\nabla}p - \kappa_T \boldsymbol{\nabla}p\,\boldsymbol{\nabla}p) \\
& + \Bigg[\eta_{11} - \underbrace{2n\left(\frac{\partial \eta_1}{\partial n}\right)_s}_{\partial \mathrm{B}^{\boldsymbol{P}}_{\boldsymbol{P}}}\Bigg](\boldsymbol{\nabla}\cdot\boldsymbol{u})\boldsymbol{S} + (\eta_{12} - 2\eta_1)\boldsymbol{S}\cdot\boldsymbol{S}^{\mathrm{T}} + 2\eta_1 \boldsymbol{\Omega}\cdot\boldsymbol{\Omega}^{\mathrm{T}} \\
& + \eta_{13}(\boldsymbol{S}^{\mathrm{T}}\cdot\boldsymbol{\Omega} + \boldsymbol{\Omega}^{\mathrm{T}}\cdot\boldsymbol{S}) \\
& + \boldsymbol{I}\Big\{\eta_4 \nabla^2 T + \eta_6 |\boldsymbol{\nabla}T|^2 \\
& + [\eta_8 - (mn)^{-2}\alpha\eta_2]\boldsymbol{\nabla}T\cdot\boldsymbol{\nabla}p - (mn)^{-1}\eta_2(\nabla^2 p - \kappa_T|\boldsymbol{\nabla}p|^2) \\
& + \Bigg[\eta_9 - \underbrace{n\left(\frac{\partial \eta_2}{\partial n}\right)_s}_{\partial \mathrm{B}^{\boldsymbol{P}}_{\boldsymbol{P}}}\Bigg](\boldsymbol{\nabla}\cdot\boldsymbol{u})^2 + (\eta_{10} - \eta_2)\boldsymbol{S} : \boldsymbol{S}^{\mathrm{T}} + \eta_2 \boldsymbol{\Omega} : \boldsymbol{\Omega}^{\mathrm{T}}\Big\}.\end{aligned} \tag{4.1}$$

Note that $\boldsymbol{S}^{\mathrm{T}} = \boldsymbol{S}$ and $\boldsymbol{\Omega}^{\mathrm{T}} = -\boldsymbol{\Omega}$. A consequence is that the trace of the $\eta_{13}$ term vanishes.

The dissipative part of the heat flux is

$$
\begin{aligned}
\boldsymbol{j}^E_{\text{diss}} &= \boldsymbol{u}\cdot\boldsymbol{\tau}_{\text{diss}} - \lambda\boldsymbol{\nabla}T \\
&\quad + \Bigg[\lambda_2 + \underbrace{(-\lambda_1)T\left(\frac{\partial p}{\partial e}\right)_n}_{\partial\mathrm{B}^E_E}\Bigg]\boldsymbol{\nabla}(\boldsymbol{\nabla}\cdot\boldsymbol{u}) + \lambda_3\nabla^2\boldsymbol{u} \\
&\quad + \Bigg(\lambda_4 - \underbrace{\left\{\lambda_1\left(\frac{\partial p}{\partial e}\right)_n + \lambda_1 T\left[\frac{\partial}{\partial T}\left(\frac{\partial p}{\partial e}\right)_n\right]_p + n\left(\frac{\partial\lambda_1}{\partial n}\right)_s\right\}}_{\partial\mathrm{B}^E_E}\Bigg)(\boldsymbol{\nabla}\cdot\boldsymbol{u})\boldsymbol{\nabla}T \\
&\quad + (\lambda_5-\lambda_1)\boldsymbol{S}\cdot\boldsymbol{\nabla}T + (\lambda_6+\lambda_1)\boldsymbol{\Omega}\cdot\boldsymbol{\nabla}T + \lambda_7\boldsymbol{S}\cdot\boldsymbol{\nabla}p \\
&\quad + \Bigg\{\lambda_8 - \underbrace{\lambda_1 T\left[\frac{\partial}{\partial p}\left(\frac{\partial p}{\partial e}\right)_n\right]_T}_{\partial\mathrm{B}^E_E}\Bigg\}(\boldsymbol{\nabla}\cdot\boldsymbol{u})\boldsymbol{\nabla}p.
\end{aligned}
\tag{4.2}
$$

In the above expressions,

$$\eta \doteq \underbrace{T^{-1}K^{\mathrm{I}}}_{\mathrm{NS}^{\boldsymbol{P}}_{\boldsymbol{P}}}, \qquad \zeta \doteq \underbrace{T^{-2}(K^{\mathrm{II}} + 2d^{-1}K^{\mathrm{I}})}_{\mathrm{NS}^{\boldsymbol{P}}_{\boldsymbol{P}}}, \tag{4.3$a$}$$

$$\eta_1 \doteq \underbrace{T^{-1}K^{\mathrm{IV}}}_{\partial\mathrm{B}^{\boldsymbol{P}}_{\boldsymbol{P}}}, \tag{4.3$b$}$$

$$\eta_2 \doteq \underbrace{T^{-1}K^{\mathrm{V}}}_{\partial\mathrm{B}^{\boldsymbol{P}}_{\boldsymbol{P}}}, \tag{4.3$c$}$$

$$\eta_3 \doteq -\underbrace{2T^{-2}K_1}_{\mathrm{B}^{\boldsymbol{P}}_E}, \tag{4.3$d$}$$

$$\eta_4 \doteq -\underbrace{T^{-2}K_2}_{\mathrm{B}^{\boldsymbol{P}}_E}, \tag{4.3$e$}$$

$$\eta_5 \doteq \underbrace{4T^{-3}K_1}_{\mathrm{B}^{\boldsymbol{P}}_E} + \underbrace{2(nT^4)^{-1}hK_3}_{\mathrm{B}^{\boldsymbol{P}}_{EN}} - \underbrace{2T^{-4}K_5}_{\mathrm{B}^{\boldsymbol{P}}_{EE}} + \underbrace{2T^{-2}\left[\frac{\partial}{\partial T}\left(\frac{h}{mn}\right)\right]_p K_{20}}_{\mathrm{B}^{\boldsymbol{P}}_{-\mathrm{P}EE}}, \tag{4.3$f$}$$

$$\eta_6 \doteq \underbrace{2T^{-3}K_2}_{\mathrm{B}^{\boldsymbol{P}}_E} + \underbrace{(nT^4)^{-2}hK_4}_{\mathrm{B}^{\boldsymbol{P}}_{EN}} - \underbrace{T^{-4}K_6}_{\mathrm{B}^{\boldsymbol{P}}_{EE}} + \underbrace{T^{-2}\left[\frac{\partial}{\partial T}\left(\frac{h}{mn}\right)\right]_p K_{21}}_{\mathrm{B}^{\boldsymbol{P}}_{-\mathrm{P}EE}}, \tag{4.3$g$}$$

$$\eta_7 \doteq -\underbrace{(nT^3)^{-1}K_3}_{\mathrm{B}^{\boldsymbol{P}}_{EN}} + \underbrace{T^{-2}\left[\frac{\partial}{\partial p}\left(\frac{h}{mn}\right)\right]_T K_{20}}_{\mathrm{B}^{\boldsymbol{P}}_{-\mathrm{P}EN}}, \tag{4.3$h$}$$

$$\eta_8 \doteq -\underbrace{(nT^3)^{-1}K_4}_{\mathrm{B}^{\boldsymbol{P}}_{EN}} + \underbrace{T^{-2}\left[\frac{\partial}{\partial p}\left(\frac{h}{mn}\right)\right]_T K_{21}}_{\mathrm{B}^{\boldsymbol{P}}_{-\mathrm{P}EN}}, \tag{4.3$i$}$$

$$\eta_9 \doteq -\underbrace{T^{-2}K_7}_{\mathrm{B}^{\boldsymbol{P}}_{\boldsymbol{PP}}} - \underbrace{T^{-1}\left(\frac{\partial p}{\partial e}\right)_n K_{21}}_{\mathrm{B}^{\boldsymbol{P}}_{-\mathrm{P}\boldsymbol{PP}}}, \tag{4.3j}$$

$$\eta_{10} \doteq -\underbrace{2T^{-2}K_8}_{\mathrm{B}^{\boldsymbol{P}}_{\boldsymbol{PP}}} + \underbrace{2T^{-1}K_{21}}_{\mathrm{B}^{\boldsymbol{P}}_{-\mathrm{P}\boldsymbol{PP}}}, \tag{4.3k}$$

$$\eta_{11} \doteq -\underbrace{2T^{-2}(K_9 + K_{10})}_{\mathrm{B}^{\boldsymbol{P}}_{\boldsymbol{PP}}} - \underbrace{2T^{-1}\left(\frac{\partial p}{\partial e}\right)_n K_{20}}_{\mathrm{B}^{\boldsymbol{P}}_{-\mathrm{P}\boldsymbol{PP}}}, \tag{4.3l}$$

$$\eta_{12} \doteq -\underbrace{4T^{-2}(K_{11} + K_{12})}_{\mathrm{B}^{\boldsymbol{P}}_{\boldsymbol{PP}}} + \underbrace{4T^{-1}K_{20}}_{\mathrm{B}^{\boldsymbol{P}}_{-\mathrm{P}\boldsymbol{PP}}}, \tag{4.3m}$$

$$\eta_{13} \doteq -\underbrace{2T^{-2}(K_{11} - K_{12})}_{\mathrm{B}^{\boldsymbol{P}}_{\boldsymbol{PP}}} + \underbrace{2T^{-1}K_{20}}_{\mathrm{B}^{\boldsymbol{P}}_{-\mathrm{P}\boldsymbol{PP}}}, \tag{4.3n}$$

and

$$\lambda \doteq T^{-2}K^{\mathrm{III}}, \tag{4.4a}$$

$$\lambda_1 \doteq \underbrace{T^{-2}K^{\mathrm{VI}}}_{\partial\mathrm{B}^{E}_{E}}, \tag{4.4b}$$

$$\lambda_2 \doteq \underbrace{-T^{-1}(K_1 + K_2)}_{\mathrm{B}^{E}_{\boldsymbol{P}}}, \tag{4.4c}$$

$$\lambda_3 \doteq \underbrace{-T^{-1}K_1}_{\mathrm{B}^{E}_{\boldsymbol{P}}}, \tag{4.4d}$$

$$\begin{aligned}\lambda_4 \doteq{}& \underbrace{T^{-2}(K_1 + K_2)}_{\mathrm{B}^{E}_{\boldsymbol{P}}} - \underbrace{T^{-3}K_{19}}_{\mathrm{B}^{E}_{\boldsymbol{P}E}} - \underbrace{T^{-3}K_{13}}_{\mathrm{B}^{E}_{E\boldsymbol{P}}} \\ &+ \underbrace{(nT^3)^{-1}h\,K_{17}}_{\mathrm{B}^{E}_{\boldsymbol{P}N}} + \underbrace{T^{-1}\left\{\left[\frac{\partial}{\partial T}\left(\frac{\partial p}{\partial n}\right)_e\right]_p K_{22} + \left[\frac{\partial}{\partial T}\left(\frac{\partial p}{\partial e}\right)_n\right]_p K_{23}\right\}}_{\mathrm{B}^{E}_{-\mathrm{P}\boldsymbol{P}N}},\end{aligned} \tag{4.4e}$$

$$\begin{aligned}\lambda_5 \doteq{}& \underbrace{T^{-2}(3K_1 + K_2)}_{\mathrm{B}^{E}_{\boldsymbol{P}}} + \underbrace{2(nT^3)^{-1}h\,K_{16}}_{\mathrm{B}^{E}_{\boldsymbol{P}N}} - \underbrace{2T^{-3}K_{18}}_{\mathrm{B}^{E}_{\boldsymbol{P}E}} \\ &- \underbrace{T^{-3}(K_{14} + K_{15})}_{\mathrm{B}^{E}_{E\boldsymbol{P}}} + \underbrace{T^{-2}\left\{\left[\left(\frac{\partial p}{\partial n}\right)_e - \frac{h}{n}\right]K_{22} + \left[1 + \left(\frac{\partial p}{\partial e}\right)_n\right]K_{23}\right\}}_{\mathrm{B}^{E}_{-\mathrm{P}E\boldsymbol{P}}},\end{aligned} \tag{4.4f}$$

$$\begin{aligned}\lambda_6 \doteq{}& \underbrace{T^{-2}(K_1 - K_2)}_{\mathrm{B}^{E}_{\boldsymbol{P}}} - \underbrace{T^{-3}(K_{14} - K_{15})}_{\mathrm{B}^{E}_{E\boldsymbol{P}}} \\ &- \underbrace{T^{-2}\left\{\left[\left(\frac{\partial p}{\partial n}\right)_e - \frac{h}{n}\right]K_{22} + \left[1 + \left(\frac{\partial p}{\partial e}\right)_n\right]K_{23}\right\}}_{\mathrm{B}^{E}_{-\mathrm{P}E\boldsymbol{P}}},\end{aligned} \tag{4.4g}$$

$$\lambda_7 \doteq \underbrace{-2(nT^2)^{-1}K_{16}}_{\mathrm{B}^E_{\boldsymbol{P}N}} - \underbrace{2(mn)^{-1}\eta_1}_{\partial\mathrm{B}^E_{\boldsymbol{P}}}, \tag{4.4$h$}$$

$$\lambda_8 \doteq -\underbrace{(nT^2)^{-1}K_{17}}_{\mathrm{B}^E_{\boldsymbol{P}N}} + \underbrace{T^{-1}\left\{\left[\frac{\partial}{\partial p}\left(\frac{\partial p}{\partial n}\right)_e\right]_T K_{22} + \left[\frac{\partial}{\partial p}\left(\frac{\partial p}{\partial e}\right)_n\right]_T K_{23}\right\}}_{\mathrm{B}^E_{-\mathrm{P}\boldsymbol{P}N}} - \underbrace{(mn)^{-1}\eta_2}_{\partial\mathrm{B}^E_{\boldsymbol{P}}}. \tag{4.4$i$}$$

This work was supported by the U. S. Department of Energy Contract DE-AC02-09CH11466.